%% file: piranha_paper.tex
\documentclass[letterpaper,11pt]{article}
\usepackage[utf8]{inputenc}

\usepackage{jheppub}
\include{includes/utils/paper_preamble}
\include{includes/utils/piranha_utils}
\include{includes/utils/comments}

\title{Pileup and Infrared Radiation Annihilation (PIRANHA{}):
A Paradigm for Continuous Jet Grooming}

\author[a,b]{Samuel Alipour-fard,}
\author[a]{Patrick T. Komiske,}
\author[a]{Eric M. Metodiev,}
\author[a,b]{and Jesse Thaler}

\affiliation[a]{
Center for Theoretical Physics, Massachusetts Institute of Technology,
\\
77 Massachusetts Avenue, Cambridge, MA 02139, U.S.A.
}
\affiliation[b]{The NSF AI Institute for Artificial Intelligence and Fundamental Interactions}

\emailAdd{samuelaf@mit.edu}
\emailAdd{pkomiske@mit.edu}
\emailAdd{metodiev@mit.edu}
\emailAdd{jthaler@mit.edu}

\preprint{MIT--CTP 5540}

\abstract{
Jet grooming is an important strategy for analyzing relativistic particle collisions in the presence of contaminating radiation.
Most jet grooming techniques introduce hard cutoffs to remove soft radiation, leading to discontinuous behavior and associated experimental and theoretical challenges.
In this paper, we introduce Pileup and Infrared Radiation Annihilation (\PIRANHA{}), a paradigm for continuous jet grooming that overcomes the discontinuity and infrared sensitivity of hard-cutoff grooming procedures.
We motivate \PIRANHA{} from the perspective of optimal transport and the Energy Mover's Distance and review Apollonius Subtraction and Iterated Voronoi Subtraction as examples of \PIRANHA{}-style grooming.
We then introduce a new tree-based implementation of \PIRANHA{}, Recursive Subtraction, with reduced computational costs.
Finally, we demonstrate the performance of Recursive Subtraction in mitigating sensitivity to soft distortions from hadronization and detector effects, and additive contamination from pileup and the underlying event.
}

\newif\ifdarkmode
\darkmodefalse

\ifdarkmode
    \pagecolor[rgb]{0,0,0} 
    \color[rgb]{0.5,0.5,0.5} 
\fi

\newif\iflistcomments
\listcommentsfalse

\newif\ifshowfinalchecks
\showfinalchecksfalse

\begin{document}
\maketitle

\iflistcomments
    \include{includes/aux/paper_comments}
\fi

\ifshowfinalchecks
    \include{includes/aux/final_checks}
\fi


\section{Introduction}

Jets of hadrons produced in high-energy particle collisions are an important guide in our understanding of the subatomic universe and quantum chromodynamics (QCD), our fundamental theory of the strong nuclear force.
Quantifying the internal structure of hadronic jets plays an important role in translating between experimental results and theoretical predictions.
The toolkit of jet substructure, including techniques from jet grooming \cite{Krohn:2009th,Ellis:2009me,Larkoski:2014wba,Dasgupta:2013ihk,Dasgupta:2013via,Tseng:2013dva,Dasgupta:2016ktv} and pileup mitigation \cite{Cacciari:2007fd,Soyez:2012hv,Komiske:2017ubm,Bertolini:2014bba,Cacciari:2008gn,Butterworth:2008iy,Chatrchyan:2012sn,Alon:2011xb,Krohn:2013lba,ATLAS-CONF-2013-083,ATLAS-CONF-2013-085,Chatrchyan:2012tt,CMS-PAS-JME-13-005,Sirunyan:2020foa,ATLAS-CONF-2012-066,Monk:2018clo,Martinez:2018fwc},
helps characterize the radiation patterns produced by jets in a wide variety of physical processes at colliders.
Jet substructure can distinguish jets produced by light quarks, gluons, and the decays of electroweak bosons and top quarks
\cite{Thaler:2008ju,Thaler:2011gf,Hook:2011cq,Gallicchio:2011xq,Soper:2012pb,Gallicchio:2012ez,CMS-PAS-JME-09-001,CMS-PAS-EXO-09-002,CMS:2013kfa,ATL-PHYS-PUB-2009-081,ATL-PHYS-PUB-2010-008,Cui:2010km,ATLAS-CONF-2011-053,Chatrchyan:2013rla,Larkoski:2013eya,Dasgupta:2012hg,Backovic:2013bga,ATLAS-CONF-2013-084,Komiske:2018vkc,Komiske:2016rsd,Metodiev:2018ftz,Krohn:2012fg,MERINO:2013tta,Bhattacherjee:2016bpy,Macaluso:2018tck,Egan:2017ojy,Kasieczka:2017nvn,Pearkes:2017hku,Butter:2017cot},
provide accurate predictions of the behavior of boosted objects in many contexts
\cite{Catani:1992ua,Dokshitzer:1998kz,Dasgupta:2001sh,Banfi:2004yd,Banfi:2005gj,Ellis:2009su,Banfi:2010pa,Walsh:2011fz,Chien:2012ur,Li:2012bw,Jouttenus:2013hs,Hatta:2013iba,Larkoski:2014tva,Procura:2018zpn,Aaboud:2017aca,Frye:2016aiz, Almeida:2008yp,Larkoski:2017iuy,Larkoski:2017cqq,Thaler:2010tr,Ellis:2009me,Abdesselam:2010pt,Katz:2010mr,Gallicchio:2010dq,Adams:2015hiv,Sirunyan:2017ezt,Moore:2018lsr,FerreiradeLima:2016gcz,Rubin:2010fc,Chatrchyan:2012sn,ATLAS:2019kwg,CMS-PAS-BTV-13-001,CMS-PAS-JME-13-006,Kribs:2009yh,Chen:2010wk,Hackstein:2010wk,Kim:2010uj,Almeida:2011aa,Pandolfi:2012ima,Vernieri:2014wfa,CMS-PAS-HIG-17-007,Procura:2014cba},
and even reveal non-perturbative features of the infrared structure of QCD
\cite{Komiske:2019jim, Komiske:2019fks, Hoang:2019ceu, Mateu:2012nk, Wobisch:1998wt}.

Jet grooming facilitates our understanding of jet substructure by removing low-energy pollution from jets.
The major goal of jet grooming is to reduce the sensitivity of jet observables to two conceptually distinct categories of pollution:
\begin{itemize}
    \item
    \textbf{Soft Distortions}
    are low-energy effects that distort the radiation coming directly from hard jet production, such as hadronization and detector effects;

    \item
    \textbf{Additive Contamination}
    denotes a backdrop of low-energy radiation that pollutes a jet but is not directly connected to jet production, including the pileup (PU) of radiation from overlapping particle collisions and an underlying event (UE) of secondary parton interactions.
\end{itemize}
The low-energy phenomena of QCD that produce both soft contamination and additive contamination are notoriously difficult to model, and jet grooming has become an important toolkit used to empower both experimental and theoretical analyses of jet substructure
\cite{Krohn:2009th,Ellis:2009me,Larkoski:2014wba,Dasgupta:2013ihk,Dasgupta:2013via,Tseng:2013dva,Dasgupta:2016ktv,Thaler:2008ju,Thaler:2011gf,Hook:2011cq,Gallicchio:2011xq,Soper:2012pb,Gallicchio:2012ez,CMS-PAS-JME-09-001,CMS-PAS-EXO-09-002,CMS:2013kfa,ATL-PHYS-PUB-2009-081,ATL-PHYS-PUB-2010-008,Cui:2010km,ATLAS-CONF-2011-053,Chatrchyan:2013rla,Larkoski:2013eya,Dasgupta:2012hg,Backovic:2013bga,ATLAS-CONF-2013-084,Komiske:2018vkc,Komiske:2016rsd,Metodiev:2018ftz,Krohn:2012fg,MERINO:2013tta,Bhattacherjee:2016bpy,Macaluso:2018tck,Egan:2017ojy,Kasieczka:2017nvn,Pearkes:2017hku,Butter:2017cot,Catani:1992ua,Dokshitzer:1998kz,Dasgupta:2001sh,Banfi:2004yd,Banfi:2005gj,Ellis:2009su,Banfi:2010pa,Walsh:2011fz,Chien:2012ur,Li:2012bw,Jouttenus:2013hs,Hatta:2013iba,Larkoski:2014tva,Procura:2018zpn,Aaboud:2017aca,Frye:2016aiz, Almeida:2008yp,Larkoski:2017iuy,Larkoski:2017cqq,Thaler:2010tr,Abdesselam:2010pt,Katz:2010mr,Gallicchio:2010dq,Adams:2015hiv,Sirunyan:2017ezt,Moore:2018lsr,FerreiradeLima:2016gcz,Rubin:2010fc,Chatrchyan:2012sn,ATLAS:2019kwg,CMS-PAS-BTV-13-001,CMS-PAS-JME-13-006,Kribs:2009yh,Chen:2010wk,Hackstein:2010wk,Kim:2010uj,Almeida:2011aa,Pandolfi:2012ima,Vernieri:2014wfa,CMS-PAS-HIG-17-007,Procura:2014cba,ATL-PHYS-PUB-2019-027,Aad:2019vyi,ATLAS:2020gwe}.

A complication of many popular jet grooming procedures is that they are \textit{discontinuous} in event space:
they may map similar ungroomed jets, which differ only by small changes in their distributions of energy, into extensively distinct groomed jets \cite{Dasgupta:2013ihk,Larkoski:2014wba}.
Discontinuous behavior in jet grooming algorithms leads in turn to undesirably sensitive responses to low-energy pollution.
Experimentally, discontinuities in grooming lead to large responses of groomed substructure to the interactions of jets with experimental detectors, and resulting difficulties in unfolding \cite{ATL-PHYS-PUB-2019-027,Aad:2019vyi,ATLAS:2020gwe}.
Theoretically, discontinuities in grooming lead to uncertainties in fixed-order calculations \cite{Larkoski:2014wba} and complications in describing the effects of hadronization on groomed jet substructure \cite{Hoang:2019ceu}.

In this paper, we introduce \textbf{P}ileup and \textbf{I}nfrared \textbf{R}adiation \textbf{A}n\textbf{n}i\textbf{h}il\textbf{a}tion (\PIRANHA{}), a paradigm for continuous jet grooming, using recent geometric perspectives in collider physics.
\PIRANHA{} applies intuition from optimal transport theory and the Energy Mover's Distance (EMD) between collider events~\cite{Komiske:2019fks,Komiske:2020qhg} to overcome the discontinuous behavior of existing jet grooming algorithms.
We introduce the new terminology of \textit{soft} and \textit{angular} \textit{discontinuities} to categorize discontinuous behavior under infinitesimal changes to the energies and angles of jet constituents, respectively.
We explore the improved continuity of \PIRANHA{} conceptually through several toy examples, and more practically by studying the responses of \PIRANHA{} grooming to several sources of soft distortions and additive contamination.

At present, \PIRANHA{} has three concrete implementations:
\begin{itemize}
    \item
    \textbf{Apollonius Subtraction} (P-AS), introduced in \Reff{Komiske:2020qhg}, is a continuous, conceptually simple application of the \PIRANHA{} paradigm with a computationally expensive implementation;

    \item
    \textbf{Iterated Voronoi Subtraction} (P-IVS), also introduced in \Reff{Komiske:2020qhg}, is a continuous, computationally efficient modification of P-AS with a simple geometric implementation;

    \item
    \textbf{Recursive Subtraction} (P-RS) is a family of nearly-continuous, computationally efficient, tree-based algorithms motivated by P-AS and P-IVS.
    We develop the concrete implementation of \textbf{Recursive Subtraction with a Fraction \(\boldsymbol{f}\)} (P-RSF\(_f\)).
    We discuss how P-RSF\(_f\) is discontinuous in suppressed regions of parameter space, and focus on \PRSF{1/2}, which is the only fully soft-continuous version of P-RSF\(_f\).
\end{itemize}

\Tab{groomerlist} expresses the infrared/collinear safety (using the definition of \Reff{Komiske:2020qhg}) and soft/angular continuity properties of the traditional and \PIRANHA{} groomers we discuss in this paper.
The soft continuity properties expressed in \Tab{groomerlist} are simple to understand by comparing the hard-cutoff framework of popular groomers to the subtractive framework of \PIRANHA{}.
The angular discontinuities expressed in \Tab{groomerlist} are more subtle and are inherited from angular-ordered jet clustering.
We note that the Recursive Subtraction algorithms we introduce are discontinuous only in a highly suppressed region of phase space, and we still find it suitable to call P-RS a family of \PIRANHA{} groomers.

\begin{table}[t]
\centering
\begin{tabular}{|c|c||c|c||c|c|}
\hhline{~~--||--}
\multicolumn{2}{c|}{}
&
\multicolumn{2}{c||}{Safety}
&
\multicolumn{2}{c|}{Continuity}
\\
\hhline{~~--||--}
\noalign{\vskip\doublerulesep
         \vskip-\arrayrulewidth}
\hhline{--||--||--}
    Groomer & Acronym &
    Infrared & Collinear &
    Soft & Angular
\\
\hhline{--||--||--}
\noalign{\vskip\doublerulesep
         \vskip-\arrayrulewidth}
\hhline{--||--||--}
Apollonius Subtraction & P-AS &
\cmark & \cmark &
\cmark & \cmark
\\
\hhline{--||--||--}
Iterated Voronoi Subtraction & P-IVS &
\cmark & \cmark &
\cmark & \cmark
\\
\hhline{--||--||--}
Recursive Sub. w/Fraction \(f = 1/2\) & \PRSF{1/2} &
\cmark & \cmark &
\cmark & \danger
\\
\hhline{--||--||--}
Recursive Sub. w/Fraction \(f \neq 1/2\) &  P-RSF\(_{f \neq 1/2}\) &
\cmark & \cmark &
\danger & \danger
\\
\hhline{--||--||--}
\noalign{\vskip\doublerulesep
         \vskip-\arrayrulewidth}
\hhline{--||--||--}
Soft Drop (SD), \(\beta_{\rm SD} > 0\) & SD\(_{\beta > 0}\) &
\cmark & \cmark &
\xmark & \danger
\\
\hhline{--||--||--}
SD (grooming mode), \(\beta_{\rm SD} \leq 0\) & SD\(_{\beta \leq 0}\) &
\cmark & \xmark &
\xmark & \danger
\\
\hline
\end{tabular}
\caption[
Safety and continuity properties of the \PIRANHA{} and Soft Drop grooming algorithms studied in this paper.
The mark \cmark~indicates that a groomer satisfies the associated condition, while the mark \xmark~indicates that it does not.
The mark \raisebox{0.15 em}{\danger} indicates that a groomer does not satisfy the condition, but only in a boundary region of phase space that is exponentially suppressed at leading order in perturbative QCD.
]{
Safety and continuity properties of the \PIRANHA{} and Soft Drop grooming algorithms studied in this paper.
The mark \cmark~indicates that a groomer satisfies the associated condition, while the mark \xmark~indicates that it does not.
The mark \raisebox{0.15 em}{\danger} indicates that a groomer does not satisfy the condition, but only in a boundary region of phase space that is exponentially suppressed at leading order in perturbative QCD.
The mark \raisebox{0.15 em}{\danger} indicates that a groomer does not satisfy the condition, but only in a boundary region of phase space that is exponentially suppressed at leading order in perturbative QCD.\footnotemark
}
\label{tab:groomerlist}
\end{table}

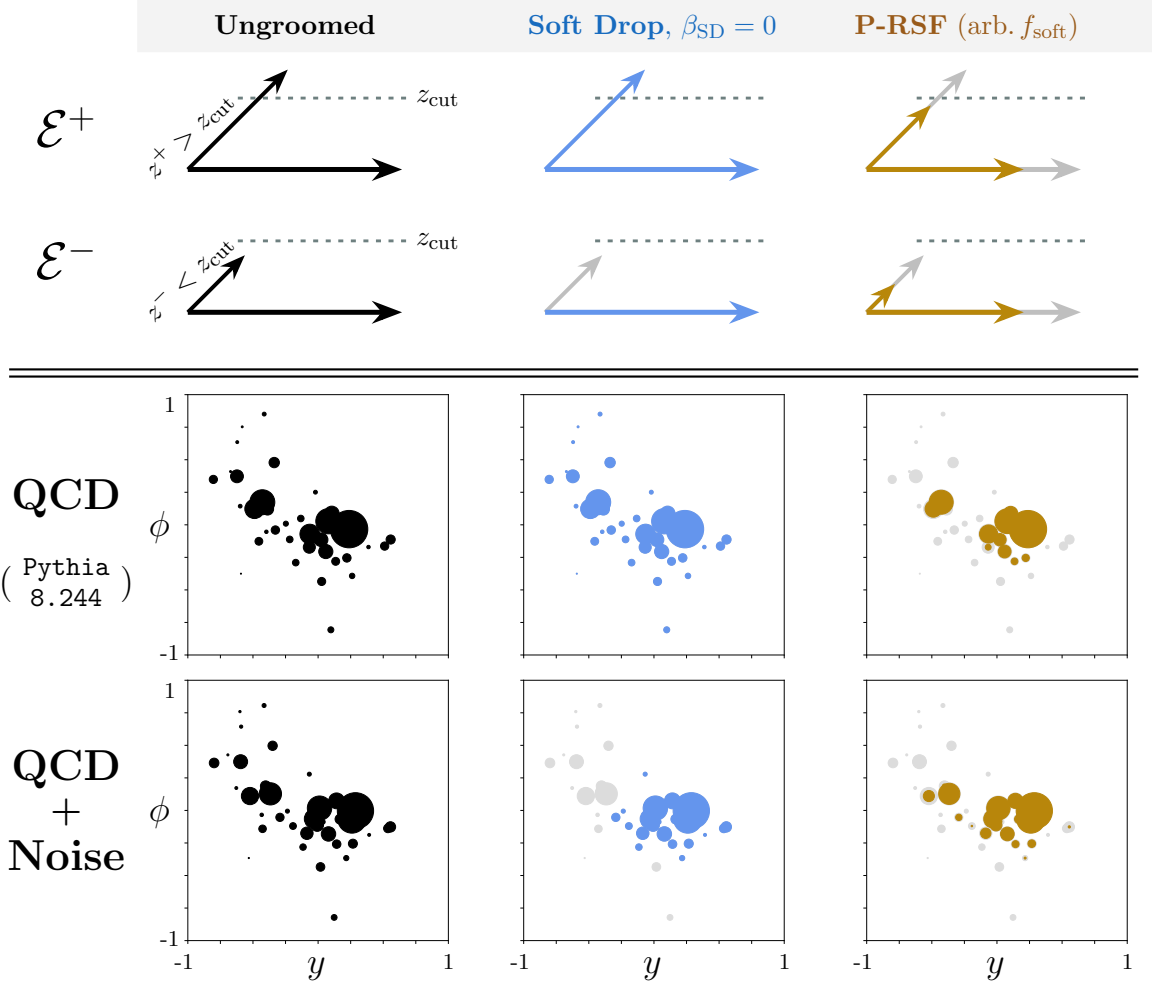
\begin{figure}[tp]
\centering
\scalebox{0.95}{
\begin{tikzpicture}[baseline=-3.5ex]

  \node at (-1.7,  .6)()
        {\huge \(\mathcal{E}^+\)};
  \node at (-1.7,  -1.25)()
        {\huge\(\mathcal{E}^-\)};

\draw [fill=black, opacity=0.05]
       (-0.7,2.4) -- (13.6,2.4) -- (13.6,1.65) -- (-0.7,1.65) -- cycle;

  \node at (1.5,  2)() {\textbf{Ungroomed}};

  \coordinate (A)  at (1.4, 1.4);
  \coordinate (O)  at (0, 0);
  \coordinate (B)  at (3, 0);
  \draw[-Stealth,line width=.55mm] (O) -> (A);
  \draw[-Stealth,line width=.7mm] (O) -> (B);

  \coordinate (A)  at (.8, -1.2);
  \coordinate (O)  at (0, -2);
  \coordinate (B)  at (3, -2);
  \draw[-Stealth,line width=.55mm] (O) -> (A);
  \draw[-Stealth,line width=.7mm] (O) -> (B);


\coordinate (I)  at (0.7, 1.0);
\coordinate (F)  at (3.15, 1.0);
\draw[dash pattern=on 2pt off 3pt, line width=.4mm, color=aurometalsaurus] (I) -> (F);
\node at (3.5,  1.0)()
        {\(\zcut\)};

\coordinate (I)  at (0.7, -1.0);
\coordinate (F)  at (3.15, -1.0);
\draw[dash pattern=on 2pt off 3pt, line width=.4mm, color=aurometalsaurus] (I) -> (F);
\node at (3.5,  -1.0)()
        {\(\zcut\)};

\node [rotate=45] at (0.0,  0.5)()
        {\(z^+ > \zcut\)};
\node [rotate=45] at (0.0,  -1.5)()
        {\(z^- < \zcut\)};

\node at (6.5,  2)() {\textcolor{dodgerblue!75!black}{\textbf{Soft Drop}, \(\beta_{\rm SD} = 0\)}};

  \coordinate (A)  at (6.4, 1.4);
  \coordinate (O)  at (5.0, 0);
  \coordinate (B)  at (8.0, 0);
  \draw[-Stealth,line width=.55mm, color=royalblue] (O) -> (A);
  \draw[-Stealth,line width=.7mm, color=royalblue] (O) -> (B);

  \coordinate (Ap)  at (5.8, -1.2);
  \coordinate (O)  at (5.0, -2);
  \coordinate (B)  at (8.0, -2);
  \draw[-Stealth, line width=.55mm, color=black!25!white] (O) -> (Ap);
  \draw[-Stealth,line width=.7mm, color=royalblue] (O) -> (B);

\coordinate (I)  at (5.7, 1.0);
\coordinate (F)  at (8.15, 1.0);
\draw[dash pattern=on 2pt off 3pt, line width=.4mm, color=aurometalsaurus] (I) -> (F);

\coordinate (I)  at (5.7, -1.0);
\coordinate (F)  at (8.15, -1.0);
\draw[dash pattern=on 2pt off 3pt, line width=.4mm, color=aurometalsaurus] (I) -> (F);

  \node at (10.9,  2)() {\textcolor{ochre!75!black}{\textbf{P-RSF} (arb.\,\(f_{\rm soft}\))}};

  \coordinate (A)  at (10.4, 0.9);
  \coordinate (Ap)  at (10.9, 1.4);
  \coordinate (O)  at (9.5, 0);
  \coordinate (B)  at (11.7, 0);
  \coordinate (Bp)  at (12.5, 0);
  \draw[-Stealth,line width=.55mm, color=black!25!white] (O) -> (Ap);
  \draw[-Stealth,line width=.55mm, color=darkgoldenrod] (O) -> (A);
  \draw[-Stealth,line width=.7mm, color=black!25!white] (O) -> (Bp);
  \draw[-Stealth,line width=.7mm, color=darkgoldenrod] (O) -> (B);

  \coordinate (A)  at (9.9, -1.6);
  \coordinate (Ap)  at (10.3, -1.2);
  \coordinate (O)  at (9.5, -2);
  \coordinate (B)  at (11.7, -2);
  \coordinate (Bp)  at (12.5, -2);
  \draw[-Stealth,line width=.55mm, color=black!25!white] (O) -> (Ap);
  \draw[-Stealth,line width=.55mm, color=darkgoldenrod] (O) -> (A);
  \draw[-Stealth,line width=.7mm, color=black!25!white] (O) -> (Bp);
  \draw[-Stealth,line width=.7mm, color=darkgoldenrod] (O) -> (B);

\coordinate (I)  at (10.2, 1.0);
\coordinate (F)  at (12.65, 1.0);
\draw[dash pattern=on 2pt off 3pt, line width=.4mm, color=aurometalsaurus] (I) -> (F);

\coordinate (I)  at (10.2, -1.0);
\coordinate (F)  at (12.65, -1.0);
\draw[dash pattern=on 2pt off 3pt, line width=.4mm, color=aurometalsaurus] (I) -> (F);

\coordinate (I)  at (-2.5, -2.8);
\coordinate (F)  at (13.3, -2.8);
\draw[line width=.3mm] (I) -> (F);

\coordinate (I)  at (-2.5, -2.9);
\coordinate (F)  at (13.3, -2.9);
\draw[line width=.3mm] (I) -> (F);

  \node at (-1.7,  -4.6)()
        {\textbf{\LARGE QCD}};
  \node at (-2.55, -5.8)(){\Large (};
  \node at (-1.7,  -5.6)()
        {\texttt{Pythia}};
  \node at (-1.7,  -6.0)()
        {\texttt{8.244}};
 \node at (-0.85, -5.8)(){\Large )};
  \node at (-1.7,  -8.4)()
         {\textbf{\LARGE QCD}};
 \node at (-1.7,  -9.0)()
         {\textbf{\Large +}};
 \node at (-1.7,  -9.6)()
        {\textbf{\LARGE Noise}};

  \node at (1.8,  -5.0)() {\includegraphics[width=.25\textwidth]{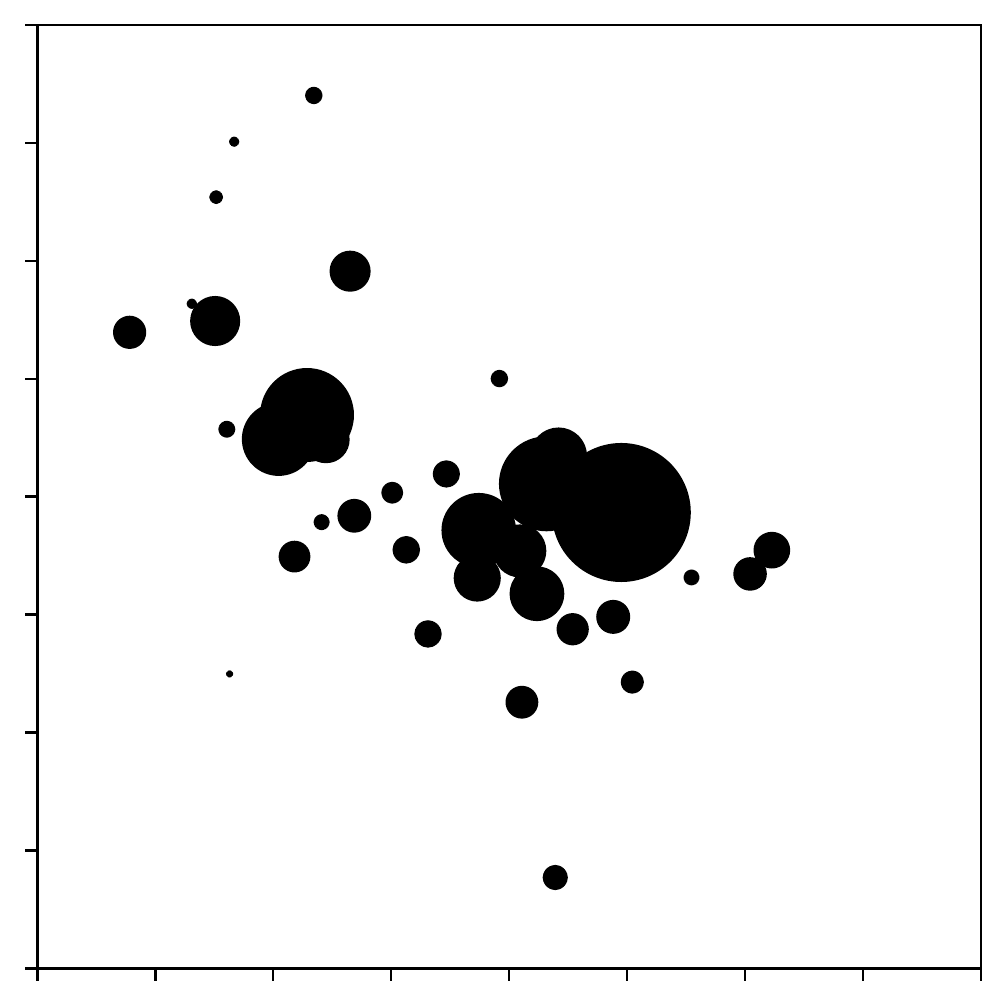}};
  \node at (1.8,  -9.0)() {\includegraphics[width=.25\textwidth]{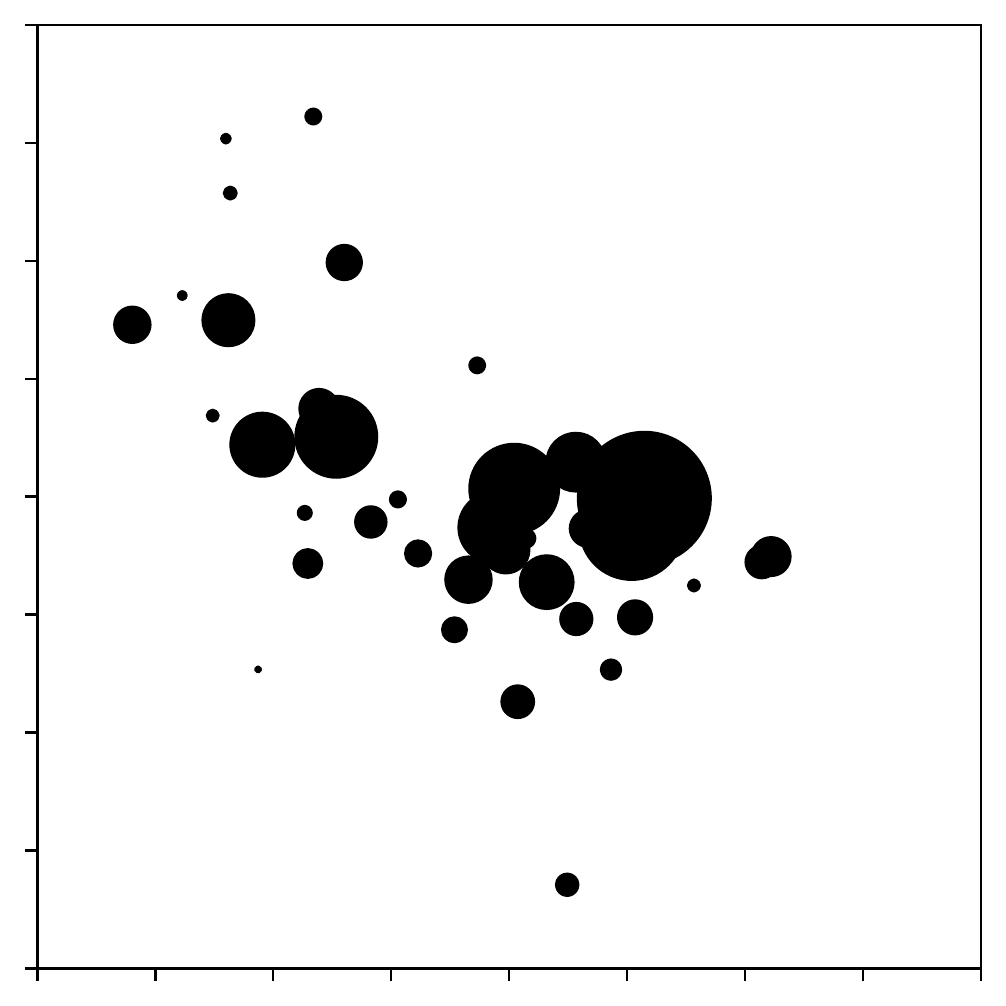}};
  \node at (6.5,  -5.0)() {\includegraphics[width=.25\textwidth]{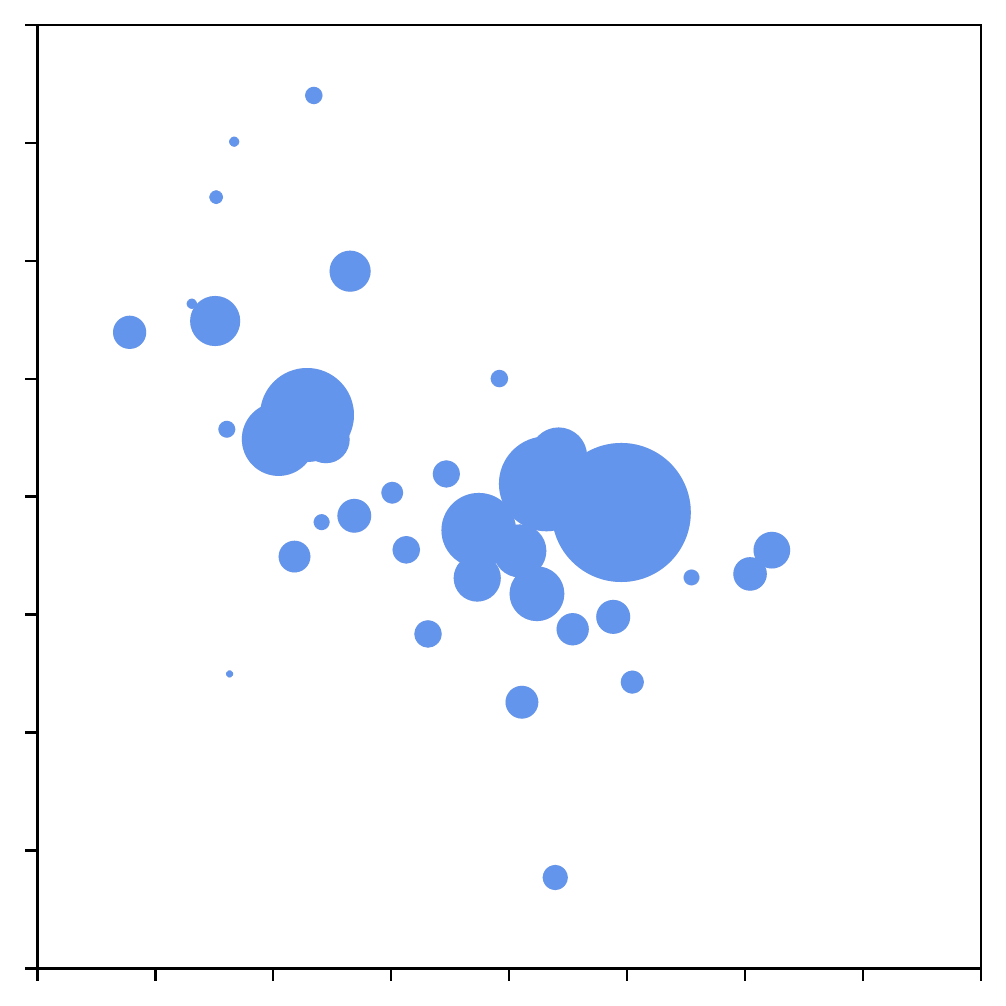}};
  \node at (6.5,  -9.0)() {\includegraphics[width=.25\textwidth]{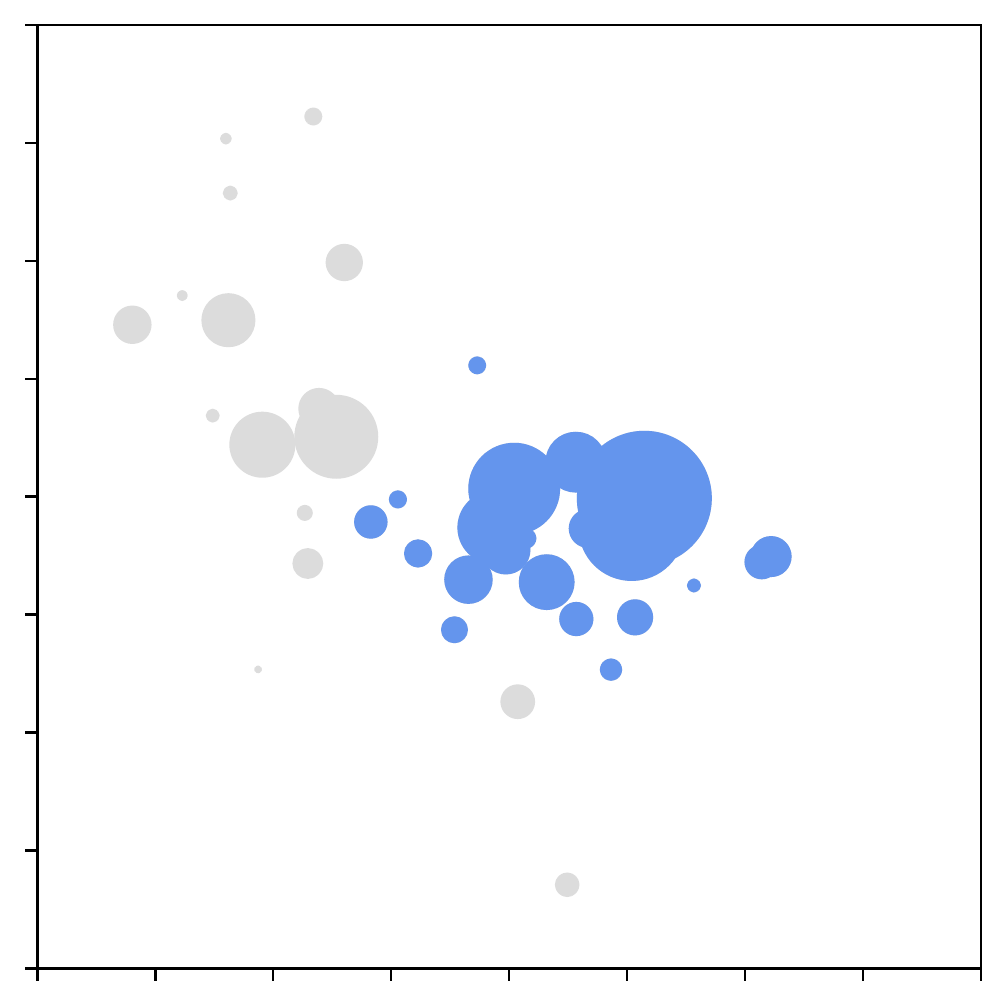}};
  \node at (11.3,  -5.0)() {\includegraphics[width=.25\textwidth]{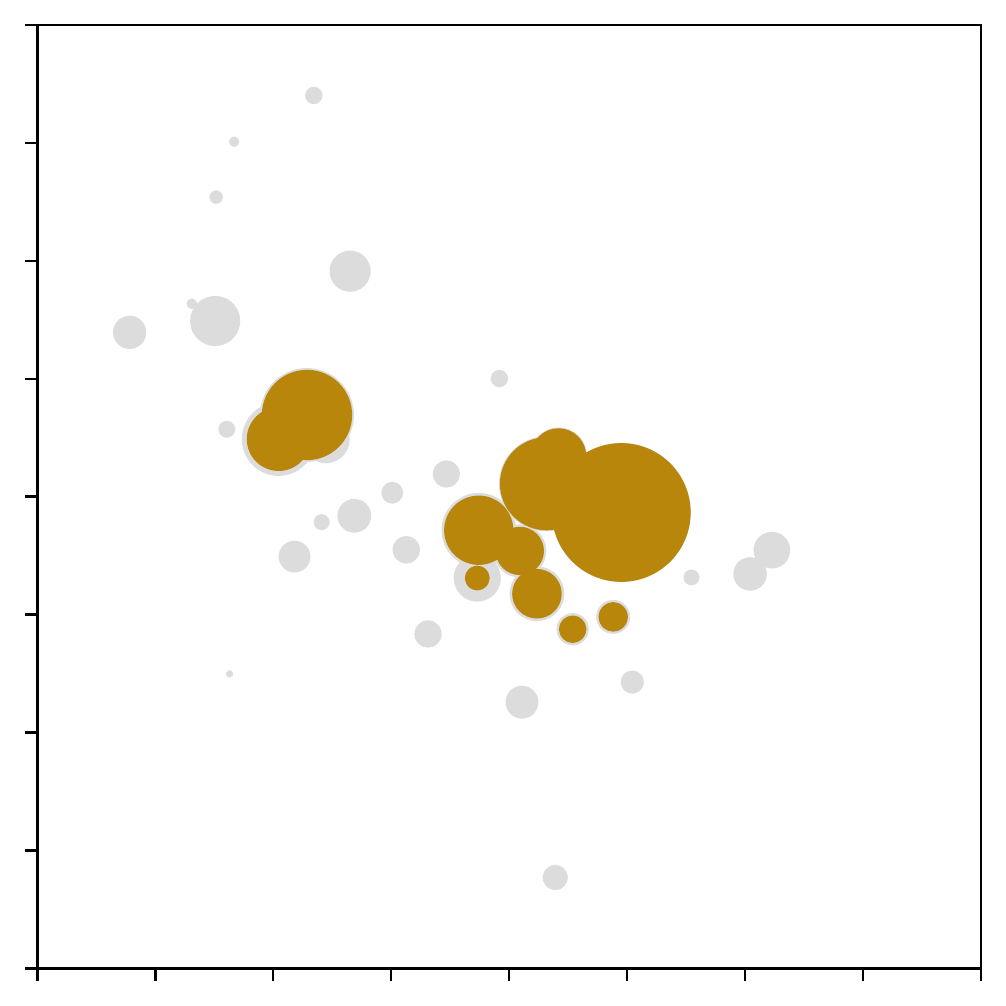}};
  \node at (11.3,  -9.0)() {\includegraphics[width=.25\textwidth]{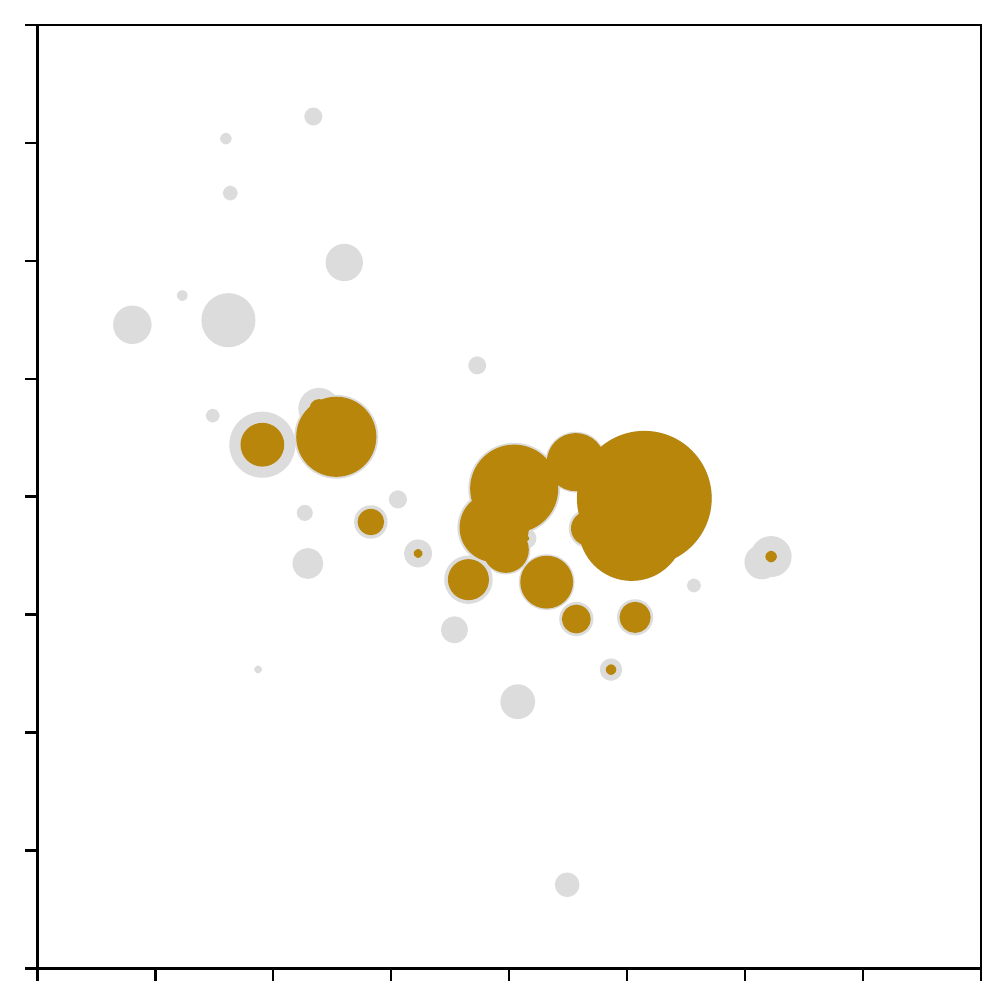}};

 \node at (-0.4,  -5.0)() {\Large \(\phi\)};
 \node at (-0.4,  -9.0)() {\Large \(\phi\)};
 \node at (1.8,  -11.2)() {\Large \(y\)};
 \node at (6.5,  -11.2)() {\Large \(y\)};
 \node at (11.3,  -11.2)() {\Large \(y\)};

 \node at (-0.25,  -10.77)() {\footnotesize -1};
 \node at (-0.25,  -7.3)() {\footnotesize 1};
 \node at (-0.25,  -6.77)() {\footnotesize -1};
 \node at (-0.25,  -3.3)() {\footnotesize 1};
 \node at (-0.05,  -11.1)() {\footnotesize -1};
 \node at (3.65,  -11.1)() {\footnotesize 1};
 \node at (4.65,  -11.1)() {\footnotesize -1};
 \node at (8.35,  -11.1)() {\footnotesize 1};
 \node at (9.45,  -11.1)() {\footnotesize -1};
 \node at (13.15,  -11.1)() {\footnotesize 1};

\end{tikzpicture}
}
\caption{
A visualization of how similar ungroomed events (first column, in black) may be mapped discontinuously to distinct groomed jets by the hard-cutoff mMDT algorithm (Soft Drop with \(\beta_{\rm SD} = 0\), or \SD{0}; middle column, in blue), while the continuous \PRSF{1/2} algorithm will map them to similar groomed results (final column, in orange).
In the first two rows, we compare the action of mMDT and \PRSF{1/2} on simple toy events \(\mathcal{E}^\pm\), with particles visualized as arrows whose size corresponds to their energy.
In the final two rows, we compare the groomers acting on a more realistic QCD event produced with \texttt{Pythia 8.244}, and on the same QCD event with additional Gaussian noise, respectively.
}
\label{fig:intro_visual}
\end{figure}

Popular jet grooming algorithms are soft discontinuous because they introduce hard cutoffs on the energy of jet constituents, removing particles that are too low-energy to be classified as important probes of hard physics;
we call such groomers \textit{hard-cutoff groomers}.
The modified Mass Drop Tagger (mMDT) \cite{Dasgupta:2013ihk}, for example, requires that certain sub-jets within a jet carry a sufficient energy fraction, \(z > z_{\rm cut}\), to survive the grooming procedure.
If a sub-jet has an energy fraction near \(\zcut\), low-energy pollution may push its energy to the other side of the cutoff, leading to soft discontinuous behavior and associated theoretical and experimental complications.
\Fig{intro_visual} demonstrates this soft discontinuous behavior, the corresponding soft continuity of \PRSF{1/2}, by comparing their action on two sets of similar events:
two toy jet events \(\mathcal{E}^\pm\) consisting of two particles each, and a QCD jet produced in \texttt{Pythia 8.244} along with the same QCD event with a small amount of gaussian noise applied to the \(p_T\), rapidity, and azimuthal angle of each particle.
We revisit the events \(\mathcal{E}^\pm\) as tools to characterize soft discontinuity more precisely in \Sec{sd_discont}.
The mMDT grooming procedure maps \(\mathcal{E}^+\) and \(\mathcal{E}^-\) to very distinct events, and similarly maps the noiseless and noisy QCD jets to very distinct groomed jets.
We chose an unusually high value of \(\zcut = 0.25\) when grooming the QCD events to make the discontinuity of mMDT more apparent.
\footnotetext{
Soft Drop with \(\beta_{\rm SD} \leq 0\) is not collinear safe in grooming mode, used in this work, where jets consisting of a single particle are retained as groomed jets.
Grooming mode is a natural choice in the context of pileup mitigation (see \Sec{sd_discont}).
Note that Soft Drop with \(\beta_{\rm SD} \leq 0\) is collinear safe in tagging mode, where jets consisting of a single particle are groomed away completely.
}
We will also use Soft Drop \cite{Larkoski:2014wba}, a generalization of mMDT reviewed in \Sec{softdrop} which reproduces mMDT when a parameter \(\beta_{\rm SD}\) is set to zero, as another example of soft discontinuous hard-cutoff grooming.


\PIRANHA{} borrows technology from optimal transport to produce soft continuous, subtractive jet grooming algorithms without hard cutoffs.
\PIRANHA{} groomers may be roughly imagined as optimally transporting hungry piranhas to eat up delicious low-energy pollution.
Given an ungroomed event with energy density \(\mathcal{E}_{0}\) in the pseudorapidity-azimuth angle plane and any ansatz \(\rhoC\) for the energy density of low-energy pollution, one particularly succinct \PIRANHA{} algorithm is
\begin{align}
    \mathcal{E}_g\left[\mathcal{E}_0\right]
    =
    \underset{\mathcal{E}'}{\arg \min}\ {\rm EMD}_{\beta, R}
    \left(\mathcal{E}_0, \mathcal{E}' + \rhoC\right)
    \label{eqn:simple_piranha_intro}
    ,
\end{align}
where \(\mathcal{E}_g\) is a groomed energy distribution, \(\rho\,\mathcal{C}\) is a free input to the algorithm indicating a model of the energy flow of the contaminating radiation (see \Sec{emd}), and explicit dependence on the EMD metric makes the continuity of \(\mathcal{E}_g\left[\mathcal{E}_0\right]\) manifest.

P-AS is a direct, continuous application of \Eq{simple_piranha_intro} with \(\beta = 1\), and for which \(\rhoC\) is uniform in the pseudorapidity-azimuth plane to approximate common sources of additive contamination such as PU and UE \cite{Soyez:2018opl,Monk:2018clo,Sjostrand:1987su,Sjostrand:2014zea,Dasgupta:2007wa,Kirchgaesser:2020poq,Moraes:2007rq,CDF:2015txs,Larkoski:2021hee,Baron:2020xoi,Marzani:2017kqd}.
P-IVS approximates P-AS and preserves its continuity.
Finally, the Recursive Subtraction with a Fraction 1/2 (\PRSF{1/2}) algorithm we introduce in \Sec{rsf} furnishes, to our knowledge, the first example of a soft continuous tree-based jet grooming algorithm;
the soft discontinuities of other P-RS algorithms occur only in highly suppressed regions of phase space.

The rest of the paper proceeds as follows:
in \Sec{PIRANHA}, we review the Energy Mover's Distance (EMD) of \Reff{Komiske:2019fks}, propose the \PIRANHA{} paradigm for continuous jet grooming, review the Apollonius Subtraction (P-AS) and Iterated Voronoi Subtraction (P-IVS) algorithms of \Reff{Komiske:2020qhg}, and introduce Recursive Subtraction (P-RS) with a Fraction (P-RSF) as a new, tree-based implementation of \PIRANHA{}.
In \Sec{traditionaldiscont}, we discuss the soft discontinuities present in hard-cutoff groomers and P-RSF, demonstrate the soft continuity of \PRSF{1/2}, and examine angular discontinuity in tree-based groomers.
In the remainder of the paper, we demonstrate the soft insensitivity enjoyed by \PIRANHA{} groomers over hard-cutoff groomers through several phenomenological studies.
In \Sec{soft_distortion}, we compare the responses of the hard-cutoff Soft Drop algorithm and \PRSF{1/2} to soft distortions from hadronization and to the removal of neutral particles, as a proxy for detector effects.
In \Sec{add_contam}, we compare the ability of Soft Drop and \PRSF{1/2} to effectively remove additive contamination from pileup and the underlying event.
We give concluding thoughts and avenues for future exploration in \Sec{Conclusions}, delegating discussions of \PIRANHA{} grooming in EMD mode, comparisons of other hard-cutoff and \PIRANHA{} groomers, and a first look at the perturbative structure of \PIRANHA{} grooming to \Appss{grooming_in_emd_mode}{feedingfrenzy}{calc}, respectively.
The \PIRANHA{} algorithms used in our analysis are available on GitHub \cite{piranhagithub}.

\section{The PIRANHA{} Paradigm for Continuous Grooming}
\label{sec:PIRANHA}

This section introduces the \PIRANHA{} paradigm for jet grooming, which uses optimal transport to continuously remove low-energy pollution from jets.
We first review the Energy Mover's Distance (EMD) of Ref.~\cite{Komiske:2019fks}, which provides a framework to introduce the principles of \PIRANHA{} groomers and to enforce their continuity.
We then discuss three implementations of \PIRANHA{}:
Apollonius Subtraction (P-AS) and Iterated Voronoi Subtraction (P-IVS) -- both introduced in \Reff{Komiske:2020qhg} -- and the new tree-based Recursive Subtraction (P-RS) family of \PIRANHA{} algorithms.
We describe a variant of \PIRANHA{} -- grooming in EMD mode -- in \App{grooming_in_emd_mode}.
We end this section with \Sec{stronger_continuity}, which may be skipped on a first reading, where we present definitions of stronger notions of continuity that could potentially pave the way for more constrained and well-behaved strategies for continuous grooming.

\subsection{Review of the Energy Mover's Distance}
\label{sec:emd}
The Energy Mover's Distance (EMD) of Ref.~\cite{Komiske:2019fks} provides a quantitative measure of the similarity between two jets.
We introduce the EMD here both to facilitate our definition of continuity in jet grooming and as a useful tool for quantifying the responses of jet groomers to low-energy pollution.
Readers familiar with the EMD should feel free to skip to \Sec{pira_intro}.

The EMD is a metric on the space of collider events.
Here and in the remainder of the paper, we borrow the terminology of \Reff{Komiske:2020qhg} and use the terms ``event'' and ``collider event'' to refer to the \textit{energy flow} of the event, or the angular distribution of energy of its outgoing radiation.\footnote{
For example, the energy flow of an event \(\mathcal E\) with \(M\) outgoing particles is the angular distribution
\begin{equation*}
    \label{eqn:energyflow}
    \mathcal E(\hat n)
    =
    \sum_{i=1}^ME_i\,\delta(\hat n - \hat n_i),
\end{equation*}
where $E_i > 0$ denotes the energy carried by particle \(i\), and $\hat n_i$ denotes its outgoing angular direction.
}
The EMD may be thought of as the amount of ``work'' required to rearrange one event into another.
For events that consist of a finite number of outgoing particles, the EMD is defined as the solution to the optimal transport problem
\begin{equation}
    \label{eqn:EMD_def_1}
    {\rm EMD}_{\beta, R}\left(\mathcal{E}, \mathcal{E}'\right)
    =
    \min_{f_{ij} > 0} \sum_{i = 1}^M \sum_{j = 1}^{M'} f_{ij} \left(\frac{\theta_{ij}}{R}\right)^\beta
    +
    \left| \sum_{i = 1}^M E_i -  \sum_{j = 1}^{M'} E'_j\right|,
\end{equation}
\begin{equation}
    \sum_{i = 1}^M f_{ij} \leq E_j',
    ~~
    \sum_{j = 1}^{M'} f_{ij} \leq E_i,
    ~~
    \sum_{i = 1}^M \sum_{j = 1}^{M'} f_{ij}
    = \min\left(\sum_{i = 1}^M E_i, \sum_{j = 1}^{M'} E'_j\right)
    \label{eqn:EMD_def_2}
    ,
\end{equation}
where \(i \in \{1, ..., M\}\) indicates a final-state particle of \(\mathcal{E}\) with energy \(E_i\), \(j \in \{1, ..., M'\}\) indicates a final-state particle of \(\mathcal{E}'\) with energy \(E'_j\), and \(\theta_{ij}\) denotes an angular distance between particles \(i\) and \(j\).
\(\beta > 0\) and \(R > 0\) are free parameters that control the behavior and relative importance of the first term on the right-hand side of \Eq{EMD_def_1}.\footnote{
Strictly speaking, the EMD defines a metric only if \(2 R\) is greater than the maximum possible angular distance \(\theta_{ij}\).
Furthermore, for \(\beta > 1\), one must raise the first term of \Eq{EMD_def_1} to the \(1/\beta\) power.
We use the EMD without a subscript, \(
{\rm EMD}\left(\mathcal{E}, \mathcal{E}'\right)
\) to denote the EMD with \(\beta = 1\) and with large enough \(R\) that it furnishes a metric.
}

In this work, we adopt the hadronic angular measure of \Reff{Komiske:2020qhg}, given by
\begin{equation}
    \theta_{ij} = \sqrt{-(n_i - n_j)^2}
    ,
    ~~~~~~
    ~~~~~~
    n_{i}^\mu = \frac{p_i^\mu}{E_{T\,i}} = (\cosh(y_i), \textbf{v}_{T\,i}, \sinh(y_i))^\mu
    .
    \label{eqn:hadronic_metric}
\end{equation}
Here, \(n_i^\mu\) is a vector describing the motion of particle \(i\), parameterized by its rapidity \(y_i\) and its velocity \(\textbf{v}_{T\, i}\) in the directions transverse to the beampipe.
This angular metric reproduces the rapidity-azimuth distance between particles \(i\) and \(j\) in the small rapidity-azimuth distance limit.\footnote{
Another popular choice for the form of the \(n_i^\mu\) is captured by
\begin{align}
    \label{eqn:ee_metric}
    n_{i,~{e^+e^-}}^\mu = \frac{p_i^\mu}{E_i} = (1, \textbf{v}_i)^\mu
    ,
\end{align}
which is a natural choice in the study of electron-positron collisions~\cite{Komiske:2020qhg}.
This choice reproduces the real-space opening angle between particles \(i\) and \(j\) in the small angle limit.
}

The EMD between a groomed jet in the presence and absence of low-energy pollution also provides an observable-independent tool for quantifying the sensitivity of jet grooming to pollution from soft distortions and additive contamination.
Since the EMD furnishes a metric on the space of events, two events are separated by zero EMD iff they have identical energy flow, and two jets separated by zero EMD will yield the same value for every infrared/collinear (IRC) safe observable (see \href{https://arxiv.org/pdf/2004.04159.pdf#page=11\&zoom=100,0,650}{Lemma 1} of \Reff{Komiske:2020qhg}).
Further, the EMD between two events also \textit{directly} bounds the differences in a large class of IRC-safe observable quantities (Lipschitz continuous observables) between the events~\cite{Komiske:2019fks}.
For example, the EMD between a parton-level jet and the same jet after hadronization (the parton-hadron EMD) directly bounds the difference between a large class of parton- and hadron-level IRC-safe observables.
The parton-hadron EMD is therefore a powerful, observable independent tool for characterizing the response of a jet to soft distortions from hadronization.
We use the EMD in our analysis of grooming sensitivity to several sources of low-energy pollution in \Secs{soft_distortion}{add_contam}.

\subsection{PIRANHA{} and Continuity}
\label{sec:pira_intro}

\PIRANHA{} grooming may be intuitively imagined as the optimal transport of a group of hungry piranhas towards the low-energy pollution within a jet.
In this analogy, henceforth the piranha analogy, a \PIRANHA{} groomer populates the \(\eta\)-\(\phi\) plane with a school of piranhas, distributed according to a particular model of the polluting energy density.
When they are given an event to groom, the piranhas discuss which piranhas will get to eat which jet constituent;
in the language of optimal transport, we say that the piranhas decide on a set of transport plans.
The piranhas then implement the plan and feed on the constituents of the jet, or subtract the \(p_T\) of each jet constituent.

\PIRANHA{} grooming uses optimal transport to design \textit{continuous} grooming procedures that have reduced sensitivity to low-energy pollution.
In particular, continuity requires that minuscule changes to the jet do not dramatically change the resulting groomed jet.
In the piranha analogy, continuity emerges when the piranhas' transport plans do not respond dramatically to small changes in the constituents of an ungroomed jet.
Since both soft distortions and fluctuations in additive contamination may lead to small changes in an event, continuity helps ensure that neither dramatically changes the information carried by groomed jets.

We may use the EMD to define continuity more precisely for maps on energy flows, in the spirit of \href{https://arxiv.org/pdf/2004.04159.pdf#page=5\&zoom=100,0,350}{Definition 1} of \Reff{Komiske:2020qhg}:
\begin{definition}\label{def:eventcontinuity}
A map \(M\) from energy flows to energy flows is \textit{continuous at an event \(\mathcal E\)} if, for any \(\varepsilon > 0\), there exists a \(\delta > 0\) such that for all \(\mathcal E'\),
\begin{equation*}\label{eqn:emdcontinuity}
    {\rm EMD}(\mathcal E,\mathcal E')<\delta
    \quad\implies\quad
    {\rm EMD}(M(\mathcal E), M(\mathcal E')) < \varepsilon.
\end{equation*}
We say that \(M\) is \textbf{continuous} if it is continuous at all events \(\mathcal{E}\).
\end{definition}
\noindent Definition~\ref{def:eventcontinuity} encodes the requirement that a continuous grooming algorithm must map jets that are infinitesimally nearby in event space to groomed results that are also infinitesimally close.\footnote{
Equivalently, we may say that a grooming procedure is continuous if it maps nearby jets to nearby groomed jets.
An event \(\mathcal{E}\) is \textbf{near} an open ball \(\mathcal{B}\) if any neighborhood of \(\mathcal{E}\) contains events in \(\mathcal{B}\).
On the real line, for example, the point \(E = 1\) is near the open ball \(B = (-1, 1)\).
A grooming procedure \(G\) is then \textbf{continuous} if for any event \(\mathcal{E}\) and open ball \(\mathcal{B}\),
    \(
    \mathcal{E} {\text ~is~ near~} \mathcal{B}
    \implies
    G(\mathcal{E}) {\text~ is~ near~} G(\mathcal{B}).
    \)
}

The EMD itself also furnishes an example of an optimal transport problem, allowing us to immediately introduce a simple \PIRANHA{} groomer.
To begin, let us model the radiation that we would like to remove from our jet as a distribution of energy \(\rhoC\) in a portion of the rapidity-azimuth (\(\eta\)-\(\phi\)) plane with area \(A\).
We use \(\rho\) to indicate the mean energy density and \(\mathcal{C}\) to indicate a distribution modeling the shape of the contaminating radiation, normalized such that
\begin{equation}
    \rho\int \dd y\,\dd\phi\,\mathcal{C}(y,\phi) = \rho A = E
\end{equation}
gives the total energy of the additive contamination we would like to remove.
Again echoing \Reff{Komiske:2020qhg}, we may phrase the subtraction of the energy distribution \(\rhoC\) from an event \(\mathcal{E}\) as the solution to the optimal transport problem
\begin{align}
    \mathcal{E}_g[\mathcal E_0] = \underset{\mathcal{E}'}{\arg \min}\ {\rm EMD}_{\beta, R}
    \left(\mathcal{E}_0, \mathcal{E}' + \rhoC\right)
    \label{eqn:simple_piranha}
    .
\end{align}
It was shown in \href{https://arxiv.org/pdf/2004.04159.pdf#page=33\&zoom=100,0,0}{Lemma 3} of \Reff{Komiske:2020qhg} that \Eq{simple_piranha} defines a subtractive algorithm that merely decreases the energy of jet constituents, and does not add new particles to an event.
In the piranha analogy, \(\rhoC\) roughly describes the distribution of our piranhas, with a ``piranha energy density'' \(\rho\), while the minimization problem of \Eq{simple_piranha} provides them with the transport plans that guide them to subtract from the \(p_T\) of jet constituents.
Furthermore, the form of \(\mathcal{E}_g[\mathcal E_0]\) in \Eq{simple_piranha} is manifestly continuous due to its dependence on the EMD metric.\footnote{
While \(\mathcal{E}_g[\mathcal E_0]\) in \Eq{simple_piranha} is continuous, manifest continuity due to dependence on the EMD should be treated skeptically.
For example, the ``grooming procedure'' associated with  \Eq{simple_piranha} may not be continuous if the \(\arg \min\) operation is restricted to a non-convex sub-space of the full space of events.
Analogously, \(
    f(x)
    =
    \underset{y}{\arg \min}\ ||x, y+1||
    =
    x-1
\)
is continuous as a map from \(\mathbb R\) to \(\mathbb R\) with the standard topology.
However, \(
    g(x)
    =
    \underset{y \in \{0, 1\}}{\arg \min}\ ||x, y+1||,
\)
which differs only by the allowed values of \(y\) in the \(\arg \min\), is not.
}

For the more complicated \PIRANHA{} groomers we introduce in this paper, it will be helpful to define two additional properties that are required by full continuity:
\begin{itemize}
    \item
    \textbf{Soft Continuity} is the requirement that a groomer is invariant under infinitesimally soft perturbations of the energies of jet constituents;

    \item
    \textbf{Angular Continuity} is the requirement that a groomer is invariant under infinitesimally small angular changes in the directions of jet constituents.
\end{itemize}

We will see that hard-cutoff groomers and even many recursive subtraction algorithms suffer from soft discontinuity in \Secs{softdrop}{rsf_discont} respectively.
In \Sec{ang_discont}, we will explore how tree-based grooming algorithms -- hard-cutoff and \PIRANHA{} alike -- may inherit angularly discontinuous behavior from angular-ordered jet clustering.\footnote{
Collinear safety, or invariance under exact collinear splittings which replace a final-state particle of momentum \(p^\mu\) by two particles with momenta \(\lambda p^\mu\) and \((1-\lambda) p^\mu\), is a weaker condition.
Exact collinear splittings do not change the energy distribution associated with an event; therefore, any well-defined map on the space of particle events must be collinear safe~\cite{Komiske:2020qhg}.
}

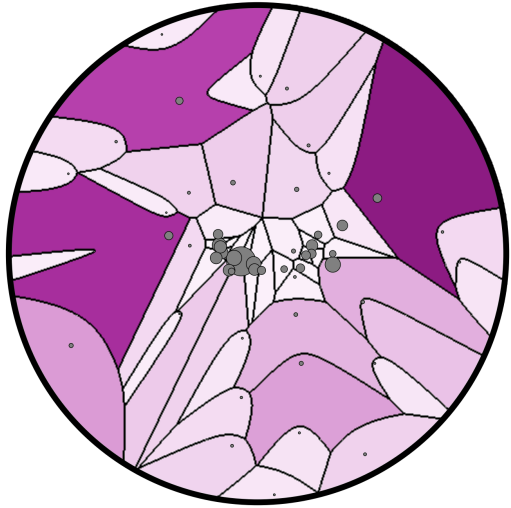
\begin{figure}[t!]
\centering

\subfloat[
]{
\begin{tikzpicture}
\node at (0,0) {
    \includegraphics[width=0.45\textwidth]{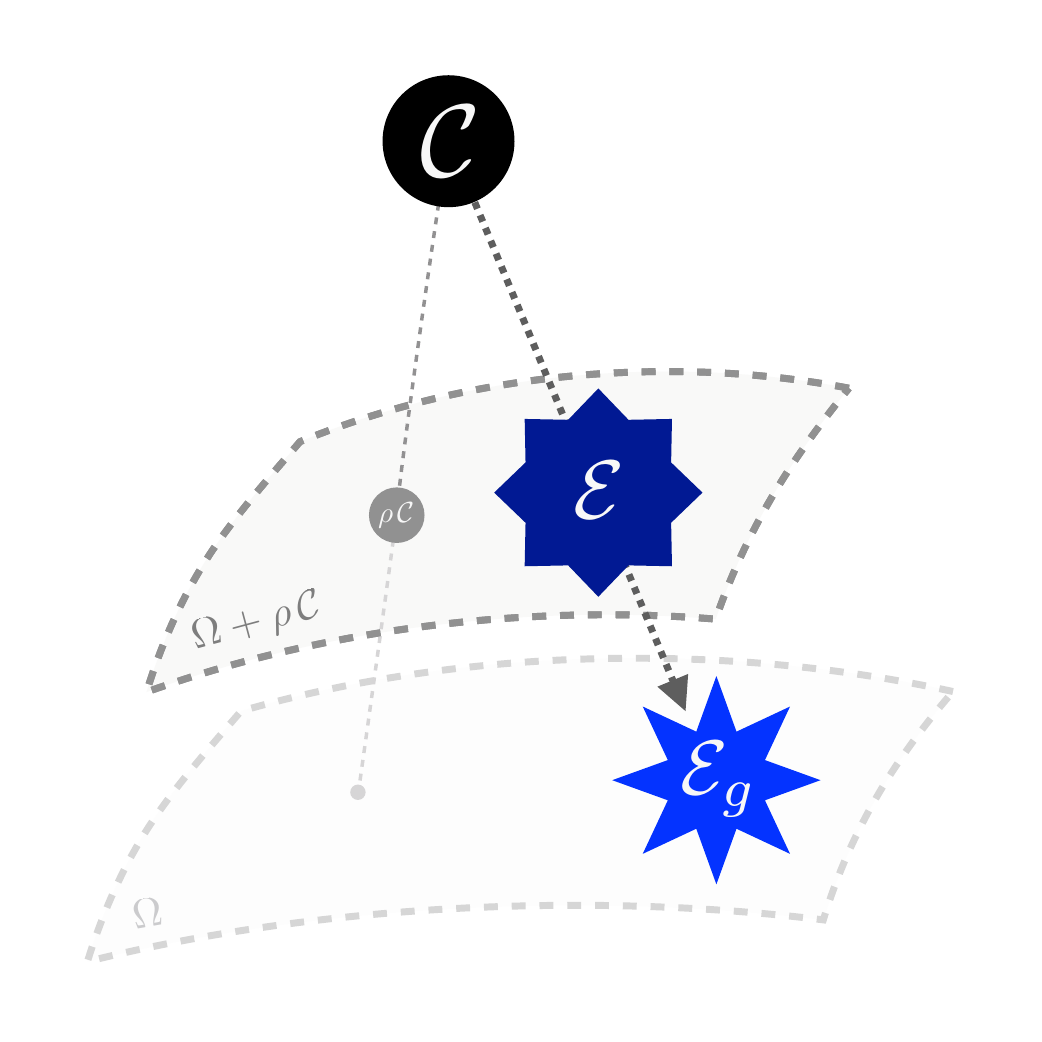}
};
\node at (0,4.25) {Cartoon of a \PIRANHA{} groomer};
\node at (0,3.75) {acting on the space of energy flows};
\end{tikzpicture}
\label{fig:piranha_projection}
}
\subfloat[
]{
\begin{tikzpicture}
\node at (0,0) {
\includegraphics[width=0.45\textwidth]{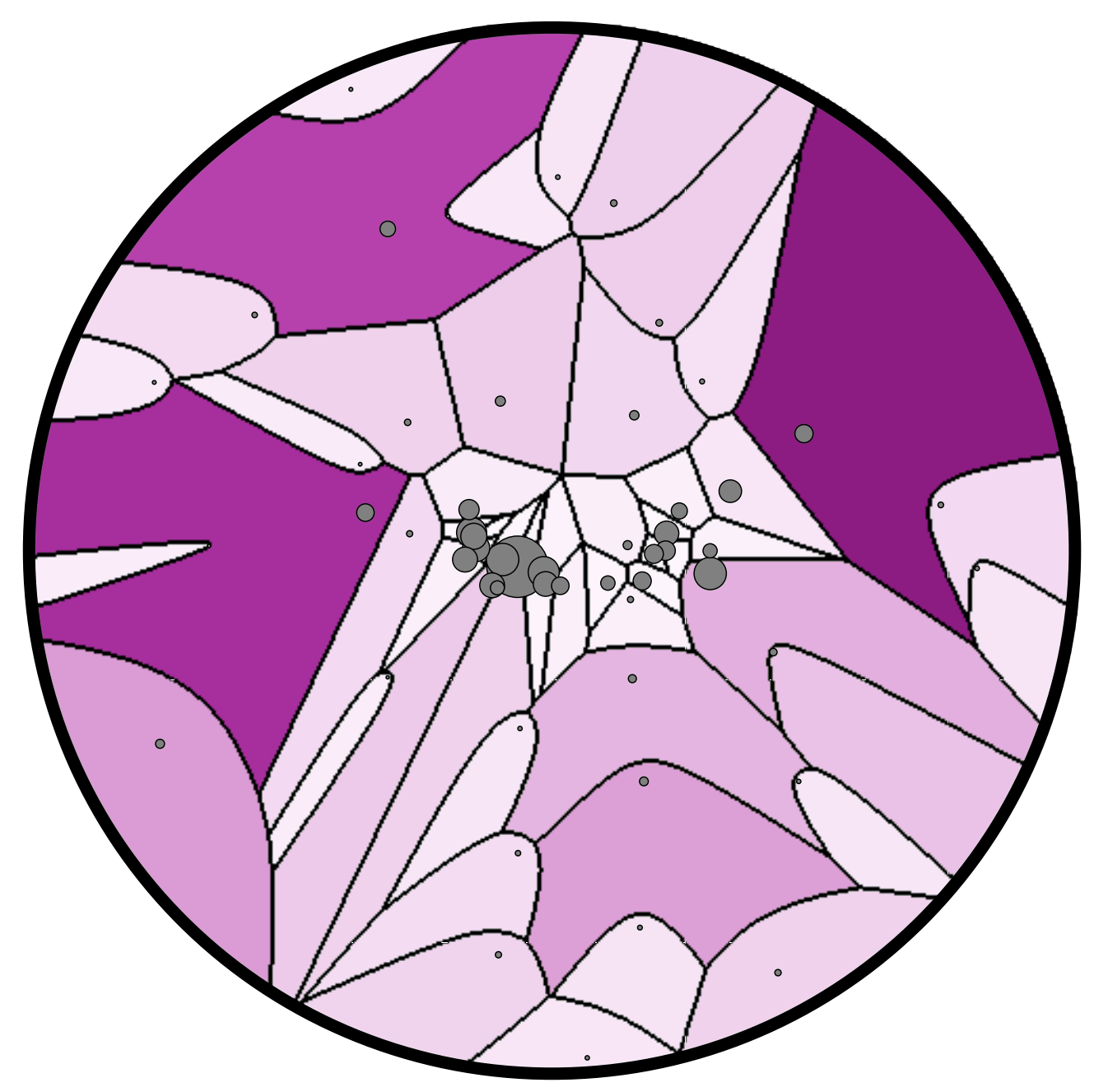}
};
\node at (0,4.35) {Visualization of P-AS acting on};
\node at (0,3.85) {an event generated in \texttt{Pythia 8.244}};
\end{tikzpicture}
\label{fig:apollonius_subtraction}
}
\caption{
    (a) A cartoon of a generic implementation of the \PIRANHA{} paradigm described by \Eq{simple_piranha}, based on a figure from \Reff{Komiske:2020qhg}.
    \(\mathcal{E} \in \Omega\) indicates a particular event \(\mathcal{E}\) in the space \(\Omega\) of events to be groomed, \(\mathcal{C}\) indicates a model energy distribution for contaminating radiation, and \(\rho\in \mathbb{R}\) parameterizes the strength of the grooming.
    \Eq{simple_piranha} uses these ingredients to produce the groomed event \(\mathcal{E}_g\).
    (b) An Apollonius diagram in the rapidity-azimuth plane, also from \Reff{Komiske:2020qhg}, used by the Apollonius Subtraction (P-AS) algorithm.
    P-AS, an implementation of the \PIRANHA{} paradigm described in \Sec{as}, models contaminating radiation as uniform in the rapidity-azimuth plane \(\mathcal C = \mathcal U\).
    The color intensity of each Apollonius region is proportional to the amount by which the corresponding particle \(i\) is groomed, and thus to the area \(A^{\rm Apoll.}_i\) of the region.
}
\label{fig:as}
\end{figure}


\subsection{Apollonius Subtraction (P-AS)}
\label{sec:as}

Apollonius Subtraction (P-AS) is a direct application of \Eq{simple_piranha}, in which we take contaminating radiation to be uniformly distributed in a region of the \(\eta\)-\(\phi\) plane bounded by a maximum pseudorapidity, \(|\eta| < \eta_{\rm max}\).
We denote the uniform event with energy density \(\rho\) by \(\rhoU\).
This has proven to be an effective model for pileup, the underlying event, and initial state radiation~\cite{Soyez:2018opl,Monk:2018clo,Sjostrand:1987su,Sjostrand:2014zea,Dasgupta:2007wa,Kirchgaesser:2020poq,Moraes:2007rq,CDF:2015txs,Larkoski:2021hee,Baron:2020xoi,Marzani:2017kqd}, and motivates our approach to jet grooming for the remainder of the paper.

More precisely, using \(\beta = 1\), \(R =1\), and replacing \(\rhoC\) with \(\rhoU\) in \Eq{simple_piranha} yields an optimal transport problem known as an \textit{Apollonius problem}.
The Apollonius problem is well studied in the optimal transport literature and can be solved with an \textit{Apollonius diagram}, or additively weighted Voronoi diagram, which assigns an \textit{Apollonius region} to each particle in the event.
The precise structure of the Apollonius diagram corresponding to an event is described in a physics context in Section 5.4 of \Reff{Komiske:2020qhg}, and in the context of optimal transport in \Reffs{hartmann2017geometrybased,hartmann2018semidiscrete,bourne2018semidiscrete}.\footnote{
\Eq{simple_piranha} with \(\rhoC = \rhoU\) and arbitrary positive \(\beta\) also describes a valid optimal transport problem and an associated \PIRANHA{} groomer.
The solution to the associated optimal transport problem is described by a \textit{generalized Laguerre diagram}~\cite{Komiske:2020qhg, bourne2018semidiscrete}.
We leave the study of \PIRANHA{} grooming motivated by generalized Laguerre diagrams for future work.
}

P-AS may then be phrased as a constituent-level area subtraction procedure that solves the Apollonius problem for a given event.
To solve the Apollonius problem associated with a particular Apollonius diagram, we subtract from the \(p_T\) of each particle an amount proportional to the area of its Apollonius region,
\(p_{T\,i}^{\rm AS} = p_{T\,i} - \rho A_i^{\rm Apoll.}\).
Letting \(\zcut = \rho \, A_{\rm tot} / p_{T\,{\rm tot}}\), in analogy to the \(\zcut\) of traditional groomers such as Soft Drop, we may equivalently write
\(p_{T\,i}^{\rm AS} = p_{T\,i} - \zcut\,p_{T\,{\rm tot}}\,A_i^{\rm Apoll.} / A_{\rm tot}\).
In the piranha analogy, the transport plans for an event are determined by its Apollonius diagram, and all of the piranhas within a given Apollonius region feed on the associated jet constituent.
A rigorous proof of the continuity of P-AS is given in \href{https://arxiv.org/pdf/1706.07403.pdf#page=15&zoom=100,0,200}{Lemma 3.3} of \Reff{hartmann2017geometrybased}, and a visual representation of P-AS, taken from \Reff{Komiske:2020qhg}, is depicted in \Fig{as}.

P-AS is closely related to constituent-level area subtraction techniques for pileup mitigation.
Indeed, P-AS was initially introduced in the context of pileup mitigation in \Reff{Komiske:2020qhg}, where it was shown that in certain limits, Apollonius Subtraction aligns with the discontinuous Voronoi Area Subtraction (VAS) procedure for pileup mitigation~\cite{Cacciari:2007fd, Cacciari:2008gn, Cacciari:2011ma}.
Thus, P-AS may be thought of as a continuous analog to existing techniques for constituent-level area subtraction.
P-AS also aligns in certain limits with continuum Constituent Subtraction (CS), a continuous but computationally expensive algorithm for the removal of pileup~\cite{Berta:2014eza, Komiske:2020qhg}.

Unfortunately, the computational cost of finding the Apollonius diagram for a given particle event is relatively high because we rely on directly solving the Apollonius problem using numerical ghosts~\cite{Komiske:2020qhg}:
we use a uniform grid of ``ghost'' particles in the \(\eta\)-\(\phi\) plane to obtain the Apollonius diagram associated with a particular event computationally.
Data collection and analysis for theoretical and experimental studies at colliders motivate the computationally efficient analogs of P-AS that we describe in the remainder of this section.\footnote{Another approach to developing analogs of P-AS that we do not pursue in this work is to use the formalism of \textit{linearized optimal transport}, explored mathematically in \Reffs{cai2022linearized,sarrazin2023linearized} and applied to particle collisions in \Reffs{Cai:2020vzx,Cai:2021hnn}, which preserves the strengths of the EMD and significantly reduces the computational costs associated with optimal transport problems.}
A comparison of the runtime for our different \PIRANHA{} algorithms is given in \Fig{runtimes} of \App{feedingfrenzy}.

\subsection{Iterated Voronoi Subtraction (P-IVS)}
\label{sec:ivs}
Iterated Voronoi Subtraction (P-IVS) is a \PIRANHA{} groomer similar to both P-AS and VAS that overcomes the computational inefficiency of P-AS.
P-IVS also uses the uniform event \(\rho\,\mathcal U\) as a model for additive contamination, but crucially uses Voronoi diagrams rather than Apollonius diagrams to subtract this contamination away.

Like P-AS, the amount of grooming performed by the P-IVS grooming procedure is encoded in a parameter \(\rho\).
We may describe P-IVS quite succinctly as the solution to the iterated series of optimal transport problems~\cite{Komiske:2020qhg}:
\begin{align}
    \mathcal{E}^{(n+1)}_{\rm IVS}
    =
    \underset{\mathcal{E}'}{\arg \min} ~ {\rm EMD}
    (\mathcal{E}^{(n)}, \mathcal{E}' + \rho^{(n)}\,\mathcal{U})
    \label{eqn:ivs1}
    ,
\end{align}
with
\begin{align}
    \rho^{(n)} =
    \min\left\{
    	\rho - \sum_{i=0}^{n-1}\rho^{(i)},
	~
	\min_i \frac{p^{(n)}_{T\,i}}{A^{(n)}_i}
    \right\}
    \label{eqn:ivs2}
    .
\end{align}
The final groomed event is given by the limit of the recursive procedure in \Eq{ivs1}, \(\mathcal{E}_{\rm IVS} = \lim_{n\to\infty} \mathcal{E}^{(n)}_{\rm IVS}\).
Here, \(\rho^{(0)} = 0\), \(p_{T\,i}^{(n)}\) is the transverse momentum of particle \(i\) after \(n\) applications of the recursive algorithm in \Eq{ivs1}, and \(A_i^{(n)}\) is the Voronoi region for the event after \(n\) applications of \Eq{ivs1}.
For example, \(p_{T\, i}^{(0)}\) and \(A_i^{(0)}\) describe the transverse momenta and Voronoi areas of the ungroomed event.

More precisely, we may describe our ungroomed event as a collection of points in the \(\eta\)-\(\phi\) plane describing the directions of outgoing momenta, each weighted by its transverse momentum.
We may then enumerate the steps of the IVS procedure as follows:
\begin{enumerate}
    \item
    P-IVS constructs a Voronoi diagram in the \(\eta\)-\(\phi\) plane for the particles in the event.
    We will label the stage of the P-IVS procedure by an integer \(n\), starting at \(n = 1\) and going up to at most the number of particles in the event.
    \label{enum:ivs_1}

    \item
    P-IVS modifies the \(p_T\) of every particle in the event, subtracting \(p_T\) proportional to the area of the particle's Voronoi region until a particle would be removed or the grooming is complete:
    \begin{align}
	\label{eqn:ivs_2}
	p_{T\,i}^{(n)} = p_{T\,i}^{(n-1)} - \rho^{(n)} A^{(n-1)}_i,
    \end{align}
    with \(\rho^{(n)}\) given by \Eq{ivs2}.
    \label{enum:ivs_2}

    \item
    If \(\rho - \sum_{i = 1}^n \rho^{(i)} > 0\), P-IVS increments \(n\) by 1 and continues from Step~\ref{enum:ivs_1} by drawing a new Voronoi diagram for the modified event with a particle removed.
    Otherwise, the grooming is complete.
    \label{enum:ivs_3}
\end{enumerate}

As with P-AS, the simple expression of P-IVS in terms of the EMD already showcases its continuity and connection to optimal transport.
One may worry that the presence of an infinitesimally soft particle in an event may dramatically change the Voronoi diagram for an event, and therefore dramatically change the result of grooming using Voronoi areas.
P-IVS overcomes this challenge by only using Voronoi areas until a particle is removed;
when a particle is removed from the event, P-IVS computes an updated set of Voronoi areas that do not rely on the removed particle, continuing recursively until it has removed the correct amount of energy.
A rigorous proof of the continuity of P-IVS is given in \href{https://arxiv.org/pdf/1706.07403.pdf#page=15&zoom=100,0,200}{Lemma 3.3} of \Reff{hartmann2017geometrybased}.

Unweighted Voronoi diagrams for the original event \(\mathcal{E}^{(0)}\) can be found efficiently, unlike the weighted Voronoi/Apollonius diagrams needed for P-AS.
Furthermore, the Voronoi diagrams for the subtracted events \(\mathcal{E}^{(n)}_{\rm IVS}\) used throughout the stages of the P-IVS algorithm do not need to be computed from scratch, and can be found in constant (amortized) time~\cite{Komiske:2020qhg}.
P-IVS thus retains the continuous grooming properties of P-AS while remaining amenable to numerical computation and data collection.

\subsection{Tree-Based PIRANHA{}: Recursive Subtraction (P-RS) with a Fraction (F)}
\label{sec:rsf}

Recursive Subtraction (P-RS) is an extension of the \PIRANHA{} paradigm into the space of tree-based grooming algorithms and draws on the strengths of both hard-cutoff and \PIRANHA{} grooming.
We use ``Recursive Subtraction'' to denote an algorithm that takes in a binary tree of emissions describing a jet and recursively subtracts from the momenta of its jet constituents.
The dependence of P-RS on binary trees is reminiscent of many hard-cutoff grooming strategies, which are computationally efficient and bear a close connection to the physics of perturbative QCD~\cite{Larkoski:2014wba,Collins:2011zzd}.
The subtractive nature of P-RS echoes the area subtraction techniques of P-AS and P-IVS, which have a close connection to optimal transport and enjoy the advantage of continuity.
P-RS combines these strengths and paves the way for \PIRANHA{} groomers that are experimentally useful and theoretically tractable.

P-RS acts subtractively on jets associated with a \textit{binary tree} or \textit{clustering history}:
a tree structure that emerges from recursively combining jet constituents, two at a time, until only the full jet remains.
In particular, P-RS is a \textit{de-clustering algorithm} on the jet tree:
it splits the full jet tree into its two sub-jets, then splits each sub-jet into two sub-jets as dictated by the clustering history, continuing until only final-state particles remain.
In the discussion that follows, we also refer to sub-jets in the binary tree as \textit{branches} or \textit{emissions}.

Following the procedures of traditional grooming algorithms such as Soft Drop~\cite{Larkoski:2014wba}, the P-RS grooming algorithms we consider in this paper are \textit{angular-ordered}:
they act on jets associated with a binary tree whose splittings are ordered by the angular separation of their sub-jets.
In particular, we consider jets are first clustered using the anti-\(k_t\) algorithm~\cite{Cacciari:2008gp} and then \textit{re-clustered} using the pair-wise angular-ordered Cambridge-Aachen (C/A) clustering algorithm~\cite{Dokshitzer:1997in}.\footnote{
Starting with the anti-\(k_t\) algorithm both mitigates fluctuations due to contamination in jet areas \cite{Cacciari:2008gp} and reduces the effects of clustering logarithms in analytic calculations \cite{Larkoski:2014wba}.
}
C/A recursively clusters the jet constituents that are closest in angle, so that as we de-cluster each branch of the C/A tree we recover narrower and narrower emissions.
Since the branching structure of an angular-ordered tree of emissions is comparable to the history of emissions within the parton model, a phenomenon known as coherent parton branching~\cite{Collins:2011zzd}, angular-ordered grooming procedures have a closer connection to the physics of perturbative QCD.

P-RS also takes in a total amount of transverse momenta to subtract from the clustered jet, analogous to the choice of \(\rho A_{\rm jet}\) in the P-AS and P-IVS algorithms.
At each splitting in the tree, P-RS assigns some amount of the grooming associated with a given branch to its two emissions.
When the de-clustering finally reaches the final-state particles of a jet, P-RS subtracts from their transverse momenta, much like P-AS and P-IVS, by an amount proportional to the assigned grooming.

\textbf{Recursive Subtraction with a Fraction \(\boldsymbol{f}_{\rm\bf soft}\)} (P-RSF\(_f\), or P-RSF) is a simple implementation of P-RS that does not depend on the kinematic information of each splitting.
P-RSF depends on two real parameters, each between 0 and 1:
\begin{itemize}
    \item
    \(\zcut\) is the fraction of transverse momentum the grooming will remove from the entire jet, \(p_{T,\,\text{P-RSF}} = (1 - \zcut) p_{T,\,\rm tot}\).
    We may write \(\zcut = \zcut^{(0)} = \rho A_{\rm jet} / p_{T,\,\rm jet}\), in analogy to the ``piranha energy density'' \(\rho\) of P-AS and P-IVS.
    We denote the total transverse momentum removed from the jet as \(\Delta^{(0)} = \zcut\,p_{T,\,\rm tot}\), and the transverse momentum removed from a branch \(i\) of the jet is \(\Delta^{(i)} = \zcut^{(i)}\,p_{T,\,\rm tot}\);

    \item
    \(f_{\rm soft}\) is the fraction of the grooming assigned to the softer sub-jet at each stage of the de-clustering, as detailed below.
    P-RSF prefers to groom softer radiation more and more as \(f_{\rm soft}\) increases.
\end{itemize}

In the steps of the algorithm below, we index the current branch of the jet on which P-RSF is acting by \(i\).
The softer sub-jet of branch \(i\) is denoted ``\(i,\,\text{soft}\)'', and the harder branch is denoted ``\(i,\,\text{hard}\)''.
If the algorithm is at its starting point and considering the entire jet, we write \(i = 0\).

The concrete prescription of P-RSF is as follows:
\begin{enumerate}
    \item
    P-RSF attempts to de-cluster sub-jet \(i\) into two sub-jets.
    If this is impossible, the branch corresponds to a final-state particle and P-RSF proceeds to Step 5.
    \label{item:rsf_initial}

    \item
    P-RSF divides the grooming assigned to branch \(i\) between the softer sub-jet (\(i,\,\text{soft}\)) and the harder sub-jet (\(i,\,\text{hard}\)):
    \begin{subequations}
    \begin{align}
    	\Delta^{(i,\,\text{soft})}
	&= 
	f_{\rm soft} \, \Delta^{(i)}
	\overset{\Delta}{=}
	\zcut^{(i,\,\text{soft})} p_{T,\, \rm tot}
	\\
        \Delta^{(i,\,\text{hard})}
	&=
	(1-f_{\rm soft}) \, \Delta^{(i)}
	\overset{\Delta}{=}
	\zcut^{(i,\,\text{hard})} p_{T,\, \rm tot}
        ,
    \end{align}
    \end{subequations}
    so that \(\Delta^{(i,\,\text{soft})} + \Delta^{(i,\,\text{hard})} = \Delta^{(i)}\).
    In words, P-RSF assigns a fraction \(f_{\rm soft}\) of the grooming to the softer sub-jet, and the remaining fraction \(1 - f_{\rm soft}\) to the harder sub-jet.
    Equivalently,
    \begin{subequations}
    \begin{align}
        \zcut^{(i,\,\text{soft})} &= f_{\rm soft} \zcut^{(i)}
        \\
        \zcut^{(i,\,\text{hard})} &= (1-f_{\rm soft}) \zcut^{(i)}
        ,
    \end{align}
    \end{subequations}
    so that \(\zcut^{(i,\,\text{soft})} + \zcut^{(i,\,\text{hard})} = \zcut^{(i)}\).
    \label{item:rsf_softhard}

    \item
    P-RSF checks if one of the sub-jets of the current branch is removed by the grooming:
    
    \begin{enumerate}
    \item
    If the grooming assigned to the softer sub-jet is greater than its actual transverse momentum, \(\Delta^{(i,\,\text{soft})} > p_T^{(\rm i,\,\text{soft})}\), P-RSF removes the softer sub-jet from the event.
    P-RSF then assigns the remaining grooming to the harder sub-jet, \(\Delta^{(i,\,\text{hard})} = \Delta^{(i)} - p_T^{(\rm i,\,\text{soft})}\).
    \label{item:remove_soft}

    \item
    Similarly, if the grooming assigned to the harder sub-jet is greater than its transverse momentum, \(\Delta^{(i,\,\text{hard})} > p_T^{(\rm i,\,\text{hard})}\), P-RSF removes the harder sub-jet from the event.
    It then assigns the remaining grooming to the softer sub-jet, \(\Delta^{(i,\,\text{soft})} = \Delta^{(i)} - p_{T}^{(i,\,\text{hard})}\).
    This is possible only if \(f_{\rm soft} < 1/2\).
    \label{item:remove_hard}
    \end{enumerate}

    \item
    P-RSF implements Step~\ref{item:rsf_initial} on the surviving sub-jets of \(i\) using the values of \(\Delta^{(i,\,\text{soft})}\) and \(\Delta^{(i,\,\text{hard})}\) calculated in the previous two steps.

    \item
    When P-RSF reaches a final-state particle (a branch that cannot be further divided), it subtracts the grooming from its transverse momentum, effectively adjusting its energy fraction.
    Explicitly,
    \begin{align}
        p^{(i)}_{T,~{\rm P-RSF}} = p^{(i)}_T - \Delta^{(i)}.
    \end{align}
    Here, \(p^{(i)}_{T,~{\rm P-RSF}}\) denotes the groomed \(p_T\) of the final-state particle associated with the current branch \(i\).
    This may also be described as a shift in the fraction of the total transverse momentum carried by the particle \(i\),
    \begin{align}
        z^{(i)}_{\text{P-RSF}} = (z^{(i)} - \zcut^{(i)})/(1-\zcut)
        ,
    \end{align}
    where the additional factor of \(1 - \zcut\) is due to normalization to the groomed transverse momentum.
    \label{item:rsf_final}
\end{enumerate}

\begin{figure}[t!]
      \centering
      \centerline{
      \subfloat[]{
      \begin{tikzpicture}
          \node at (0,0) {
              \includegraphics[width=.4\textwidth]{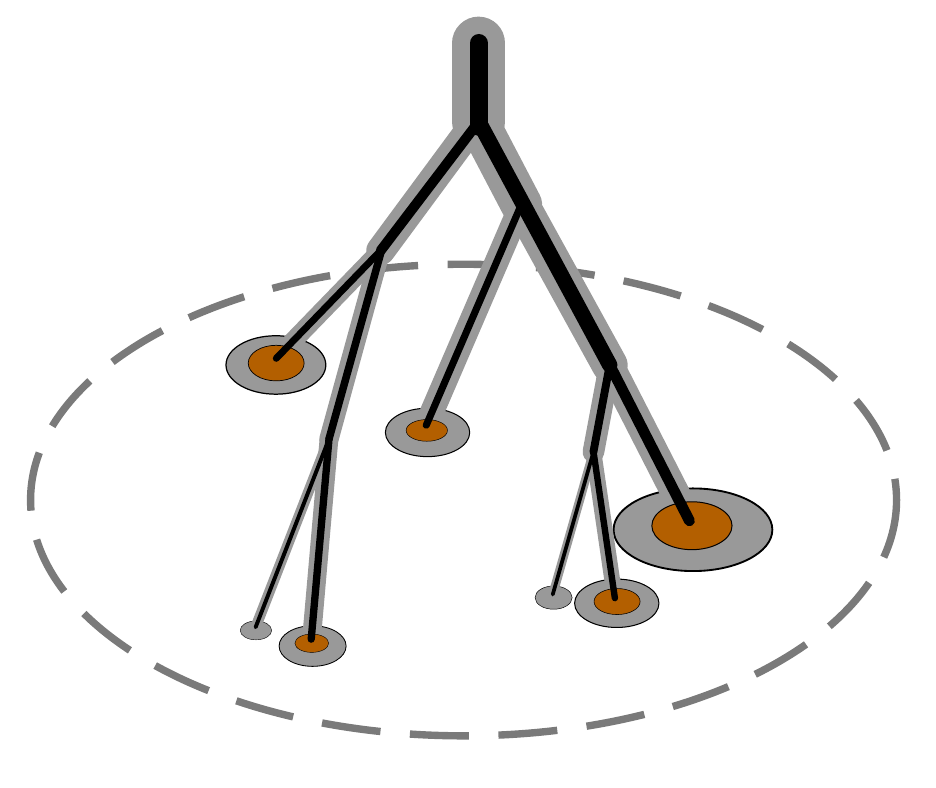}
          };
          \node at (0, 2.8) {\normalsize \PRSF{1/2} (Balanced P-RSF)};
      \end{tikzpicture}
      \label{fig:rsf_half_vis}
      }
      ~~~~
      \subfloat[]{
      \begin{tikzpicture}
          \node at (0,0) {
              \includegraphics[width=.4\textwidth]{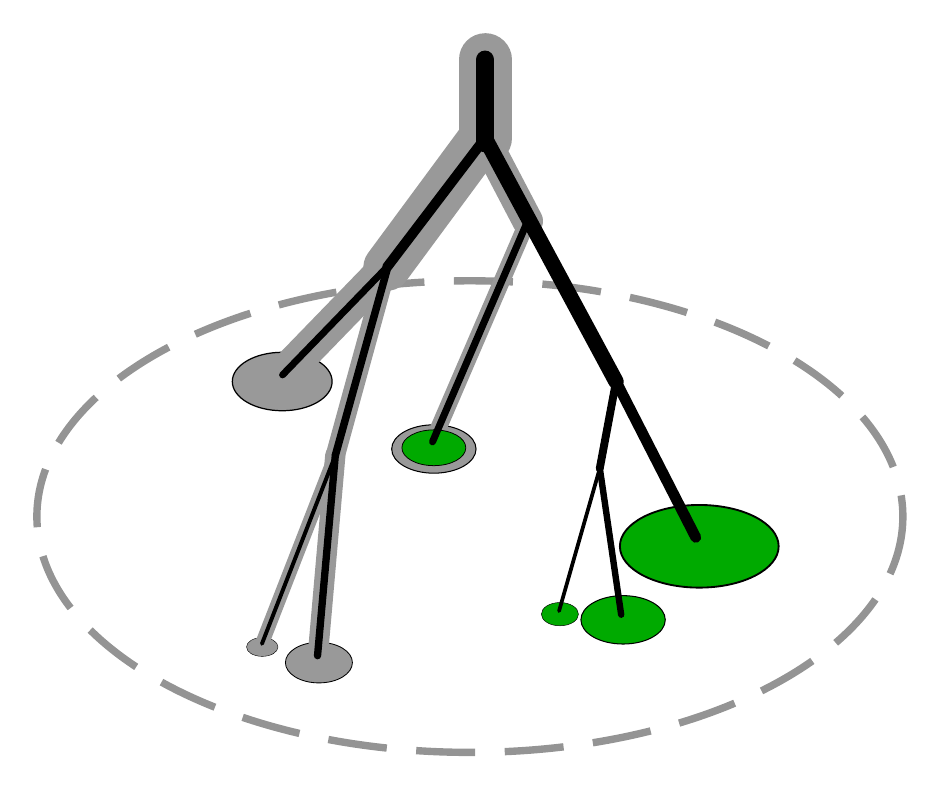}
          };
          \node at (0, 2.8) {\normalsize \PRSF{1} (example of Unbalanced P-RSF)};
      \end{tikzpicture}
      \label{fig:rsf_1_vis}
      }
      } 
\caption{
    Visualization of (a) \PRSF{1/2} and (b) \PRSF{1} acting on a toy jet depicted as an angular-ordered tree of emissions.
    The jet cone is indicated by the dashed ellipse, and the groomed final-state particles are indicated by colored ellipses, in dark orange (\PRSF{1/2}) and green (\PRSF{1});
    the size of each circle is proportional to the groomed \(p_T\) of the corresponding particle.
    The grey highlights behind each emission in the tree indicate the transverse momentum subtracted from each branch, and eventually from the final-state particles of that branch, as described in \Sec{rsf}.
    The size of the grey highlights behind each final-state particle indicates the amount of transverse momentum subtracted from that particle.
}
\label{fig:rsf_tree}
\end{figure}

P-RSF builds on the strengths of the P-RS framework, providing a simple implementation that does not rely on detailed kinematic information.
By combining the advantages of tree-based hard-cutoff and continuous \PIRANHA{} grooming paradigms, P-RS and P-RSF offer a promising path for experimentally useful and theoretically tractable continuous groomers.
A visualization of P-RSF acting on a tree of emissions for \(f_{\rm soft} = 1/2\) (which we also call \textbf{Balanced P-RSF}) and \(f_{\rm soft} = 1\) (an example of P-RSF\(_{f_{\rm soft}\neq1/2}\) or \textbf{Unbalanced P-RSF}) is shown in \Fig{rsf_tree}.

\subsection{P-RSF is (Almost) Continuous}
P-RSF is well-defined on the space of energy distributions of particle events because it is invariant under exactly collinear splittings.
To see this, we first notice that exact collinear splittings are infinitely narrow, and always correspond to branchings at the final layer of an angular-ordered tree of emissions.
With this in mind, let us implement an exact collinear splitting, replacing a final-state particle \(i\) within a jet by two exactly collinear final-state particles, ``\(i,\,\text{soft}\)'' and ``\(i,\,\text{hard}\)''.
Following the presentation of \Sec{rsf}, the transverse momentum subtracted from the new final state particles adds up to the transverse momentum subtracted from \(i\), \(\Delta^{(i,\,\text{soft})} + \Delta^{(i,\,\text{hard})} = \Delta^{(i)}\).
Since \(i,\,\text{soft}\) and \(i,\,\text{hard}\) are exactly collinear, the energy distribution of \(i\) after grooming is equal to the sum of the energy distributions for \(i,\,\text{soft}\) and \(i,\,\text{hard}\) after grooming.
The result of the P-RSF grooming procedure is therefore robust against exactly collinear splittings for any value of \(f_{\rm soft}\).

While the subtractive algorithm of P-RSF cannot be expressed simply in terms of geometry, P-RSF echoes the features of P-AS and P-IVS that grant their continuity.
For example, Steps 3(a) and 3(b)
of the P-RSF algorithm ensure that P-RSF does not assign an area to any final-state particle that is larger than its transverse momentum.

However, P-RSF still suffers from discontinuities in suppressed regions of parameter space.
We show in \Sec{rsf_discont} that Unbalanced P-RSF, or P-RSF with \(f_{\rm soft} \neq 1/2\), suffers from soft discontinuities:
small changes to the energy of the jet have the potential to change which emissions of the jet are softer.
The balanced recursive subtraction procedure, \PRSF{1/2}, overcomes this weakness and is soft-continuous.
Furthermore, \Sec{ang_discont} discusses how recursive tree-based grooming algorithms suffer from discontinuities inherited from pairwise clustering, leading to angular discontinuities in both Balanced and Unbalanced P-RSF as well as traditional grooming algorithms such as Soft Drop.

\subsection{Stronger Notions of Continuity}
\label{sec:stronger_continuity}
As a final note for this section, we point out that there are analogs of continuity that provide stronger constraints.
One example of a stronger form of continuity is \textit{uniform continuity}:
\begin{definition}\label{def:eventuniformcontinuity}
A map \(M\) from energy flows to energy flows is \textit{uniformly continuous} if, for any \(\varepsilon > 0\), there exists a \(\delta > 0\) such that for all \(\mathcal{E}\) and \(\mathcal E'\),
\begin{equation*}\label{eqn:emduniformcontinuity}
    {\rm EMD}(\mathcal E,\mathcal E')<\delta
    \quad\implies\quad
    {\rm EMD}(M(\mathcal E), M(\mathcal E')) < \varepsilon.
\end{equation*}
Roughly, we might say that uniform continuity requires that when we pick \(\varepsilon\), \(M\) must be ``continuous with the same \(\delta\) for every event''.
\end{definition}

\noindent
An even stronger condition is that of \textit{H\"older continuity}:
\begin{definition}\label{def:eventholdercontinuity}
A map \(M\) from energy flows to energy flows is \textit{H\"older continuous} with exponent \(\alpha \in \mathbb R\) if, for all \(\mathcal{E}\) and \(\mathcal E'\),
\begin{equation*}\label{eqn:emdholdercontinuity}
    {\rm EMD}(M(\mathcal E), M(\mathcal E'))
    <
    K \cdot \big({\rm EMD}(\mathcal E, \mathcal E')\big)^\alpha,
\end{equation*}
with \(K\in \mathbb R\) a constant.
The special case \(\alpha = 1\) is called \textit{Lipschitz continuity}.
\end{definition}

Placing stronger constraints on grooming, such as uniform continuity and H\"older continuity, has the potential to further constrain the effects of soft contamination and fluctuations in additive radiation on groomed results.
For example, in which regions of parameter space are the hard-cutoff or \PIRANHA{} groomers presented in this paper uniformly or H\"older continuous?
Are other methods for continuous grooming, such as EMD-mode \PIRANHA{} introduced in \App{grooming_in_emd_mode} or \PIRANHA{} groomers designed using linearized optimal transport \cite{Cai:2020vzx,Cai:2021hnn,cai2022linearized,sarrazin2023linearized}, more amenable to strongly continuous generalizations?
Can this knowledge help us identify obstacles to achieving these stronger types of continuity and guide us in designing more powerful, robust observables and groomers?

Though we do not investigate grooming methods that utilize these more restrictive analogs of continuity in this work, the invention and understanding of such groomers offer an intriguing and promising avenue for future research.
Instead, the remainder of this work is devoted to introducing hard-cutoff grooming, examining the types of discontinuities that may emerge in jet grooming, and studying the sensitivity of the continuous groomers we introduced in this section to important examples of low-energy pollution that appear in the study of particle collision data.

\section{Discontinuities in Grooming}
\label{sec:traditionaldiscont}
In this section, we begin by reviewing the hard-cutoff Soft Drop grooming algorithm (\Sec{softdrop}).
We then compare the leading order (LO) soft continuity properties of Soft Drop and Recursive Subtraction both conceptually (\Sec{sd_discont}) and through a perturbative calculation (\Sec{sd_discont_lo}).

In \Secs{rsf_discont}{ang_discont}, we explore potential challenges to continuity that can arise in more intricate scenarios.
Specifically, we discuss suppressed soft discontinuities that emerge in Unbalanced Recursive Subtraction (\Sec{rsf_discont}) and angular discontinuities induced by angular-ordered jet clustering that emerge beyond LO in tree-based grooming algorithms (\Sec{ang_discont}).
We do not delve into the mathematical intricacies or perturbation theory of these examples, and they may be skipped on an initial reading.

\subsection{Review of Hard-Cutoff Grooming: Soft Drop}
\label{sec:softdrop}
A common and elegant approach to jet grooming is to provide a hard cutoff on the energy of the radiation within a jet;
this approach retains high-energy information, removes radiation that may be sensitive to soft distortions or due to additive contamination, and even facilitates perturbative calculations of jet substructure observables.
In this section, we review a commonly used hard-cutoff grooming algorithm:
the Soft Drop de-clustering algorithm \cite{Larkoski:2014wba}.
Readers familiar with Soft Drop may skip to \Sec{sd_discont}, where we discuss the soft discontinuities of the Soft Drop algorithm.

The Soft Drop de-clustering algorithm is tree-based, computationally efficient, and serves as a representative example of a hard-cutoff grooming algorithm.
Much like P-RSF, Soft Drop begins with an angular ordered tree of emissions and grooms at each branch of the tree.
For Soft Drop, however, the parameter \(\zcut\) provides a hard cutoff, and Soft Drop sometimes maps similar events into groomed events that are very distinct.
The resulting discontinuity in the grooming procedure leads to theoretical challenges and uncertainties, unpredictable and non-linear responses to non-perturbative physics, and difficulties in interpreting experimental data \cite{ATL-PHYS-PUB-2019-027,Aad:2019vyi,ATLAS:2020gwe}.

The Soft Drop algorithm depends on two parameters.
The first, \(\zcut\), is a hard cutoff that parameterizes the strength of the grooming.
The second, \(\beta_{\rm SD}\), is a parameter that controls the angular dependence of the grooming procedure.
Soft Drop preferentially removes soft radiation at wider and wider angles as \(\beta_{\rm SD}\) is increased.

Soft Drop acts on a jet with characteristic radius \(R_0\) that is first clustered using an algorithm such as the anti-\(k_t\) algorithm, and then re-clustered with the Cambridge-Aachen (C/A) algorithm.
As in \Sec{rsf}, this produces a binary tree of clustered particles within the jet, such that particles deeper in the tree tend to be closer in angle.
Soft Drop then implements the following algorithm:
\begin{enumerate}
    \item
    \label{item:declustering}
    The most recent stage of the C/A clustering is undone, breaking the jet into two sub-jets.

    \item
    The softer sub-jet is removed from the jet if it is not energetic enough to be considered relevant to the hard physics under study;
    in particular, let \(z_{\rm soft} = \min(p_{T,1}, p_{T,2})/(p_{T,1} + p_{T,2})\) denote the fraction of the \(p_T\) carried by the softer sub-jet, and let \(\theta\) denote the angular distance between the two sub-jets.
    If the two sub-jets pass the Soft Drop grooming criterion:
     \begin{align}
        z_{\rm soft} > \zcut \left(\frac{\theta}{R_0}\right)^{\beta_{\rm SD}}
        \label{eqn:softdropcriterion}
        ,
    \end{align}
    the grooming procedure is stopped.
    The two sub-jets are then re-merged and labeled as the groomed jet without further modification.

    \item
    If the inequality of \Eq{softdropcriterion} is not satisfied, the softer sub-jet is not energetic enough to be considered relevant to the hard physics of jet production.
    In this case, the softer of the two sub-jets is removed from the event and the procedure is repeated recursively by de-clustering the harder of the two sub-jets.
\end{enumerate}
In this work, we use Soft Drop in grooming mode:
if Soft Drop reaches a final-state particle that cannot be de-clustered as required by Step \ref{item:declustering}, we keep it as the groomed jet.
Tagging mode would instead remove it from the jet entirely and leave a groomed event with no particles \cite{Larkoski:2014wba}.

Grooming mode is a natural choice in the context of pileup mitigation because of the extreme behavior of tagging mode, which completely eliminates a single particle jet even when \(\zcut = 0\).
In particular, if we have an event in which we estimate there is \textit{no} contamination from pileup (see \Sec{pileup}), we want to avoid grooming the jet at all and correspondingly set \(\zcut = 0\).
For a tree-level jet, consisting of a single quark, grooming mode respects our wish to leave the jet untouched, while tagging mode would nonetheless remove it entirely.

Soft Drop de-clustering is a generalization of the modified Mass Drop Tagger (mMDT) algorithm \cite{Dasgupta:2013ihk}, and reproduces mMDT when \(\beta_{\rm SD} = 0\).
Soft Drop has seen success in a wide variety of phenomenological applications, such as the characterization of boosted objects \cite{Thaler:2008ju,Thaler:2011gf,Hook:2011cq,Gallicchio:2011xq,Soper:2012pb,Gallicchio:2012ez,CMS-PAS-JME-09-001,CMS-PAS-EXO-09-002,CMS:2013kfa,ATL-PHYS-PUB-2009-081,ATL-PHYS-PUB-2010-008,ATLAS:2019kwg,Cui:2010km,ATLAS-CONF-2011-053,Chatrchyan:2013rla,Larkoski:2013eya,Dasgupta:2012hg,Backovic:2013bga,ATLAS-CONF-2013-084,Komiske:2018vkc,Komiske:2016rsd,Metodiev:2018ftz,Krohn:2012fg,MERINO:2013tta,Bhattacherjee:2016bpy,Macaluso:2018tck,Egan:2017ojy,Kasieczka:2017nvn,Pearkes:2017hku,Butter:2017cot,Catani:1992ua,Dokshitzer:1998kz,Dasgupta:2001sh,Banfi:2004yd,Banfi:2005gj,Ellis:2009su,Banfi:2010pa,Walsh:2011fz,Chien:2012ur,Li:2012bw,Jouttenus:2013hs,Hatta:2013iba,Larkoski:2014tva,Procura:2018zpn,Aaboud:2017aca,Frye:2016aiz, Almeida:2008yp,Larkoski:2017iuy,Larkoski:2017cqq,Thaler:2010tr,Ellis:2009me,Abdesselam:2010pt,Katz:2010mr,Gallicchio:2010dq,Adams:2015hiv,Sirunyan:2017ezt,Moore:2018lsr,FerreiradeLima:2016gcz,Rubin:2010fc,Chatrchyan:2012sn,CMS-PAS-BTV-13-001,CMS-PAS-JME-13-006,Kribs:2009yh,Chen:2010wk,Hackstein:2010wk,Kim:2010uj,Almeida:2011aa,Pandolfi:2012ima,Vernieri:2014wfa,CMS-PAS-HIG-17-007,Procura:2014cba} and the extraction of parameters of the Standard Model, such as the top quark mass \cite{Hoang:2017kmk,ATLAS:2021urs,Negrini:2022gec} and the strong coupling constant \cite{Marzani:2019evv}, from particle collision data.

Soft Drop with \(\beta_{\rm SD} > 0\) is collinear safe (invariant under exactly collinear splittings), and is therefore well-defined on the space of energy flows.
However, when \(\beta_{\rm SD} \leq 0\), Soft Drop in grooming mode is not collinear safe.
The lack of collinear safety in grooming mode can be quickly derived when considering a collinear splitting with \(z < \zcut\) which fails the Soft Drop criterion.
A similar problem emerges if one makes appropriate definitions for the algorithm in the limit \(\theta \to 0\), and Soft Drop in grooming mode is also not collinear safe when \(\beta_{\rm SD} < 0\).
Therefore, Soft Drop in grooming mode with \(\beta_{\rm SD} \leq 0\) is not well-defined as a map on the space of energy distributions, and we do not consider it in the remainder of the paper.

\subsection{An Invitation to Soft Discontinuities: Soft Drop Versus P-RS}
\label{sec:sd_discont}

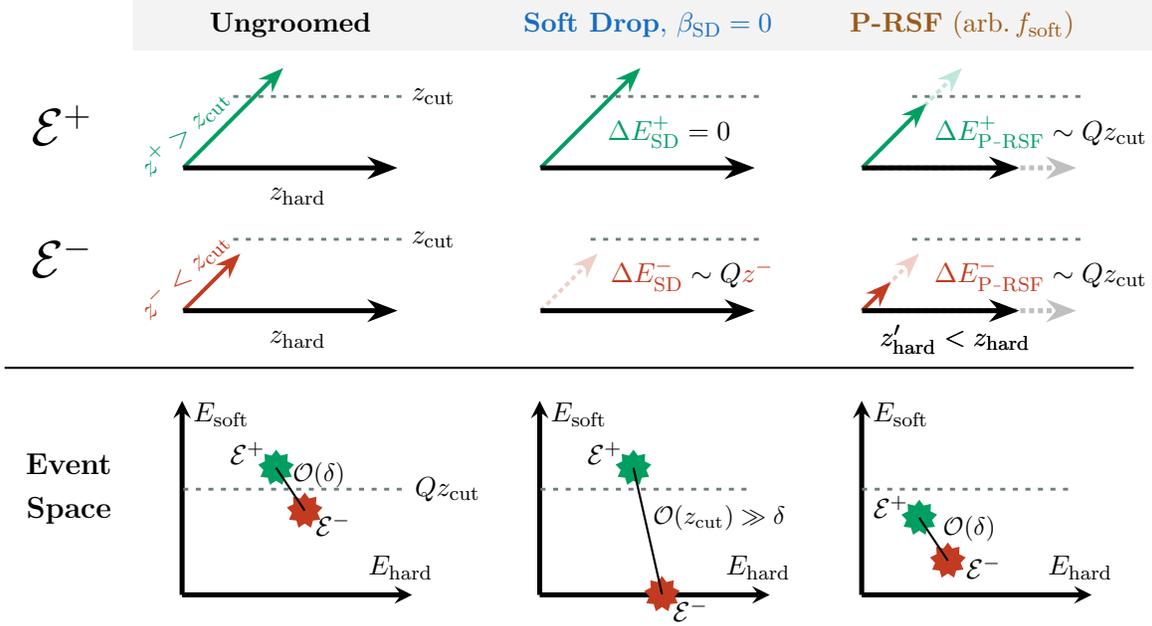
\begin{figure}[]

\centering
\scalebox{0.95}{
\begin{tikzpicture}[baseline=-3.5ex]

  \node at (-1.7,  .6)()
        {\huge \(\mathcal{E}^+\)};
  \node at (-1.7,  -1.25)()
        {\huge\(\mathcal{E}^-\)};

\draw [fill=black, opacity=0.05]
       (-0.7,2.4) -- (13.6,2.4) -- (13.6,1.65) -- (-0.7,1.65) -- cycle;

  \node at (1.5,  2)() {\textbf{Ungroomed}};

  \coordinate (A)  at (1.4, 1.4);
  \coordinate (O)  at (0, 0);
  \coordinate (B)  at (3, 0);
  \draw[-Stealth,line width=.55mm, color=Epluscolor] (O) -> (A);
  \draw[-Stealth,line width=.7mm] (O) -> (B);

  \coordinate (A)  at (.8, -1.2);
  \coordinate (O)  at (0, -2);
  \coordinate (B)  at (3, -2);
  \draw[-Stealth,line width=.55mm, color=Eminuscolor] (O) -> (A);
  \draw[-Stealth,line width=.7mm] (O) -> (B);

  \coordinate (I)  at (3.0, -1.7);
  \coordinate (F)  at (3.0, -2.3);

  \coordinate (I)  at (3.0, 0.3);
  \coordinate (F)  at (3.0, -0.3);

\coordinate (I)  at (0.7, 1.0);
\coordinate (F)  at (3.15, 1.0);
\draw[dash pattern=on 2pt off 3pt, line width=.4mm, color=aurometalsaurus] (I) -> (F);
\node at (3.5,  1.0)()
        {\(\zcut\)};

\coordinate (I)  at (0.7, -1.0);
\coordinate (F)  at (3.15, -1.0);
\draw[dash pattern=on 2pt off 3pt, line width=.4mm, color=aurometalsaurus] (I) -> (F);
\node at (3.5,  -1.0)()
        {\(\zcut\)};

\node [rotate=45] at (0.0,  0.5)()
        {\(\color{Epluscolor} z^+ > \zcut\)};
\node [rotate=45] at (0.0,  -1.5)()
        {\(\color{Eminuscolor}     z^- < \zcut\)};

\node at (1.6,  -0.4)()
        {\(z_{\rm hard}\)};
\node at (1.6,  -2.4)()
        {\(z_{\rm hard}\)};

\node at (6.5,  2)() {\textcolor{dodgerblue!75!black}{\textbf{Soft Drop}, \(\beta_{\rm SD} = 0\)}};

  \coordinate (A)  at (6.4, 1.4);
  \coordinate (O)  at (5.0, 0);
  \coordinate (B)  at (8.0, 0);
  \draw[-Stealth,line width=.55mm, color=Epluscolor] (O) -> (A);
  \draw[-Stealth,line width=.7mm] (O) -> (B);

  \coordinate (Ap)  at (5.8, -1.2);
  \coordinate (O)  at (5.0, -2);
  \coordinate (B)  at (8.0, -2);
  \draw[-Stealth, dash pattern=on 2pt off 1.5pt, line width=.55mm, color=Eminuscolor!25!white] (O) -> (Ap);
  \draw[-Stealth,line width=.7mm] (O) -> (B);

  \coordinate (I)  at (8.0, -1.7);
  \coordinate (F)  at (8.0, -2.3);

  \coordinate (I)  at (8.0, 0.3);
  \coordinate (F)  at (8.0, -0.3);

\coordinate (I)  at (5.7, 1.0);
\coordinate (F)  at (8.15, 1.0);
\draw[dash pattern=on 2pt off 3pt, line width=.4mm, color=aurometalsaurus] (I) -> (F);

\coordinate (I)  at (5.7, -1.0);
\coordinate (F)  at (8.15, -1.0);
\draw[dash pattern=on 2pt off 3pt, line width=.4mm, color=aurometalsaurus] (I) -> (F);

\node at (6.8,  0.5)()
        {\( {\color{Epluscolor} \Delta E^+_{\rm SD}} = 0 \)};
\node at (7.12,  -1.5)()
        {\( {\color{Eminuscolor} \Delta E^-_{\rm SD}} \sim  Q  {\color{Eminuscolor} z^-} \)};

\node at (10.8,  -2.4)()
        {\(z_{\rm hard}' < z_{\rm hard}\)};

  \node at (10.9,  2)() {\textcolor{ochre!75!black}{\textbf{P-RSF} (arb.\,\(f_{\rm soft}\))}};

  \coordinate (A)  at (10.4, 0.9);
  \coordinate (Ap)  at (10.9, 1.4);
  \coordinate (O)  at (9.5, 0);
  \coordinate (B)  at (11.7, 0);
  \coordinate (Bp)  at (12.5, 0);
  \draw[-Stealth,line width=.55mm, dash pattern=on 2pt off 1.5pt, color=Epluscolor!25!white] (O) -> (Ap);
  \draw[-Stealth,line width=.55mm, color=Epluscolor] (O) -> (A);
  \draw[-Stealth,line width=.7mm, dash pattern=on 2pt off 1.5pt, color=black!25!white] (O) -> (Bp);
  \draw[-Stealth,line width=.7mm] (O) -> (B);

  \coordinate (A)  at (9.9, -1.6);
  \coordinate (Ap)  at (10.3, -1.2);
  \coordinate (O)  at (9.5, -2);
  \coordinate (B)  at (11.7, -2);
  \coordinate (Bp)  at (12.5, -2);
  \draw[-Stealth,line width=.55mm, dash pattern=on 2pt off 1.5pt, color=Eminuscolor!25!white] (O) -> (Ap);
  \draw[-Stealth,line width=.55mm, color=Eminuscolor] (O) -> (A);
  \draw[-Stealth,line width=.7mm, dash pattern=on 2pt off 1.5pt, color=black!25!white] (O) -> (Bp);
  \draw[-Stealth,line width=.7mm] (O) -> (B);

  \coordinate (I)  at (12.5, -1.7);
  \coordinate (F)  at (12.5, -2.3);

  \coordinate (I)  at (12.5, 0.3);
  \coordinate (F)  at (12.5, -0.3);

\coordinate (I)  at (10.2, 1.0);
\coordinate (F)  at (12.65, 1.0);
\draw[dash pattern=on 2pt off 3pt, line width=.4mm, color=aurometalsaurus] (I) -> (F);

\coordinate (I)  at (10.2, -1.0);
\coordinate (F)  at (12.65, -1.0);
\draw[dash pattern=on 2pt off 3pt, line width=.4mm, color=aurometalsaurus] (I) -> (F);

\node at (12.0,  0.5)()
        {\( {\color{Epluscolor} \Delta E^+_{\rm P\mhyphen RSF}} \sim Q \zcut \)};
\node at (12.0,  -1.5)()
        {\( {\color{Eminuscolor} \Delta E^-_{\rm P\mhyphen RSF}} \sim Q \zcut \)};

\node at (10.8,  -2.4)()
        {\(z_{\rm hard}' < z_{\rm hard}\)};

\coordinate (I)  at (-2.5, -2.8);
\coordinate (F)  at (13.3, -2.8);
\draw[line width=.3mm] (I) -> (F);

\node at (-1.6,  -4.15)()
        {\large \textbf{Event}};
\node at (-1.6,  -4.8)()
        {\large \textbf{Space}};

\begin{axis}
[at={(-.5, -220)}, width=4.8cm, height=4.3cm,
xlabel=\(E_{\rm hard}\), ylabel=\(E_{\rm soft}\),
axis line style={line width=2pt},
axis y line*=left,
axis x line*=bottom,
axis lines = middle,
x label style={at={(axis description cs:0.95,0.25)},anchor=north},
y label style={at={(axis description cs:0.17,0.82)},anchor=south},
ticks=none]
\end{axis}
\coordinate (I)  at (0.0, -4.5);
\coordinate (F)  at (3.15, -4.5);
\draw[dash pattern=on 2pt off 3pt, line width=.4mm, color=aurometalsaurus] (I) -> (F);
\node at (3.7,  -4.5)()
        {\(Q \zcut\)};
\node at (2.1,  -5)()
        {\(\mathcal{E}^-\)};
\node[star,star points=9,minimum width=.01cm,star point ratio=1.4,fill=Eminuscolor] at (1.7, -4.8) {};
\filldraw[fill=Epluscolor,draw=Epluscolor] (1.3, -4.2) circle(0.15);
\node at (0.9,  -4)()
        {\(\mathcal{E}^+\)};
\node[star,star points=9,minimum width=.01cm,star point ratio=1.4,fill=Epluscolor] at (1.3, -4.2) {};
\coordinate (I)  at (1.3, -4.2);
\coordinate (F)  at (1.7, -4.8);
\draw[line width=.3mm] (I) -> (F);
\node at (1.9,  -4.3)() {\small \(\mathcal{O}(\delta)\)};

\begin{axis}
[at={(155, -220)}, width=4.8cm, height=4.3cm,
xlabel=\(E_{\rm hard}\), ylabel=\(E_{\rm soft}\),
axis line style={line width=2pt},
axis y line*=left,
axis x line*=bottom,
axis lines = middle,
x label style={at={(axis description cs:0.95,0.25)},anchor=north},
y label style={at={(axis description cs:0.17,0.82)},anchor=south},
ticks=none]
\end{axis}
\coordinate (I)  at (5.0, -4.5);
\coordinate (F)  at (7.95, -4.5);
\draw[dash pattern=on 2pt off 3pt, line width=.4mm, color=aurometalsaurus] (I) -> (F);
\node at (7.1, -6.2)()
        {\(\mathcal{E}^-\)};
\node[star,star points=9,minimum width=.01cm,star point ratio=1.4,fill=Eminuscolor] at (6.7, -5.97) {};
\node at (5.9,  -4)()
        {\(\mathcal{E}^+\)};
\node[star,star points=9,minimum width=.01cm,star point ratio=1.4,fill=Epluscolor] at (6.3, -4.2) {};
\coordinate (I)  at (6.7, -5.97);
\coordinate (F)  at (6.3, -4.2);
\draw[line width=.3mm] (I) -> (F);
\node at (7.5,  -4.9)() {\small \(\mathcal{O}(\zcut)\gg\delta\)};

\begin{axis}
[at={(295, -220)}, width=4.8cm, height=4.3cm,
xlabel=\(E_{\rm hard}\), ylabel=\(E_{\rm soft}\),
axis line style={line width=2pt},
axis y line*=left,
axis x line*=bottom,
axis lines = middle,
x label style={at={(axis description cs:0.95,0.25)},anchor=north},
y label style={at={(axis description cs:0.17,0.82)},anchor=south},
ticks=none]
\end{axis}
\coordinate (I)  at (9.5, -4.5);
\coordinate (F)  at (12.45, -4.5);
\draw[dash pattern=on 2pt off 3pt, line width=.4mm, color=aurometalsaurus] (I) -> (F);
\node at (11.2, -5.6)()
        {\(\mathcal{E}^-\)};
\node[star,star points=9,minimum width=.01cm,star point ratio=1.4,fill=Eminuscolor] at (10.7, -5.5) {};
\node at (9.9,  -4.75)()
        {\(\mathcal{E}^+\)};
\node[star,star points=9,minimum width=.01cm,star point ratio=1.4,fill=Epluscolor] at (10.3, -4.9) {};
\coordinate (I)  at (10.7, -5.5);
\coordinate (F)  at (10.3, -4.9);
\draw[line width=.3mm] (I) -> (F);
\node at (11.0,  -5.05)() {\small \(\mathcal{O}(\delta)\)};
\end{tikzpicture}
}

\caption{
A cartoon comparing the discontinuous action of hard-cutoff grooming to the continuous action of \PIRANHA{} on the nearly identical two-particle events with energy \(Q\), \(\mathcal{E}^+\) and \(\mathcal{E}^-\).
We use Soft Drop (with \(\beta_{\rm SD} = 0\)) and P-RSF (with arbitrary \(f_{\rm soft}\)) with a grooming parameter \zcut{} as representative examples of the hard-cutoff and \PIRANHA{} paradigms, respectively.
The events \(\mathcal E^\pm\) differ only by the energy fraction of the softer particle,
\(
z^\pm = \zcut \pm \delta/2
\),
and are separated by an infinitesimal EMD,
EMD\(
(\mathcal{E}^+, \mathcal{E}^-)
\sim \delta \ll \zcut
\).
Despite their similarity, Soft Drop maps the events to distinct groomed results separated by a large EMD:
%
%
%
EMD\(
(\mathcal{E}_{\rm SD}^+, \mathcal{E}_{\rm SD}^-)
\sim \zcut \gg \delta
\).
P-RSF instead
maps the events continuously to infinitesimally similar groomed events:
EMD\(
(\mathcal{E}_{\rm P\mhyphen RSF}^+,
\mathcal{E}_{\rm P\mhyphen RSF}^-)
\sim \delta
\).
\(\Delta E^\pm_G\) indicates the amount of energy removed from the event \(\mathcal E^\pm\) by the groomer \(G\).
}
\label{fig:groomcartoon}
\end{figure}

Traditional groomers are discontinuous in the regions of parameter space near a hard cutoff.
For example, Soft Drop is discontinuous in the region of parameter space where
\(
    z_{\rm soft} = \zcut \theta^{\beta_{\rm SD}} / R_0^{\beta_{\rm SD}}
\)
due to the Soft Drop criterion of \Eq{softdropcriterion}.
In this subsection, we discuss the discontinuity of hard-cutoff grooming and the relative continuity of \PIRANHA{}.

The following discussion centers around the illustrative example shown in \Fig{groomcartoon}, which shows the simplest instance of discontinuous behavior near a hard cutoff as well as the resolution offered by \PIRANHA{}.
In this example, we consider two nearly identical events, \(\mathcal{E}^+\) and \(\mathcal{E}^-\), that each contain two particles.
The softer constituent of \(\mathcal E^+\) has an energy fraction \(z^+ =  \zcut + \delta\) slightly above the grooming parameter \zcut, while the softer constituent of \(\mathcal E^-\) has an energy fraction \(z^- =\zcut - \delta\) slightly below \zcut.
The energy flows of the two events are nearly indistinguishable, in the sense that \(
{\rm EMD}(\mathcal{E}^+, \mathcal{E}^-)\sim\delta
\),
and we choose \(\delta\) for consistency with Definition~\ref{def:eventcontinuity}.
We use Soft Drop with \(\beta_{\rm SD} = 0\), or mMDT, as a representative example of hard-cutoff grooming, and P-RSF with arbitrary \(f_{\rm soft}\) as a representative example of the \PIRANHA{} paradigm;
for each, we use the parameter \(\zcut\).

Soft Drop treats \(\mathcal E^+\) and \(\mathcal E^-\) very differently:
it does not modify \(\mathcal E^+\), but completely removes the softer particle of \(\mathcal E^-\).
The energy flow of \(\mathcal E^+\) is unchanged, while that of \(\mathcal E^-\) is changed dramatically, and the EMD between the groomed jets is relatively large:
EMD\(
(\mathcal{E}^+_{\rm SD},
\mathcal{E}^-_{\rm SD})
\sim \zcut
\gg \delta\).
Direct application of Definition~\ref{def:eventcontinuity} to our example shows that Soft Drop is discontinuous.\footnote{
We note that Soft Drop with \(\beta \neq 0\) is also soft discontinuous.
Indeed, the above arguments still hold for the case \(\beta \neq 0\), verbatim, when the angle between the particles of \(\mathcal E^\pm\) is fixed to \(\theta = R_0\).
As expressed in \Eq{softdropcriterion}, the Soft Drop criterion at this fixed angle is still \(z_{\rm soft} > \zcut\).
However, since Soft Drop with \(\beta \neq 0\) drops the softer particle if it does not satisfy \(z / \theta^{\beta} > \zcut/ R_0^\beta\), we may also say that Soft Drop with \(\beta \neq 0\) is ``\(z / \theta^{\beta}\)-discontinuous'', where \(z\)-discontinuity indicates the soft discontinuity for the case of \(\beta = 0\) that we discuss above.
}
Since the events \(\mathcal E^+\) and \(\mathcal E^-\) differ by small changes to the energy of jet constituents, we see that Soft Drop is soft discontinuous in the region of parameter space near the cutoff \zcut.

The procedure of P-RSF is quite different.
When grooming both events, \PRSF{1/2} grooms the energy of both emissions by the same amount, leading to no large discrepancy in the two groomed results.
Indeed, after the \PRSF{1/2} grooming procedure, these two events are still separated by an infinitesimal EMD:
EMD\((\mathcal{E}^+_{\rm RSF},
\mathcal{E}^-_{\rm RSF})\sim\delta\).
This simple example demonstrates how hard-cutoff grooming methods may discontinuously map two nearly identical events into vastly different groomed results, and showcases the soft continuous resolution of the \PRSF{1/2} grooming procedure and other \PIRANHA{} groomers.


\subsection{Soft Discontinuities in Perturbation Theory: Soft Drop Versus P-RS}
\label{sec:sd_discont_lo}

Next, we examine the manifestations of the soft discontinuous behavior of hard-cutoff grooming more quantitatively by comparing Soft Drop and P-RSF at leading order in perturbation theory (LO).
In particular, we examine the LO effects of discontinuity on the distributions of the two-prong generalized energy correlation functions (ECFs) of \Reff{Larkoski:2013eya}.
For jets that are central (\(y = 0\)) and narrow (\(R_0 \ll 1\)), we may write the ECFs in the form
\begin{equation}
    C_1^{(\varsigma)}
    \simeq
    \frac{1}{2}\sum_{i=1}^M\sum_{j=1}^M
    z_i z_j \left(\frac{\theta_{ij}}{R_0}\right)^\varsigma
    \label{eqn:ECFdefn_lo}
    ,
\end{equation}
up to non-singular corrections in powers of the jet radius, which we neglect \cite{Larkoski:2014wba}.
In \Eq{ECFdefn_lo}, \(z_i\) represents the fraction of the jet energy carried by particle \(i\), \(\theta_{ij}\) indicates the angle between particles \(i\) and \(j\), \(R_0\) indicates the jet radius, and the sum is over all particles in the groomed jet.

In our discussion below, we focus on the calculation of LO substructure because it demonstrates the effects of the discontinuity of Soft Drop and the continuity of \PIRANHA{} in a simple context.
We provide a more detailed description of our LO calculation of \PIRANHA{}-groomed substructure in \App{LO_RSF}, and extend our substructure analysis to leading logarithmic (LL) accuracy in \App{resumresults}.
The resummed analysis of \App{resumresults} does not change the qualitative conclusions of our discussion below.
However, it highlights subtleties in systematically improving our resummed substructure calculations due to the global, subtractive nature of \PIRANHA{} grooming.
In \App{groomedenergyfraction}, we also discuss the calculation of the groomed energy fraction \(z_g\) -- another common observable used in the study of jet grooming -- which exhibits similar subtleties.
Therefore, while our resummation does not change the qualitative conclusions of the following discussion, it indicates that the computation of resummed \PIRANHA{} observables at higher accuracy may require new calculational tools.

At LO, the emission of a single parton with energy fraction \(z\) and angle \(\theta\) is described approximately by the pseudo-probability distribution
\begin{align}
    \frac{\alpha_s}{\pi} p_i(z) \frac{1}{\theta}
    \approx
    \frac{2 C_{R_i} \alpha_s}{\pi}~\frac{1}{z}~\frac{1}{\theta},
    \label{eqn:dglap_approx_lo}
\end{align}
where \(p_i(z)\) is a Dokshitzer-Gribov-Lipatov-Altarelli-Parisi (DGLAP) splitting function \cite{Gribov:1972ri,Dokshitzer:1977sg,Altarelli:1977zs} describing the splitting of a mother parton \(i\), and \(C_{R_i}\) is the quadratic Casimir for the SU(3) color representation \(R_i\) of the mother parton.
\(C_F = (N_C^2 - 1)/(2 N_C) = 4/3\) for quarks and \(C_A = N_C = 3\) for gluons.
We are interested in particular in the \textit{reduced} splitting functions,
\begin{align}
    \overline{p}_i(z) = p_i(z) + p_i(1-z)
    ,
\end{align}
where \(z \in (0, 1/2)\).
Reduced splitting functions are relevant for our groomed calculations because they describe the energy distribution of the softer of the two partons after the splitting.

The probability distribution of \(C_1^{(\varsigma)}\) for Soft Drop was studied in \Reff{Larkoski:2014wba}, which found, for \(\beta_{\rm SD} > 0\), the LO result
\begin{align}
    \label{eqn:sd_lo}
    \rho_{i,\,\,\text{SD}}(C_1^{(\varsigma)})
    \approx
    \frac{2\alpha_s C_{R_i}}{\pi \varsigma} \frac{1}{C_1^{(\varsigma)}}
    \times
    \begin{cases}
        -\log C_1^{(\varsigma)} + B_i,
        &
        C_1^{(\varsigma)} > \zcut;
        \\
        -
        \frac{\beta_{\rm SD}}{\varsigma+\beta_{\rm SD}} \log C_1^{(\varsigma)}
        -
        \frac{\varsigma}{\varsigma+\beta_{\rm SD}}\log \zcut
        +
        B_i,
        &
        C_1^{(\varsigma)} < \zcut,
    \end{cases}
\end{align}
away from \(C_1^{(\varsigma)} = 0\), up to \(\mathcal{O}(\alpha_s^2)\) and terms that are power suppressed in \(C_1^{(\varsigma)}\), \zcut, or both.
\(B_i\) is a factor due to hard-collinear pieces of the splitting function that are not singular as \(z\) approaches 0:
$B_q = -3/4$ for quarks and $B_g = -11/12+n_f/(6 C_A)$ for gluons, where $n_f$ is the number of active quark flavors.
The case \(\beta_{\rm SD} < 0\) leads to an identical result up to another constraint on \(C_1^{(\varsigma)}\), which simply multiplies the expression above by \(\Theta(C_1^{(\varsigma)} > \zcut^{\varsigma/ |\beta_{\rm SD}|})\).

The piece-wise behavior of \Eq{sd_lo} and the associated kink in the Soft Drop \(C_1^{(\varsigma)}\) distribution are due to the discontinuous behavior of Soft Drop.
As noted by \Reff{Larkoski:2014wba}, when \(\beta_{\rm SD} \geq 0\) any two-parton configurations with \(C_1^{(\varsigma)} = z \, (\theta/R_0)^\varsigma \, > \, \zcut\) must have \(z \, (\theta/R_0)^{\beta_{\rm SD}} \, > \, \zcut\), and are therefore not affected by Soft Drop.
On the other hand, configurations with \(C_1^{(\varsigma)} < \zcut\) may be Soft Dropped, leading to the piece-wise change in groomed substructure in this region of phase space.

We discuss the calculation of the \(C_1^{(\varsigma)}\) distribution for P-RSF groomed jets in \App{LO_RSF}, where we find the LO result
\begin{equation}
    \rho_{i,\,\,\text{P-RSF}}(C_1^{(\varsigma)}; f)
    \approx
    \frac{2\alpha_s C_{R_i}}{\varsigma~\pi}
    \frac{1}{C_1^{(\varsigma)}}
    \left(
        -\log\left(C_1^{(\varsigma)} + f\,\zcut\right)
        + B_i
    \right),
    \label{eqn:prsf_LO}
\end{equation}
away from \(C_1^{(\varsigma)} = 0\), using the same factors of \(B_i\) as for Soft Drop, and up to terms that are power-suppressed in \(C_1^{(\varsigma)}\), \(\zcut\), or both.\footnote{
We include some additional power-suppressed terms and additional contributions when \(C_1^{(\varsigma)} = 0\) in the LO discussion of \App{LO_RSF}.
}

Notably, the P-RSF distribution for \(C_1^{(\varsigma)}\) is smooth and does not need to be defined in a piece-wise fashion.
Even at LO, the subtractive nature of P-RSF leads to a smooth interpolation between double-logarithmic behavior at large \(C_1^{(\varsigma)}\) and single-logarithmic behavior at small \(C_1^{(\varsigma)}\).\footnote{The terms ``double logarithmic'' and ``single logarithmic'' here refer to the behavior of the LO cumulative distribution function.
In particular, we note that when \(C_1^{(\varsigma)} \gg f\,\zcut\), the pseudo-probability distribution \(\rho \propto \log\left(C\right)/C\) demonstrates double-logarithmic behavior.
On the other hand, when \(0 < C_1^{(\varsigma)} \ll f\,\zcut\), the distribution \(\rho \propto \log\left(f\,\zcut\right)/C\) demonstrates single-logarithmic behavior.
}
Much as the kink in the Soft Drop distribution can be attributed to the sudden activation of the grooming procedure in certain regions of phase space, the smoothness of the P-RSF distribution reflects in part that the grooming procedure is always active.\footnote{
We say that the smoothness of P-RSF only reflects the global activity of the grooming \textit{in part} because we expect that \PIRANHA{} groomers that turn on gradually will also have smooth substructure distributions.}
While the piece-wise behavior of Soft Drop observable distributions is smoothed out by all-orders effects \cite{Benkendorfer:2021unv}, the smoothness of P-RSF observable distributions at LO is a manifestation of the subtractive behavior of P-RSF that ensures its continuity on two-parton jets.

We also note that the substructure distributions of other \PIRANHA{} grooming procedures can be calculated at LO with a similar method by upgrading \(f_{\rm soft}\) to a function of \(z\) and \(\theta\).
Further, we expect that such an \(f(z, \theta)\) may be well approximated by \(f(0,0)\), with sub-leading corrections proportional to \(\mathcal{O}(z, \theta, \zcut)\).

In our discussion above, we emphasized that the kinks in Soft Drop groomed substructure distributions are manifestations of the discontinuity of Soft Drop itself.
Generalizations of these arguments, however, must be verified carefully.
In particular, there is not generically a one-to-one correspondence between kinks in distributions and discontinuities in the associated observable.
As a toy one-dimensional example, consider a random variable \(X\) that is uniformly distributed in \((-1/2,\,1/2)\).
The discontinuous, ``zig-zagging'' function \(f(X) = X \pm 1/2\), choosing the upper sign for \(X < 0\) and the lower sign for \(X > 0\), is uniformly distributed.
Therefore, discontinuous functions on our toy ``phase space'' -- the domain of \(X\) -- do not necessarily have kinks in their distributions.
Conversely, there is a kink in the distribution of \(f(X) = \mp (1 - \sqrt{1 \pm 2X})/2\), where we take the upper sign when \(X < 0\) and the lower sign for \(X \geq 0\), even though \(f(x)\) is a continuous function of \(x\) and has a continuous first derivative.


\subsection{Suppressed Soft Discontinuities in Unbalanced P-RS}
\label{sec:rsf_discont}

Unbalanced P-RSF algorithms are also discontinuous in the suppressed region of parameter space where \(z = 1/2\).\footnote{
The probability density of seeing a softer sub-jet with energy fraction \(z\) scales as \(1/z\) at leading logarithmic accuracy in perturbative QCD.
Therefore, the region with \(z \sim 1/2\) at the boundary of the jet phase space is far less populated than the regions of phase space near a hard-cutoff with \(z \sim \zcut < 1/2\).
}
The discontinuous behavior emerges because Unbalanced P-RSF algorithms treat harder and softer sub-jets differently;
since an infinitesimally soft perturbation can lead to either sub-jet having more energy when \(z = 1/2\), an infinitesimal change in an ungroomed energy flow can lead to macroscopic differences in groomed results.
In the language of \Sec{pira_intro}, we say that Unbalanced P-RSF is soft discontinuous in the suppressed, measure zero region of parameter space where \(z = 1/2\), and that Balanced P-RSF is the only soft continuous P-RSF algorithm.

In the following discussion, we explore the soft discontinuity of Unbalanced P-RS in the context of a jet containing only two partons.
We find that the discontinuity at \(z = 1/2\) can lead to macroscopic changes to the groomed jet axis.
However, only the orientation of the jet is affected by the soft discontinuity in our two-parton example, and the soft discontinuity of Unbalanced P-RS is not reflected at LO by traditional jet substructure variables.
Nonetheless, important features of a jet, such as its thrust axis \cite{Brandt:1964sa,Farhi:1977sg}, are susceptible to discontinuities at LO.
Furthermore, even traditional jet substructure observables are impacted by the soft discontinuity of Unbalanced P-RS in three-parton jet configurations that emerge beyond LO.

\begin{figure}[t!]
      \centering
\scalebox{0.95}{
\begin{tikzpicture}[baseline=-3.5ex]

  \node at (-1.7,  .6)()
        {\textcolor{black!40!E1color}{\huge \(\mathcal{E}^{(\rm high)}\)}};
  \node at (-1.7,  -1.25)()
         {\textcolor{black!20!E2color}{\huge\(\mathcal{E}^{(\rm low)}\)}};

\draw [fill=black, opacity=0.05]
       (-0.7,2.6) -- (8.65,2.6) -- (8.65,1.85) -- (-0.7,1.85) -- cycle;

  \node at (1.5,  2.2)() {\textbf{Ungroomed}};

  \coordinate (O)  at (0.5, .5);
  \coordinate (A)  at (2.4, 1.3);
  \coordinate (B)  at (2.0, -0.1);
  \coordinate (M) at (2.4, 0.65);

  \draw[-Stealth,line width=.8mm, color=aurometalsaurus!40!E1color, dash pattern=on 5pt off 2pt] (O) -> (M);
  \draw[-Stealth,line width=.9mm] (O) -> (A);
  \draw[-Stealth,line width=.75mm] (O) -> (B);

    \node [rotate=25] at (1.1,  1.2)()
        {\(z_{\rm high} = \frac{1}{2} + \delta\)};

  \coordinate (O)  at (0.5, -1.5);
  \coordinate (A)  at (2.4, -2.3);
  \coordinate (B)  at (2.0, -0.9);
  \coordinate (M) at (2.4, -1.65);

  \draw[-Stealth,line width=.8mm, color=aurometalsaurus!40!E2color, dash pattern=on 5pt off 2pt] (O) -> (M);
  \draw[-Stealth,line width=.9mm] (O) -> (A);
  \draw[-Stealth,line width=.75mm] (O) -> (B);

    \node [rotate=-23] at (1.0,  -2.2)()
        {\(z_{\rm low} = \frac{1}{2} + \delta\)};

  \node at (6.25,  2.2)() {\textcolor{ochre!75!black}{\textbf{P-RSF} (arb.\,\(f_{\rm soft}\neq 1/2\))}};

  \coordinate (O)  at (5.0, .5);
  \coordinate (A)  at (6.9, 1.3);
  \coordinate (Ap)  at (6.6, 1.176);
  \coordinate (B)  at (6.5, -0.1);
  \coordinate (Bp)  at (5.8, 0.18);
  \coordinate (M) at (6.9, 0.95);

  \draw[-Stealth,line width=.7mm, color=aurometalsaurus!40!E1color, dash pattern=on 5pt off 2pt] (O) -> (M);
  \draw[-Stealth,line width=.6mm, dash pattern=on 2pt off 1.5pt, color=black!25!white] (O) -> (A);
  \draw[-Stealth,line width=.9mm] (O) -> (Ap);
  \draw[-Stealth,line width=.6mm, dash pattern=on 2pt off 1.5pt, color=black!25!white] (O) -> (B);
  \draw[-Stealth,line width=.6mm] (O) -> (Bp);

  \coordinate (O)  at (5.0, -1.5);
  \coordinate (A)  at (6.9, -2.3);
  \coordinate (Ap)  at (6.6, -2.176);
  \coordinate (B)  at (6.5, -0.9);
  \coordinate (Bp)  at (5.8, -1.18);
  \coordinate (M) at (6.9, -1.95);

  \draw[-Stealth,line width=.7mm, color=aurometalsaurus!40!E2color, dash pattern=on 5pt off 2pt] (O) -> (M);s
  \draw[-Stealth,line width=.6mm, dash pattern=on 2pt off 1.5pt, color=black!25!white] (O) -> (A);
  \draw[-Stealth,line width=.9mm] (O) -> (Ap);
  \draw[-Stealth,line width=.6mm, dash pattern=on 2pt off 1.5pt, color=black!25!white] (O) -> (B);
  \draw[-Stealth,line width=.6mm] (O) -> (Bp);

\coordinate (I)  at (-2.5, -2.9);
\coordinate (F)  at (8.8, -2.9);
\draw[line width=.3mm] (I) -> (F);

\node at (-1.575,  -3.7)()
        {\large \textbf{Jet}};
\node at (-1.6,  -4.35)()
        {\large \textbf{Axis}};

\coordinate (I)  at (-0.5, -4.25);
\coordinate (F)  at (2.8, -4.25);
\draw[-Stealth, line width=.45mm] (I) -> (F);

  \node[star,star points=9,minimum width=.01cm,star point ratio=1.4,fill=E2color] at (1.0, -4.25) {};
  \node[star,star points=9,minimum width=.01cm,star point ratio=1.4,fill=E1color] at (1.3, -4.25) {};

  \coordinate (I)  at (0.9, -3.9);
  \coordinate (F)  at (1.4, -3.9);
  \draw[Stealth-Stealth, line width=.3mm] (I) -> (F);
  \node at (1.15,  -3.5)()
        {\(\Delta\theta_0 / R \sim \mathcal O(\delta)\)};

\coordinate (I)  at (4.5, -4.25);
\coordinate (F)  at (7.8, -4.25);
\draw[-Stealth, line width=.45mm] (I) -> (F);

  \node[star,star points=9,minimum width=.01cm,star point ratio=1.4,fill=E2color] at (5.25, -4.25) {};
  \node[star,star points=9,minimum width=.01cm,star point ratio=1.4,fill=E1color] at (7.05, -4.25) {};

  \coordinate (I)  at (5.15, -3.9);
  \coordinate (F)  at (7.15, -3.9);
  \draw[Stealth-Stealth, line width=.3mm] (I) -> (F);
  \node at (6.15,  -3.5)()
        {\(
          \Delta\theta_g / R
          \sim
          2\,\zcut\,\left|f - \frac{1}{2}\right|
          + \mathcal O(\delta)
        \)};

\end{tikzpicture}
}
\caption{
A cartoon demonstrating the soft discontinuity of P-RSF for \(f_{\rm soft} \neq 1/2\) acting on the nearly identical events \(\mathcal E^{(\rm high, low)}\), each with two particles with nearly identical energy fractions, \(z \sim 1/2\), and with opening angle \(R\).
Red denotes the event \(\mathcal E^{(\rm high)}\), blue denotes \(\mathcal E^{(\rm low)}\), and a thick dotted arrow indicates the jet axis for each event.
The hardest particle of \(\mathcal E^{(\rm high)}\) is its upper particle, while the hardest particle of \(\mathcal E^{(\rm low)}\) is its lower particle.
Since P-RSF with \(f_{\rm soft} \neq 1/2\) preferentially subtracts radiation from the softer particle of each event, the events are mapped discontinuously to distinct groomed results.
The discontinuous behavior of the groomed jet axis is one manifestation of this soft discontinuous behavior:
while the ungroomed energy-weighted jet axes of \(\mathcal E^{(\rm high)}\) and \(\mathcal E^{(\rm low)}\) differ by an infinitesimal amount (\(\Delta\theta_0 \sim R \epsilon\)), the groomed jet axes are widely separated for any \(f_{\rm soft} \neq 1/2\).
}
\label{fig:rsf_discont}
\end{figure}
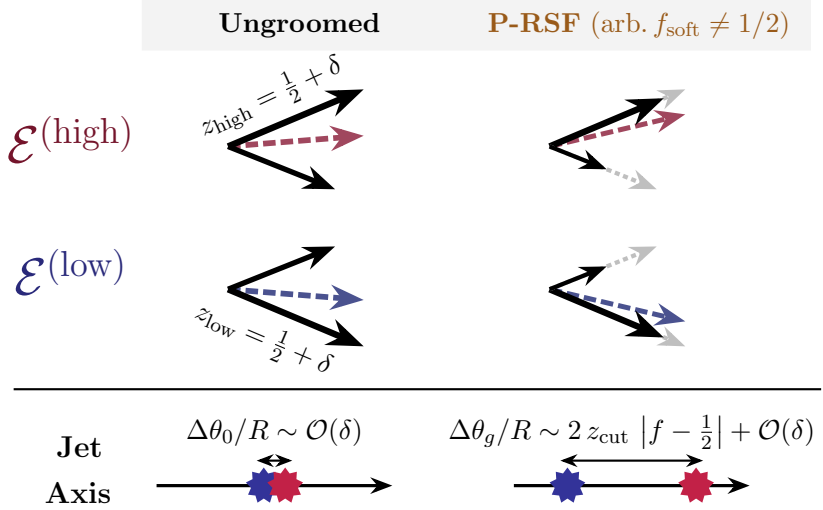

In the example of \Fig{rsf_discont}, we explore the effect of the soft discontinuity of Unbalanced P-RSF on the axis of a jet.
In particular, we examine nearly identical two-parton jets in the region of phase space near \(z \sim 1/2\) that differ only by an infinitesimally soft perturbation.
We denote these events by \(\mathcal{E}^{(i)}\), where \(i \in \{{\rm high, low}\}\).
The \(\mathcal E^{(i)}\) each consist of two particles: a ``high'' and ``low'' particle separated by an angular distance \(R\).
In \(\mathcal{E}^{(i)}\), particle \(i\) has just slightly more energy, \(z_i = 1/2 + \delta\).

For simplicity, we focus on the discontinuous behavior of the energy-weighted jet axis,
\begin{equation}
    \hat n_{\rm jet} \propto z_{\rm high} \hat n_{\rm high} +  z_{\rm low} \hat n_{\rm low}
    ,
\end{equation}
for the groomed events \(\mathcal E^{(i)}\), and the proportionality holds up to normalization.
Since the ungroomed \(\mathcal{E}^{(i)}\) are nearly identical, the angle between the ungroomed jet axes is infinitesimally small:
\begin{subequations}
\begin{align}
    \Delta\theta_{0}
   &=
    2\,R\,\delta
    +
    \mathcal{O}(R^3,\,\delta^3)
    ,
    \\
    \lim_{\delta\to 0}\Delta\theta_0 &= 0
    ,
\end{align}
\end{subequations}
where \(\Delta\theta_{0}\) indicates the angle between the jet axes of \(\mathcal{E}^{(\rm high)}\) and \(\mathcal{E}^{(\rm low)}\) before grooming.
After the application of P-RSF, however, the softer branch of each event will be dramatically diminished, and there is a large angle between the two groomed jet axes:
\begin{equation}
    \lim_{\delta\to 0}\Delta \theta_{g}
    =
    2\,R
    \,
    \frac{\zcut}{1 - \zcut}
    \,
    \left|f_{\rm soft} - 1/2\right|
    +
    \mathcal{O}(R^3)
    \neq 0
    \label{eqn:prsf_soft_discont}
    ,
\end{equation}
where \(\Delta\theta_{g}\) indicates the angle between the jet axes of \(\mathcal{E}^{(\rm high)}\) and \(\mathcal{E}^{(\rm low)}\) after grooming with P-RSF.
The continuous behavior indicated by \(\lim_{\delta\to 0} \Delta\theta_0 = 0\) before grooming is in sharp contrast to the discontinuous behavior after grooming, \(\lim_{\delta\to 0} \Delta\theta_g \neq 0\).
In \Eq{prsf_soft_discont}, we have assumed that \(\max(f,\,1-f)\,\zcut < 1/2 - \delta\), so that neither particle is fully removed from the event.
Of course, when \(f_{\rm soft} = 1/2\), the grooming algorithm treats the harder and softer sub-jets equally, so that we once again have \(\Delta \theta_g \to \mathcal{O}(\delta)\) and the soft discontinuous behavior fades away.

This simple example demonstrates a general principle for designing Recursive Subtraction algorithms:
if they are to be soft-continuous, they must treat the softer and harder sub-jets identically in the limit \(z \to 1/2\).
For example, let us imagine a P-RSF--type grooming algorithm with \(f_{\rm soft}\) upgraded to a function of the energy fraction \(z\) and angle \(\theta\) of a branch, \(f_{\rm soft} \to f(z,\theta)\).
Any P-RSF-type groomer with \(f(1/2,\, \theta) = 1/2\) overcomes the soft discontinuity discussed of the discussion above.
We call such P-RS groomers that treat emissions in the region \(z = 1/2\) identically ``Hard-Balanced'' Recursive Subtractors, as they are balanced in the region of phase space where two sub-jets are equally hard.
A P-RSF algorithm with \(f(z, \theta) = 1 - z\), for example, is Hard-Balanced, and still preferentially grooms infinitesimally soft radiation.
We leave the study of Hard-Balanced Recursive Subtractors to future work.

Finally, we note that while the soft discontinuity we detailed above is an important formal feature of P-RSF, the regions of phase space where \(z \sim 1/2\) contribute far less to observable distributions than those for which \(z \sim \zcut\).
In particular, the pseudo-probability distribution for \(z\) scales as \(1/z\) in perturbative QCD, implying that the soft discontinuity of Unbalanced P-RSF that we explore above is suppressed relative to the soft discontinuities of hard-cutoff groomers such as Soft Drop.
Indeed, the soft discontinuities of P-RSF do not manifest in leading logarithmic substructure distributions, and we may even use P-RSF with \(f_{\rm soft} \neq 1/2\) to gain perturbative insight into the soft continuous analog, \PRSF{1/2}, and into \PIRANHA{} grooming in general;
we explain this reasoning in more depth and pursue this goal in \App{calc}.

On the other hand, soft discontinuities associated with hard cutoffs in traditional grooming procedures lie on the interior of the jet phase space, straddled by the regions \(0 < z < \zcut\) and \(\zcut < z < 1/2\).
As we saw in \Sec{sd_discont_lo}, the associated soft discontinuities lead to piece-wise definitions of groomed observable distributions even at LO to isolate the behavior of radiation above and below the associated hard cutoff.

\subsection{Clustering Discontinuities Beyond Leading Order in Tree-Based Grooming}
\label{sec:ang_discont}

Recursive tree-based grooming algorithms also suffer from \textit{clustering discontinuities}, or discontinuities inherited from the procedure of clustering the jet into a tree of emissions (as discussed in \Sec{rsf}).
Clustering discontinuities are inevitable in tree-based grooming because any clustering algorithm relies on the discontinuous notion of which particles are ``closest'' to one another.
However, since there is only one way to cluster a two-parton jet, clustering discontinuities only emerge beyond LO and when a jet contains three or more partons.

Soft Drop and P-RSF in particular suffer from an \textit{angular clustering discontinuity}.
Both rely on an angular-ordered jet clustering history obtained from the angular-ordered C/A clustering algorithm;
the notion of the C/A clustering history is discontinuous because small changes to the angles of jet constituents may lead to distinct clustering histories.
In the language of \Sec{pira_intro}, we say angular-ordered clustering algorithms and angular-ordered, tree-based grooming algorithms are angularly discontinuous in a measure zero region of phase space.

\begin{figure}[t!]
      \centering
      \subfloat[]{
          \includegraphics[width=.4\textwidth]{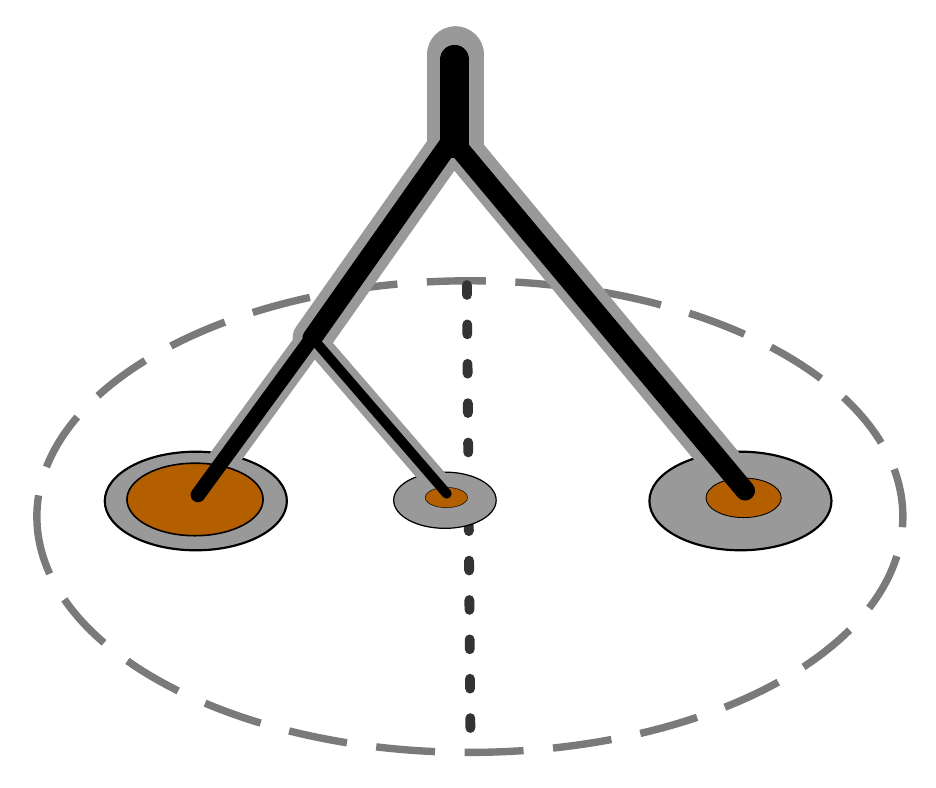}
          \label{fig:ang_discont_left}
      }
      ~~~~
      \subfloat[]{
          \includegraphics[width=.4\textwidth]{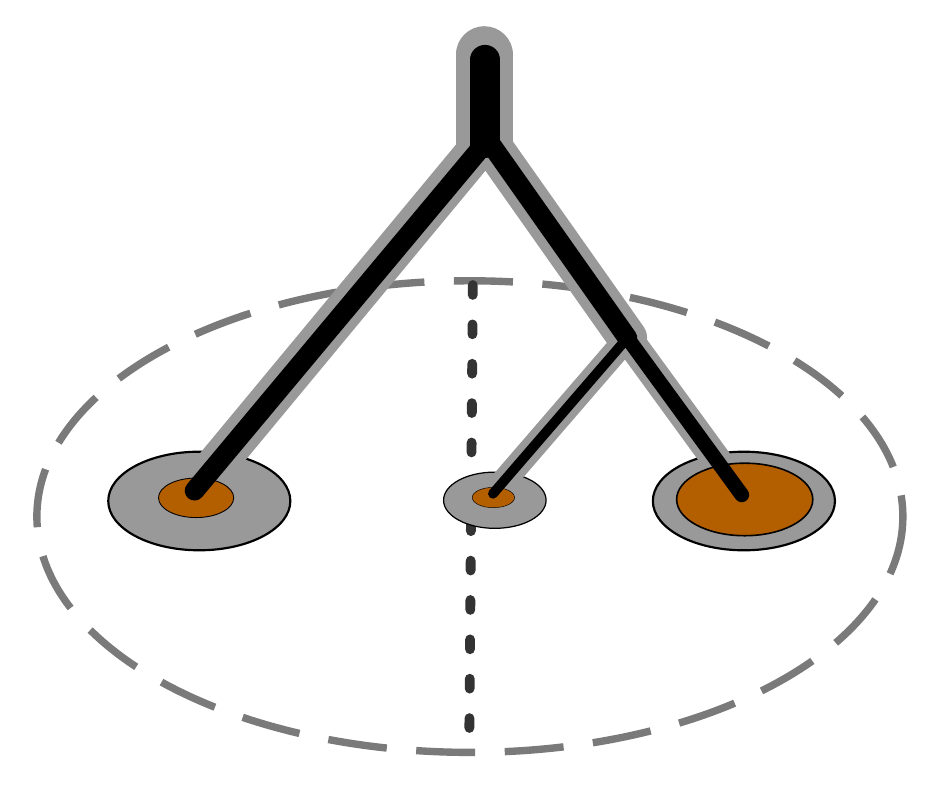}
          \label{fig:ang_discont_right}
      }
\caption{
    A visualization of angular discontinuity in tree-based grooming in the style of \Fig{rsf_tree}, with the angular-ordered \PRSF{1/2} algorithm as a representative example.
    The thick dashed line in the center of the jet is equidistant in angle from the left and right sub-jets of the event, and the middle emission lies nearly on this line.
    In (a), the middle emission lies closer to the left sub-jet, and the right sub-jet is therefore groomed more.
    In (b), the middle emission is closer to the right sub-jet, and the left sub-jet is groomed more.
    We see that a small change in the angle of the middle emission may drastically change the clustering history of the jet, and the result of a tree-based grooming procedure.
}
\label{fig:angdiscont}
\end{figure}

In our discussion, we focus on a simple manifestation of angular clustering discontinuities in the three-parton jet shown in \Fig{angdiscont}.
In particular, we consider the region of phase space where a particular sub-jet is nearly equidistant from a sub-jet on the left and another on the right.
We call the left and right sub-jets \(L\) and \(R\), and we denote the equidistant emission as \(M\), for ``middle''.

A small change in the location of emission \(M\) may lead to distinct C/A clustering histories.
If \(M\) is slightly closer to the left emission \(L\), then \(L\) will receive a quarter of the total grooming while \(R\) will receive half (as in \Fig{ang_discont_left});
the result is that \(L\) is groomed less than \(R\), and much more of the structure of the left side of the jet is preserved after the grooming.
Alternatively, if \(M\) is slightly closer to the right, then \(R\) will be groomed less than \(L\);
the right side of the jet will be preserved while much of the left side will be groomed away (as in \Fig{ang_discont_right}).
Despite the similarity of the ungroomed events, the resulting groomed events will have macroscopic differences and be separated by a relatively large EMD.

To emphasize this point quantitatively, let us consider P-RSF acting on the simple example where all three emissions lie in a plane, the left and right emissions each have an energy fraction \(z_\ell = z_r = z_0\), and the middle emission -- nearly equidistant from the left and right emissions -- has \(z_m = 1-2z_0\).\footnote{We focus on P-RSF in this example to avoid the more complex analysis required for Soft Drop that may obfuscate the physics at hand.
Soft Drop also exhibits an angular clustering discontinuity that can be seen in the context of this simple example, but that is complicated by the possible behaviors of the grooming:
\(\zcut\) and \(z_0\) may conspire such that zero, one, or two emissions are groomed from the event.
The effects of the clustering discontinuity on the energy-weighted jet axis must therefore be considered piece-wise in the \((\zcut,\,z_0)\) parameter space.
}
We use \(R\) to denote the angle between the left and right emissions, and \(\delta\) to denote a small perturbation in the angle of the middle emission:
\(\theta_{\ell m} = R/2 + \delta\) in one event, while \(\theta_{\ell m} = R/2 - \delta\) in the other.
We again examine the discontinuity in the energy-weighted jet axis, which takes the more general form
\begin{align}
    \hat{n}_{\rm jet} \propto \sum_i z_i \hat{n}_i
    ,
\end{align}
where the proportionality holds up to normalization.
A simple computation then yields
\begin{subequations}
\begin{align}
    \Delta\theta_0
    &=
    2\, z_m \, \delta
    +
    \mathcal{O}\left(R^2, \delta^3\right)
    ,
    \\
    \lim_{\delta\to 0}\Delta\theta_0 &= 0
    ,
\end{align}
\end{subequations}
for the angle \(\Delta\theta_0\) between the two jet axes before grooming, while
\begin{align}
    \lim_{\delta\to 0}\Delta\theta_g
    &=
    R\,\frac{\zcut}{1-\zcut}\,\left|f(f-3)+1\right|
    +
    \mathcal{O}(R^3)
    \neq 0
    ,
\end{align}
for the angle between the two jet axes after the application of P-RSF, where we have assumed that \(\zcut\) is sufficiently small that all particles survive the grooming.
We again see a discontinuous behavior in the energy-weighted axis of the groomed jets, \(\lim_{\delta\to 0}\Delta\theta_g \neq 0\), despite the fact that the ungroomed events were nearly identical and \(\lim_{\delta\to 0}\Delta\theta_0 = 0\).

Fortunately, the angular discontinuities associated with tree-based clustering algorithms are also suppressed in the phase space of perturbative QCD.
First, we note that the angular discontinuities in angular-ordered clustering algorithms emerge only at higher orders of perturbative accuracy for which a jet has three or more constituents.
These regions of phase space are generically suppressed in high-energy QCD calculations, where the emission of extra final-state partons is suppressed by the strong coupling constant \(\alpha_s\).
Furthermore, perturbative QCD predicts a strong hierarchy in the angles between different emissions in a jet, \(\theta_1 \gg \theta_2 \gg \cdots\) \cite{Collins:2011zzd}.
Therefore, even among phase space configurations with three or more constituents, it is very unlikely in perturbative QCD to find a configuration in which a particular final-state particle lies nearly equidistant in angle between two others.

We also emphasize that the angular discontinuity of the C/A clustering algorithm is representative of more general discontinuities in tree-based clustering and in jet algorithms.\footnote{
To mitigate the discontinuities of tree-based clustering, one can introduce a continuous weighting procedure for clustering histories.
One way to assign continuous weights to different clustering histories is the \textsc{Q-jet} scheme \cite{Ellis:2012sn,Ellis:2014eya}, which notably leads to reduced statistical fluctuations in observable distributions compared to the standard scheme of using a single clustering history for jet substructure calculations.
While the \textsc{Q-jet} scheme has not been applied directly to grooming in the context of energy flows, to our knowledge, it provides a particularly simple way to overcome the clustering discontinuities we discuss here:
we may simply define an overall groomed energy flow -- devoid of clustering discontinuities -- as the weighted average of the groomed energy flows associated with each possible clustering history.
}
For example, the \(k_t\) and anti-\(k_t\) algorithms exhibit a similar \(k_t\) discontinuity by identical arguments.
Discontinuities of different clustering algorithms may be more or less suppressed in the phase space of perturbative QCD, or more or less sensitive to particular sources of low-energy pollution.
Therefore, a more detailed discussion of clustering discontinuities and re-clustering schemes that minimize their effects is a potential direction for future research.

\section{Responses of Grooming to Soft Distortions}
\label{sec:soft_distortion}

We now compare the responses of \PIRANHA{} and traditionally groomed jets to soft distortions in more detail, focusing first on hadronization and then on the exclusion of neutral particles as a simplified probe of smearing effects.
In particular,
\begin{itemize}
    \item We use \PRSF{1/2} as a representative of \PIRANHA{} grooming procedures;
    \item We use Soft Drop with \(\beta_{\rm SD} = 2\) as a representative of traditional grooming procedures.
\end{itemize}
For details regarding other \PIRANHA{} grooming procedures and Soft Drop with different \(\beta_{\rm SD}\), see \App{feedingfrenzy}.

\subsection{Samples and Observables}
In the studies of this section, we use \texttt{Pythia 8.244} \cite{Sjostrand:2014zea} with the default 4C tune \cite{Corke:2010yf}, and work with samples of QCD quark and gluon jets without multiple parton interactions.
We consider jets with transverse momentum \(p_T > 500\) GeV, maximum pseudorapidity \(|\eta| < 4\), and clustered with the anti-\(k_T\) algorithm \cite{Cacciari:2008gp} with \(R = 1\);
we then re-cluster using the Cambridge-Aachen algorithm \cite{Dokshitzer:1997in}.
Our analyses focus on Soft Drop with \(\beta_{\rm SD} = 2\) as a representative for traditional grooming procedures and on Balanced Recursive Subtraction (\PRSF{1/2}) as a representative for \PIRANHA{} grooming;
the Balanced Recursive Subtraction algorithm is available on GitHub \cite{piranhagithub}.

We study the responses of groomed jets to soft distortion through three qualitatively different lenses:
\begin{itemize}
    \item \textbf{Energy Flow: EMD}
    \\
    To understand the overall response of groomed jet energy flow to soft radiation, we use the EMD of \Sec{emd}.
    The EMD between two jets bounds the difference of a large class of IRC-safe observables between the jets and provides an observable-independent tool to study the robustness of different grooming procedures to the presence of soft contamination.\footnote{
    In particular, the EMD of \Sec{emd} tends to reflect the response of extensive properties of jets, as we see concretely in our examples;
    a less concrete way to see this is by directly looking at \Eq{EMD_def_1}, and noticing that the EMD has units of energy.
    }

    \item \textbf{Extensive Properties: \(p_T\)}
    \\
    To understand how the extensive properties of groomed jets are modified by soft contamination, we examine the responses of the transverse momentum of groomed jets.
    We are concerned in particular with the amount by which transverse momentum in the presence of contamination, \(\widetilde{p_T}\), is shifted relative to the associated un-contaminated value, \(p_T\):
    \begin{align}
        \Delta p_T = \widetilde{p_T}- p_T
        .
        \label{eqn:pt_response}
    \end{align}
   \(p_T\) indicates transverse momentum in the absence of distortion or contamination and \(\widetilde{p_T}\) captures the response of transverse momentum to different sources of low-energy physics.
   This is specified more precisely in each of the subsequent sections.

    \item \textbf{Substructure: \(C_1^{(\varsigma)}\)}
    \\
    We use jet substructure to capture the response of the intensive properties of jets to contaminating radiation.
    In particular, we study the response of the two-prong energy correlation functions (ECFs) of \Reff{Larkoski:2013eya}:
    \begin{equation}
        C_1^{(\varsigma)}
        =
        \frac{1}{2}\frac{\sum_{i=1}^M\sum_{j=1}^M p_{T,\,i}\ p_{T,\,j}\ \left(R_{ij}/R_0\right)^\varsigma}{p_{T,\,{\rm tot.}}^2}
        \label{eqn:ECFdefn_pheno}
        .
    \end{equation}
    \(p_{T,\,i}\) represents the fraction of the jet transverse momentum carried by particle \(i\), \(R_{ij}\) indicates the rapidity-azimuth distance between particles \(i\) and \(j\), and the sum is over all particles in the jet.
    In the limit of jets that are central (\(y = 0\)) and narrow (have jet radius \(R_0 \ll 1\)), we can write the ECFs in the approximate form of \Eq{ECFdefn_lo}.
    As a concrete example of the response of substructure to soft radiation, we focus on \(\Delta C_1^{(2)} = \widetilde{C_1^{(2)}} - C_1^{(2)} \approx \Delta\left(m^2 / p_T^2\right)\), where a tilde again indicates the associated observable in the presence of a particular source of low-energy physics, while the absence of a tilde indicates the absence of low-energy effects.
\end{itemize}

We emphasize that in all of our phenomenological studies (except our studies of the underlying event in \Secs{ue}{massres}), we examine how each source of contamination leads to changes on a ``per-jet'' level, rather than at the level of distributions of observables.
In particular, we examine the EMD between each jet before and after contamination, and the changes in \(p_T\) and \(C_1^{(2)}\) induced by contamination for individual jets.
To the best of our knowledge, the effects of hadronization on a per-jet basis have not been studied in this way before, as the sensitivity of groomed jets to hadronization is often studied instead at a statistical level by studying the distributions of groomed jet observables.
Therefore, the results we present offer a more detailed examination of the effects of the hadronization model of \texttt{Pythia 8.244};
however, these results may not be precisely reflected solely by the statistical changes in distributions of groomed jet observables that are induced by hadronization.

\subsection{Hadronization}
\label{sec:hadronization}

We first provide phenomenological evidence that continuous grooming is less sensitive than hard-cutoff grooming to the physics of hadronization.
A deep understanding of the physics of traditional jets requires careful consideration of hadronization corrections, such as in jet masses \cite{Hoang:2019ceu, Marzani:2017kqd, Benkendorfer:2021unv} and hadronic event shapes \cite{Dokshitzer:1995zt, Baron:2020xoi}.
%
Hard-cutoff groomed jet observables in particular can undergo large corrections from the hadronization process due to the discontinuity of the hard-cutoff paradigm:
hadronization can transform partons with energy below a hard cutoff into hadrons with energy above the hard cutoff, and vice versa.
These subtleties lead to additional complications in the theoretical calculation of non-perturbative effects on traditionally groomed jet observables \cite{Hoang:2019ceu}.

\begin{figure}[p]
\centering
\subfloat[][]{
      \includegraphics[width=.32\textwidth]
      {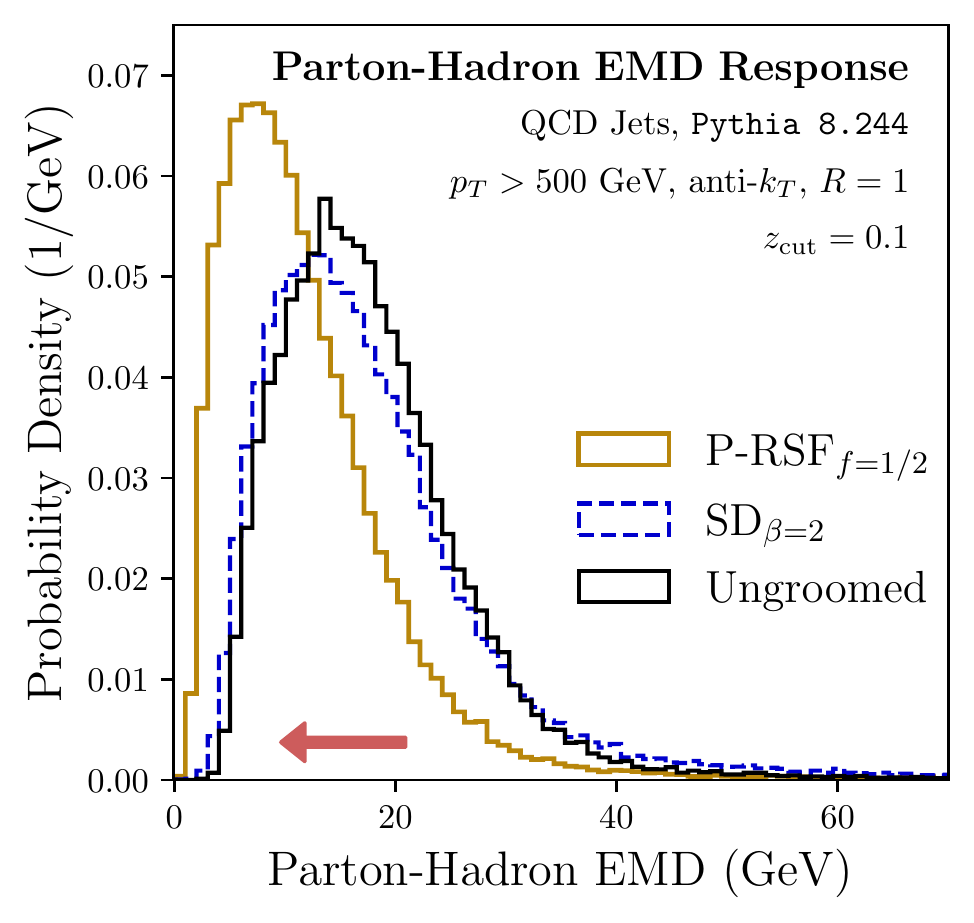}
      \label{fig:ph_emd_dist}
}
\subfloat[][]{
      \includegraphics[width=.32\textwidth]
      {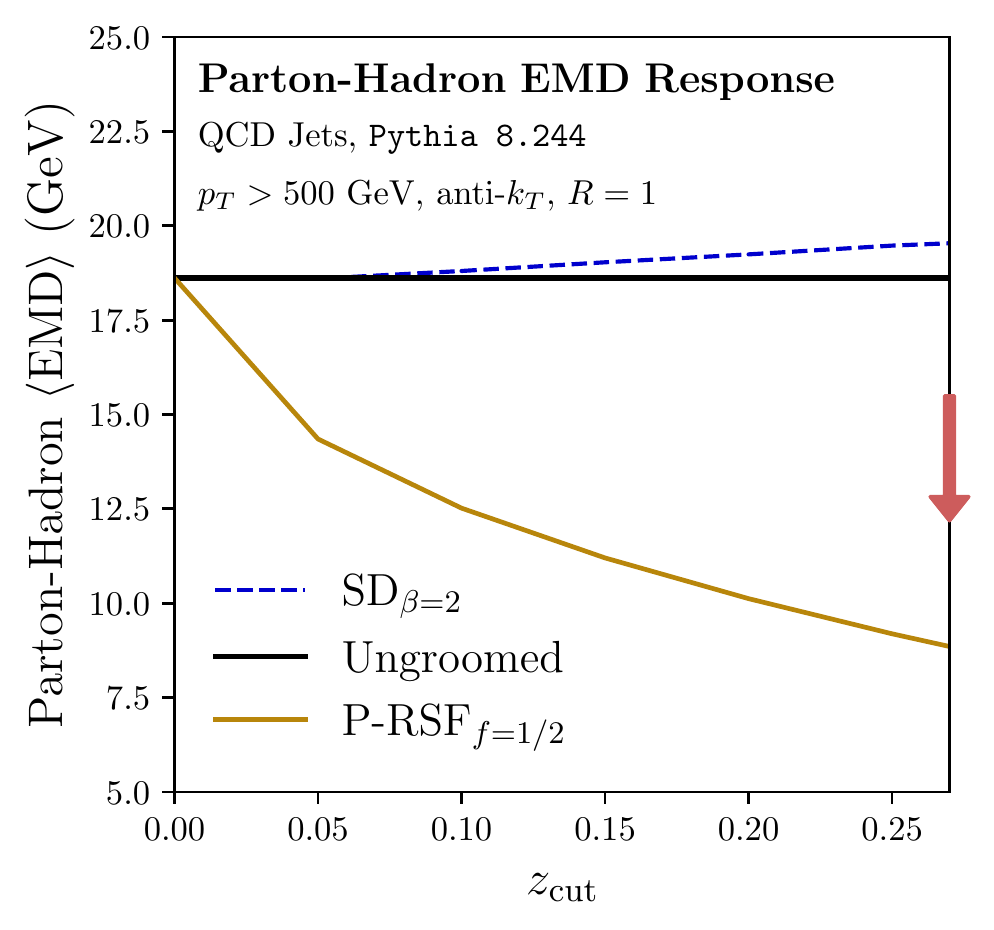}
      \label{fig:ph_emd_delta}
}
\subfloat[][]{
      \includegraphics[width=.32\textwidth]
      {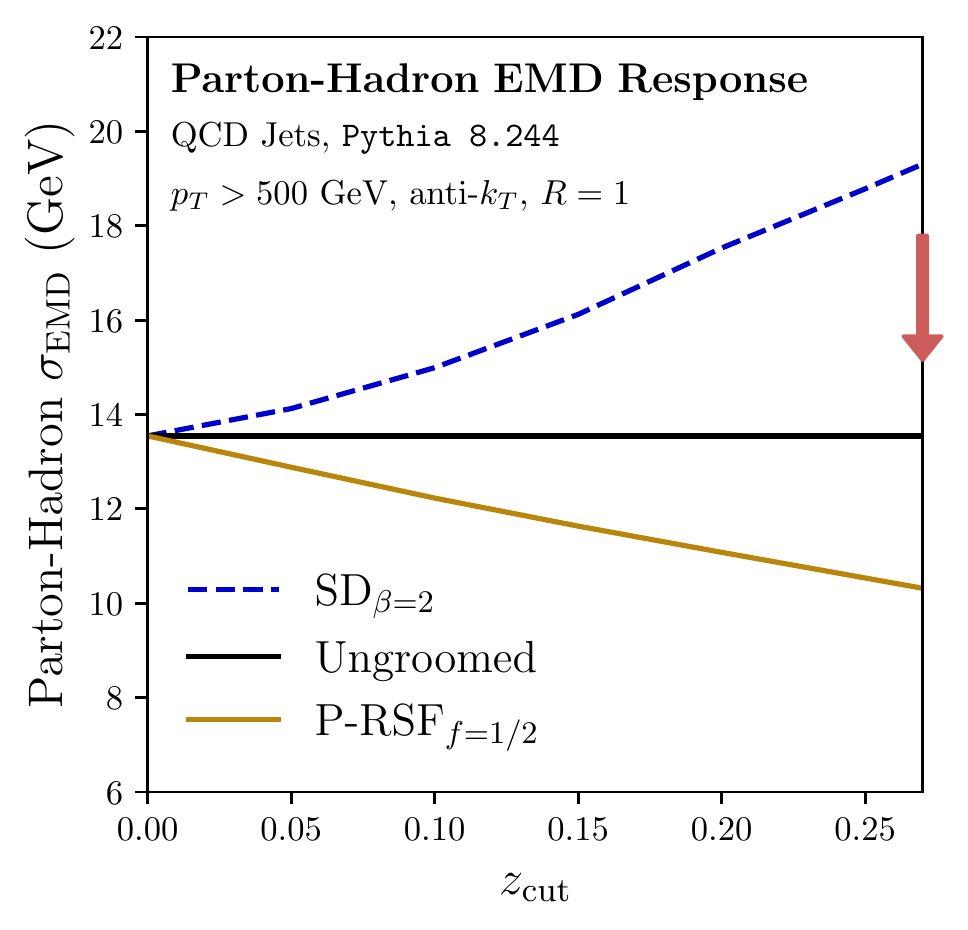}
      \label{fig:ph_emd_sigma}
}
\\
\subfloat[][]{
      \includegraphics[width=.32\textwidth]
      {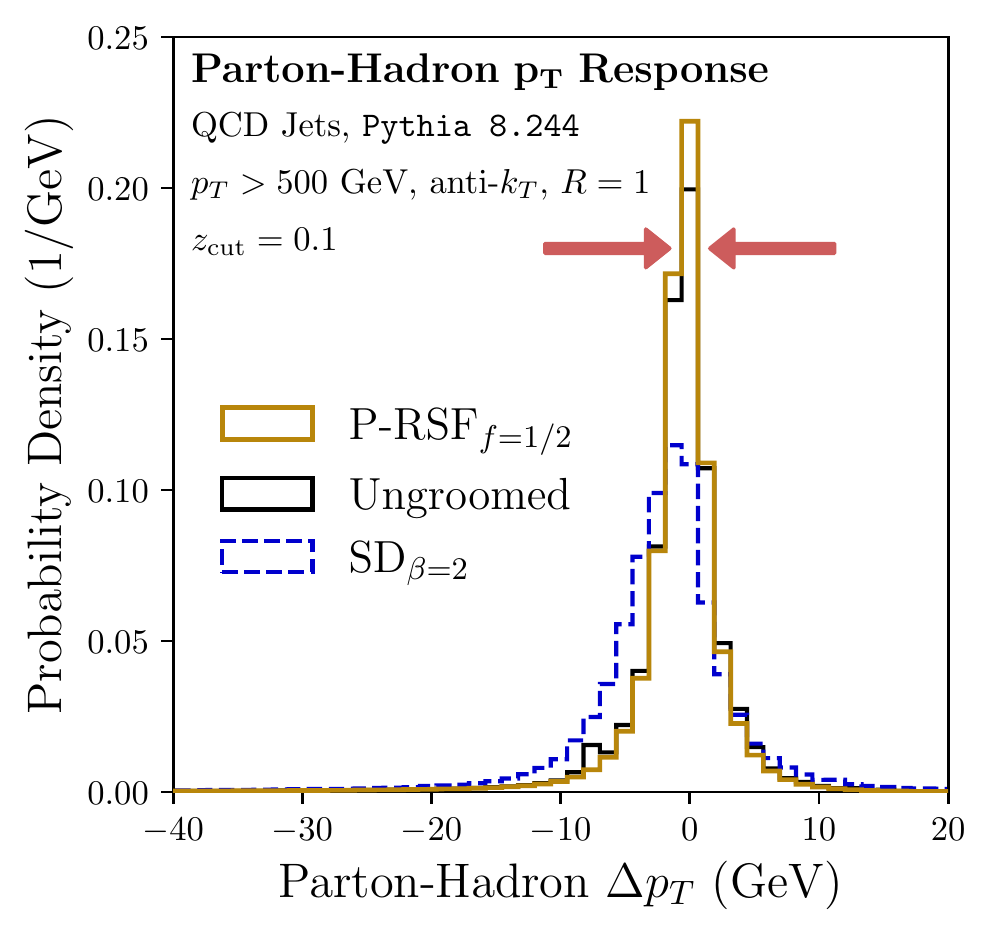}
      \label{fig:ph_pt_dist}
}
\subfloat[][]{
      \includegraphics[width=.32\textwidth]
      {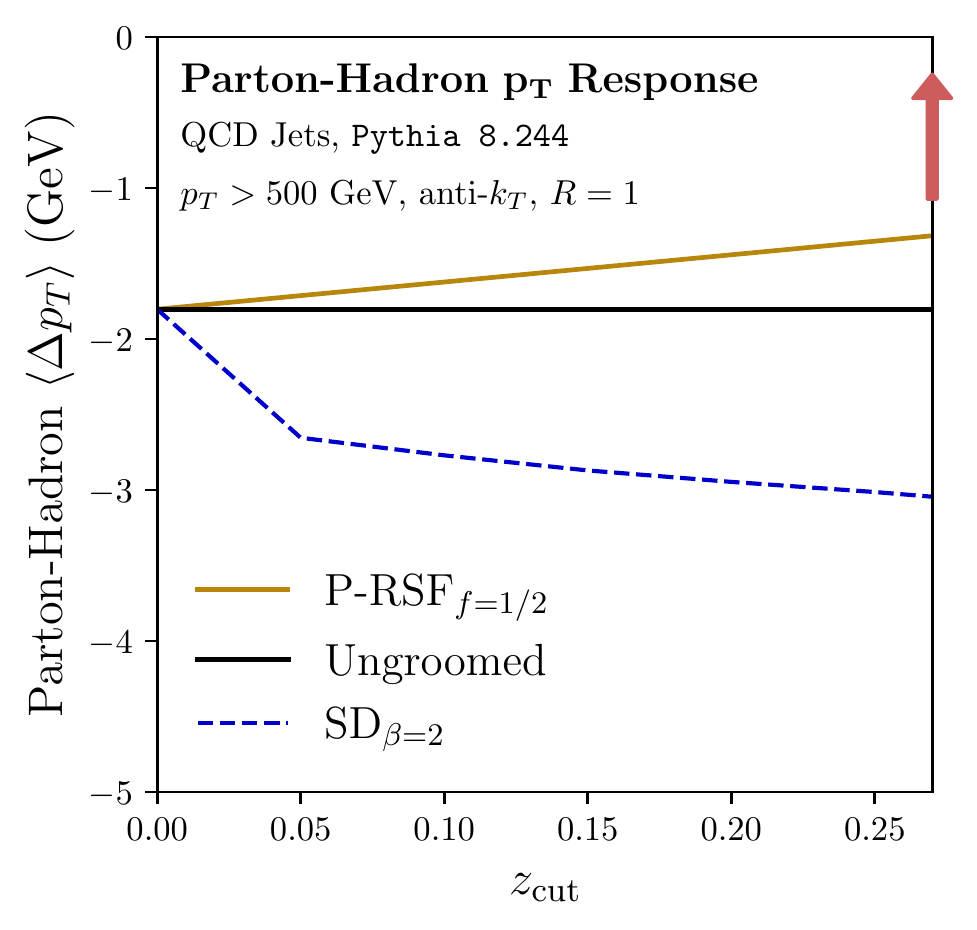}
      \label{fig:ph_pt_delta}
}
\subfloat[][]{
      \includegraphics[width=.32\textwidth]
      {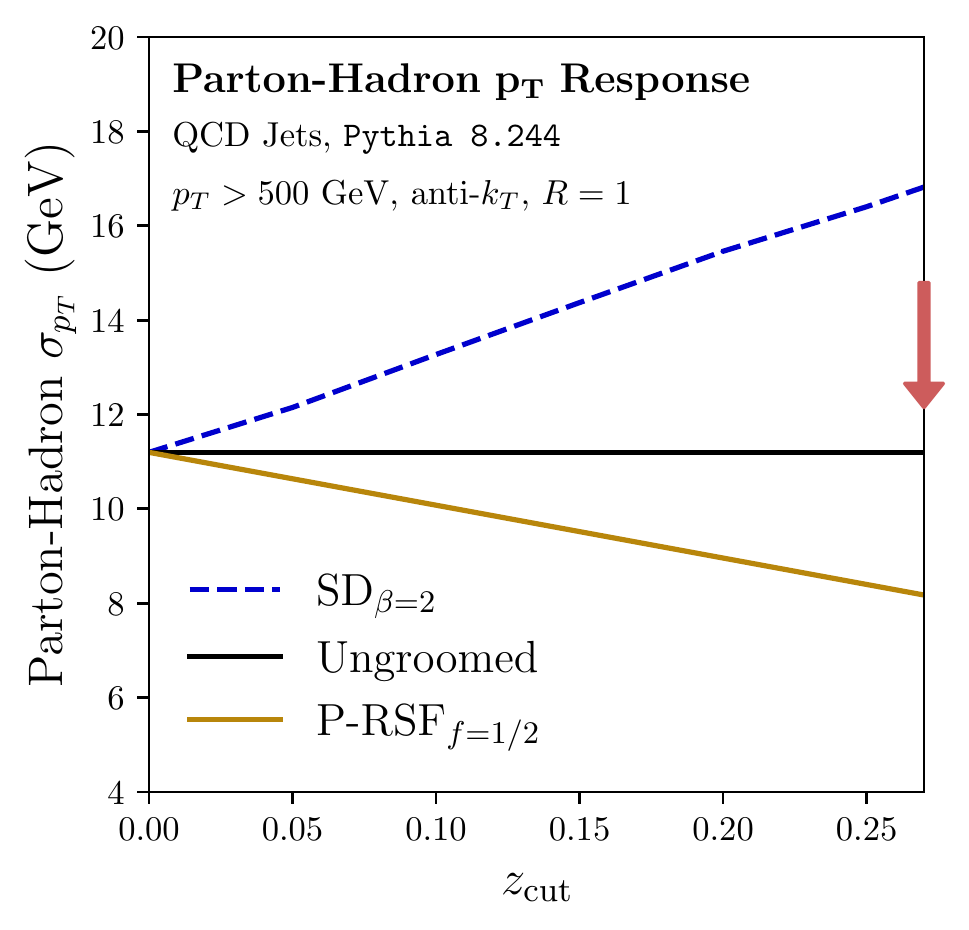}
      \label{fig:ph_pt_sigma}
}
\\
\subfloat[][]{
      \includegraphics[width=.32\textwidth]
      {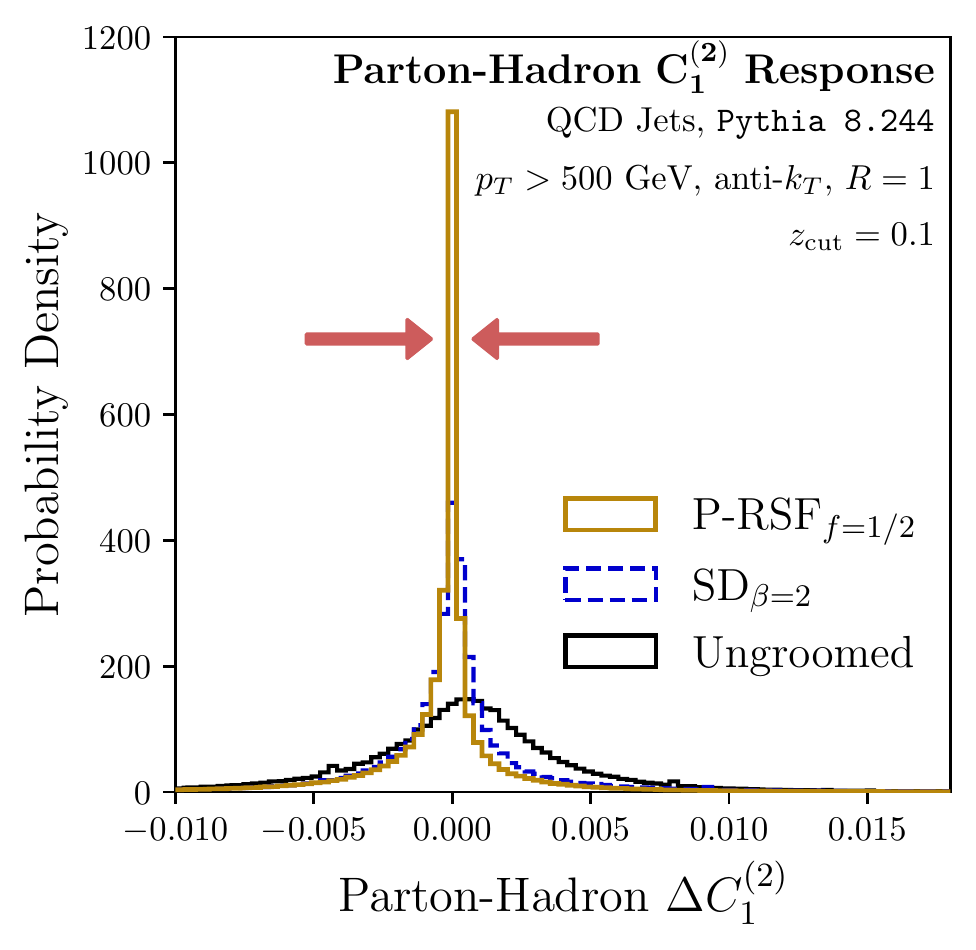}
      \label{fig:ph_c12_dist}
}
\subfloat[][]{
      \includegraphics[width=.32\textwidth]
      {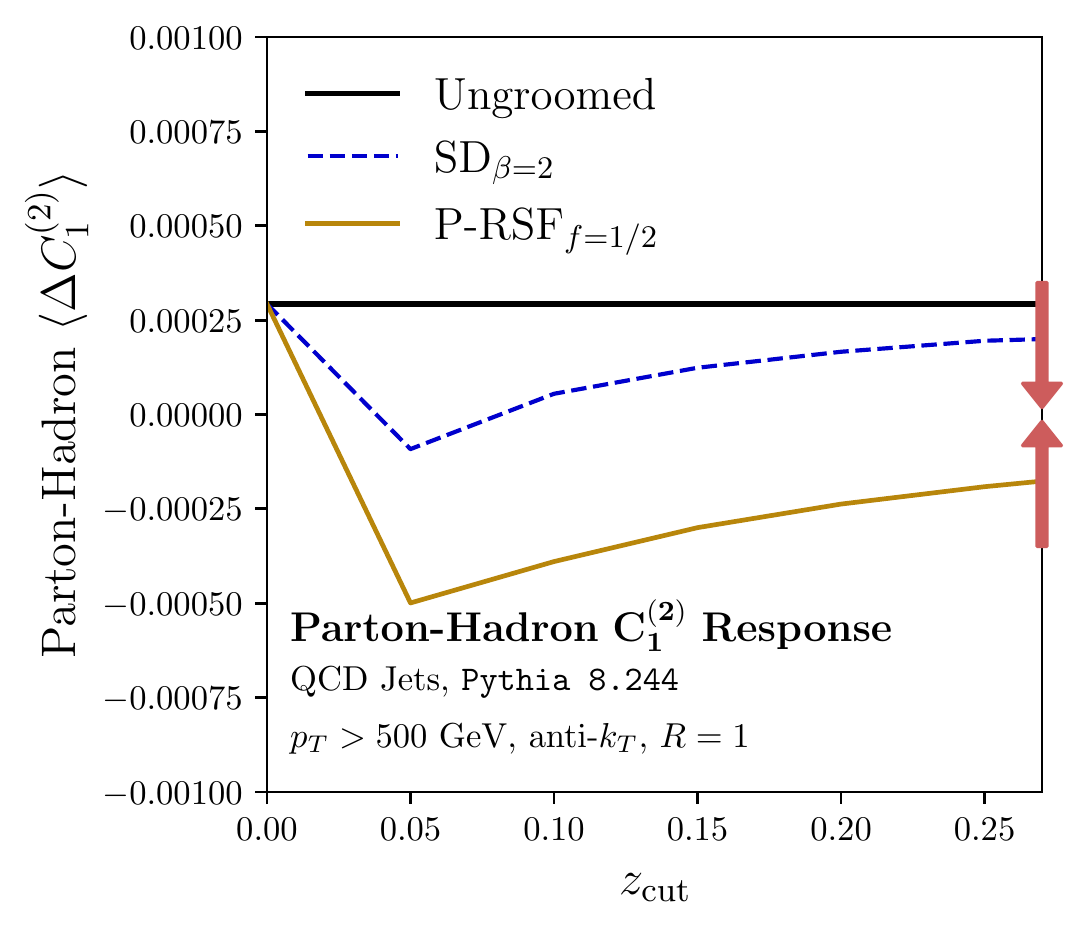}
      \label{fig:ph_c12_delta}
}\subfloat[][]{
      \includegraphics[width=.32\textwidth]
      {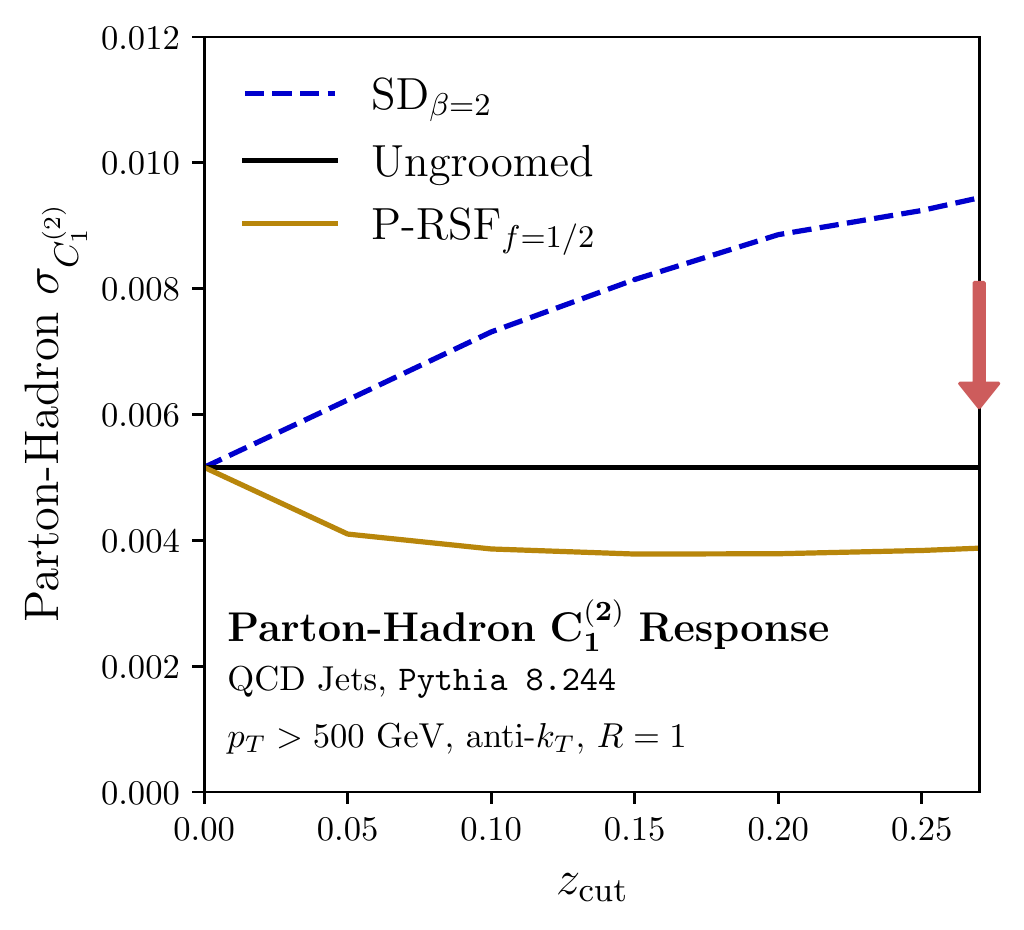}
      \label{fig:ph_c12_sigma}
}
\caption{
    Per-jet hadronization responses of (top row) EMD, (middle row) \(\Delta p_T\), and (bottom row) \(\Delta C_1^{(2)}\) using Balanced Recursive Subtraction (\PRSF{1/2}, orange) and Soft Drop with \(\beta_{\rm SD} = 2\) (\SD{2}, blue).
    We display (left column) the distribution of the response for \zcut=0.1, (middle column) the mean response as a function of \zcut, and (right column) the standard deviation of the response as a function of \zcut.
The red arrows indicate the direction corresponding to better performance.
}
\label{fig:parton_hadron_response}
\end{figure}

In \Fig{parton_hadron_response}, we compare how ungroomed jets, traditionally groomed jets, and \PIRANHA{}-groomed jets respond to hadronization in the context of energy flow, extensive properties, and substructure.
Our studies focus on the parton-hadron EMD (the EMD between parton-level jets and their hadron-level counterparts), the parton-hadron \(\Delta p_T\), and the parton-hadron \(\Delta C_1^{(2)}\).
\PIRANHA{}-groomed jets typically exhibit smaller and more predictable responses to hadronization, evinced by the reduced variance in their hadronization response.
One key factor in the improved responses of \PIRANHA{}-groomed jets is the scaling-down of distortions due to hadronization in \PIRANHA{}-groomed jets:
the subtractive approach of \PIRANHA{} (and in particular P-RSF when \(f_{\rm soft} \neq 0, 1\)) removes energy from every particle in the event, resulting in a controlled and uniform hadronization response.
On the other hand, Soft Drop grooming does not implement continuous, event-wide subtraction.
Jets groomed with Soft Drop either retain or remove particles as a result of distortions due to hadronization, resulting in a larger variance in the properties of jets groomed with Soft Drop.

We plot distributions of the parton-hadron EMD for groomed QCD jets for the benchmark value of $\zcut{}=0.1$ in \Fig{ph_emd_dist}.
We see a sharper peak at a smaller EMD in the parton-hadron EMD distributions for \PRSF{1/2} and longer tails for Soft Drop, already indicating that hadronization is more likely to dramatically change the result of Soft Drop.
\Figs{ph_emd_delta}{ph_emd_sigma} show that the average parton-hadron EMD and the variance in the parton-hadron EMD are both smaller for \PRSF{1/2} groomed jets than for Soft Drop groomed jets for a wide range of \zcut{} values, again evincing that \PIRANHA{}-groomed jets have less sensitive responses to hadronization.

The extensive observable \(p_T\) and the substructure observable \(C_1^{(2)}\) also reflect the increased robustness of \PIRANHA{}-groomed jets to hadronization.
In our discussion of hadronization, we characterize the response of \(p_T\) by using \Eq{pt_response} with \(p_T = p_T^{\rm(parton)}\) indicating the parton-level transverse momentum, before the addition of model-dependent hadronization effects, and \(\widetilde{p_T} = p_T^{\rm(hadron)}\).
In \Fig{ph_pt_dist}, we see that for \(\zcut = 0.1\), \PIRANHA{}-groomed \(p_T\) again tends to have a sharper response to hadronization, while \Figs{ph_pt_delta}{ph_pt_sigma} indicates that this sharper response holds for a wide range of \zcut{} values.
The behavior of the \(p_T\) closely mimics that of the parton-hadron EMD, evincing our arguments that the parton-hadron EMD provides a probe of generic jet observables to hadronization.
In \Figss{ph_c12_dist}{ph_c12_delta}{ph_c12_sigma}, we make similar conclusions for \(C_1^{(2)} \approx m^2 / p_T^2\).
We first note that the parton-hadron shift in the \(C_1^{(2)}\) substructure observable is larger for \PRSF{1/2}, at least for fixed \zcut{}, and the substructure of \PRSF{1/2} groomed jets may undergo larger changes due to hadronization than that of Soft Drop groomed jets.
However, the variance in the shift in \PRSF{1/2} groomed substructure is smaller than that of Soft Drop, indicating that the response of \PIRANHA{}-groomed substructure to hadronization is more predictable than that of Soft Drop.

Overall, we find that \PIRANHA{}-groomed jets have more predictable responses to hadronization than traditionally groomed jets, supported by the observable independent EMD, as well as by extensive observables and jet substructure.
We have provided evidence that IRC-safe observables of jets groomed with Soft Drop have generically larger responses to hadronization, while \PIRANHA{}-groomed observables are generally less affected by the physics of hadronization.
We hope that the stability of \PIRANHA{}-groomed jets to the non-perturbative physics of hadronization may facilitate even further communication between theoretical predictions and experimental results for groomed jet observables.

\subsection{All Versus Charged}
\label{sec:all_v_charged}
\begin{figure}[p]
\subfloat[][]{
      \includegraphics[width=.32\textwidth]
      {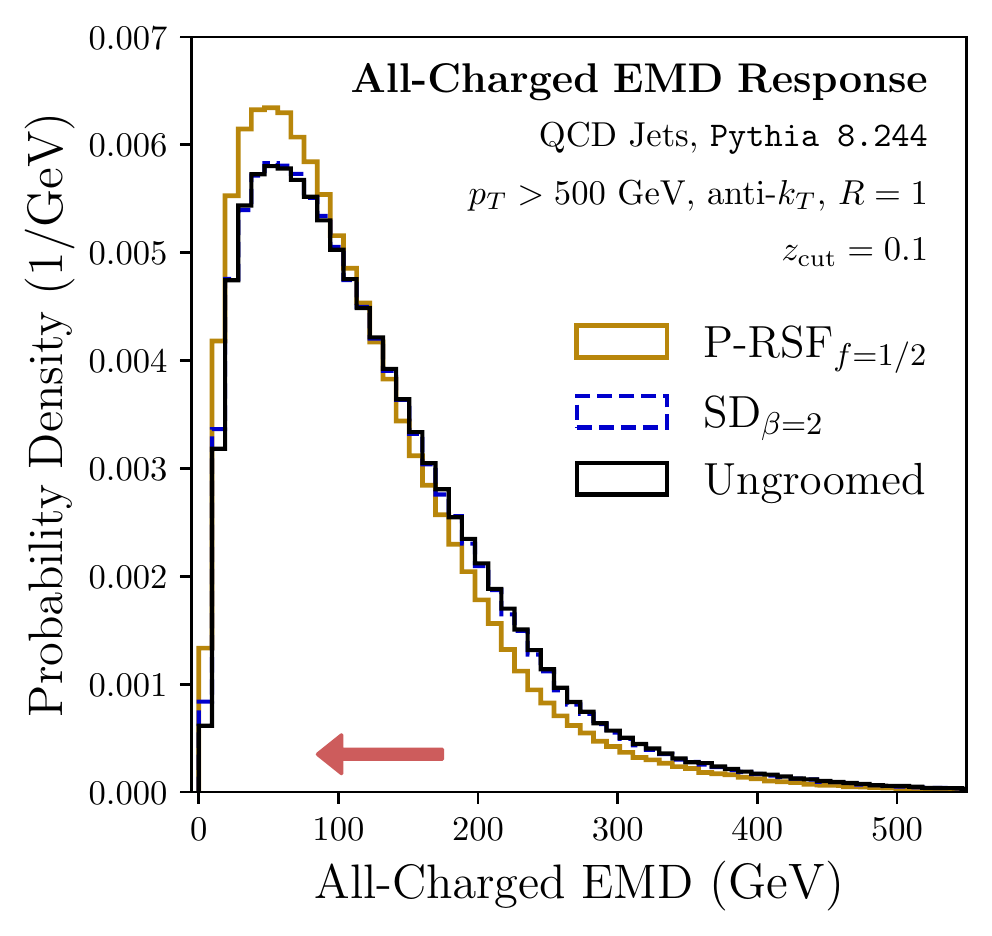}
      \label{fig:avc_emd_dist}
}
\subfloat[][]{
      \includegraphics[width=.32\textwidth]
      {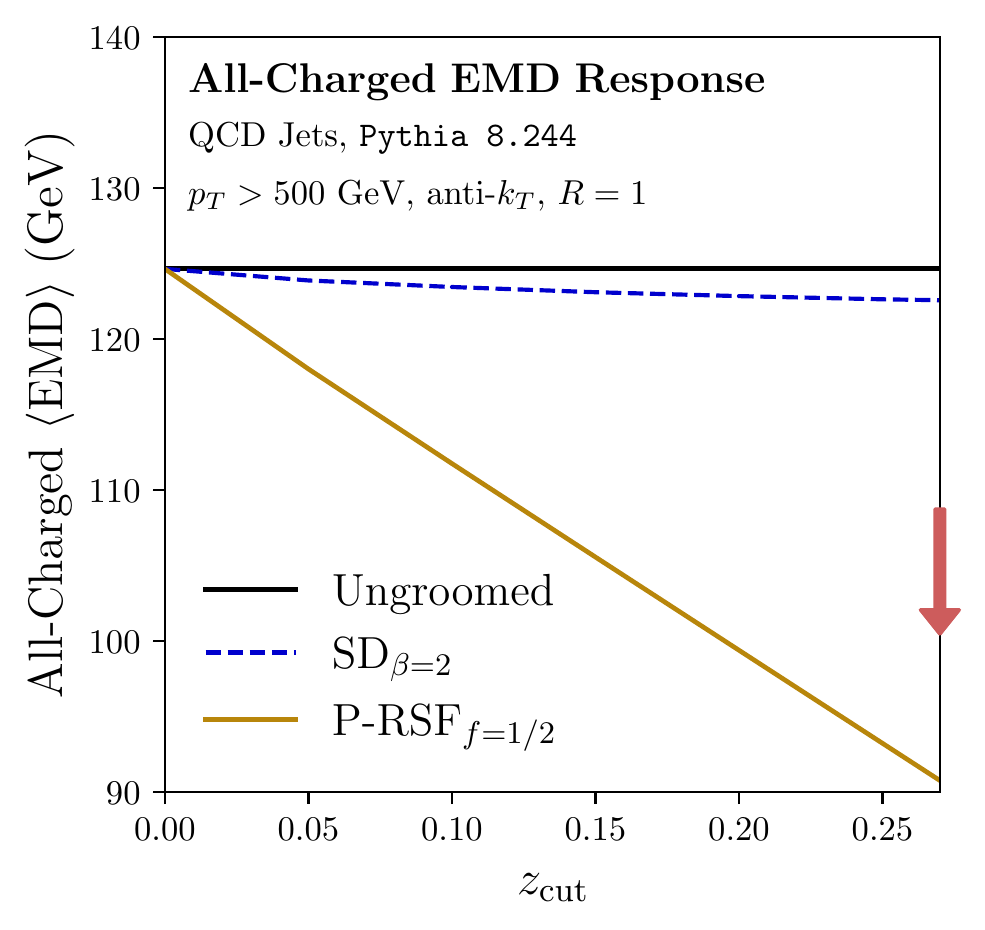}
      \label{fig:avc_emd_delta}
}
\subfloat[][]{
      \includegraphics[width=.32\textwidth]
      {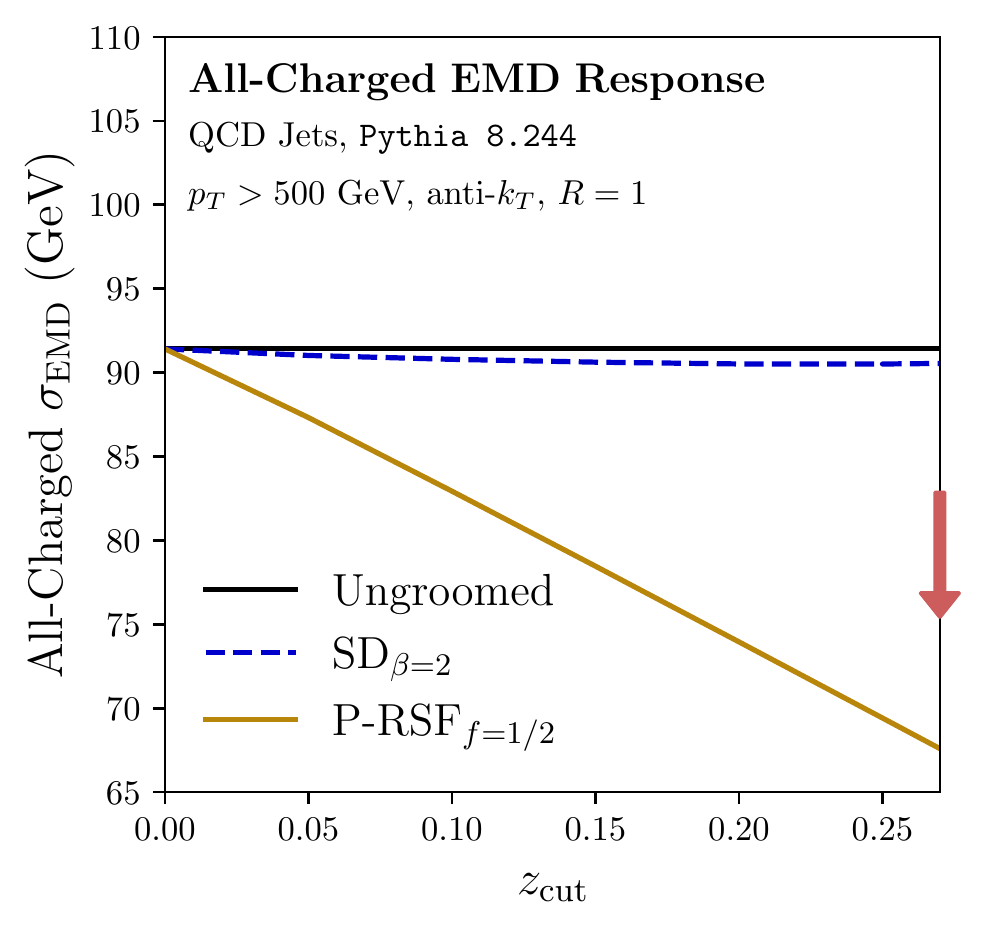}
      \label{fig:avc_emd_sigma}
}
\\
\subfloat[][]{
      \includegraphics[width=.32\textwidth]
      {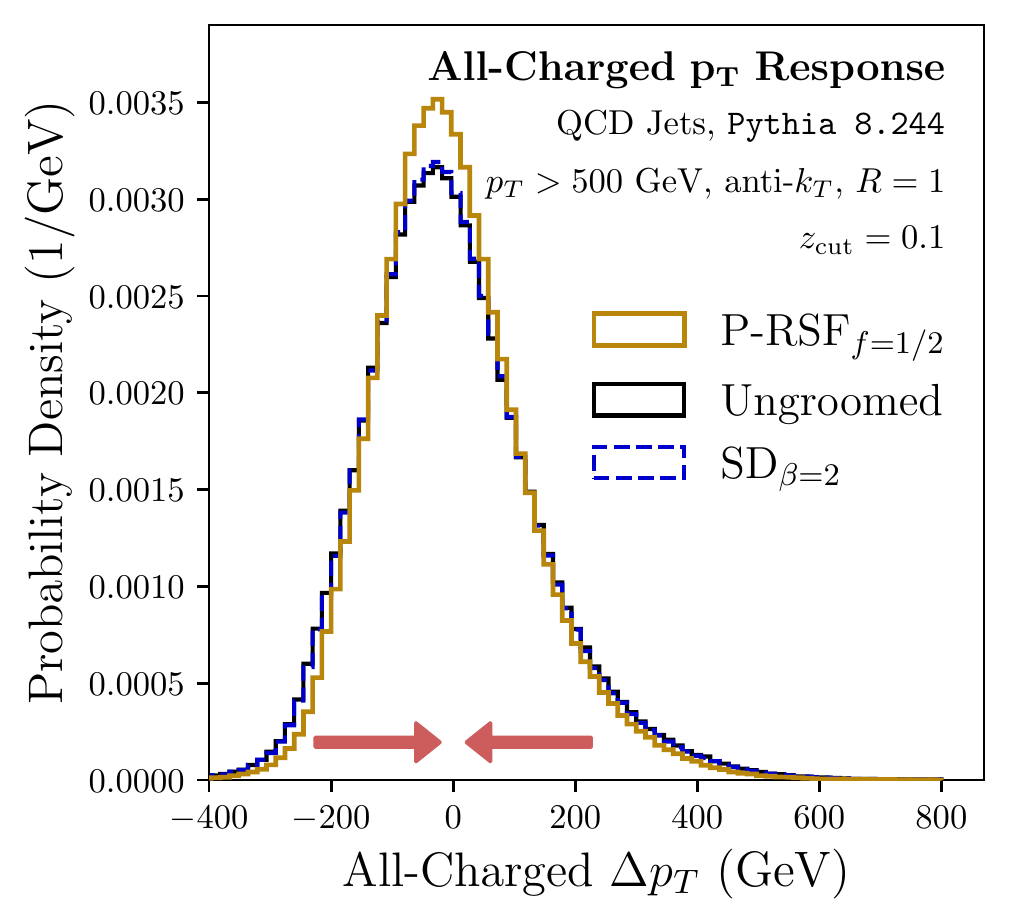}
      \label{fig:avc_pt_dist}
}
\subfloat[][]{
      \includegraphics[width=.32\textwidth]
      {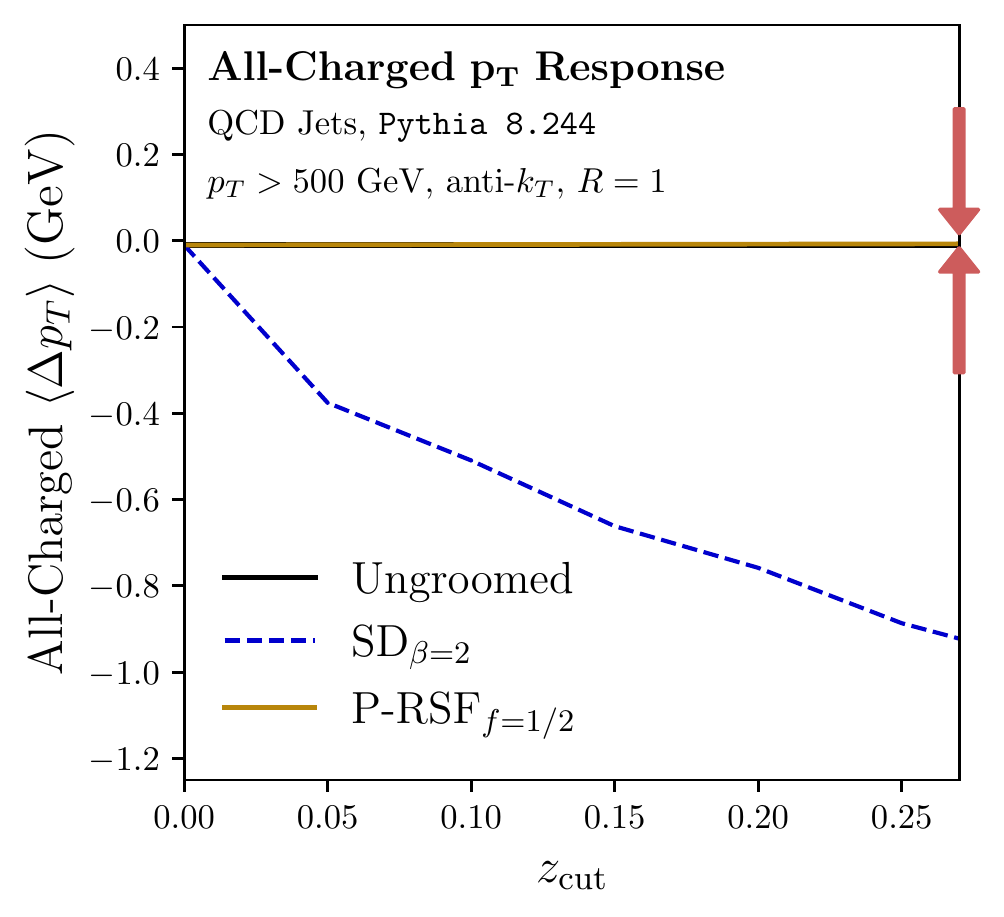}
      \label{fig:avc_pt_delta}
}
\subfloat[][]{
      \includegraphics[width=.32\textwidth]
      {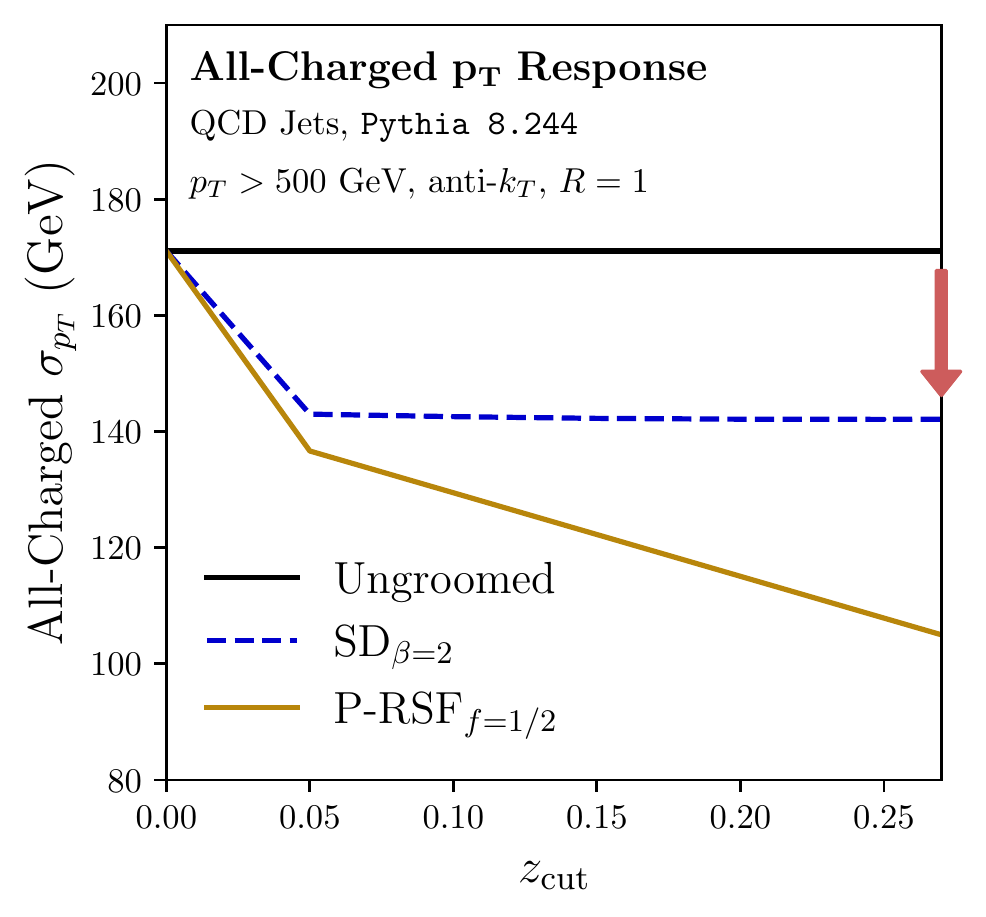}
      \label{fig:avc_pt_sigma}
}
\\
\subfloat[][]{
      \includegraphics[width=.32\textwidth]
      {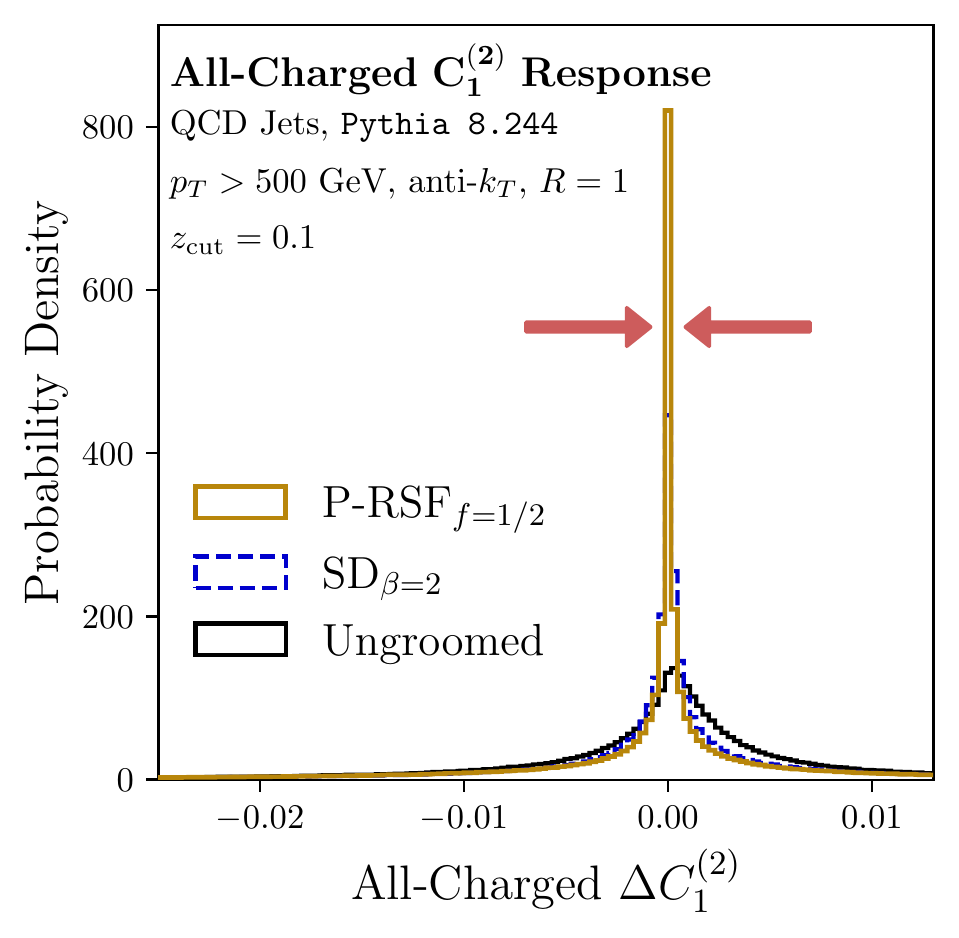}
      \label{fig:avc_c12_dist}
}
\subfloat[][]{
      \includegraphics[width=.32\textwidth]
      {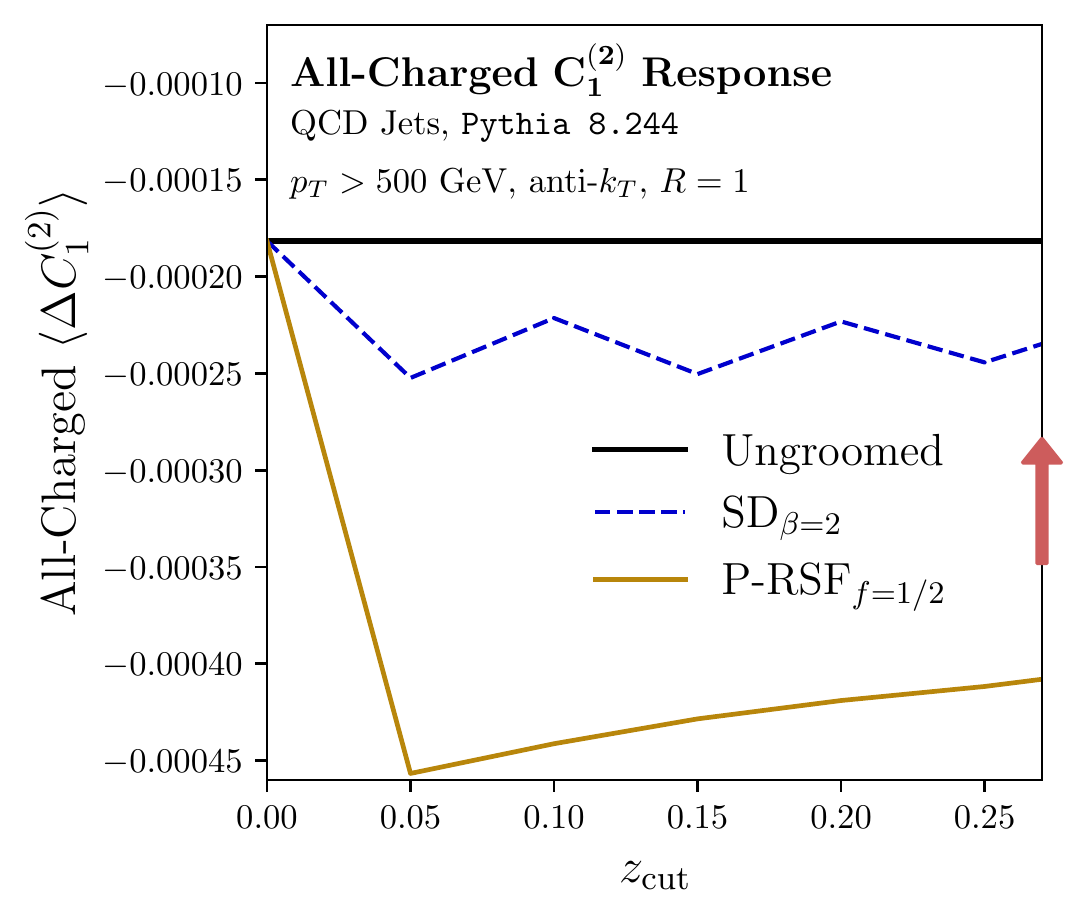}
      \label{fig:avc_c12_delta}
}
\subfloat[][]{
      \includegraphics[width=.32\textwidth]
      {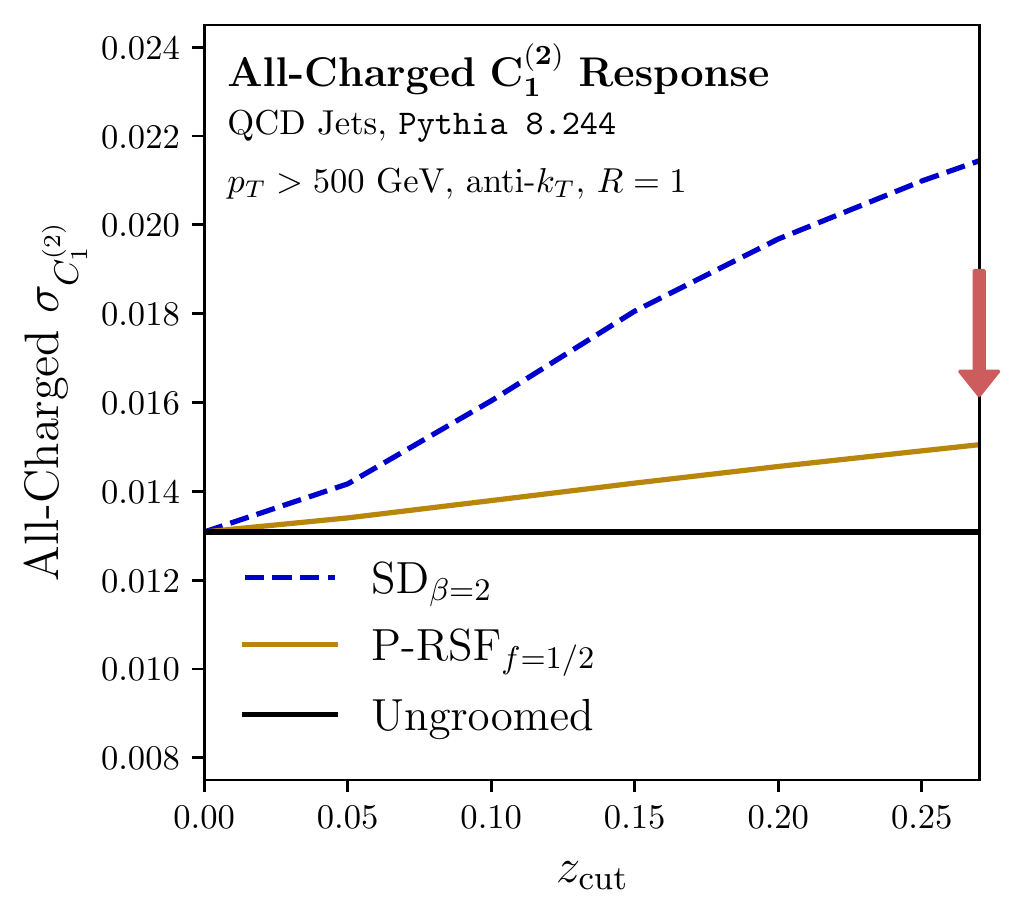}
      \label{fig:avc_c12_sigma}
}
\caption{
    Same as \Fig{parton_hadron_response}, but for per-jet responses to the exclusion of neutral particles.
    We rescale the \(p_T\) of the charged particles in order to eliminate the leading $p_T$ bias due to the loss of neutral energy.
}
\centering
\label{fig:avc_response}
\end{figure}

We next compare the response of hard-cutoff and \PIRANHA{} grooming to the exclusion of neutral particles, as a rough analog of the smearing of neutral particles due to detector effects.
Small tweaks in experimental signatures due to detector effects may produce large changes in hard-cutoff groomed observables, providing another obstacle in extracting fundamental physics from traditionally groomed jets.
The continuity of \PIRANHA{} again leads us to expect that \PIRANHA{}-groomed observables are more robust to detector effects than traditionally groomed jet observables.

To counteract the leading bias due to the loss of neutral energy in the jet, we apply a rescaling to the \(p_T\) of the charged-only jet constituents:
\begin{align}
    \tilde p_T^i =
    \begin{cases}
        0,
        &\text{particle }i\text{ is neutral}
        \\
        p_T^i
        \frac{\left\langle p_T^\text{(all)}\right\rangle}{\left\langle p_T^\text{(charged)}\right\rangle},
        &\text{particle }i\text{ is charged}
    \end{cases}
    ,
    \label{eqn:all-charged_rescaling}
\end{align}
where \(p_T^i\) is the transverse momentum of particle \(i\) in the original jet, \(\mathcal{J}\), containing both charged and neutral particles, and \(\tilde p_T^i\) is the transverse momentum we assign to particle \(i\) in the charged-only, rescaled jet, \(\tilde{\mathcal{J}}\), with neutral particles excluded.
Here, \(p_T^{\rm(all)} = \sum_i p_T^i\) is the total transverse momentum including both charged and neutral particles, while \(p_T^{\rm(charged)} = \sum_{i\text{ charged}} p_T^i\) includes only charged constituents.
The rescaling factor \(\big\langle p_T^\text{(all)}\big\rangle/\big\langle p_T^\text{(charged)}\big\rangle\)\(\approx\)\(1.47\) is defined such that the average $p_T$ is preserved after removing neutral particles.%
\footnote{
We can gain intuition for this rescaling factor by considering the isospin-preserving limit, where we expect similar numbers of \(\pi^+\), \(\pi^-\), and \(\pi^0\) mesons to be produced.
Events for which we discard the \(\pi^0\) particles should have roughly \(2/3\) the total transverse momentum, ignoring subtleties associated with kaons and other heavier states.
This leads to an estimate of \(\big\langle p_T^\text{(all)}\big\rangle/\big\langle p_T^\text{(charged)}\big\rangle \approx 3/2\) -- remarkably close to the numerical value of 1.47 found in \texttt{Pythia 8.244}.
}

In our discussion of EMD, we compute the response \(\text{EMD}\left(G(\mathcal{J}), G(\tilde{\mathcal{J}})\right)\), where \(G\) indicates the grooming algorithm under consideration;
we call this the \textit{all-charged EMD}.
Similarly, in our discussion of \Eq{pt_response}, we take \(p_T = p_T^{\rm(all)}\) to be the transverse momentum including both charged and neutral particles and \(\widetilde{p_T} = \sum_i \tilde p_T^i = p_T^{\text{(charged, rescaled)}}\) to be the rescaled contribution from charged particles only.
We refer to the difference as the all-charged \(p_T\) shift, or the all-charged \(\Delta p_T\).
Note that the rescaling procedure does not impact dimensionless charged-only substructure observables like \(C_1^{(2)}\), since they are normalized by the jet momentum.

Our comparison of the all-charged response of \PIRANHA{} grooming to that of traditional grooming procedures is shown in \Fig{avc_response}.
As in our study of hadronization, we begin our study with a discussion of the EMD.
The all-charged EMD bounds changes in IRC-safe observables due to the exclusion of neutral particles and corresponding rescaling of charged particles.
Since neutral particles are the most susceptible to smearing effects due to detector responses, the all-charged EMD furnishes an observable-independent probe for the effects of experimental detectors on groomed jets.
We demonstrate distributions of the all-charged EMD for groomed QCD jets with several choices of grooming in \Fig{avc_emd_dist}.
The contrast between \PIRANHA{} and traditionally groomed all-charged EMD is not as sharp as for the parton-hadron EMD.
However, \PIRANHA{} grooming again enjoys smaller and more sharply peaked EMD responses to the exclusion of neutral particles.
\Fig{avc_emd_delta} shows that the all-charged EMD is smaller for P-RSF\(_{1/2}\) groomed jets than for Soft Drop groomed jets for a wide range of \zcut{} values.
\Fig{avc_emd_sigma} shows that the fluctuations in the all-charged EMD, and therefore the fluctuations in the response of grooming to our naive model of smearing, are noticeably smaller for \PIRANHA{} groomers than for traditional groomers.

Our results for the all-charged \(p_T\) shifts are similar to our results for the all-charged EMD.
\Fig{avc_pt_dist} demonstrates that the \(p_T\) response of \PRSF{1/2} is smaller and more sharply peaked for $\zcut = 0.1$, while \Figs{avc_pt_delta}{avc_pt_sigma} respectively show that the \PRSF{1/2} all-charged response remains small and sharply peaked for a wide range of \zcut{} values.
Indeed, the fluctuations of the all-charged \(p_T\) response are nearly identical for each \PIRANHA{} groomer.

\Figss{avc_c12_dist}{avc_c12_delta}{avc_c12_sigma} compare the robustness of \PRSF{1/2} and Soft Drop for the substructure observable \(C_1^{(2)}\).
\Fig{avc_c12_dist} again shows that the all-charged response of \PRSF{1/2} is more sharply peaked than that of Soft Drop.
Similarly, \Figs{avc_c12_delta}{avc_c12_sigma} demonstrate that the all-charged shifts to \PIRANHA{}-groomed jet substructure are generally larger but more predictable than those of Soft Drop for a wide range of \zcut{} values.
The increased robustness of \PIRANHA{} grooming procedures compared to Soft Drop is less sharp than in the case of hadronization, but the \PIRANHA{}-groomed observables nonetheless have less spread and a stronger linear correlation in response to our naive model of smearing.

\section{Responses of Grooming to Additive Contamination}
\label{sec:add_contam}

We examine the ability of \PIRANHA{} grooming to mitigate the effects of additive contamination from pileup (PU) and the underlying event (UE), which both consist of \textit{extra} soft radiation that sits on top of a hard process.
\begin{itemize}
    \item We use \PRSF{1/2} as a representative of \PIRANHA{} grooming procedures;
    \item We use \SD{2} as a representative of traditional grooming procedures;
    \item In our PU studies, we use Constituent Subtraction (CS), a constituent-level area subtraction method for pileup mitigation \cite{Berta:2014eza}, as a representative of PU mitigation procedures.
\end{itemize}
The case of additive contamination is qualitatively different than that of soft distortions;
we do not want to examine the robustness of the grooming procedure to additive contamination, but rather the ability of the grooming to \textit{remove} the contamination.
In our PU mitigation studies, we find that \PRSF{1/2} -- which behaves comparably to CS -- is a more effective tool than \SD{2}.
In our UE studies, we find that \PRSF{1/2} and \SD{2} are comparable tools for the subtraction of UE effects from jet substructure.

We simulate PU using the dijet and minimum bias samples from \Reff{Soyez:2018opl}, produced using \texttt{Pythia 8.185} with tune 4C for proton-proton collisions at \(\sqrt{s}\) = 14 TeV.
We layer minimum bias events on top of the hard dijet events, taking the number of pileup events to be Poisson distributed with a mean of \(\langle n_{\rm PU}\rangle\) = 50 events.

We simulate UE by turning on multiple parton interactions in \texttt{Pythia 8.244} \cite{Sjostrand:2014zea} with the default 4C tune \cite{Corke:2010yf}.
We begin with a study of QCD jets and, to roughly echo similar studies in the original Soft Drop paper \cite{Larkoski:2014wba}, also explore the groomed mass resolution of jets produced by boosted \(W\) bosons and top quarks in the presence of UE.

\subsection{Minimum Bias Events and Pileup}
\label{sec:pileup}

\begin{figure}[p]
\centerline{
\subfloat[][]{
      \includegraphics[width=.32\textwidth]
      {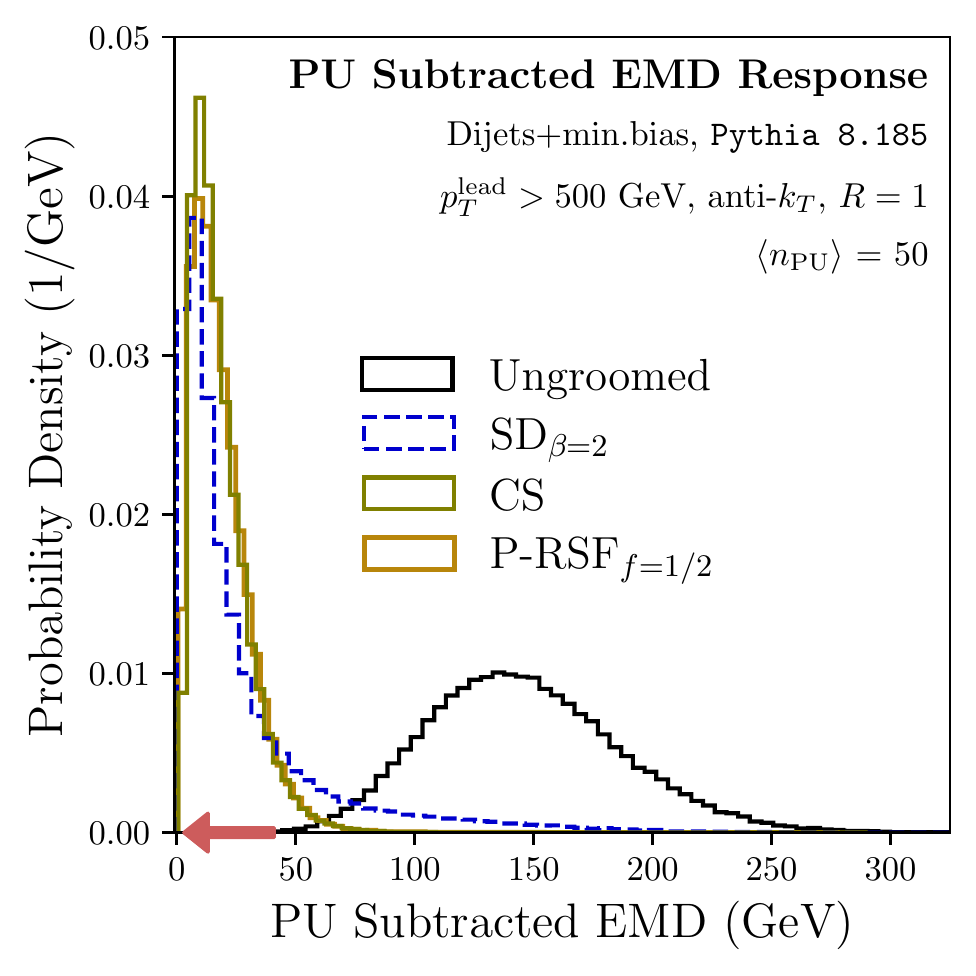}
      \label{fig:pu_emd_dist}
}
\subfloat[][]{
      \includegraphics[width=.32\textwidth]
      {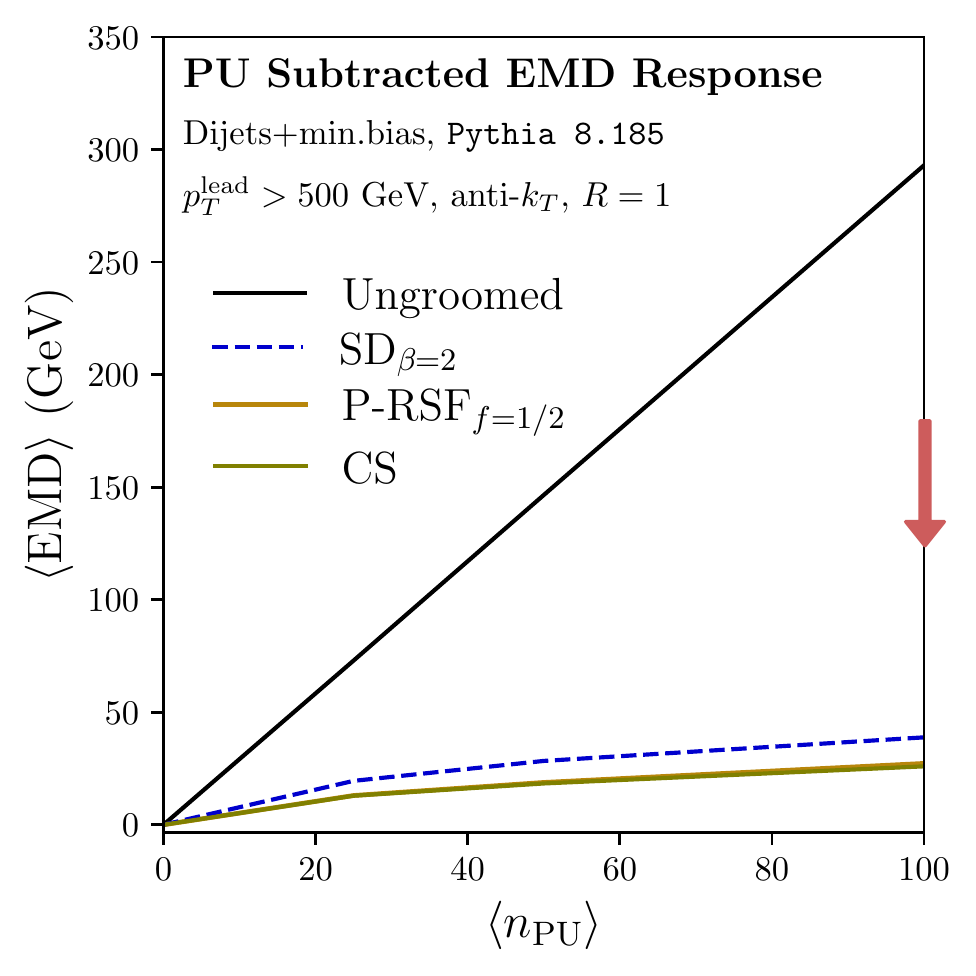}
      \label{fig:pu_emd_shift}
}
\subfloat[][]{
      \includegraphics[width=.32\textwidth]
      {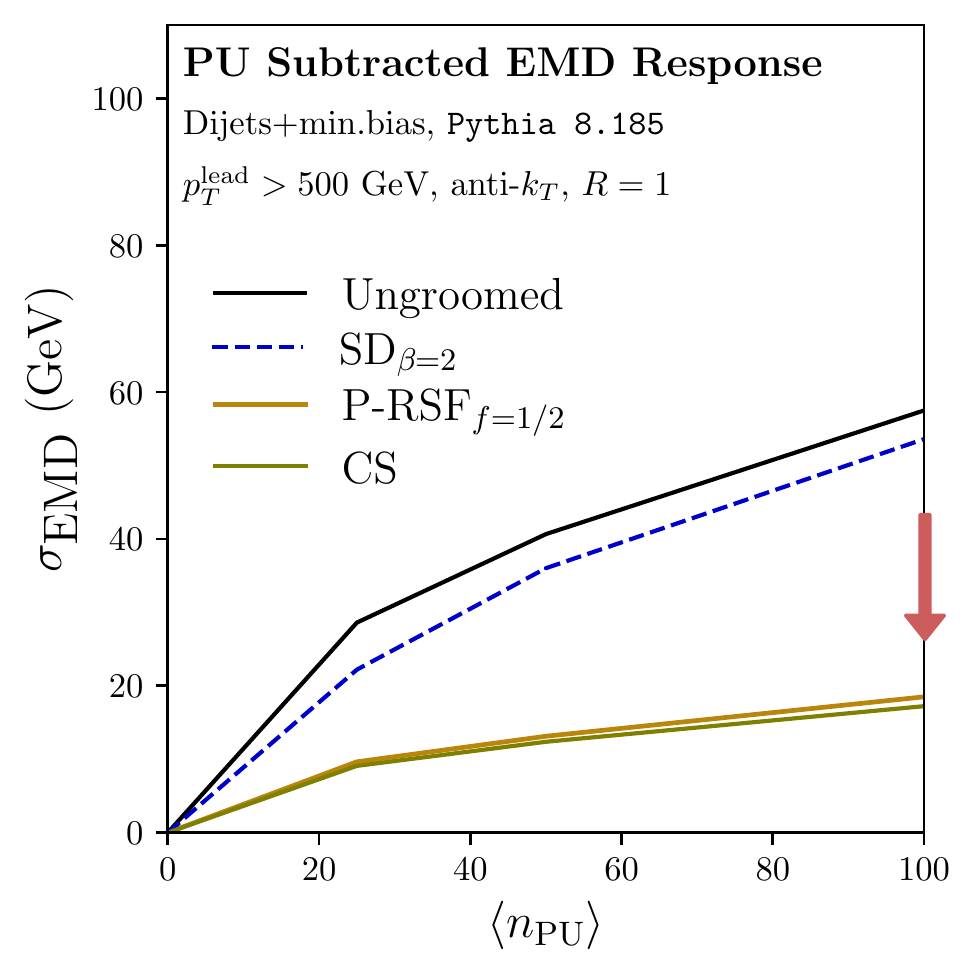}
      \label{fig:pu_emd_std}
}
} 
~\\
\centerline{
\subfloat[][]{
      \includegraphics[width=.32\textwidth]
      {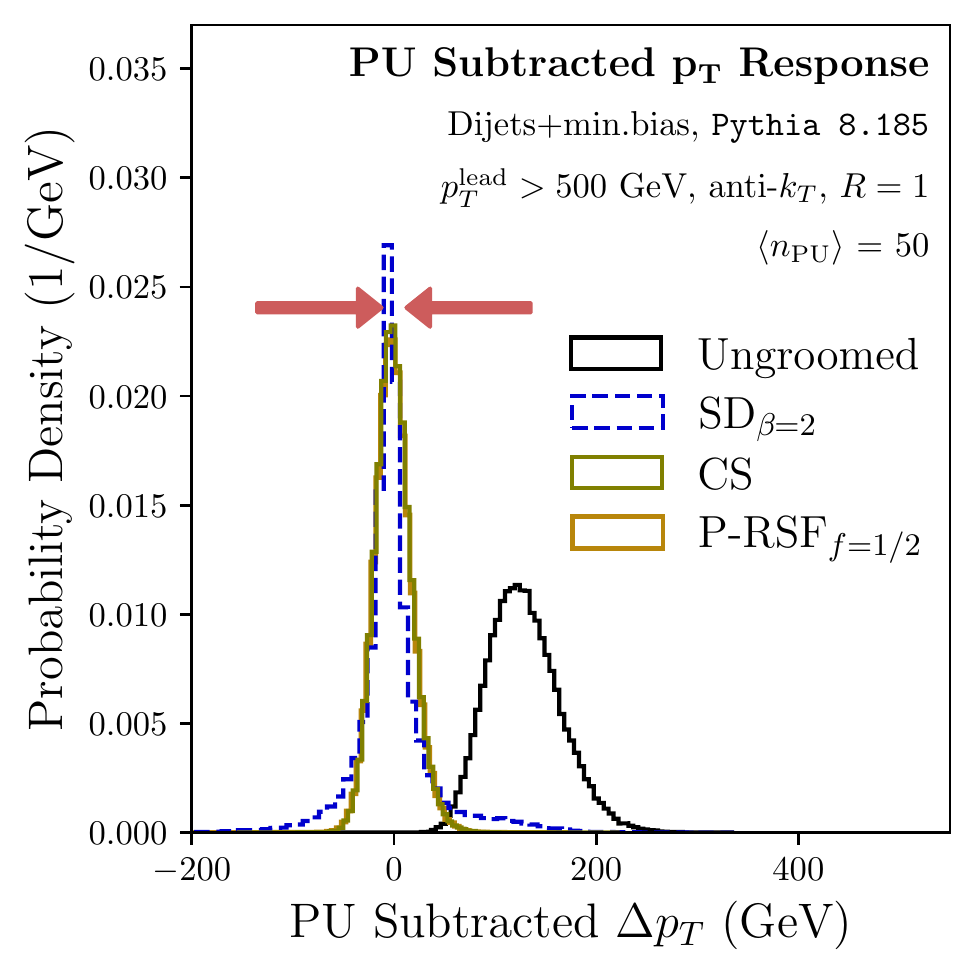}
      \label{fig:pu_pt_dist}
}
\subfloat[][]{
      \includegraphics[width=.32\textwidth]
      {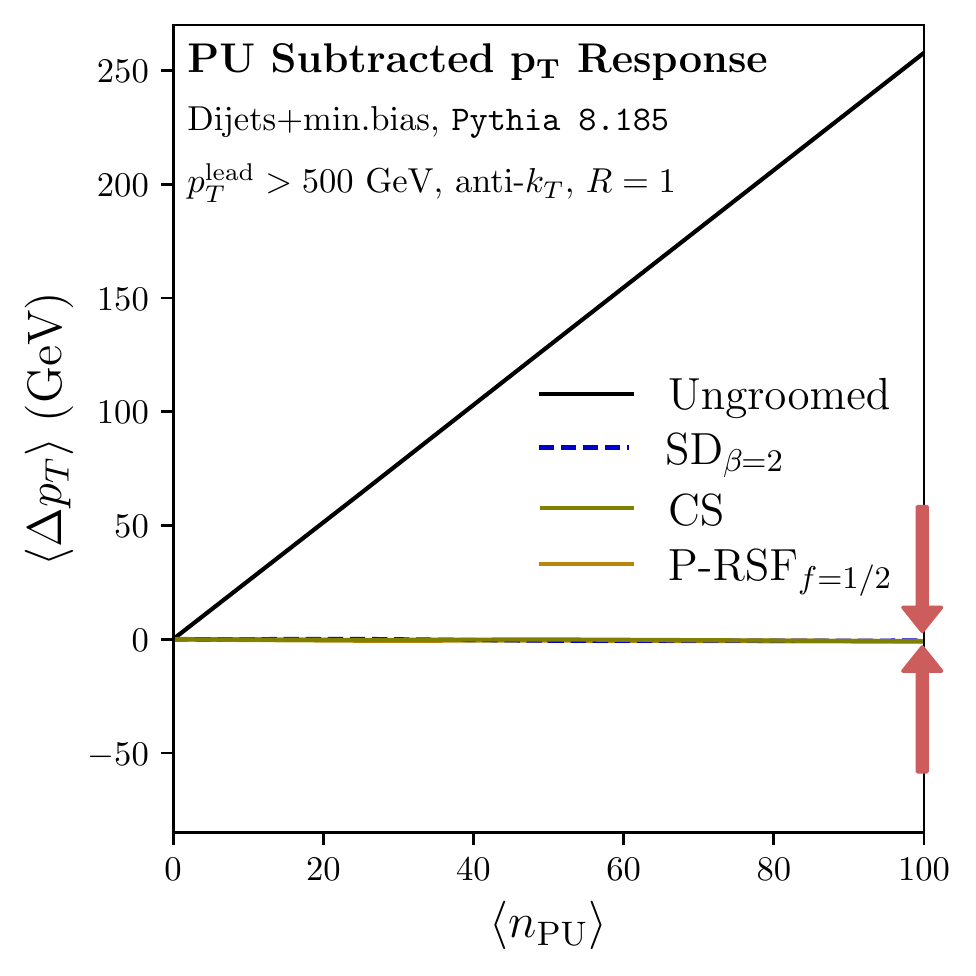}
      \label{fig:pu_pt_shift}
}
\subfloat[][]{
      \includegraphics[width=.32\textwidth]
      {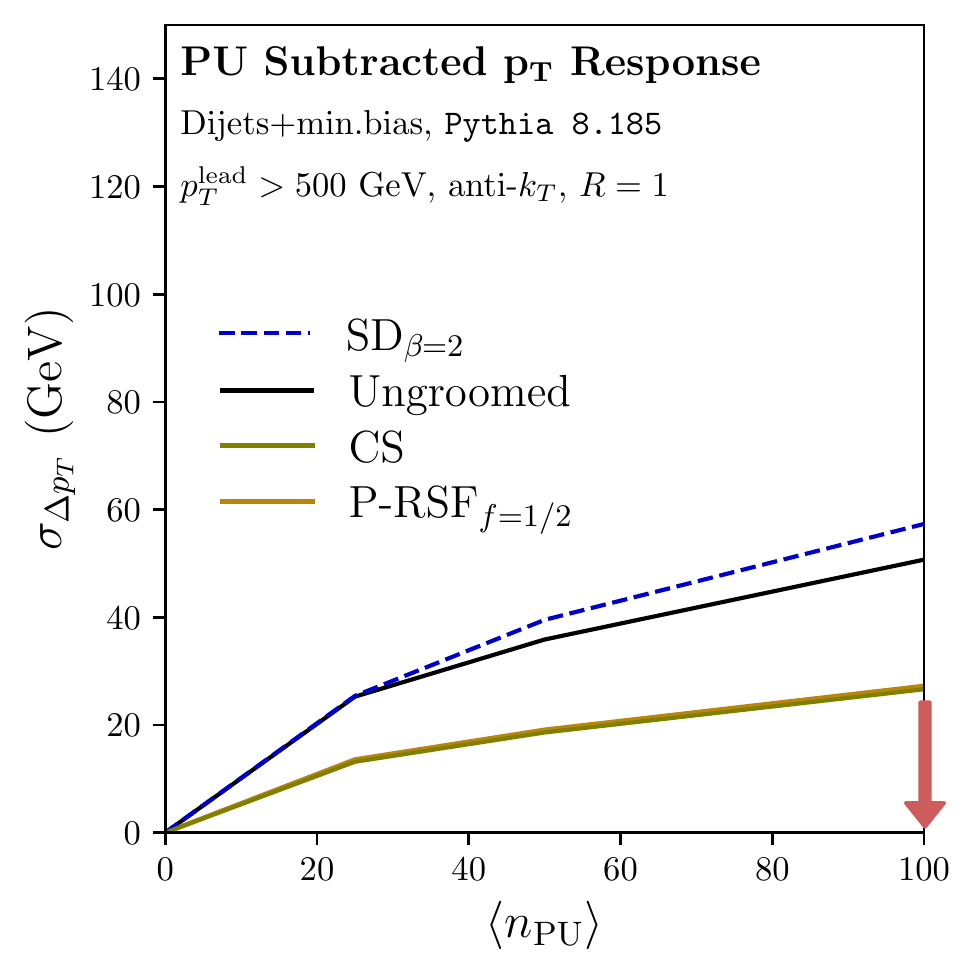}
      \label{fig:pu_pt_std}
}
} 
~\\
\centerline{
\subfloat[][]{
      \includegraphics[width=.32\textwidth]
      {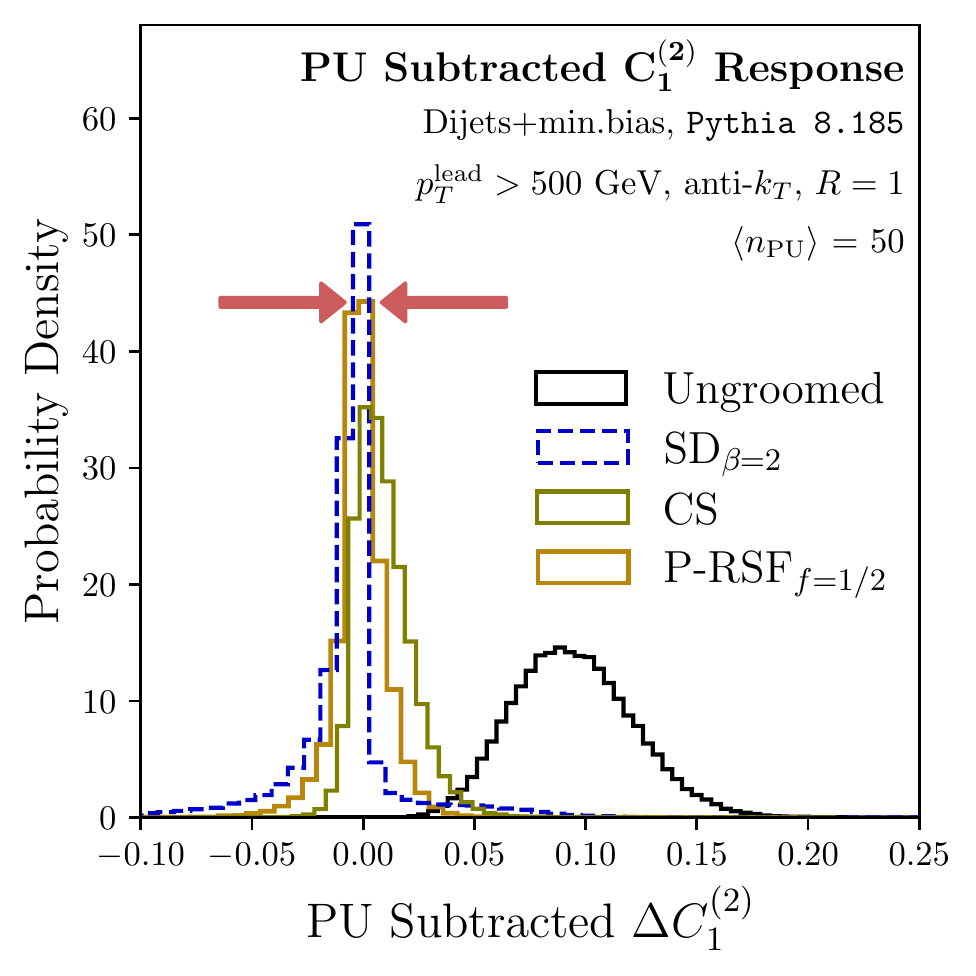}
      \label{fig:pu_c12_dist}
}
\subfloat[][]{
      \includegraphics[width=.32\textwidth]
      {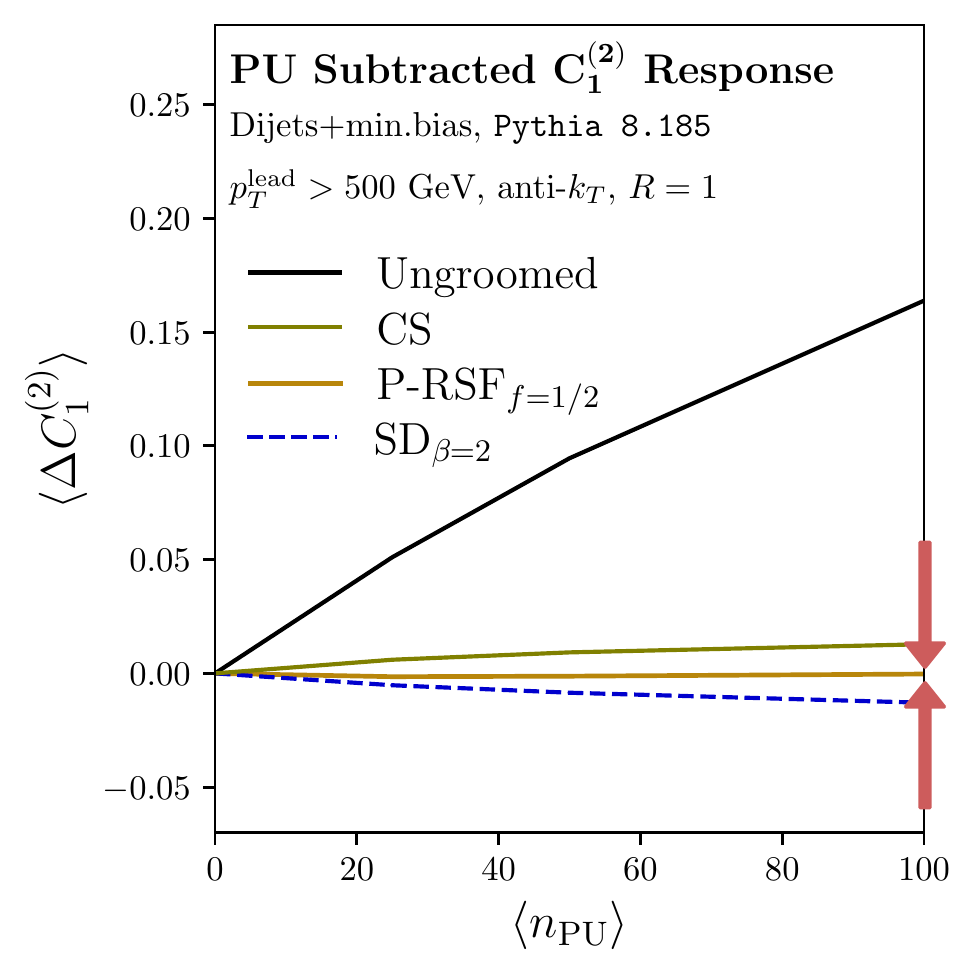}
      \label{fig:pu_c12_shift}
}
\subfloat[][]{
      \includegraphics[width=.32\textwidth]
      {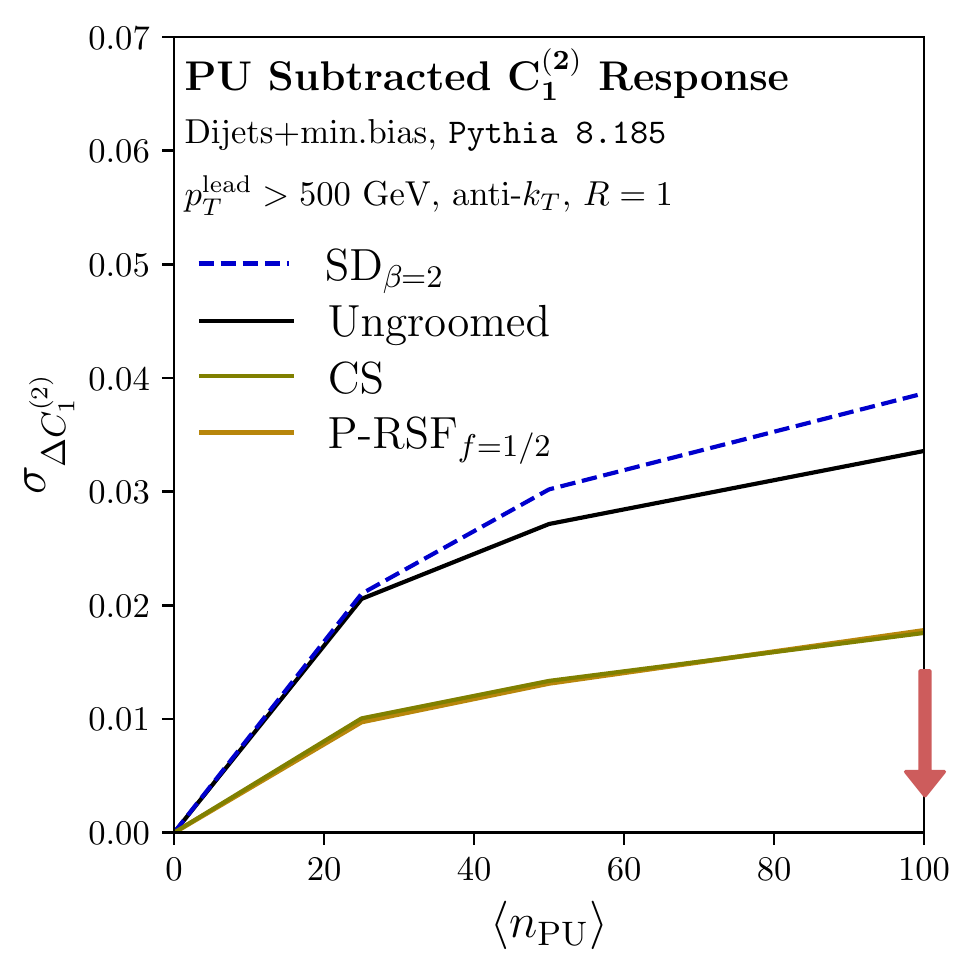}
      \label{fig:pu_c12_std}
}
} 
\caption{
    Properties of per-jet PU-subtracted (top row) EMD, (middle row) \(\Delta p_T\), and (bottom row) \(\Delta C_1^{(2)}\), discussed in \Sec{pileup} for Balanced Recursive Subtraction (\PRSF{1/2}, orange), Soft Drop with \(\beta_{\rm SD} = 2\) (\SD{2}, blue), and Constituent Area Subtraction (CS, green).
    We display (left column) the distribution of the each PU-subtracted observable for \(\langle n_{\rm PU}\rangle = 50\), (middle column) the mean PU-induced shift in each observable as a function of \(\langle n_{\rm PU}\rangle\), and (right column) the standard deviation of the PU-induced shifts in each observable as a function of \(\langle n_{\rm PU}\rangle\).
    The red arrows indicate the direction corresponding to better performance.
}
\label{fig:pu_response}
\centering
\end{figure}

We first examine the effectiveness of different groomers in mitigating the effects of PU on the dijet events of \Reff{Soyez:2018opl}.
To quantify the ability of different groomers to remove PU, we perform an event-by-event comparison of \textit{hard events} to the associated \textit{PU-subtracted events}:
\begin{itemize}
    \item
    A \textbf{hard event} is a dijet event without any PU, representing the physics of a hard process;

    \item
    A \textbf{PU-subtracted event} is produced by simulating the additive contamination due to PU on top of a hard event and then attempting to groom the PU away.
\end{itemize}
To produce a PU-subtracted event, we first simulate PU by adding the energy distributions of a Poisson-distributed number of minimum bias events to the energy distribution of a given hard event.
    We then estimate the contaminating energy density due to PU using the \texttt{GridMedianBackgroundEstimator} (GMBE) method of \texttt{FastJet} \cite{Cacciari:2011ma}.
    We tune an additional correction factor that scales our estimate \(\rho_{\rm est.}\) relative to the GMBE value \(\rho_{\rm GMBE}\) for each grooming algorithm to subtract pileup more effectively, as discussed in more detail in \App{feedingfrenzy}.
    Finally, we produce the PU-subtracted event by removing a fraction \(\zcut\) of the jet \(p_T\) consistent with the estimated pileup density:
\begin{align}
   \zcut\,p_T^{\rm(jet)}
   =
   \rho_{\rm est.}\,A_{\rm jet}
   \approx
   p_T^{\rm(PU~in~jet)}
   .
\end{align}

We examine the EMD, \(\Delta p_T\), and \(\Delta C_1^{(2)}\) between the \textit{leading jets} of the hard and pileup-events in \Fig{pu_response}.\footnote{
In some situations, the pileup contributes enough energy that the leading jet after the addition of pileup corresponds to the direction of the sub-leading jet before the addition of pileup.
This phenomenon occurs only for a small fraction of events, and we do not include these events in the following discussion.
}
In \Figss{pu_emd_dist}{pu_emd_shift}{pu_emd_std}, we examine the EMD between the leading jet in the hard dijet event and the leading jet in the subtracted event with pileup.
\Fig{pu_emd_dist} demonstrates that when \(\langle n_{\rm PU} \rangle=50\), \PRSF{1/2} tends to subtract pileup more accurately and predictably, evinced by a more sharply peaked EMD distribution and a smaller average PU-subtracted EMD than found for CS and Soft Drop.
This strength of \PRSF{1/2} persists as the amount of pileup is increased, as suggested by \Figs{pu_emd_shift}{pu_emd_std}.

\Figss{pu_pt_dist}{pu_pt_shift}{pu_pt_std} offer similar conclusions for the jet \(p_T\).
In our study of pileup responses, \(p_T = p_T^{\rm(hard)}\) indicates the transverse momentum of the hard event, before the layering of minimum bias/pileup events, and \(\widetilde{p_T} = p_T^{\rm (groomed~PU)}\) indicates the transverse momentum after pileup has been added and then groomed away.
RSF\(_{1/2}\) again reproduces the extensive quantity \(p_T\) of the jet in the absence of PU for a wide range of \(\langle n_{\rm PU} \rangle\) values.
Identical conclusions hold for the substructure of the jet, as suggested by the behavior of \(C_1^{(2)}\) shown in \Figss{pu_c12_dist}{pu_c12_shift}{pu_c12_std}.

\Figs{pufrenzy_ave}{pufrenzy_stddev} of \App{feedingfrenzy} evince further that with our tuned procedure for PU mitigation, \PIRANHA{} groomers and CS perform more reliably than traditional groomers in the removal of pileup.
Furthermore, P-RSF algorithms are tree-based and offer orders-of-magnitude faster PU mitigation over CS, P-AS, and P-IVS (see \Fig{runtimes} of \App{feedingfrenzy}).

\subsection{Underlying Event}
\label{sec:ue}

We next examine the ability of grooming to eliminate additive contamination from UE.
Since we model UE by generating events in the presence of multiple parton interactions in \texttt{Pythia 8.244}, it is less straightforward to study the effects of UE on an event-by-event basis.
We instead work at the level of \textit{distributions}:
we tune both \PRSF{1/2} and Soft Drop so that, when acting on events with UE, they most accurately reproduce the substructure distribution of events without UE.
We focus on substructure alone in our discussion of UE correction because we cut on jets with \(p_T>500\) GeV both for events with and without UE, and it is less meaningful to focus on the associated \(p_T\) distributions;
we leave a more detailed study of the ability of \PIRANHA{} groomers to remove UE at an event-by-event level for future work.

We characterize the ability of groomers to correct for the presence of UE by comparing \textit{base distributions} to \textit{UE-corrected} distributions:
\begin{itemize}
    \item
    A \textbf{base distribution} is a substructure distribution with a fixed grooming parameter, \(\zcut^{(\rm base)}\);

    \item
    A \textbf{UE-corrected distribution} is a substructure distribution in the presence of UE and with slightly more grooming tuned to soak up the additional energy contribution from UE:
    \(\zcut^{(\rm corr.)} = \zcut^{(\rm base)} + \delta\zcut\).
\end{itemize}
We tune \(\delta \zcut\) by hand, for every \(\zcut^{(\rm base)}\), to optimally correct for the presence of UE\footnote{
We note that the optimized \(\delta\zcut\) is energy-dependent and process-dependent.
For example, we expect that \(\delta\zcut\) should be parametrically smaller for processes at higher energies since the energy removed from a jet scales as \(Q \zcut\), where \(Q\) is the energy scale of the hard process, while the energy scales associated with UE are only weakly correlated with the energy of the hard process (see, for example, Fig.~3 of \Reff{Field:2011iq} or Section 7.2 of \Reff{Campbell:2017hsr}).
} by minimizing the \textit{Wasserstein distance} between the two distributions.
The Wasserstein distance is a metric on the space of distribution, much in the same way that the EMD is a metric on the space of energy flows, and provides a useful and quantitative measure of the ability of the grooming to correct for UE.

\begin{figure}[p]
\centering
    \subfloat[][]{
        \includegraphics[width=.45\textwidth]{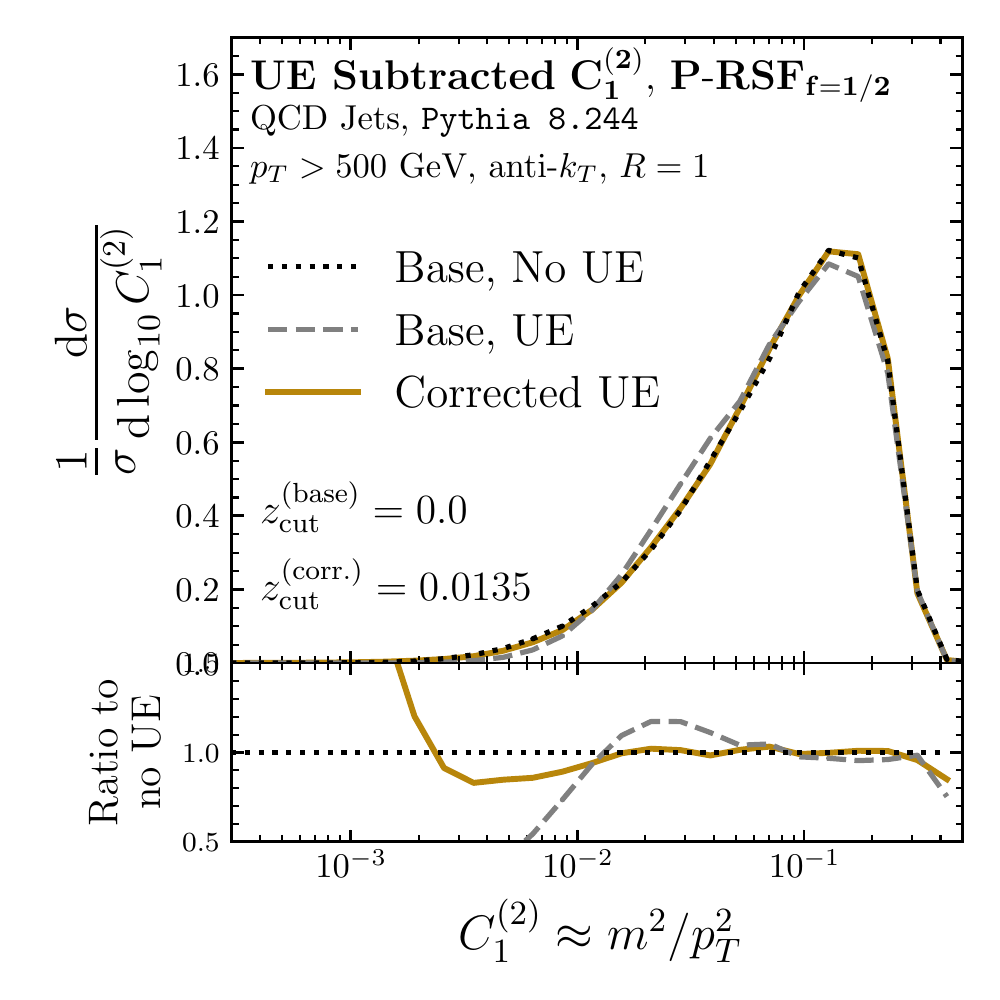}
\label{fig:rss_ue_0}
    }
    \subfloat[][]{
        \includegraphics[width=.45\textwidth]{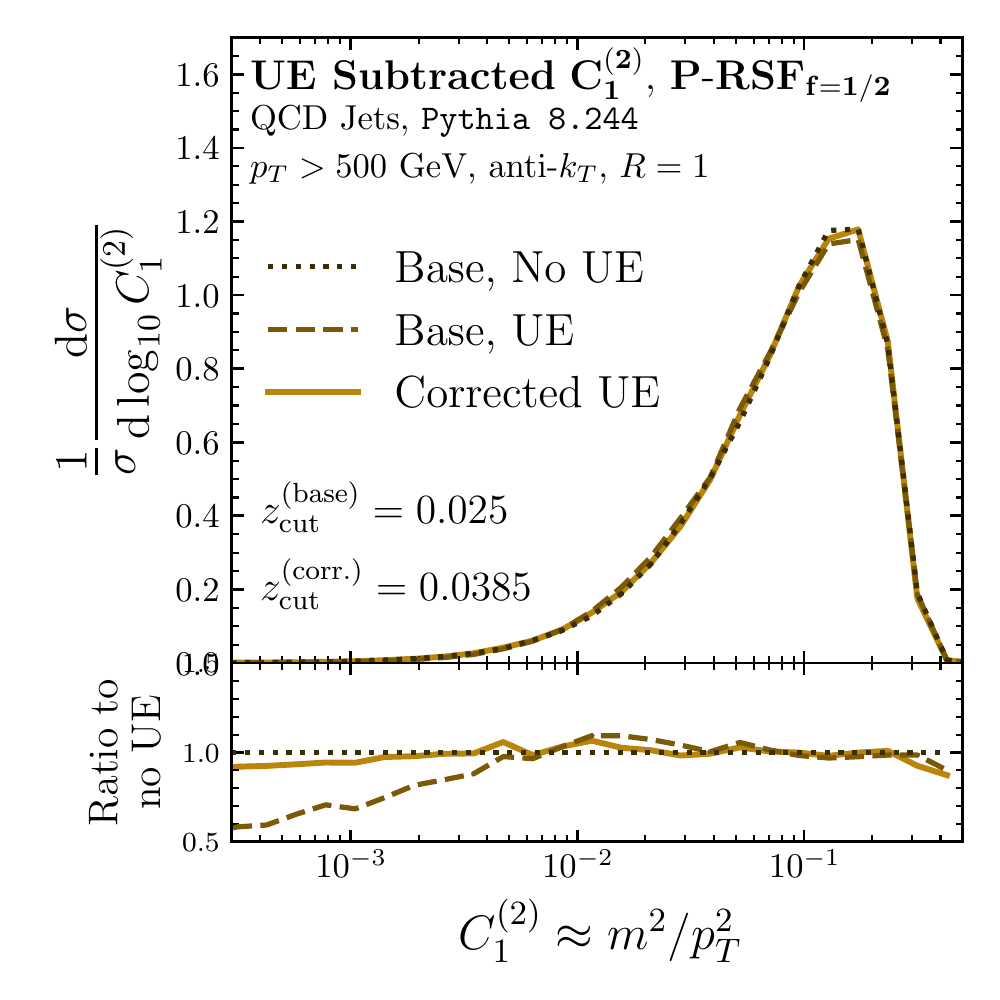}
\label{fig:rss_ue_025}
    }
    \\
    \subfloat[][]{
        \includegraphics[width=.45\textwidth]{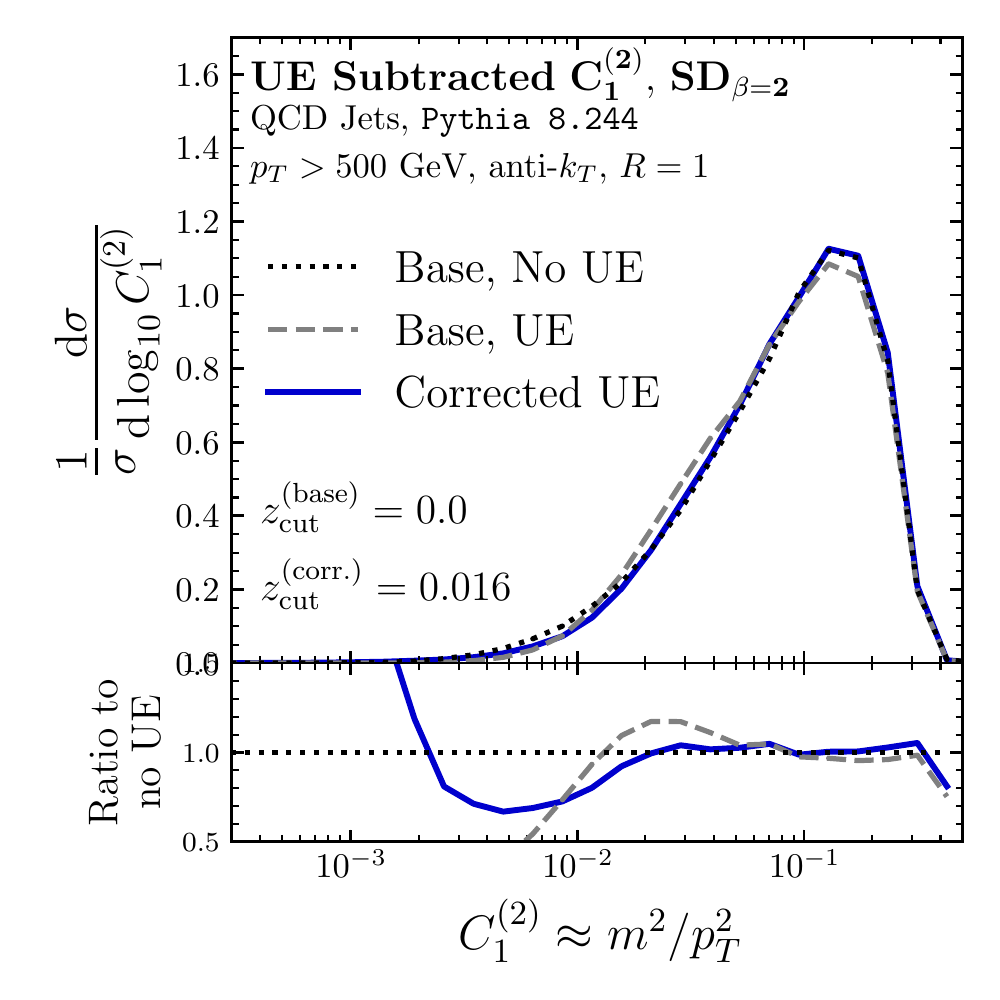}
\label{fig:sd_ue_0}
    }
    \subfloat[][]{
        \includegraphics[width=.45\textwidth]{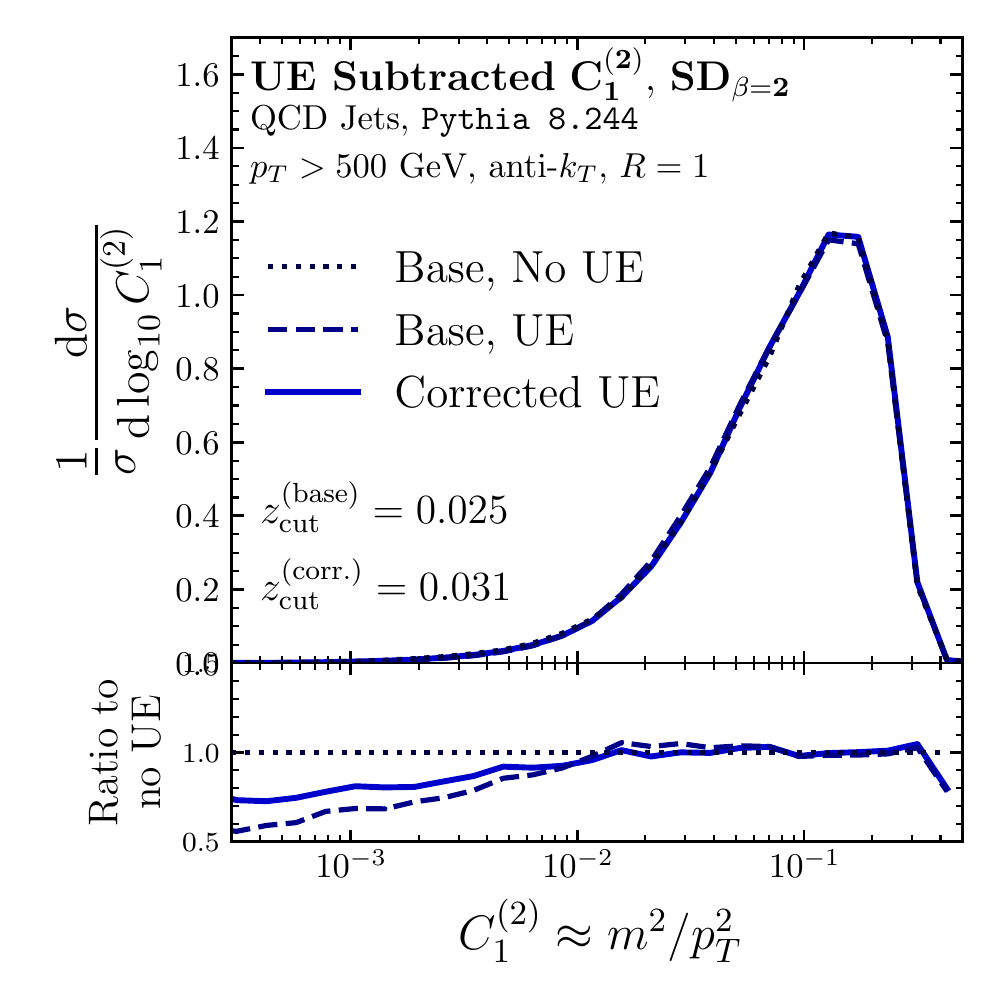}
\label{fig:sd_ue_025}
    }
\caption{
The response of groomed substructure distributions of QCD jets to the underlying event (UE), using \PRSF{1/2} (top row) and Soft Drop with \(\beta_{\rm SD} = 2\) (bottom row).
We examine the robustness of groomed substructure to the presence of UE by comparing ``Base'' substructure distributions with and without UE for fixed \(\zcut = z^{(\rm base)}_{\rm cut}\).
We also examine the possibility of adding \textit{additional} grooming to remove the effects of UE, picking a \(z^{(\rm corr.)}_{\rm cut}\) that removes UE to most accurately reproduce substructure distributions using \(\zcut = z^{(\rm base)}_{\rm cut}\) in the absence of UE.
The left column shows the substructure distributions for \(z^{(\rm base)}_{\rm cut} = 0\), while the right column shows the substructure distributions for \(z^{(\rm base)}_{\rm cut} = 0.025\).
}
\label{fig:ue}
\end{figure}

\Fig{ue} compares base distributions to UE-corrected distributions for \PRSF{1/2} and Soft Drop.
%
\Figs{rss_ue_0}{sd_ue_0} examine the ability of the groomers to reproduce ungroomed base distributions, with \(\zcut^{(\rm base)} = 0\).
\Figs{rss_ue_025}{sd_ue_025} examine instead groomed base distributions with \(\zcut^{(\rm base)} = 0.025\).

\begin{figure}[]
\centerline{
    \subfloat[][]{
        \includegraphics[width=.48\textwidth]{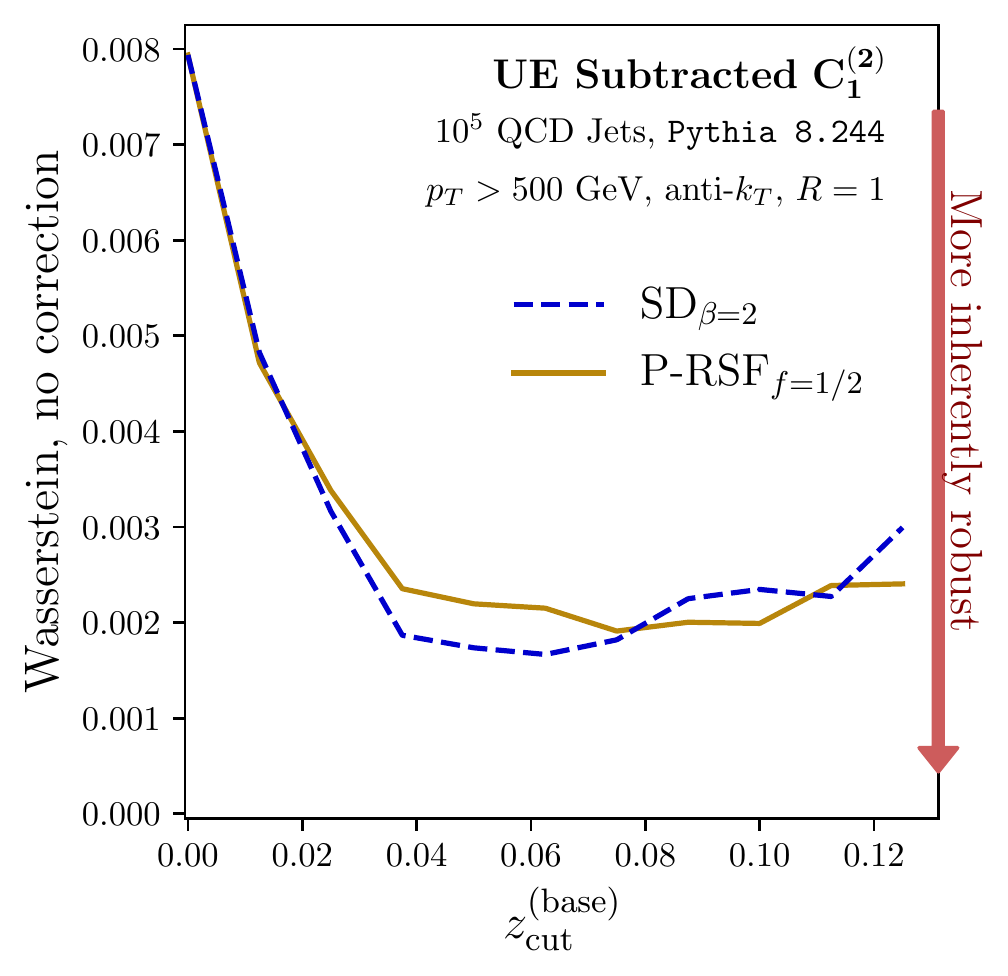}
\label{fig:ue_wasser_nocorrection}
    }
    \subfloat[][]{
        \includegraphics[width=.48\textwidth]{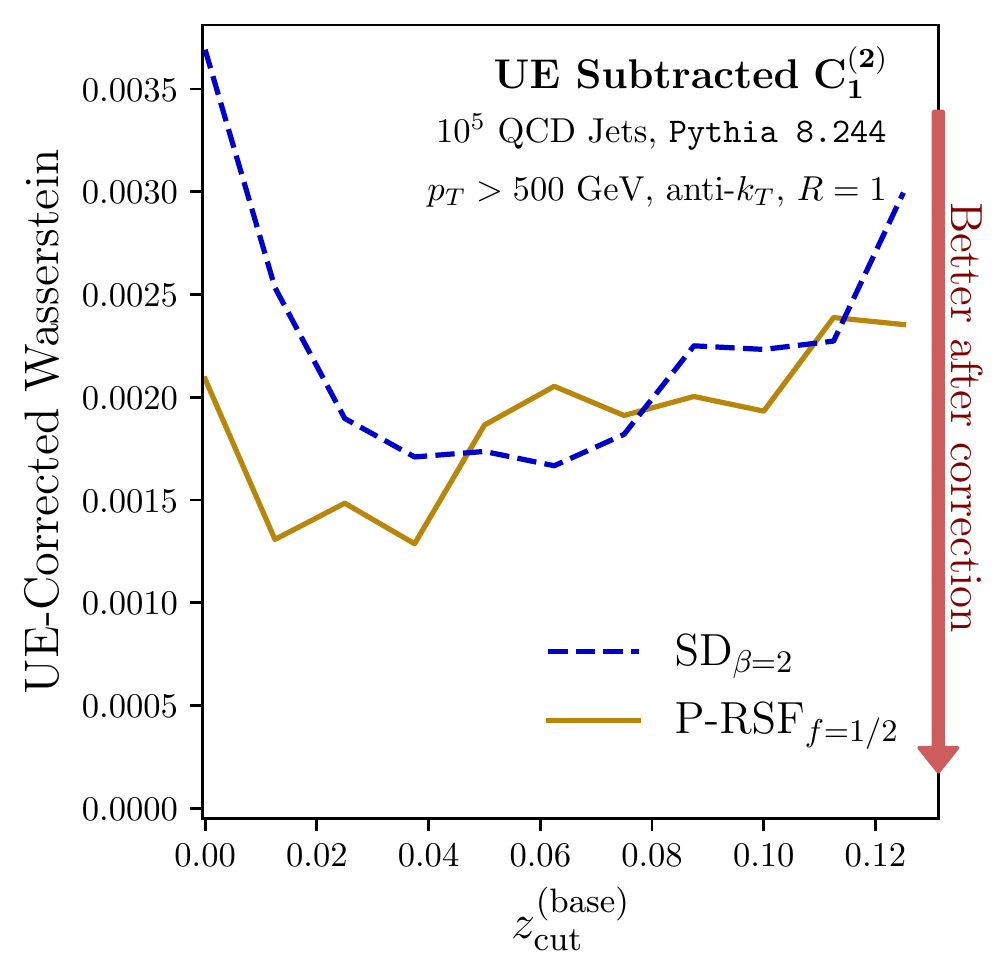}
\label{fig:ue_wasser_uecorrection}
    }
}
\caption{
    The behavior of the Wasserstein distance, in arbitrary units, between the \(C_1^{(2)}\) distribution: (a) before UE correction, and (b) after UE correction, for \SD{2} and \PRSF{1/2} as a function of the base amount of grooming \(\zcut^{(\rm base)}\).
    In (b), a lower UE-corrected Wasserstein distance quantitatively indicates better correction against the effects of UE:
    the substructure distribution of UE-corrected jets more closely resembles the substructure distribution of jets without UE.
}
\label{fig:ue_2}
\centering
\end{figure}

\Figs{ue_wasser_nocorrection}{ue_wasser_uecorrection} quantify the effectiveness and robustness of \PRSF{1/2} and Soft Drop as a function of \(\zcut^{(\rm base)}\) by studying the un-corrected Wasserstein distance and the UE-corrected Wasserstein distance, respectively.
\Fig{ue_wasser_uecorrection} shows the UE-corrected Wasserstein distance as a function of \zcut, and suggests that \PRSF{1/2} is better at correcting for UE when \(\zcut^{(\rm base)} \lesssim .05\), and that \SD{0} and \PRSF{1/2} have similar UE correction ability for values of \(\zcut^{(\rm base)}\) between 0.05 and .1.

We note that we expect hard-cutoff groomers to be generically more \textit{robust} to the presence of UE because even at fixed \zcut, they can remove an arbitrary amount of energy due to soft, wide-angle radiation in an event.
We correspondingly expect that hard-cutoff groomers require less additional grooming to remove the effects of UE.
\PIRANHA{} groomers, on the other hand, cannot remove arbitrary amounts of energy from an event;
we expect that we must always slightly increase the strength of \PIRANHA{} grooming to soak up additive contamination from UE.
We demonstrate that hard-cutoff groomers indeed require less additional \(\delta\zcut\) for optimal UE correction in \Fig{uefrenzy} of \App{uefrenzy}.

    We also point out that Recursive Subtraction techniques that are more geared towards the removal of the soft and wide-angle radiation which is characteristic of UE may perform even better in UE correction.
    In hard-cutoff grooming, for example, Soft Drop with \(\beta_{\rm SD} > 0\) preferentially grooms wide-angle radiation, Soft Drop with positive \(\beta_{\rm SD}\) is particularly well suited to remove UE (see, for example, \Fig{uefrenzy} of \App{uefrenzy}).
   Perhaps P-RS grooming procedures that upgrade the parameter \(f_{\rm soft}\) of P-RSF into a function of the soft energy fraction \(z\) and the splitting angle \(\theta\) of a branch  (such as Hard-Balanced P-RSF, as in \Sec{rsf_discont}) may be more suited to the removal of soft wide-angle radiation from UE.

   We have argued that \PRSF{1/2} and \SD{2} are comparable tools for the subtraction of UE effects in jet substructure, though \PRSF{1/2} requires more tuning to achieve the same level of UE correction.
   We emphasize that further investigation of such P-RS algorithms might improve the ability of \PIRANHA{} in the tagging of boosted objects and beyond.

\subsection{Case Study: Mass Resolution for \(W\) and Top Jets}
\label{sec:massres}

    As a final case study, we explore the use of \PIRANHA{} grooming strategies in the tagging of boosted objects, another useful application of traditional grooming strategies such as Soft Drop.
   We must once again overcome the subtlety that \PIRANHA{} grooming strategies remove energy proportional to the grooming parameter (such as \(\zcut\) in Recursive Safe Subtraction).
   To successfully tag boosted objects decaying into jets, we must tune the grooming parameter for \PIRANHA{} strategies more than we would for Soft Drop.
   Nonetheless, we examine the mass resolution of \PIRANHA{}-groomed jets from the decays of boosted \(W\) bosons and top quarks and discover that \PIRANHA{} may offer some advantages that are complementary to those of traditional grooming strategies.
   The tagging procedure we implement in the following discussion is quite simple, however, and the applications of \PIRANHA{} in boosted object tagging should be subjected to a more complete analysis.

\begin{figure}[p]
\centering
\subfloat[][]{
\includegraphics[width=.45\textwidth]{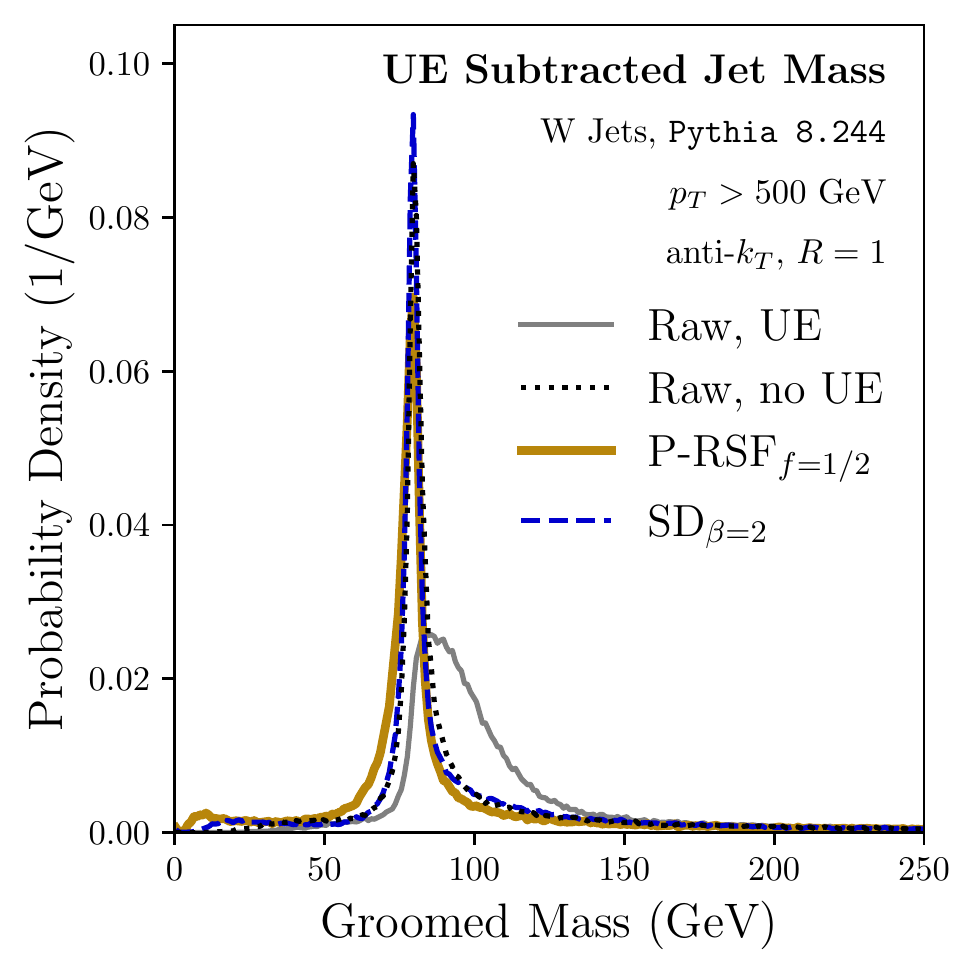}
\label{fig:w_tagging_dist}
}
\subfloat[][]{
\includegraphics[width=.45\textwidth]{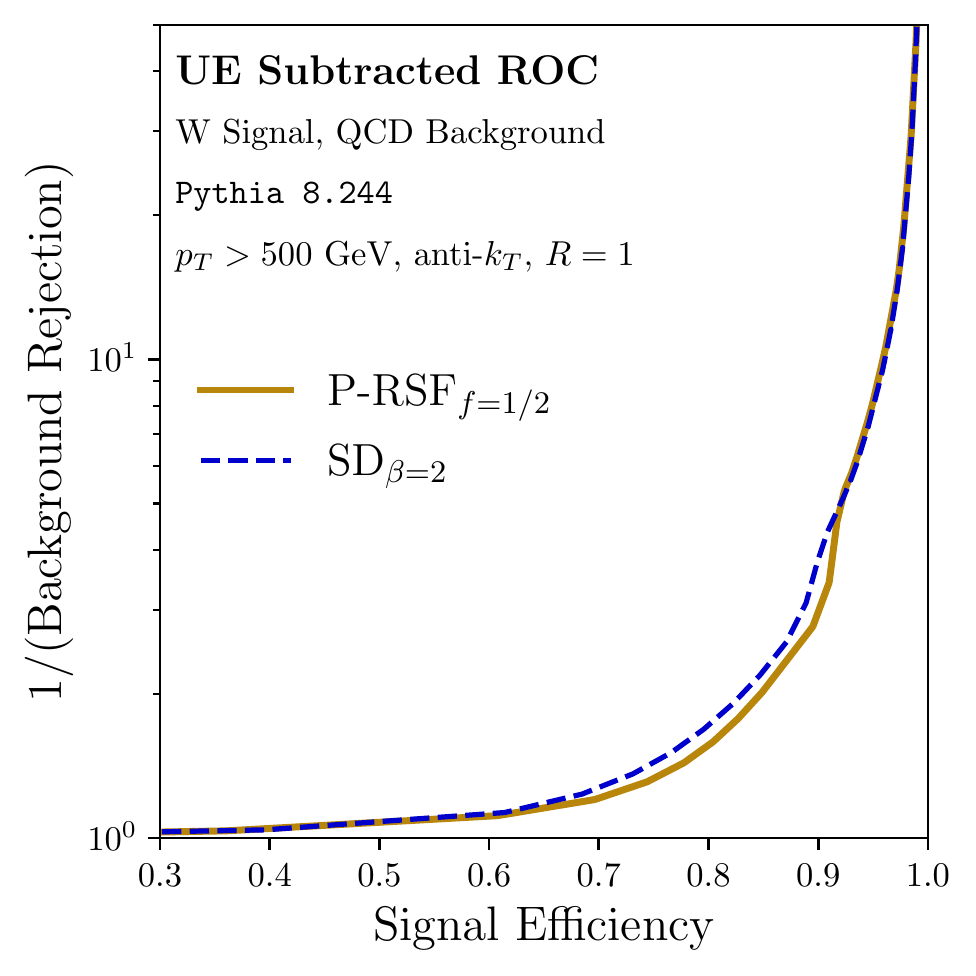}
\label{fig:w_tagging_roc}
}
\\
\subfloat[][]{
\includegraphics[width=.45\textwidth]{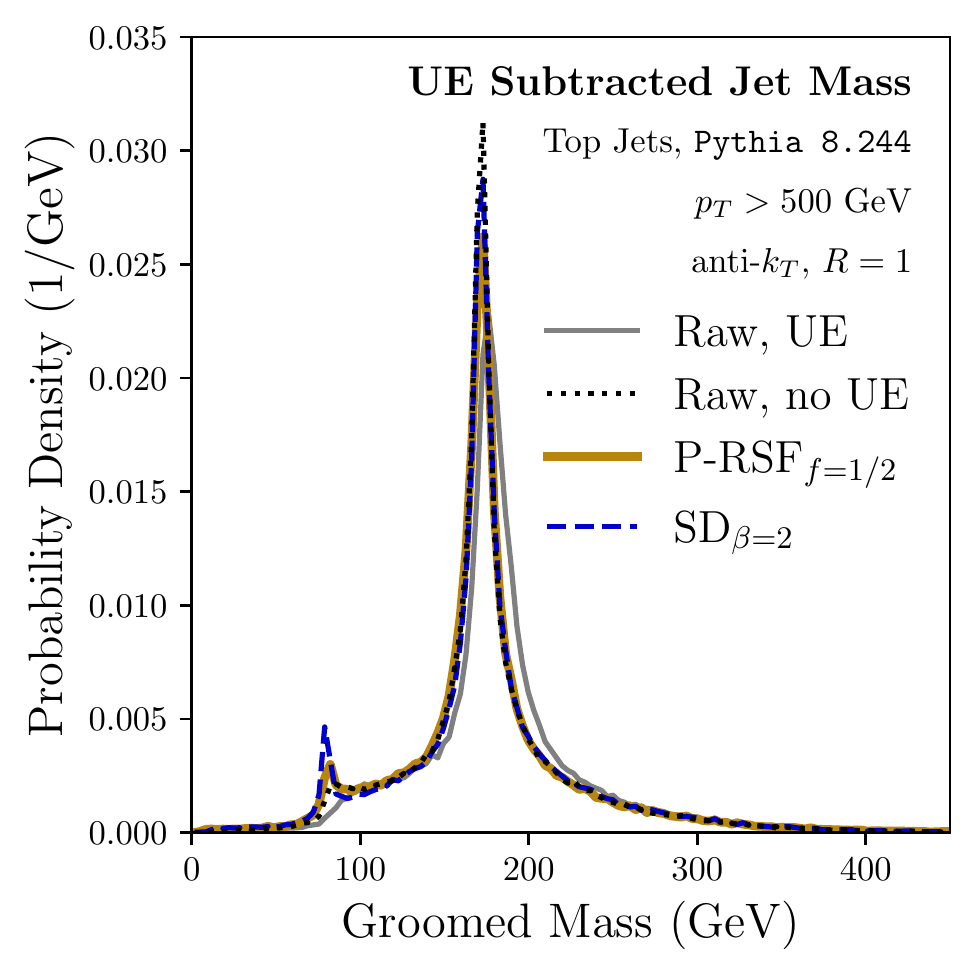}
\label{fig:top_tagging_dist}
}
\subfloat[][]{
\includegraphics[width=.45\textwidth]{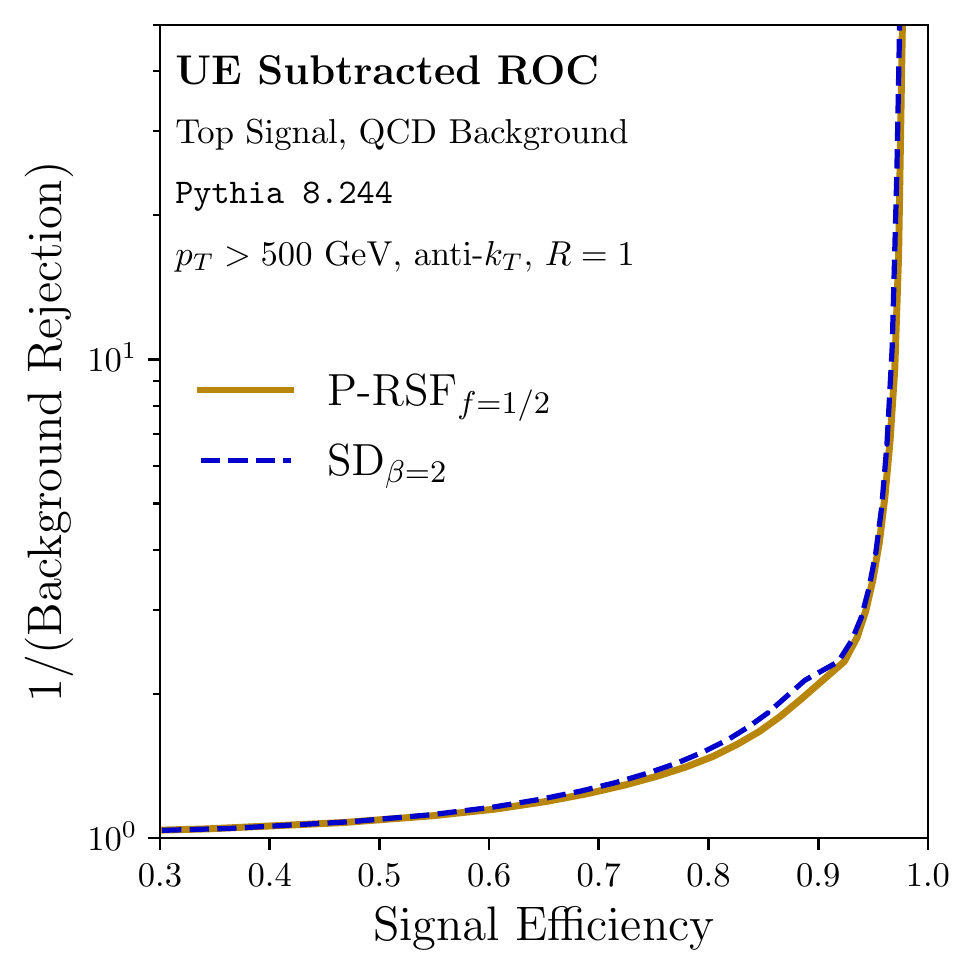}
\label{fig:top_tagging_roc}
}
\caption{
     A comparison of traditional and \PIRANHA{} grooming procedures in the tagging of (top row) boosted \(W\) bosons and (bottom row) top quarks.
      The data in the plots correspond to 100,000 jet events with \(p_T > 500\) GeV in \texttt{Pythia 8.244}.
      The left column displays the mass distributions of \(W\) and top jets with and without UE, and with UE groomed away using either Soft Drop with \(\beta_{\rm SD} = 2\) or \PRSF{1/2} with additional post-processing discussed in the text.
      The right column displays the relationship between background rejection and signal efficiency associated with \(W\) and top tagging, respectively, produced by considering symmetric windows around the mass of the relevant boosted object and comparing with a QCD background.
      }
\label{fig:tagging}
\end{figure}

   In \Fig{tagging}, we assess the behavior of \PIRANHA{} in tagging boosted objects.
   We focus on jets of \(p_T \geq 500\) GeV produced from the decays of \(W\) bosons and top quarks in \Figs{w_tagging_dist}{top_tagging_dist}, respectively.
   We then plot the associated background rejection as a function of signal efficiency in \Figs{w_tagging_roc}{top_tagging_roc}.

To produce the groomed mass distributions of \Figs{w_tagging_dist}{top_tagging_dist}, we groom jets in the presence of UE.
 Afterward, we rescale the mass of the jet to correct for the shift of the mass peak due to the removal of energy from the jet during grooming.
   This rescaling procedure is necessary to reproduce the correct mass peak for \PRSF{1/2};
   however, rescaling Soft Drop groomed mass distributions has a minimal effect since Soft Drop tends to remove wide-angle soft radiation without severely affecting radiation from the hard event.

   \Figs{w_tagging_dist}{top_tagging_dist} show rescaled mass distributions for each grooming algorithm for the value of \zcut{} that most precisely removes UE, in the sense that it most closely reproduces the mass distribution of jets without UE.
   We find that the best performance is achieved by \(\zcut^{(\text{P-RSF}_{1/2})}=0.034\) and \(\zcut^{(\text{SD}_2)}=0.025\) when grooming W jets, and \(\zcut^{(\text{P-RSF}_{1/2})}=0.023\) and \(\zcut^{(\text{SD}_2)}=0.035\) when grooming top jets.
   In each case, Soft Drop does not require any additional rescaling, while we find the best performance if we rescale the energy of \PRSF{1/2} groomed jet constituents by about \(+4\%\).

   In this simple context, \PRSF{1/2} offers slightly better background rejection for reasonable signal efficiencies.
   We evince this in \Figs{w_tagging_roc}{top_tagging_roc}, where we show the signal efficiency versus background rejection curves associated with comparing signal W or top jets to QCD background, again with \(p_T \geq 500\) GeV.
   Our tagging procedure is minimal:
   we accept events within a symmetric window of width \(\delta m\) around the mass of the \(W\) boson or top quark, respectively, and \Figs{w_tagging_roc}{top_tagging_roc} are produced by varying the window size \(\delta m\).
   For reasonable values of signal efficiency, between around 60\% and 90\%, the \PRSF{1/2} algorithm gives greater background rejection than Soft Drop.

   Again, the results in \Figs{w_tagging_dist}{top_tagging_dist} and \Figs{w_tagging_roc}{top_tagging_roc} suggest that \PRSF{1/2} and \SD{2} behave comparably as tools to remove the effects of UE in jet mass distributions, and to tag underlying boosted object.
   These results are preliminary, and to state these conclusions with more confidence will require more detailed tagging procedures than the simple algorithm described above.
   We leave a more detailed study of the tagging potential of \PRSF{1/2} and \PIRANHA{} to future work.

\section{Discussion and Conclusions}
\label{sec:Conclusions}

In this work, we proposed the paradigm of Pileup and Infrared Radiation Annihilation (\PIRANHA{}) for continuous jet grooming.
Motivated by optimal transport theory and the Energy Mover's Distance (EMD) of \Reff{Komiske:2019fks}, we re-framed the Apollonius Subtraction (P-AS) and Iterated Voronoi Subtraction (P-IVS) procedures of \Reff{Komiske:2020qhg} as implementations of the \PIRANHA{} paradigm.
We additionally introduced Recursive Subtraction with a Fraction (P-RSF) as a family of new groomers motivated by the \PIRANHA{} paradigm.
We showed that a particular Recursive Subtractor, \PRSF{1/2}, overcomes the soft discontinuities of traditional hard-cutoff grooming procedures, and that general P-RSF algorithms only have soft discontinuities in suppressed regions of phase space.
We highlighted the unprecedented robustness of \PRSF{1/2} to hadronization, detector effects, and pileup.
We showed also that \PRSF{1/2} may be able to correct for the presence of the underlying event.
Though hard-cutoff groomers may be more robust against effects from UE without additional tuning, \PIRANHA{} groomers can be tuned to remove additive contamination from the underlying event.
We used the example of additive contamination from UE to argue that \PIRANHA{} may also have applications in the tagging of boosted objects.

   There are several immediately evident avenues for future phenomenological and theoretical exploration.
    While we argued that P-RSF had more robust responses to soft distortions, it will be interesting to quantify this including theoretical, model-dependent uncertainties.
   For example, it will be interesting to study the robustness of \PRSF{1/2} to hadronization when using different models of color recombination, with different \texttt{Pythia} tunes, and from the perspective of effective field theory as in \Reff{Hoang:2019ceu}.
   Similarly, the effects of experimental detectors on \PRSF{1/2} groomed quantities must be explored with more realistic models of detector responses and compared in more detail with traditional grooming techniques.
   More detailed studies of these model-dependent uncertainties may facilitate a more precise, and even process-dependent method to tune the parameters of \PIRANHA{} groomers such as P-RSF to the removal of PU and UE in more realistic scenarios.

   Designing and studying variants of P-RSF in which the amount of grooming for a particular emission depends on the angle of the emission, as in Soft Drop with \(\beta_{\rm SD} \neq 0\), and even the energy fraction, may be another easy way to improve the robustness and precision of P-RS algorithms in removing specific models of soft contamination.
   Indeed, any method that varies \(f_{\rm soft}\), \(\zcut\), \(\rho\), or other parameters of the jet grooming procedure in a way that depends on more detailed information within the jet could be useful in more precise removal of contaminating radiation, such as the removal of PU, UE, or the thermal noise that contaminates jets originating in heavy ion collisions.
   
       A first-principles calculation of Recursive Subtraction groomed jet observables will rely on analytic exploration of the elaborately correlated emissions of P-RS groomed jets, as well as contributions from non-global configurations.
    We begin this journey in \App{calc}, but further exploration may depend on a more precise application of existing tools for jet substructure or even on newer tools in perturbative QCD, such as the techniques of \Reff{Larkoski:2015zka}.
    
   The study of EMD-mode \PIRANHA{} grooming, introduced in \App{grooming_in_emd_mode}, is another interesting avenue for the development of continuous grooming techniques.
   Unlike the \PIRANHA{} groomers explored in the main text, EMD-mode \PIRANHA{} groomers have the potential to remove an arbitrary amount of \(p_T\) from an event.
   We look forward to future studies of EMD-mode grooming for its potential to address stochastic fluctuations in the levels of jet contamination.
   
   Finally, the use of alternative methods for optimal transport, such as the EMD without the restriction of \(\beta = 1\) that we chose in this work, or the flexible and computationally efficient formalism of linearized optimal transport \cite{Cai:2020vzx,Cai:2021hnn,cai2022linearized,sarrazin2023linearized}, may offer new tools and additional insights into continuous grooming.
   
The \PIRANHA{} grooming strategy has intricate geometric origins, but its goal is simple: the optimal and continuous removal of contaminating low-energy radiation.
We hope that the simple strategies and examples of continuous grooming discussed in this work may provide and inspire new tools for clear communication between experimental results and theoretical predictions regarding our microscopic universe.

\acknowledgments

The authors would like to express their gratitude to Matthew LeBlanc, Jennifer Roloff, and the ATLAS Jet Definitions subgroup for valuable discussions, and to Nima Zardoshti for raising insightful questions during LHCP 2022.
We would also like to thank Pier Monni for helpful discussions regarding resummation, and Sean Benevedes and Rikab Gambhir for engaging conversations and careful readings of the manuscript.
Finally, we extend our appreciation to an anonymous referee whose thoughtful comments and feedback contributed to improving the quality and clarity of this work and especially of \Sec{traditionaldiscont}.
This work was supported by the Office of Nuclear Physics of the U.S. Department of Energy (DOE) under grant DE-SC-0011090, by the DOE Office of High Energy Physics under grants DE-SC0012567 and DE-SC0019128, and by the National Science Foundation under Cooperative Agreement PHY-2019786 (The NSF AI Institute for Artificial Intelligence and Fundamental Interactions, \url{http://iaifi.org}).

\appendix


\section{Grooming in EMD Mode}
\label{app:grooming_in_emd_mode}

In this appendix, we describe another route for developing \PIRANHA{} groomers:
\textit{EMD mode}.
EMD-mode grooming produces a groomed event with a fixed EMD relative to its ungroomed counterpart, and furnishes a complementary approach to ``\(p_T\)-mode'' groomers.
We provide a conceptual introduction to EMD-mode grooming below, leaving an exploration of the phenomenological implications of EMD mode to future work.

The \PIRANHA{} groomers described above all subtract a fixed amount of transverse momentum from the energy flow of an event.
Let us call them \textit{\(p_T\)-mode groomers}.
\(p_T\)-mode grooming gives us a great deal of control over the amount of energy we remove from an event.
When using \PIRANHA{} groomers in \(p_T\) mode to remove additive contamination, however, this precise control over \(\Delta p_T\) leads to additional complications;
since additive contamination may add arbitrary amounts of contaminating soft radiation to an event, removing a fixed amount of \(p_T\) from the contaminated event may not be the best strategy for reproducing the un-contaminated event.
These complications lead to important considerations when using \PIRANHA{} for pileup mitigation, as in \Sec{pileup} and \App{pufrenzy}, and they are especially disadvantageous when using \PIRANHA{} to correct for the presence of the underlying event, as in \Sec{ue} and \App{uefrenzy}.

Grooming in EMD mode is a complementary approach that allows \PIRANHA{} groomers to remove arbitrary amounts of soft radiation from an event.
\PIRANHA{} in EMD mode therefore furnishes a conceptually interesting alternative for continuous grooming.

First, let us briefly review how the \(p_T\)-mode algorithms we have presented for P-AS, P-IVS, and P-RS all subtract a fixed amount of transverse momentum, \(\Delta p_T = \rho A_{\rm tot}\), or \(\Delta p_T = \zcut p_T\), from the event under consideration.
\begin{itemize}
\item
P-AS subtracts the pre-specified \(p_T\) all at once by finding the Apollonius regions associated with the event, as in \Sec{as}.
\item
P-IVS subtracts the pre-specified \(p_T\) by subtracting \(p_T\) from each particle in the event until it removes a particle, and then continues recursively until the specified \(p_T\) has been removed, as in \Sec{ivs}.
\item
P-RS subtracts the pre-specified \(p_T\) by recursively assigning fractions of the total \(p_T\) to be removed to each branch of the jet tree until it reaches the final-state particles of the jet, as in \Sec{rsf}.
\end{itemize}

The approach for EMD-mode grooming is quite similar.
In EMD mode, however, each algorithm continues until a specific amount of EMD has been subtracted from the event, often in an iterative procedure.
As a concrete example, let us construct the algorithm for P-IVS in EMD mode, or P-IVS\(^{(\rm EMD)}\):
\begin{enumerate}
    \item
    P-IVS\(^{(\rm EMD)}\) finds the Voronoi diagram for the event, as discussed in \Sec{ivs}, and indexes the steps of the algorithm by \(n\), starting at \(n=1\).
    We define the energy flow of the algorithm after step \(n\) is completed by \(\mathcal{E}_n\), where \(\mathcal{E}_0\) is the energy flow of the original event.
    We also define the EMD between \(\mathcal{E}_{n-1}\) and \(\mathcal{E}_n\) as \({\rm EMD}_n\).

    \item
    P-IVS\(^{(\rm EMD)}\) attempts to modify the \(p_T\) of every particle in the event in the same way as P-IVS in \(p_T\)-mode:
    \begin{align}
	p_{T\,i}^{(n)} = p_{T\,i}^{(n-1)} - \rho^{(n)} A^{(n-1)}_i,
	\label{eqn:ivs_emd}
    \end{align}
    where the \(p_{T\,i}^{(n-1)}\) and \(A^{(n-1)}\) are the transverse momenta/Voronoi areas after step \(n-1\), and
    \begin{align}
        \rho^{(n)}
        =
        \min_i p^{(n)}_{T\,i}/A^{(n)}_i
        ,
    \end{align}
    in analogy with \Eq{ivs2}.
    As in \Sec{ivs}, \(\rho^{(n)}\) describes the maximum \(p_T\) that may be subtracted from an event before one of the particles in the event is groomed away entirely.

    \item
    P-IVS\(^{(\rm EMD)}\) next calculates \({\rm EMD}_n\) for this proposed groomed event, and asks whether it has subtracted a total EMD of \( {\rm EMD}_{\rm cut}\) from the event.
    If \(\sum_{k = 1}^n {\rm EMD}_n < {\rm EMD}_{\rm cut}\), P-IVS\(^{(\rm EMD)}\) has not yet subtracted the full \({\rm EMD}_{\rm cut}\) from the event.
    In this case, we have more grooming to do:
    P-IVS\(^{(\rm EMD)}\) continues recursively, going back to the first step of the algorithm.

    \item
    If \(\sum_{k = 1}^n {\rm EMD}_k > {\rm EMD}_{\rm cut}\), P-IVS\(^{(\rm EMD)}\) has subtracted too much EMD from the event.
    In this case, P-IVS\(^{(\rm EMD)}\) revises its proposed groomed event at this step in the algorithm by evaluating the value of \(\rho^{(n)}\) that would give \(\sum_{k = 1}^n {\rm EMD}_k > {\rm EMD}_{\rm cut}\), and finding the associated groomed event by using \Eq{ivs_emd}.
    It returns the resulting event as the final groomed event.
\end{enumerate}
As in P-IVS, P-IVS\(^{(\rm EMD)}\) does not need to re-compute the Voronoi diagram from scratch at each step of the algorithm.
Similarly, our procedure for finding the value of \(\rho^{(n)}\) that subtracts the correct amount of total EMD in the final step of the algorithm is computationally efficient, and does not rely on a complete re-evaluation of the EMD at each step of the algorithm.

At present, we do not have an implementation of P-RS in EMD mode.
That said, we expect that the generalization of EMD-mode grooming to P-RS and P-RSF will be similar:
at step \(n\) of the EMD mode algorithm, we perform grooming with the maximum value of \(z_{\rm cut}\) until a particle would be removed, and define EMD\(_n\) as the EMD between the energy flows at steps \(n-1\) and \(n\) of the grooming.
If the sum of the subtracted EMDs is greater than EMD\(_{\rm cut}\), \(\sum_{k = 1}^n {\rm EMD}_k > {\rm EMD}_{\rm cut}\),
we must compute the value of \(z_{\rm cut}\) that fixes \({\rm EMD}_n > {\rm EMD}_{\rm cut} - \sum_{k = 1}^{n-1} {\rm EMD}_k \), groom with this value of \(z_{\rm cut}\), and return the resulting groomed event.
Otherwise, if \(\sum_{k = 1}^n {\rm EMD}_k < {\rm EMD}_{\rm cut}\), we would continue grooming recursively.

We hope that future exploration of EMD-mode grooming may offer additional utility in the use of continuous grooming in the subtraction of additive contamination, and in the mitigation of obfuscating radiation in general.


\section{Feeding Frenzy: Comparing a Plethora of Grooming Options}
\label{app:feedingfrenzy}
In this appendix, we present a brief collection of additional results regarding the responses of \PIRANHA{} and Soft Drop groomers to soft distortions and additive contamination.
We extend the results in the main text by comparing \PRSF{1/2}, the focus of our phenomenological studies in the main text, to three groups of grooming algorithms:
\begin{itemize}
    \item
    \textbf{Hard-Cutoff Groomers:}
    \SD{0}, \SD{1}, and \SD{2};

    \item
    \textbf{Fully Continuous Groomers:}
    P-AS and P-IVS;

    \item
    \textbf{Recursive Subtractors:}
    \PRSF{0}, \PRSF{1/2}, \PRSF{3/4}, and \PRSF{1}.
\end{itemize}
In our pileup studies, we also count Constituent Subtraction (CS) \cite{Berta:2014eza} among the continuous groomers due to its continuity properties in the continuum limit.
The comparisons of this appendix demonstrate that the conclusions we drew in the main text by comparing \PRSF{1/2} to \SD{2} seem to hold in greater generality:
\PIRANHA{} groomers tend to have more robust responses to soft distortions and additive contamination than traditional grooming methods.
Furthermore, the responses of \PRSF{1/2} are similar to the responses of the fully continuous groomers P-AS and P-IVS (and CS, in the case of pileup mitigation);
each of the \PIRANHA{} algorithms we explore in this appendix are available on GitHub \cite{piranhagithub}.

As in the main text, we first extend our results involving hadronization corrections (\Sec{hadronization}) and all-charged corrections (\Sec{all_v_charged}) to explore how several choices of grooming and grooming parameters may respond to soft distortions.
We then begin our discussion of additive contamination by providing more detail regarding our pileup studies:
we explain in greater detail our procedure for pileup mitigation used in \Sec{pileup} and extend our results to several groomers and values of \(\langle n_{\rm PU}\rangle\).
Finally, we similarly explore how different groomers respond to the underlying event, extending the results of \Sec{ue}.

\begin{figure}[p]
    \centering
    \subfloat[]{
        \includegraphics[width=.32\textwidth]{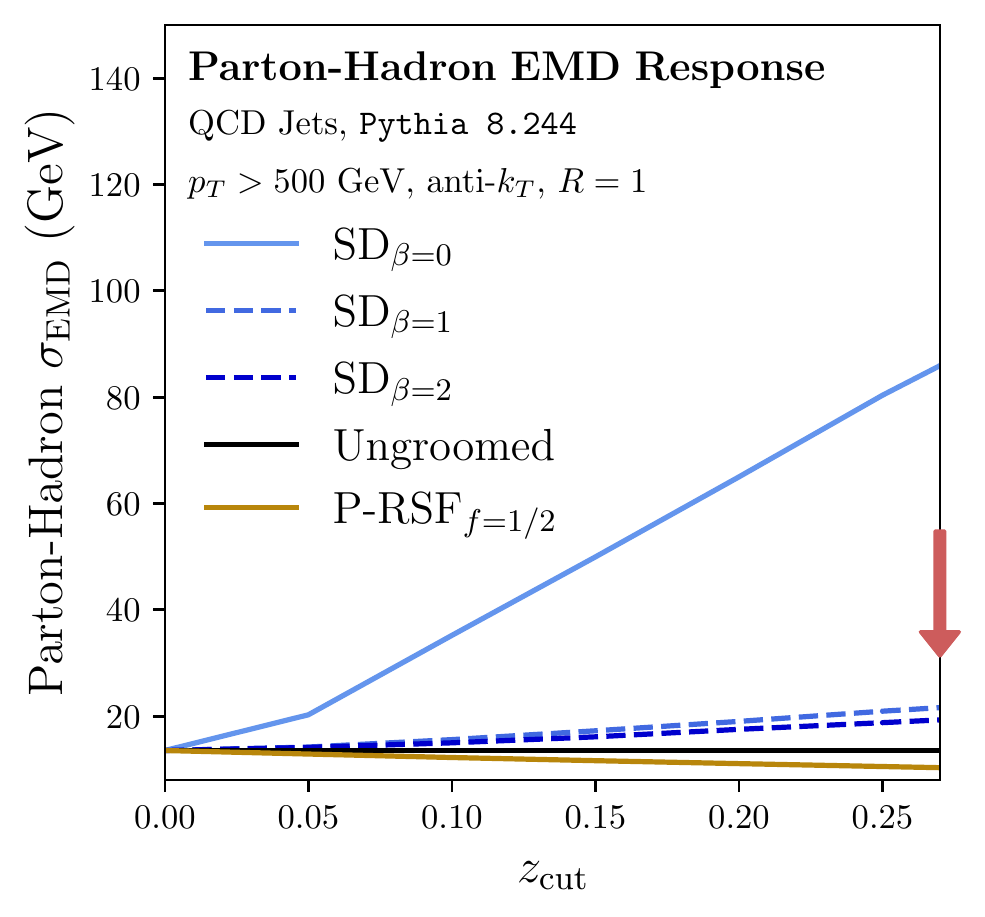}
    }
    \subfloat[]{
        \includegraphics[width=.32\textwidth]{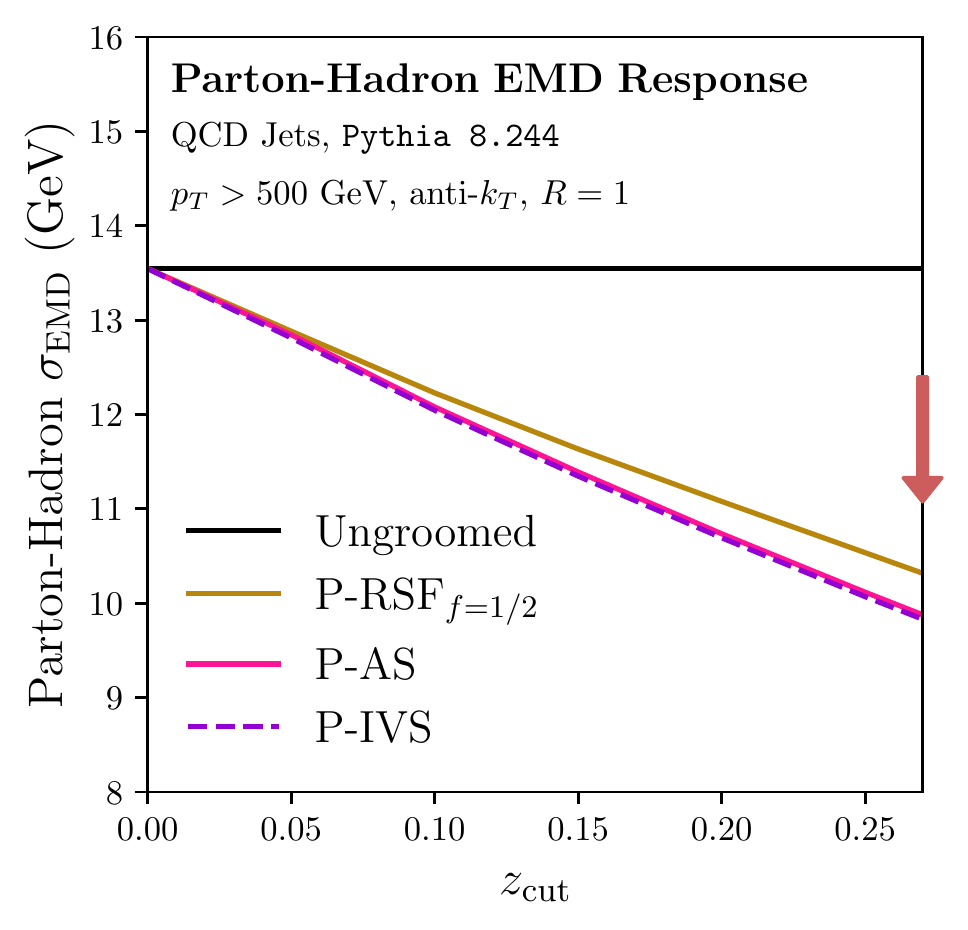}
    }
    \subfloat[]{
        \includegraphics[width=.32\textwidth]{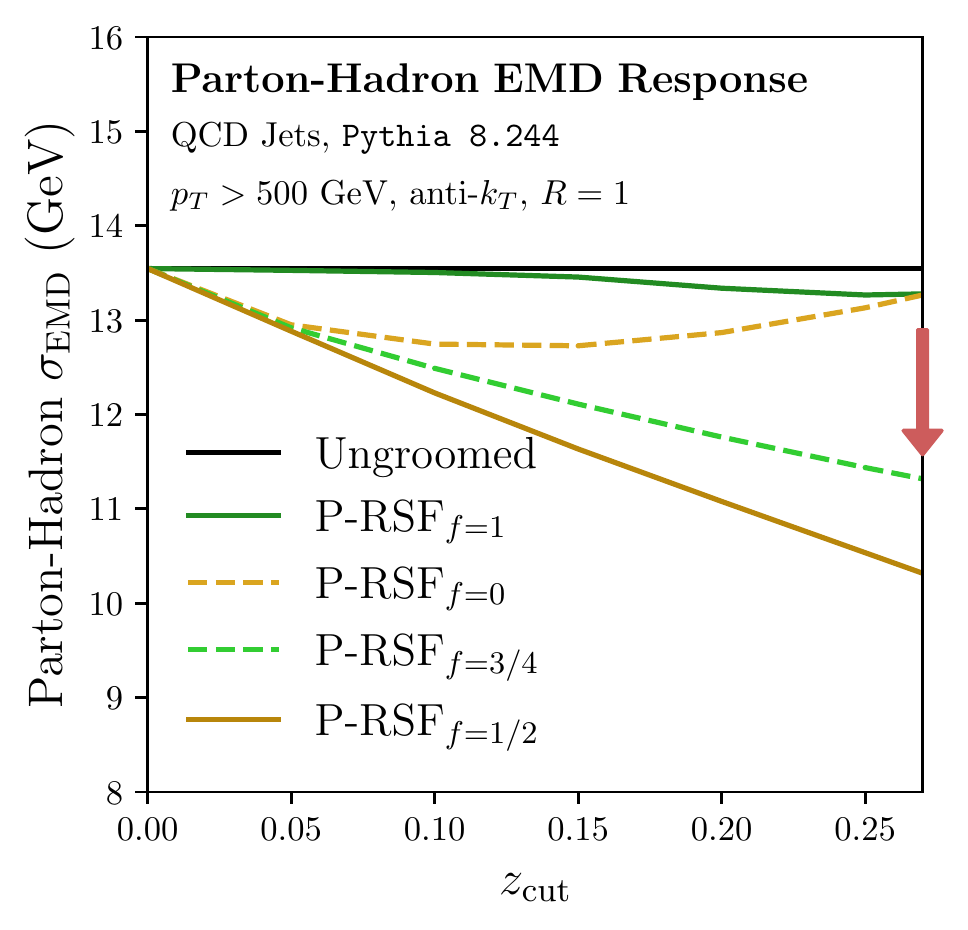}
    }
    \\
    \subfloat[]{
        \includegraphics[width=.32\textwidth]{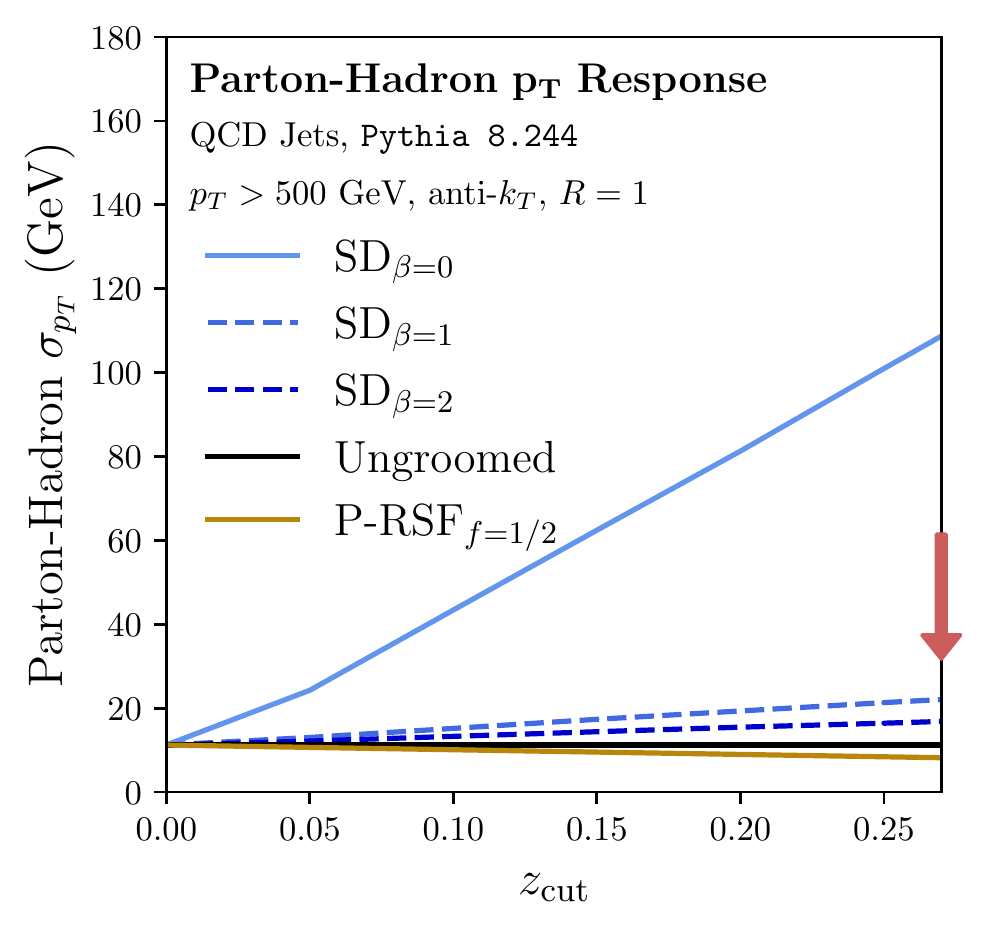}
    }
    \subfloat[]{
        \includegraphics[width=.32\textwidth]{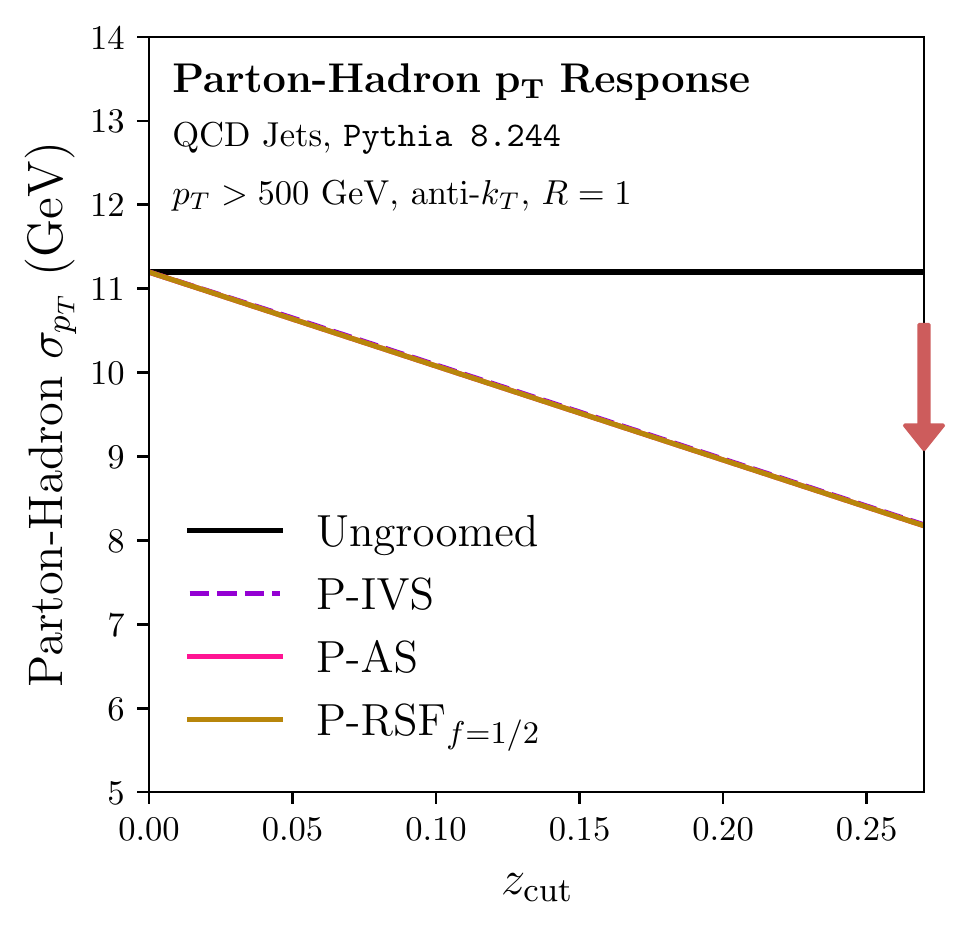}
    }
    \subfloat[]{
        \includegraphics[width=.32\textwidth]{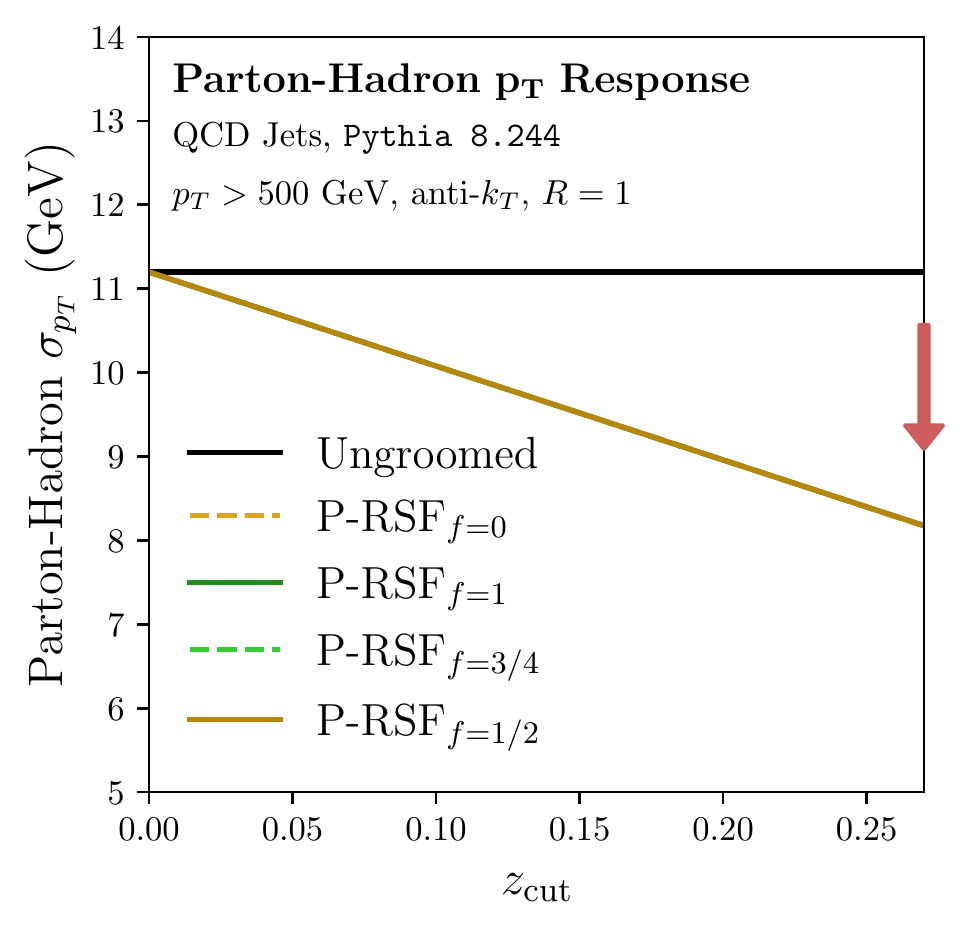}
    }
    \\
    \subfloat[]{
        \includegraphics[width=.32\textwidth]{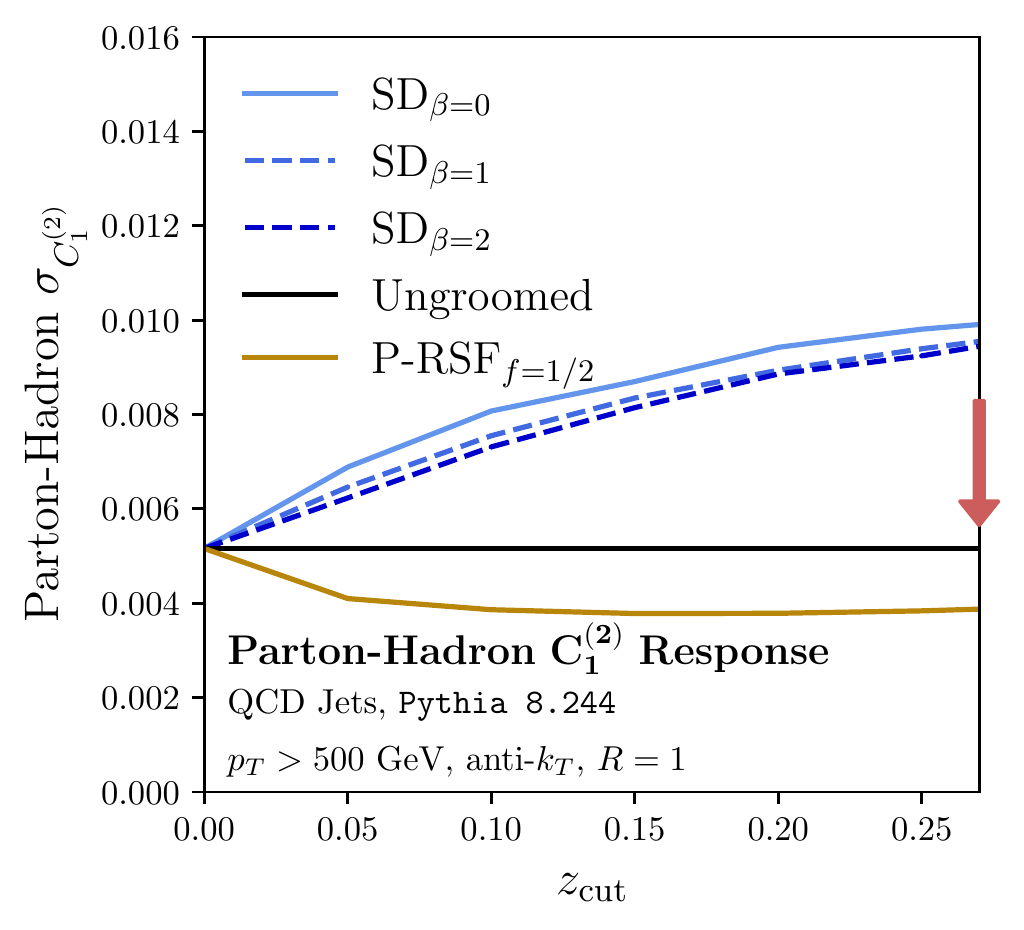}
    }
    \subfloat[]{
        \includegraphics[width=.32\textwidth]{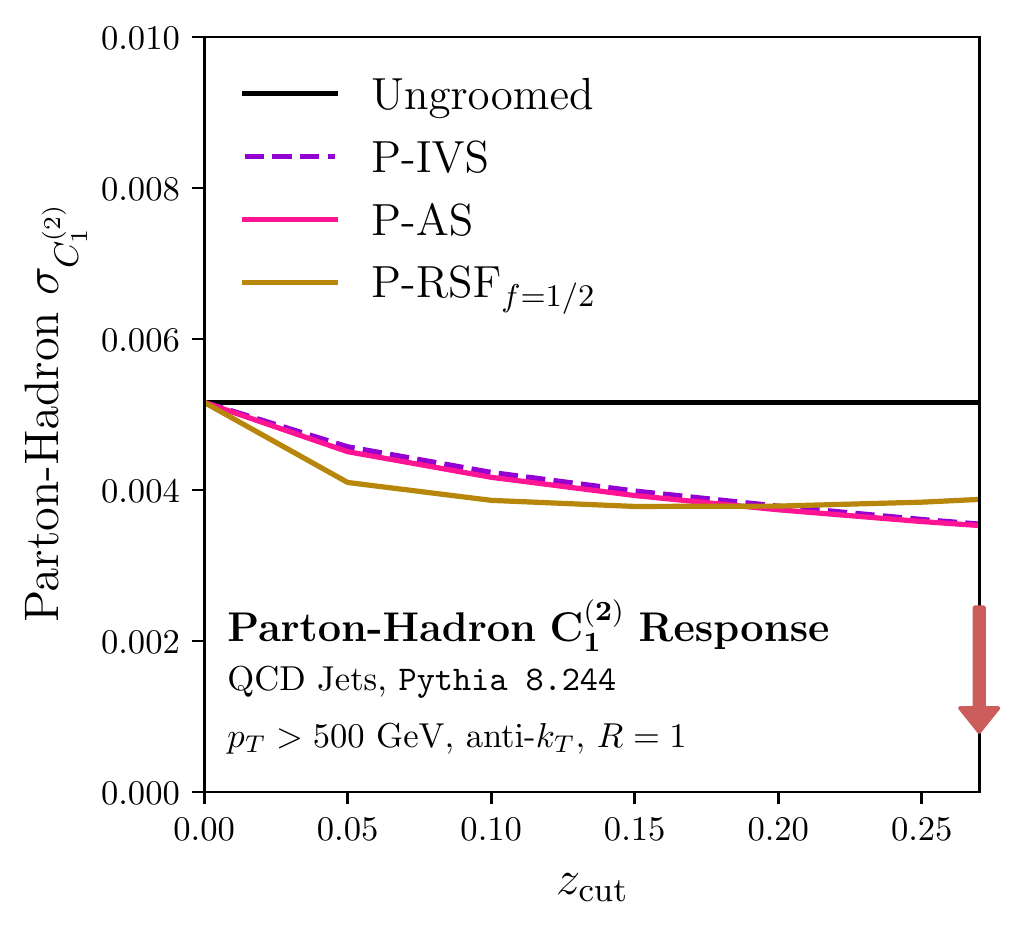}
    }
    \subfloat[]{
        \includegraphics[width=.32\textwidth]{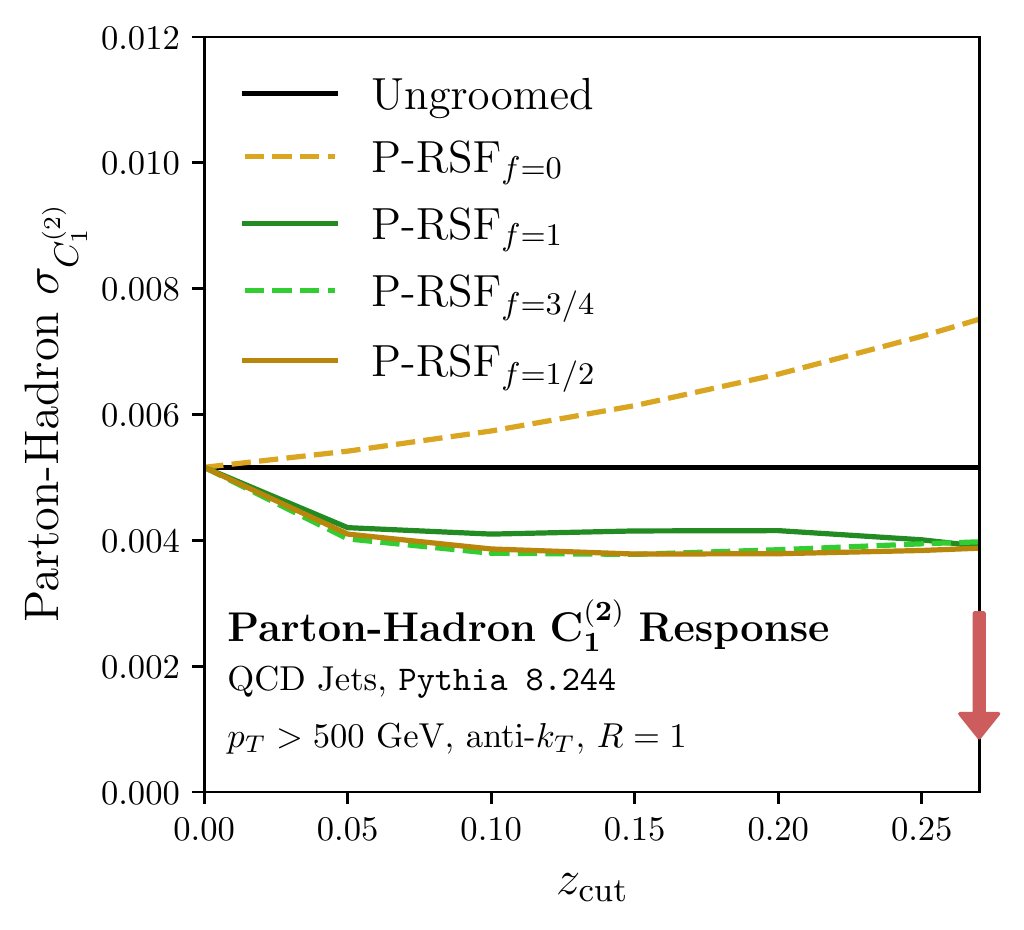}
    }
\caption{
    Per-jet hadronization responses of \PIRANHA{} and traditional groomers for (top row) EMD, (middle row) \(\Delta p_T\), and (bottom row) \(C_1^{(2)}\);
    we compare \PRSF{1/2} to (left column) hard-cutoff groomers, (middle column) fully continuous groomers, and (right column) recursive subtractors.
    For brevity in this appendix, we focus on the variance of the shifts in each observable due to hadronization in groomed jets.
    The red arrows indicate the direction corresponding to better performance.
}
\label{fig:pvhfrenzy}
\end{figure}

\begin{figure}[p]
    \subfloat[]{
        \includegraphics[width=.32\textwidth]{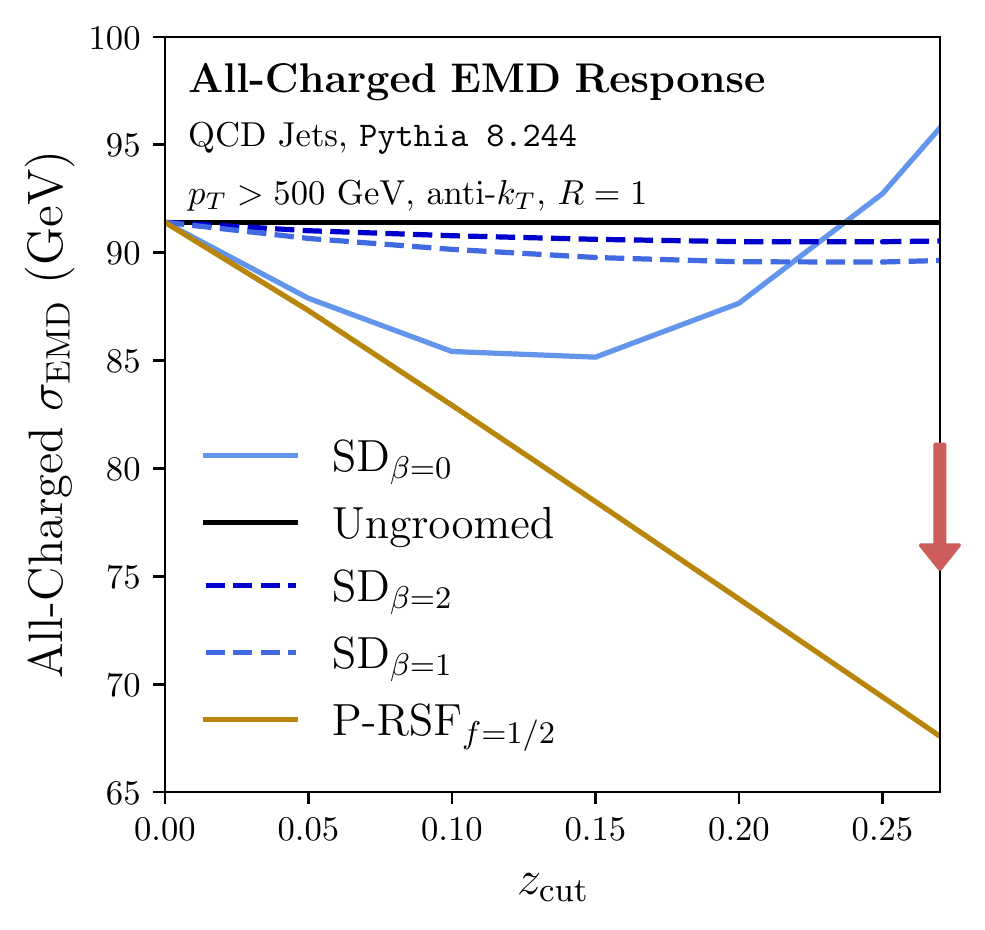}
    }
    \subfloat[]{
        \includegraphics[width=.32\textwidth]{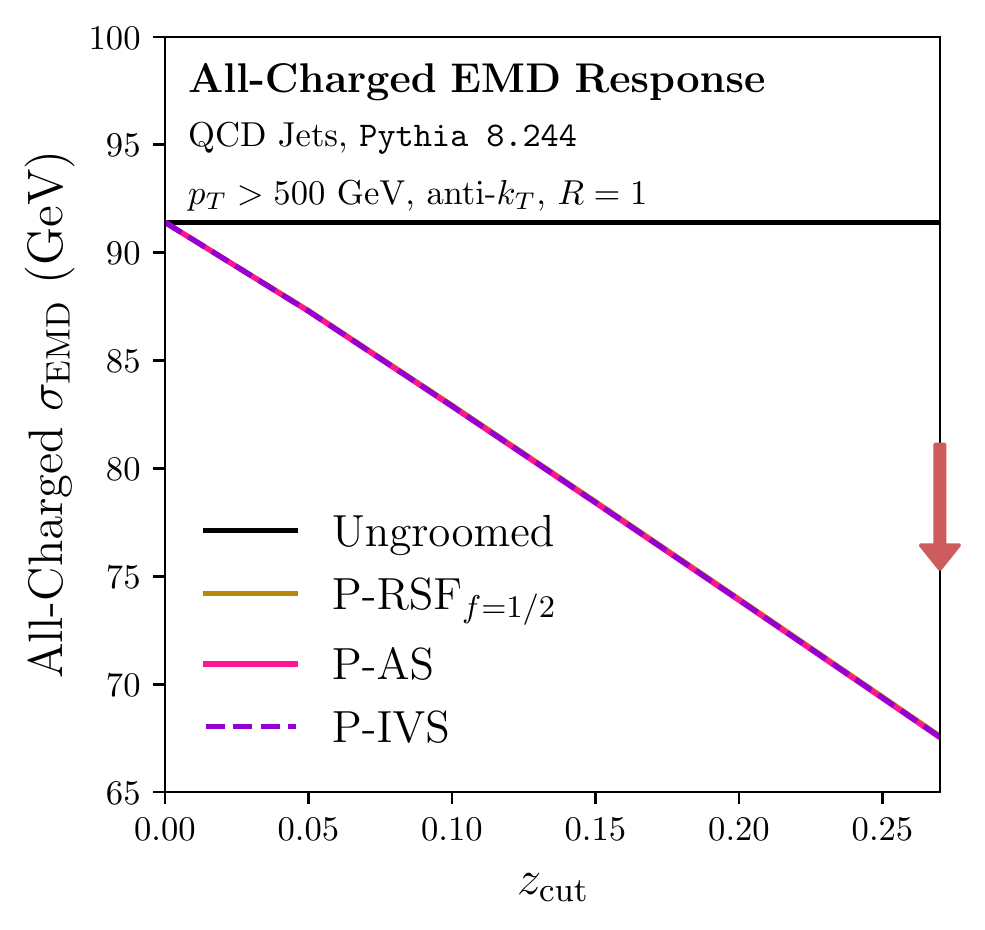}
    }
    \subfloat[]{
        \includegraphics[width=.32\textwidth]{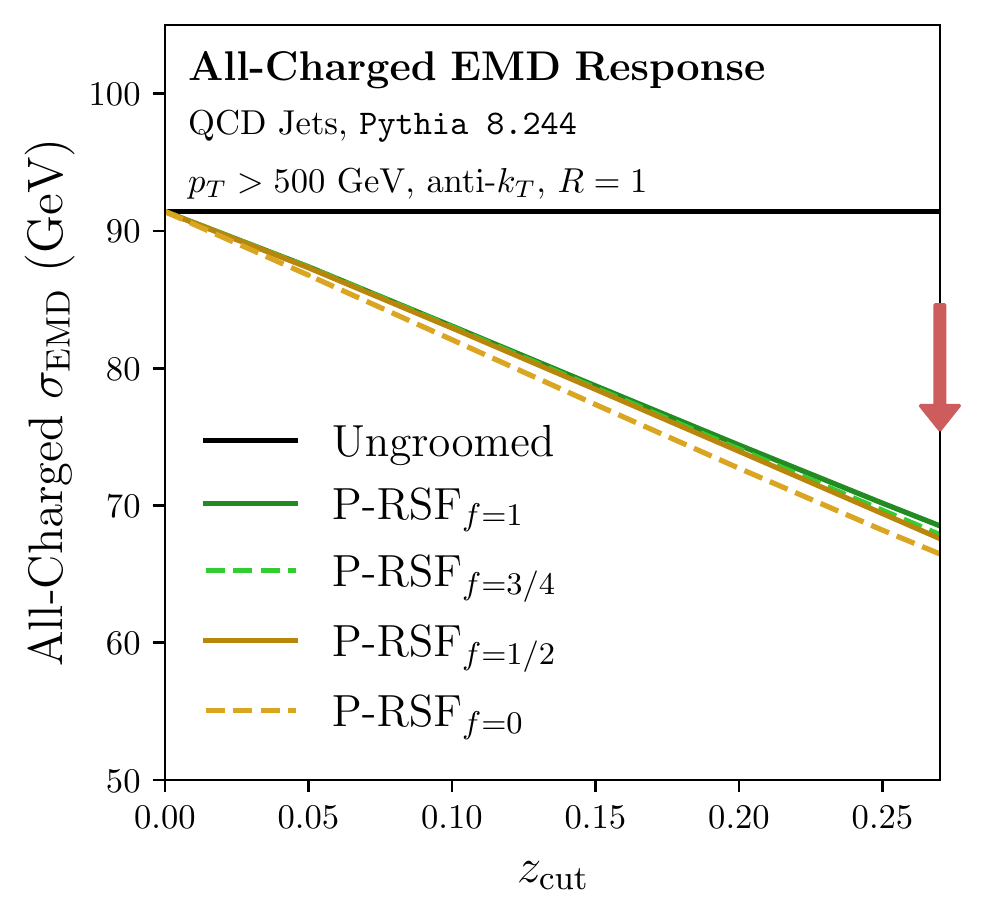}
    }
    \\
    \subfloat[]{
        \includegraphics[width=.32\textwidth]{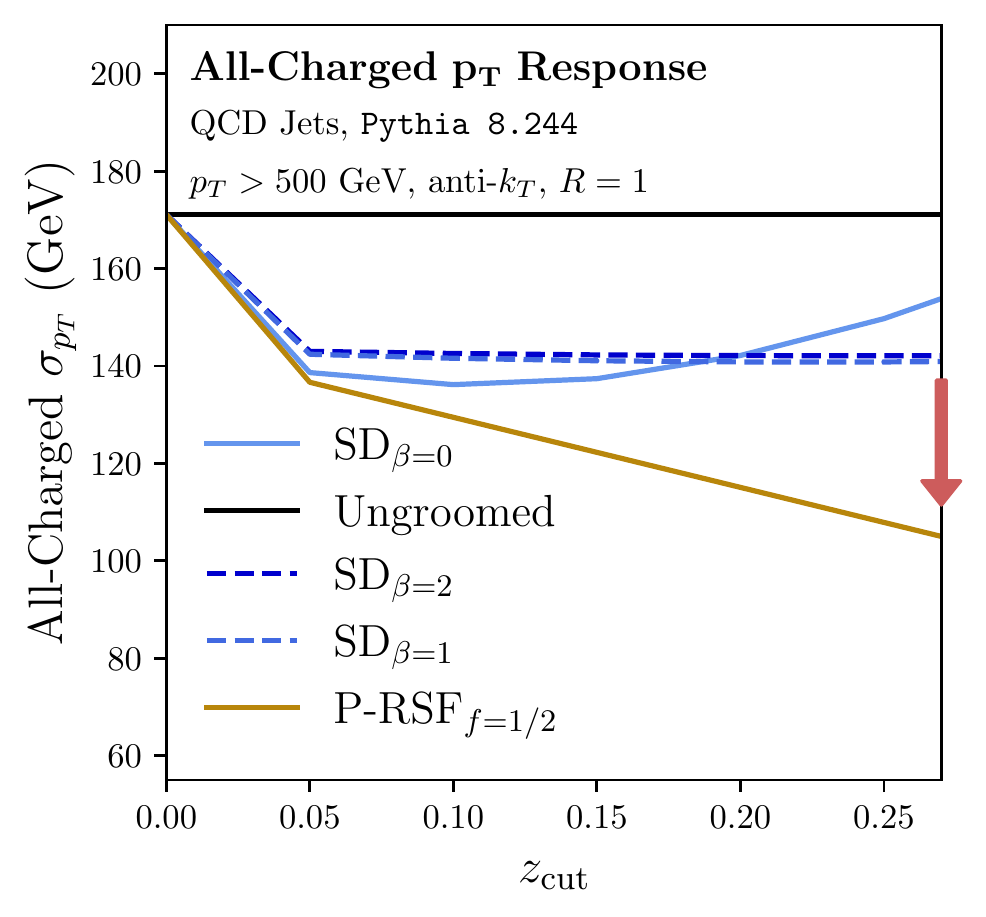}
    }
    \subfloat[]{
        \includegraphics[width=.32\textwidth]{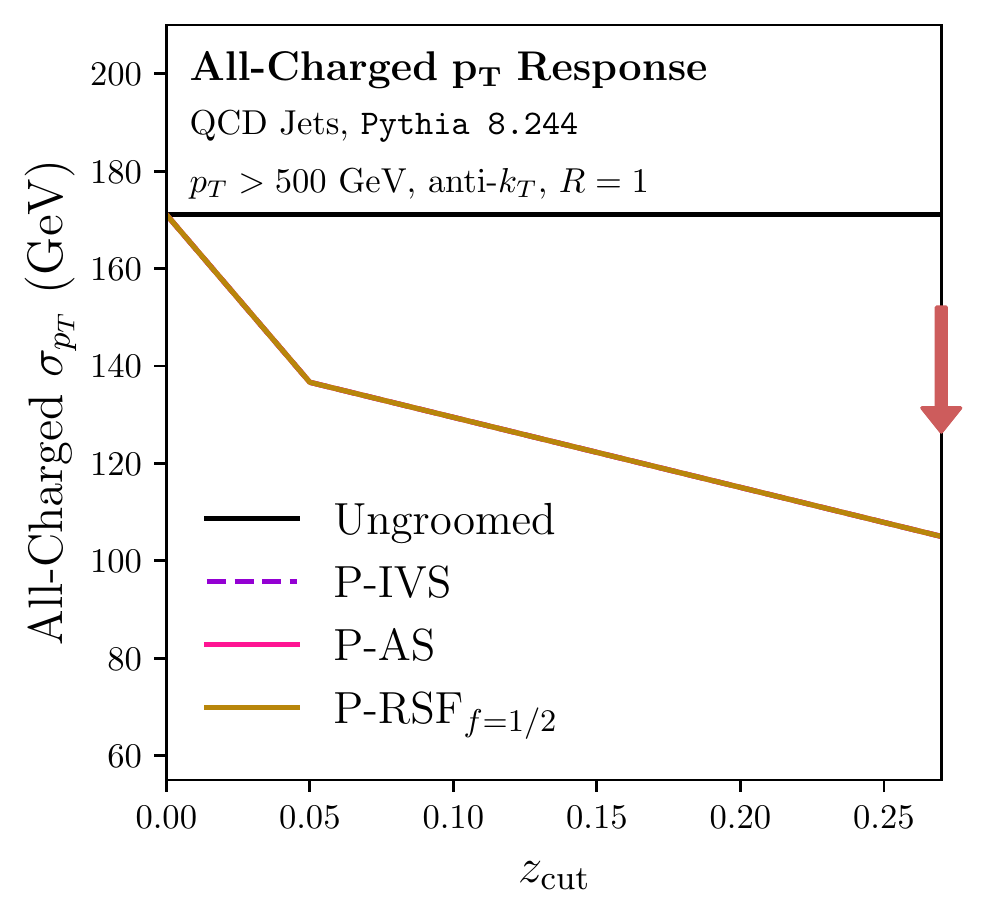}
    }
    \subfloat[]{
        \includegraphics[width=.32\textwidth]{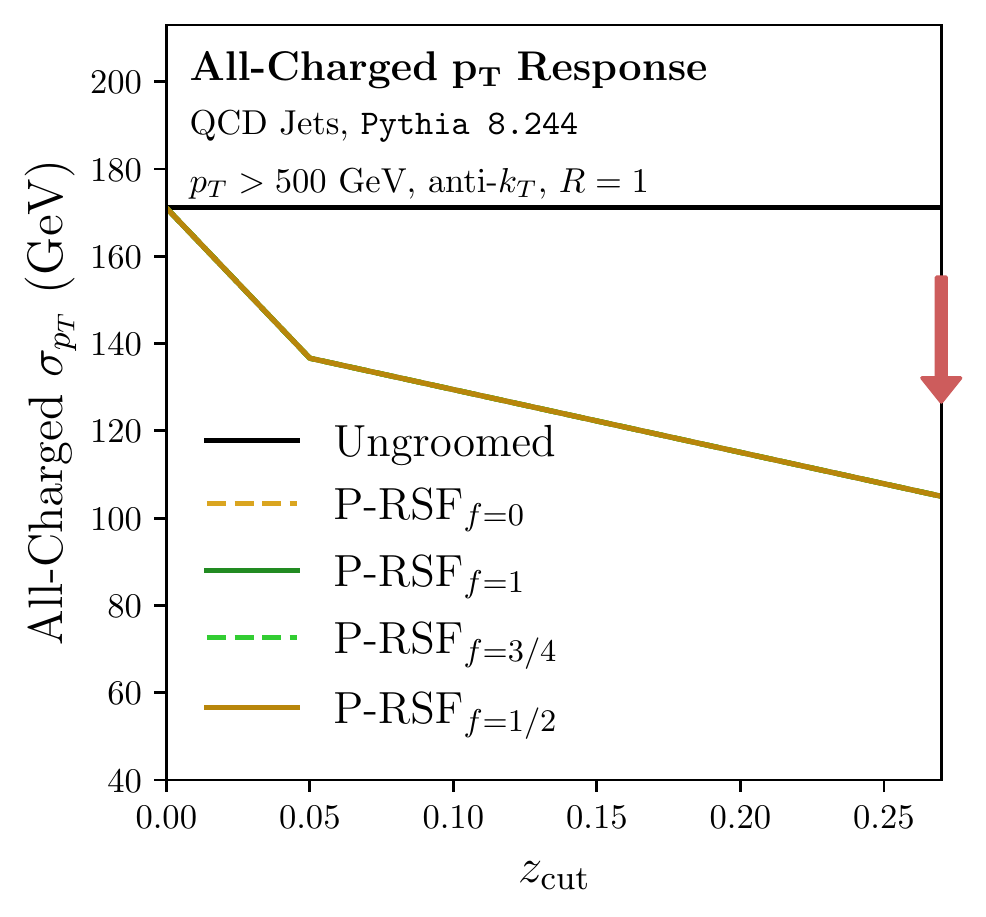}
    }
    \\
    \subfloat[]{
        \includegraphics[width=.32\textwidth]{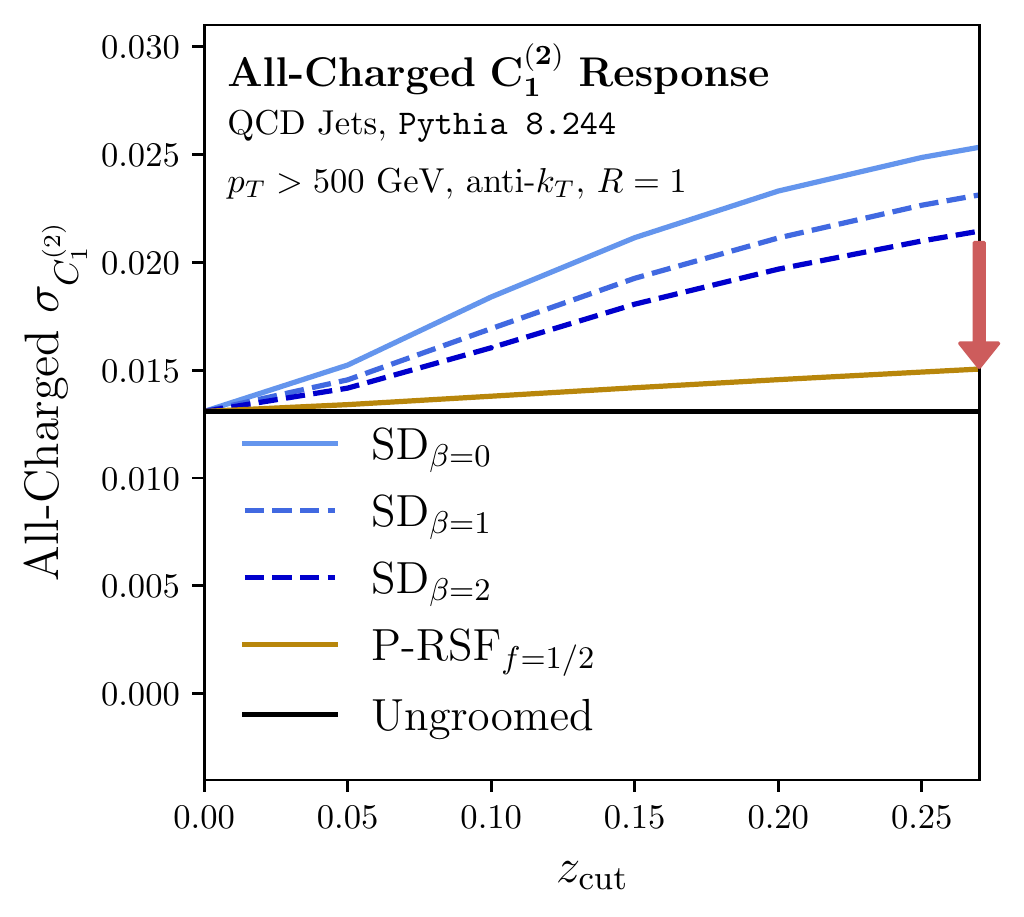}
    }
    \subfloat[]{
        \includegraphics[width=.32\textwidth]{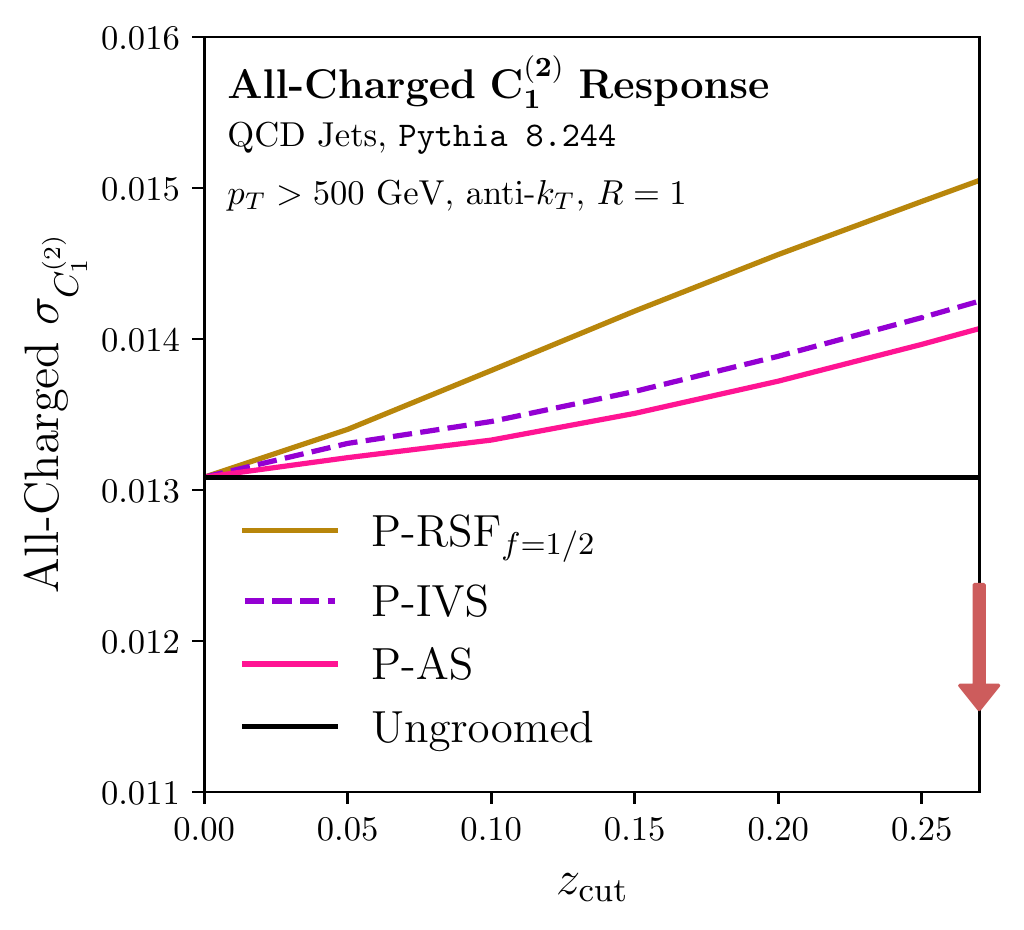}
    }
    \subfloat[]{
        \includegraphics[width=.32\textwidth]{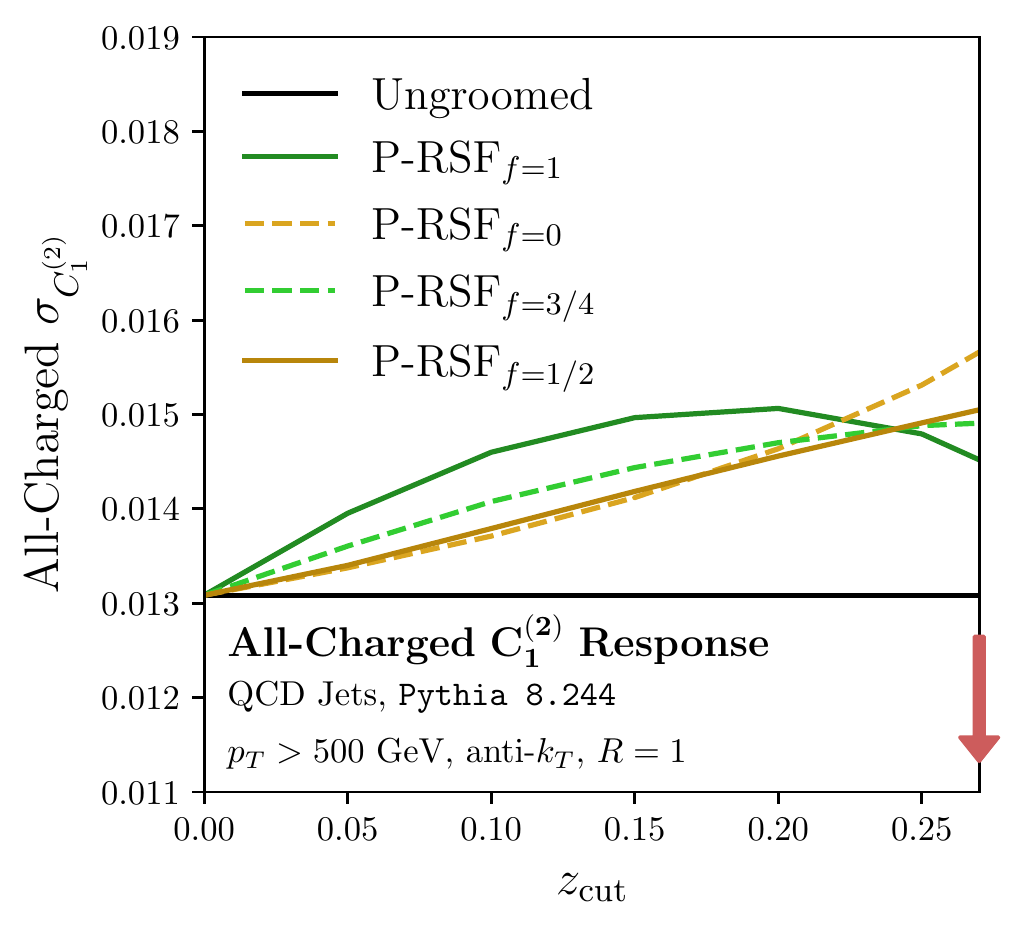}
    }
\caption{
    Same as \Fig{pvhfrenzy}, but for jet responses to the exclusion of neutral particles.
}
\label{fig:avcfrenzy}
\end{figure}

\subsection{Soft Distortions}
In the main text, we made the argument that continuity provides more predictable responses to soft distortions;
we therefore focus on the fluctuations induced by soft distortions by studying the standard deviations of the groomed EMD and \(C_1^{(2)}\) as a function of \zcut.

\Fig{pvhfrenzy} shows the fluctuations in the parton-hadron groomed EMD and the parton-hadron \(\Delta C_1^{(2)}\), as discussed in \Sec{hadronization}, for a variety of grooming options.
The first column of \Fig{pvhfrenzy} demonstrates that \PRSF{1/2} groomed jets tend to have significantly smaller fluctuations due to hadronization, echoing the conclusions of \Sec{hadronization}.
We note that \SD{2} is significantly more robust to hadronization than \SD{0} when considering extensive observables (EMD and \(p_T\)).
The second two columns of \Fig{pvhfrenzy} show that the \PIRANHA{} groomers we consider all have similar hadronization responses in the considered range of \zcut.

\Fig{avcfrenzy} presents the all-charged groomed EMD and \(\Delta C_1^{(2)}\) as discussed in \Sec{all_v_charged}.
The first column again shows that \PRSF{1/2} experiences significantly smaller fluctuations due to the exclusion of neutral particles than Soft Drop, and the second two columns show similar responses for all \PIRANHA{} groomers.

\subsection{Pileup}
\label{app:pufrenzy}

We now provide more detail regarding our PU mitigation procedure.
As discussed in \Sec{pileup}, we use the dijet and minimum bias events of \Reff{Soyez:2018opl} to study the effects of PU.
We use the \texttt{GridMedianBackgroundEstimator} (GMBE) method of \texttt{FastJet} \cite{Cacciari:2011ma} to provide a rough estimate, \(\rho_{\rm GMBE}\), of the energy density of additive contamination due to pileup in each jet we would like to groom.
However, since GMBE is designed to provide an estimate of the contaminating energy in an entire event,\footnote{
Note that \texttt{FastJet} also supports \textit{Jet} Median Background Estimation (JMBE), which is more suited to the task of estimating the contaminating energy density in each jet.
We were unable to find \texttt{Python} support for JMBE and used the method detailed here as a replacement.
} we improve upon the GMBE procedure by optimizing a correction factor that scales the GMBE estimate for each groomer.
We optimize the correction factor for each groomer by considering its action on 100,000 events, throwing away events where the presence of PU changes which jet within the dijet has more energy, and implementing the following procedure:

\begin{enumerate}
    \item
    For each epoch \(i\) during our optimization, estimate the contaminating energy density due to PU within a jet,
    \begin{align}
        \rho_{\rm PU}[\mathcal E] = \frac{p_T^{\rm (PU~in~jet)}}{A_{\rm jet}}
        ,
    \end{align}
    that needs to be applied by a groomer \(g\) to accurately remove PU contamination as
    \begin{align}
        \rho^{{\rm(epoch}\,i)}_{\rm est.}[\mathcal E ; g)
        =
        \rho_{\rm GMBE}\,[\mathcal E] f_i(g)
        ,
    \end{align}
    where we use parentheses to indicate that \(\rho_{\rm est.}\) and \(f\) are functions of the groomer \(g\), and square brackets to indicate that the energy densities \(\rho\) are functionals of the energy flow \(\mathcal E\).
    Since we are using jets clustered using the anti-\(k_t\) algorithm, we may take the jet area to be \(\pi R_{\rm jet}^2\) \cite{Cacciari:2008gp}.

    \item
    Introduce a new epoch every 1,000 events we consider, by adjusting our proposed correction factor \(f_i\) to
    \begin{align}
        f_{i + 1}
        =
        f_i + r_{\rm learn}
        \left\langle \frac{\Delta \rho}{\rho} \right\rangle_i
        \label{eqn:epochcorrectionfactor}
        ,
    \end{align}
    where the angular brackets indicate an average over the jets considered during epoch \(i\), \(\Delta \rho\) indicates the difference in \(p_T\) density in a pileup subtracted event and a hard event,
    \begin{align}
        \Delta \rho
        =
        \frac{\Delta p_T}{A_{\rm jet}}
        =
        \frac{p_T^{\rm(groomed~PU)} - p_T^{\rm(hard)}}{A_{\rm jet}}
        ,
    \end{align}
    and \(r_{\rm learn}\) is a learning rate which we set to 1/3.

    \item
    At the end of the final epoch, save the final correction factor \(f_{\rm learned}(g)\) for each groomer.

    \item
    Perform a new analysis of the ability of each groomer to remove PU by estimating the optimal \(\rho\) required for PU-subtraction as
    \begin{align}
        \rho_{\rm est.}[\mathcal E ; g)
        =
        \rho_{\rm GMBE}\,[\mathcal E] f_{\rm learned}(g)
        .
    \end{align}
    We recall for completeness that \(
    	z_{\rm cut}
	=
	\rho_{\rm est.} A_{\rm jet}/p_T^{(\rm jet)}
    \).

\end{enumerate}

\begin{figure}[p]
\centering
    \subfloat[]{
        \includegraphics[width=.32\textwidth]
        {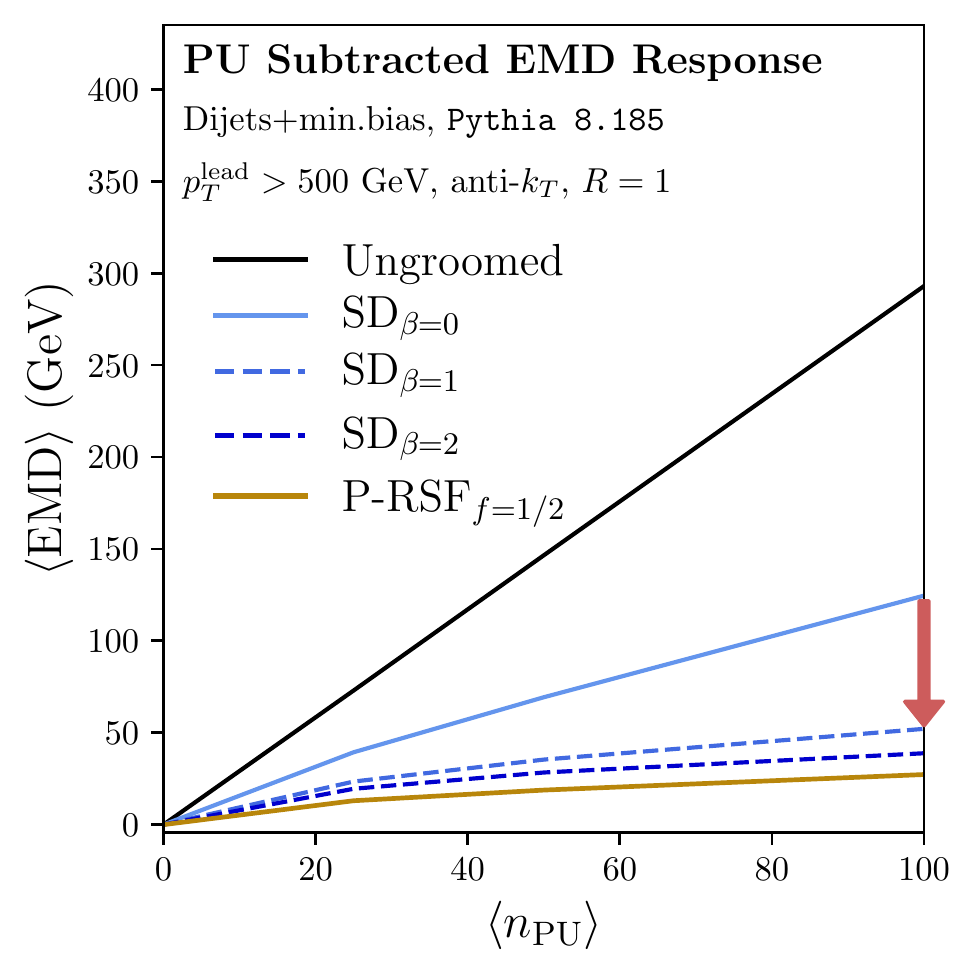}
    }
    \subfloat[]{
        \includegraphics[width=.32\textwidth]
        {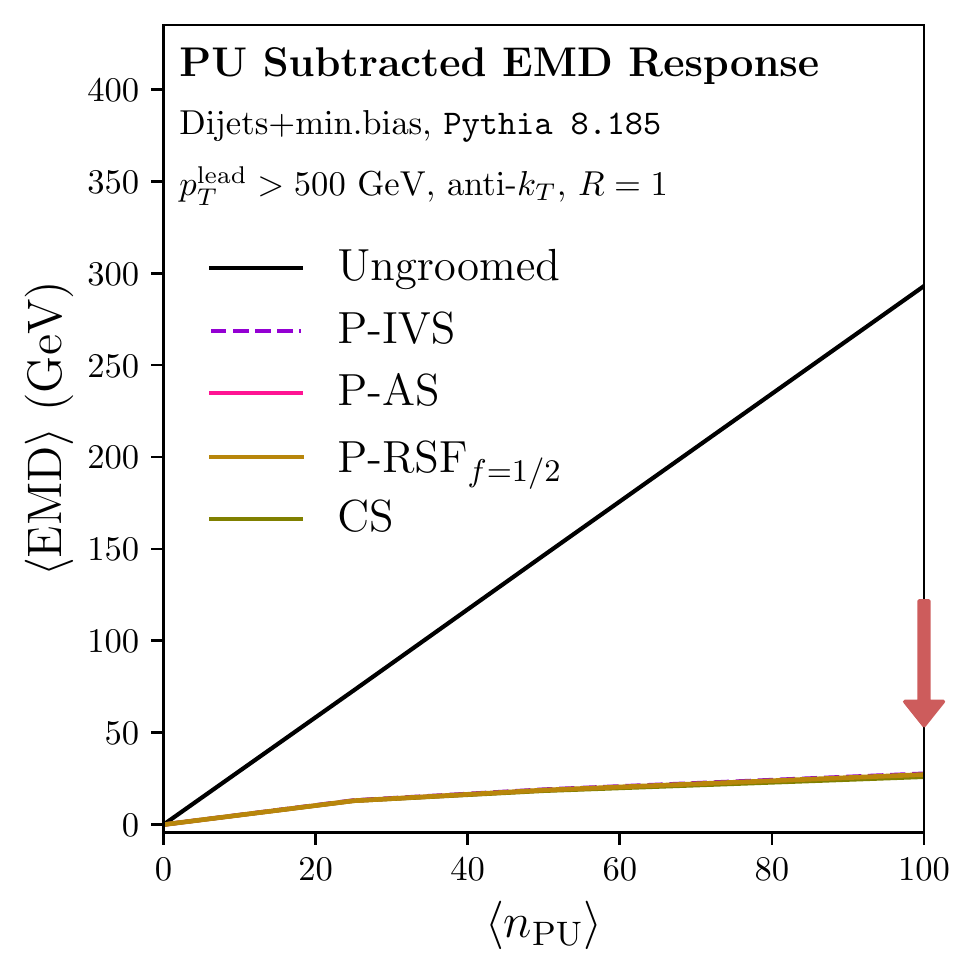}
    }
    \subfloat[]{
        \includegraphics[width=.32\textwidth]
        {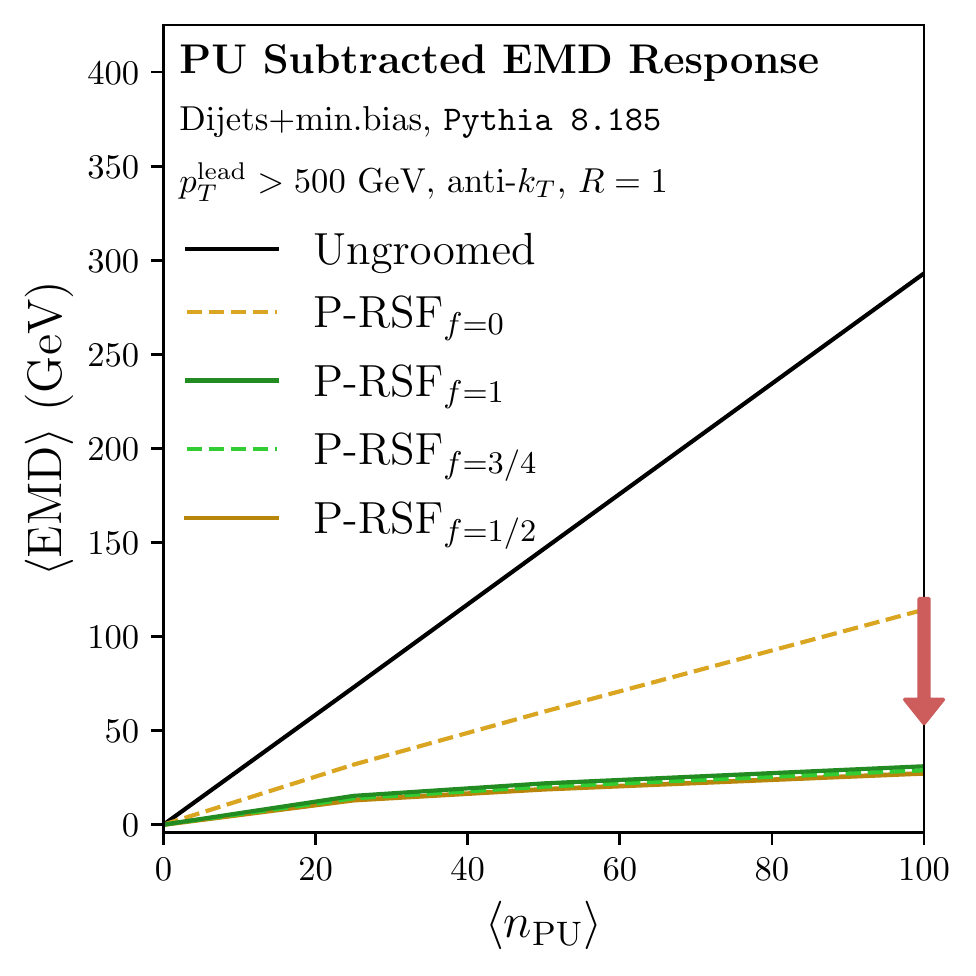}
    }
    \\
    \subfloat[]{
        \includegraphics[width=.32\textwidth]
        {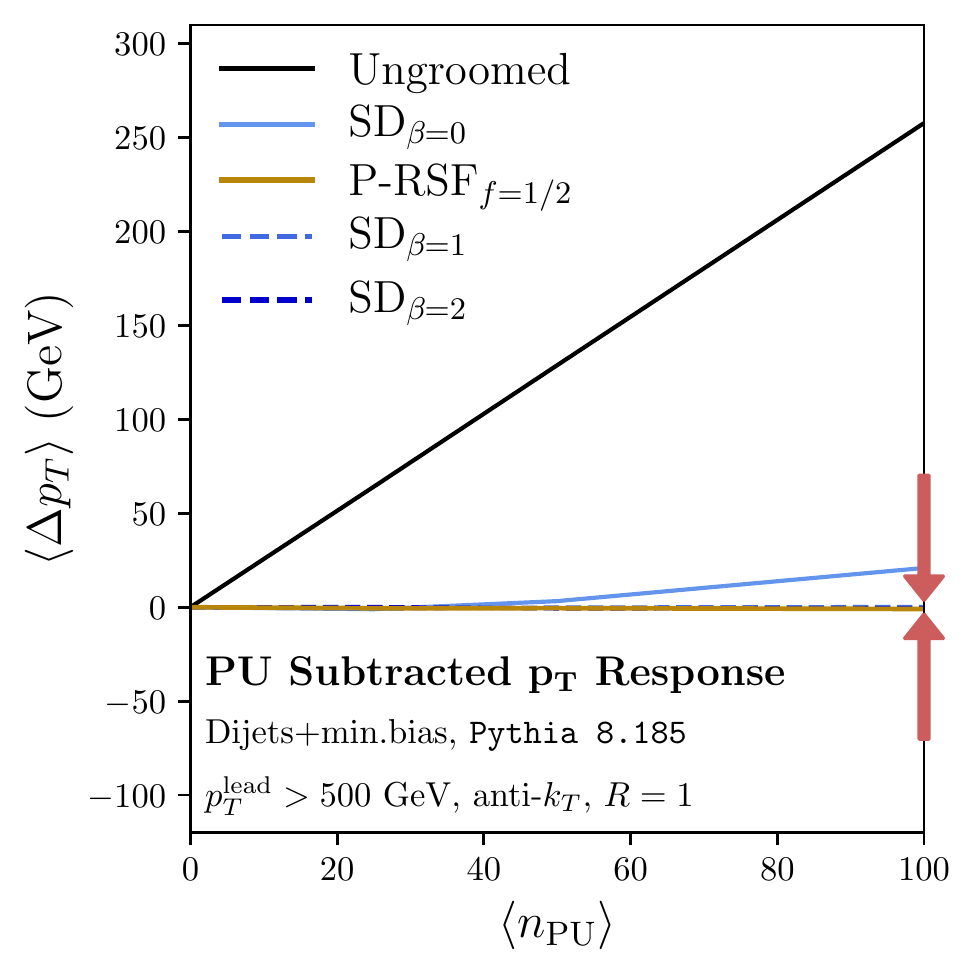}
    }
    \subfloat[]{
        \includegraphics[width=.32\textwidth]
        {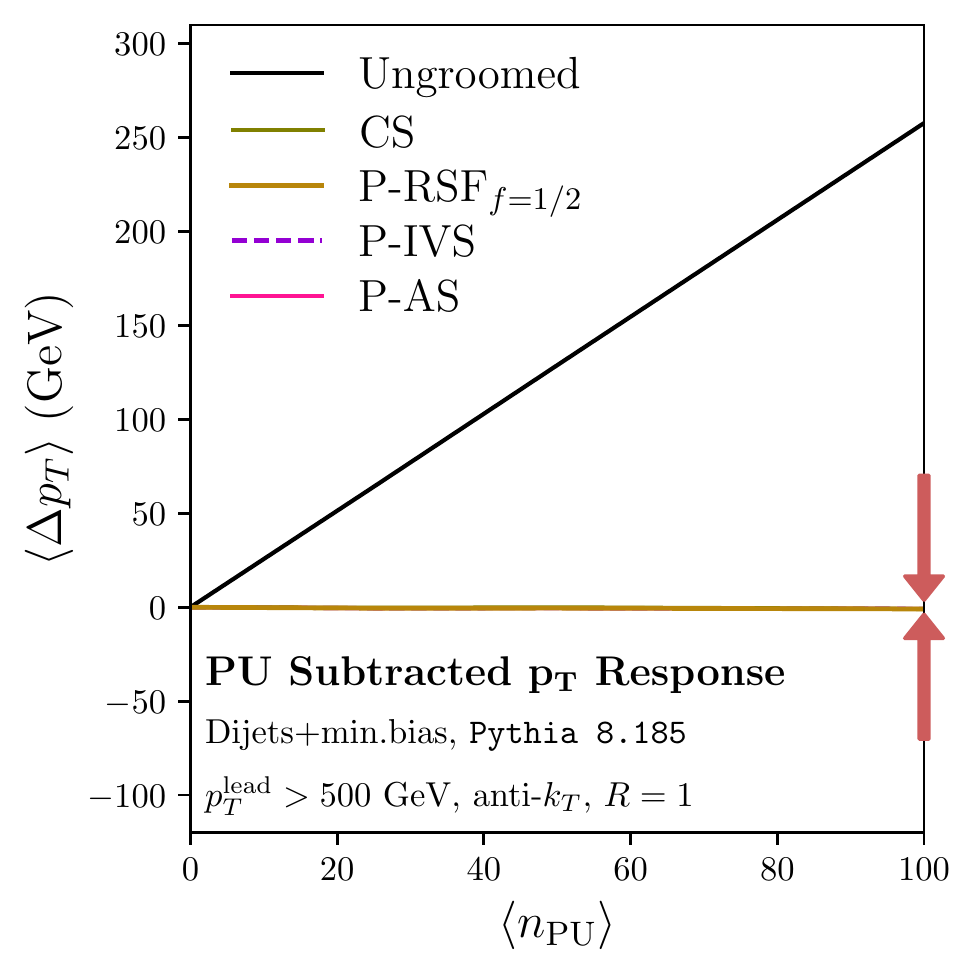}
    }
    \subfloat[]{
        \includegraphics[width=.32\textwidth]
        {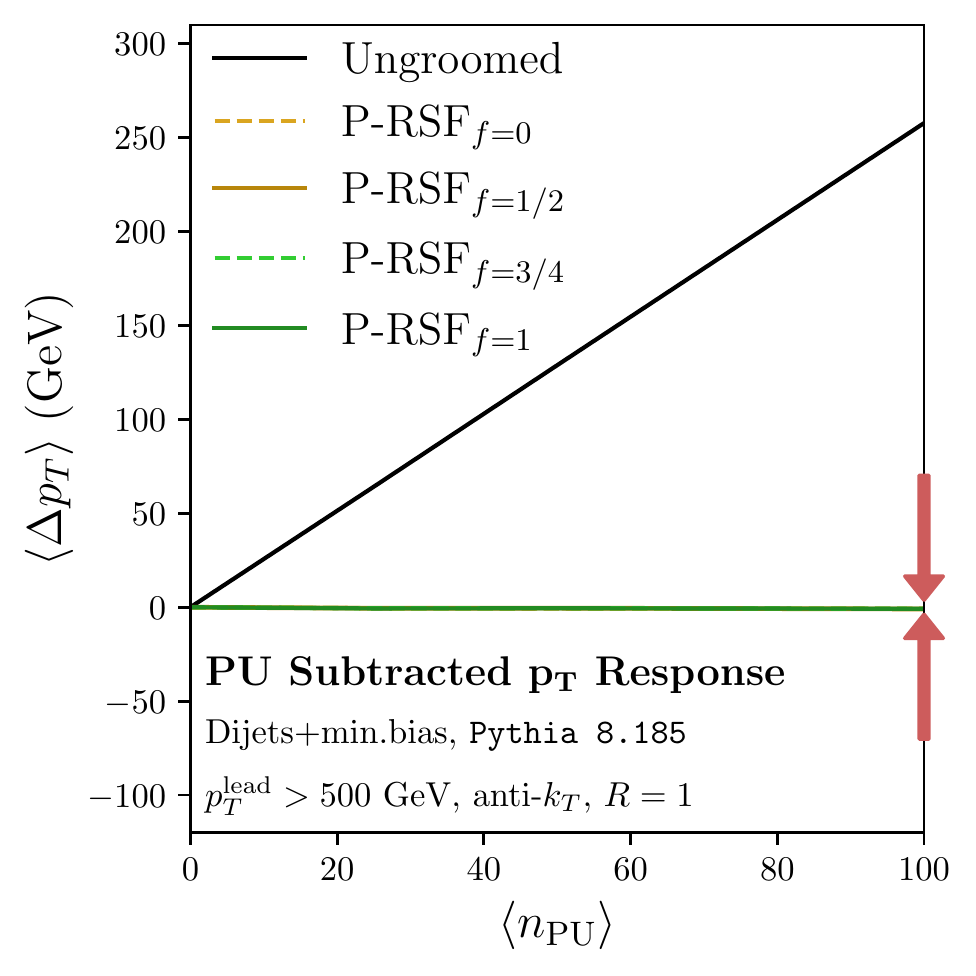}
    }
    \\
    \subfloat[]{
        \includegraphics[width=.32\textwidth]
        {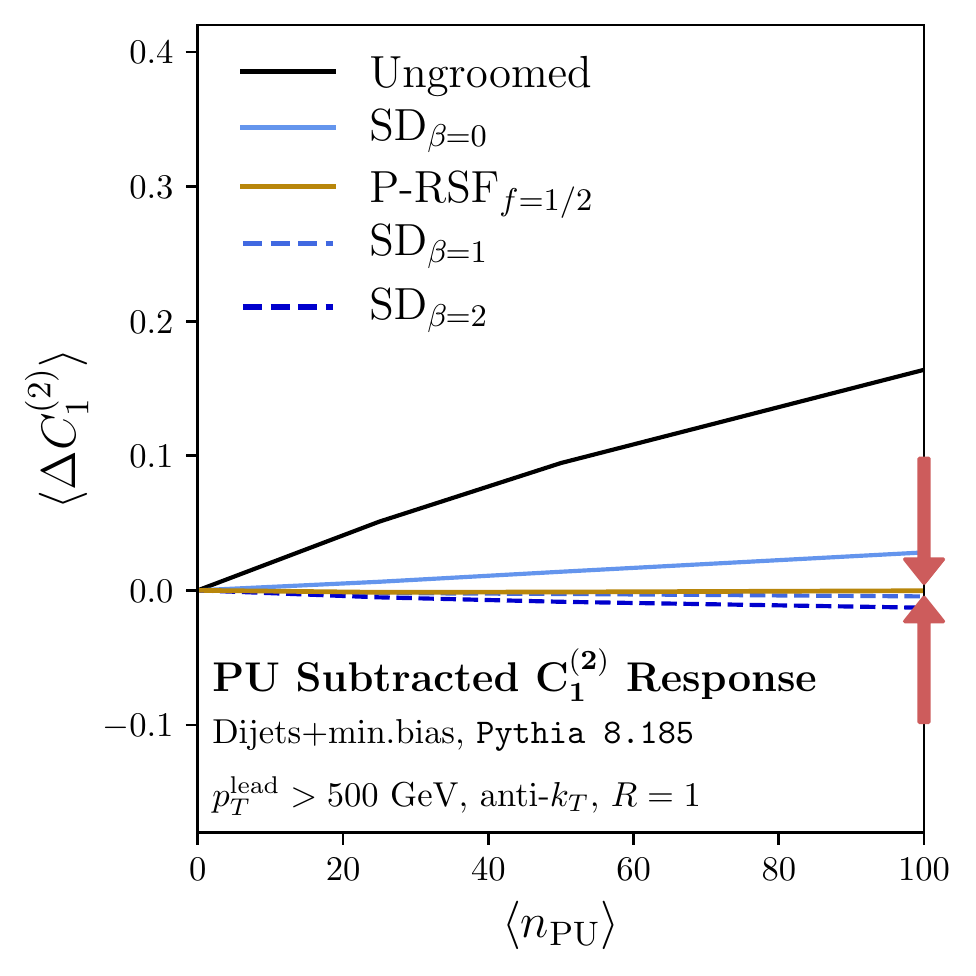}
    }
    \subfloat[]{
        \includegraphics[width=.32\textwidth]
        {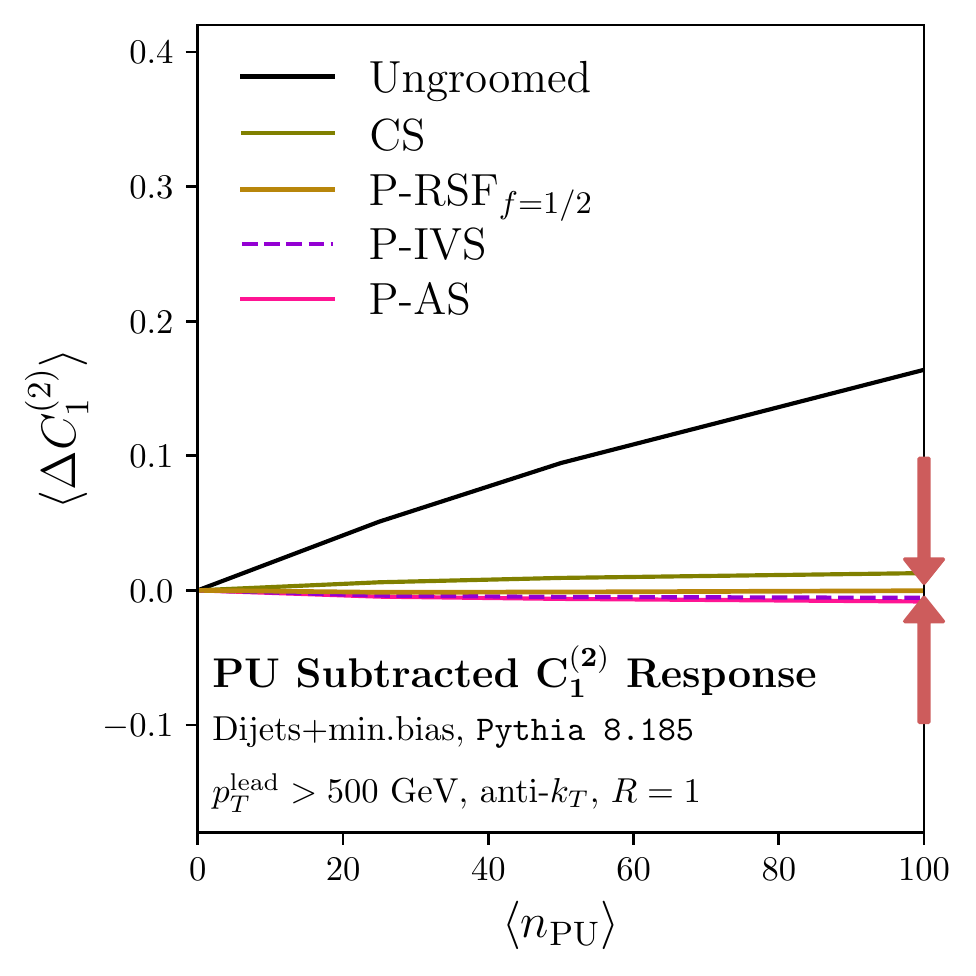}
    }
    \subfloat[]{
        \includegraphics[width=.32\textwidth]
        {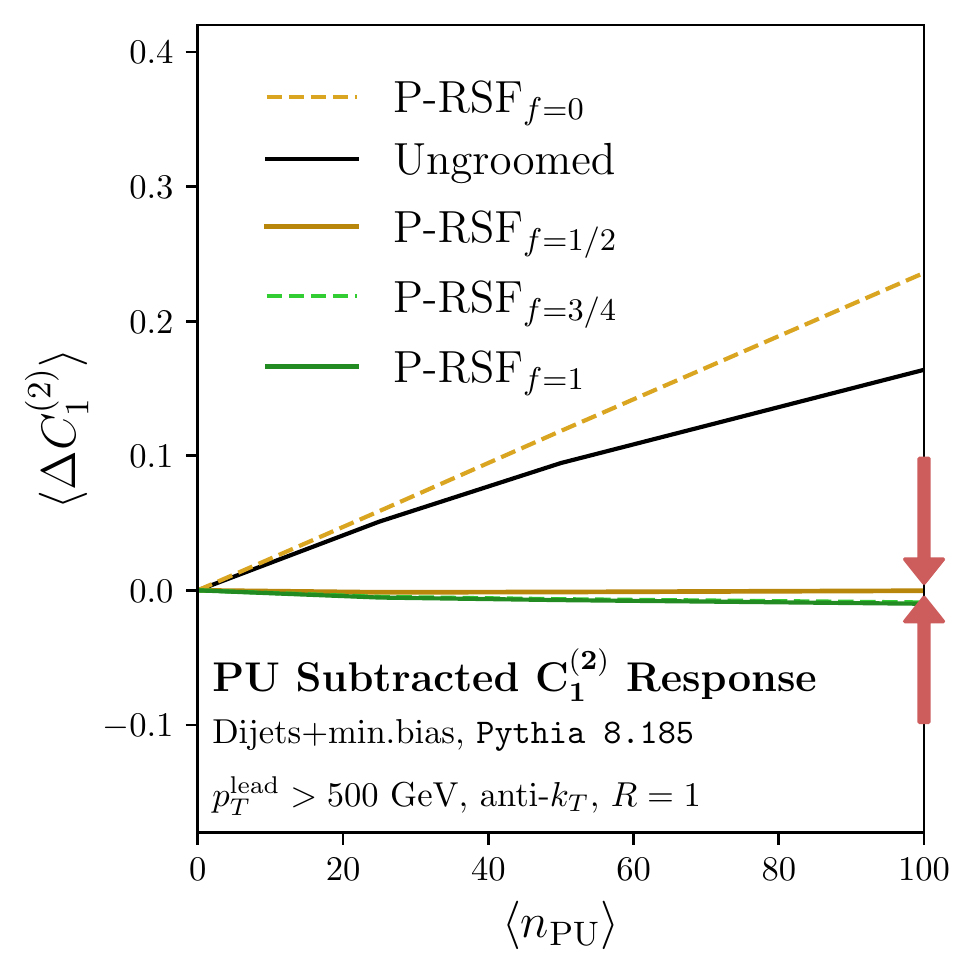}
    }
\caption{
    Average per-jet pileup-induced shifts in (top row) EMD, (middle row) \(\Delta p_T\), and (bottom row) \(C_1^{(2)}\);
    we compare \PRSF{1/2} to (left column) hard-cutoff groomers, (middle column) fully continuous groomers, and (right column) recursive subtractors, as discussed in \Sec{pileup} and \App{feedingfrenzy}.
    The red arrows indicate the direction corresponding to better performance.
}
\label{fig:pufrenzy_ave}
\end{figure}

\begin{figure}[p]
\centering
    \subfloat[]{
        \includegraphics[width=.32\textwidth]
        {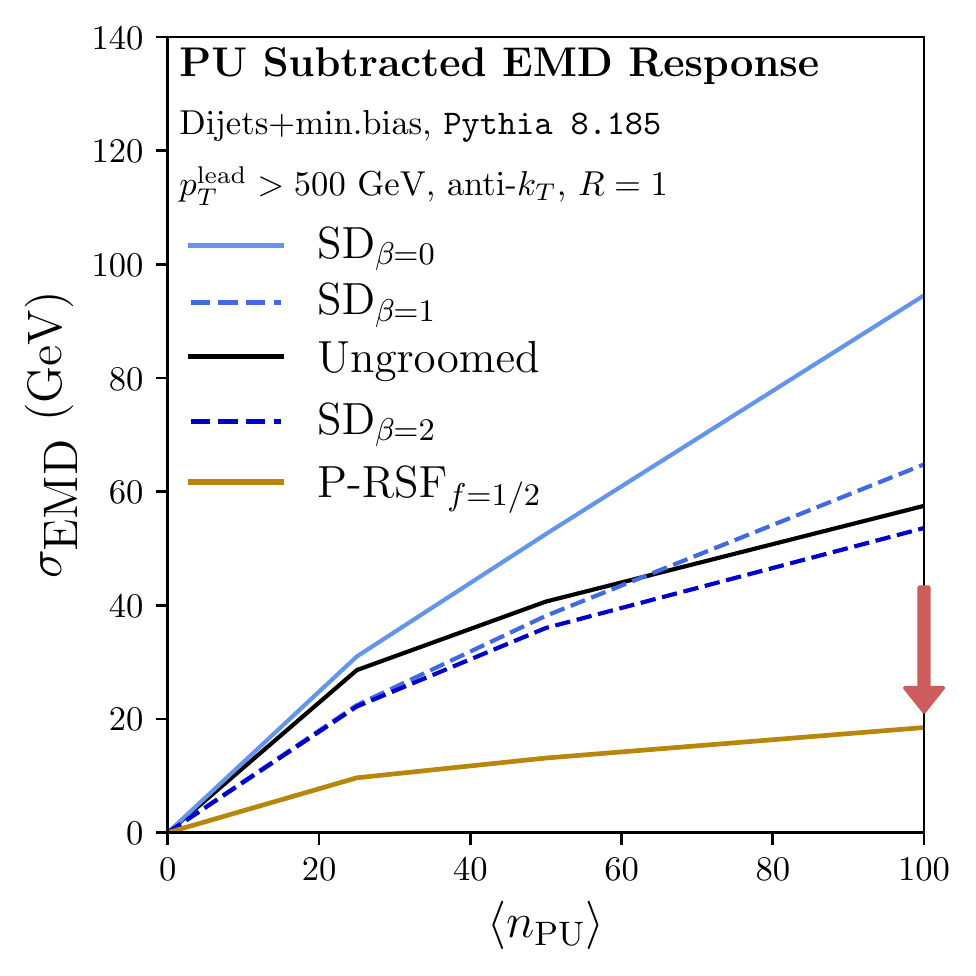}
    }
    \subfloat[]{
        \includegraphics[width=.32\textwidth]
        {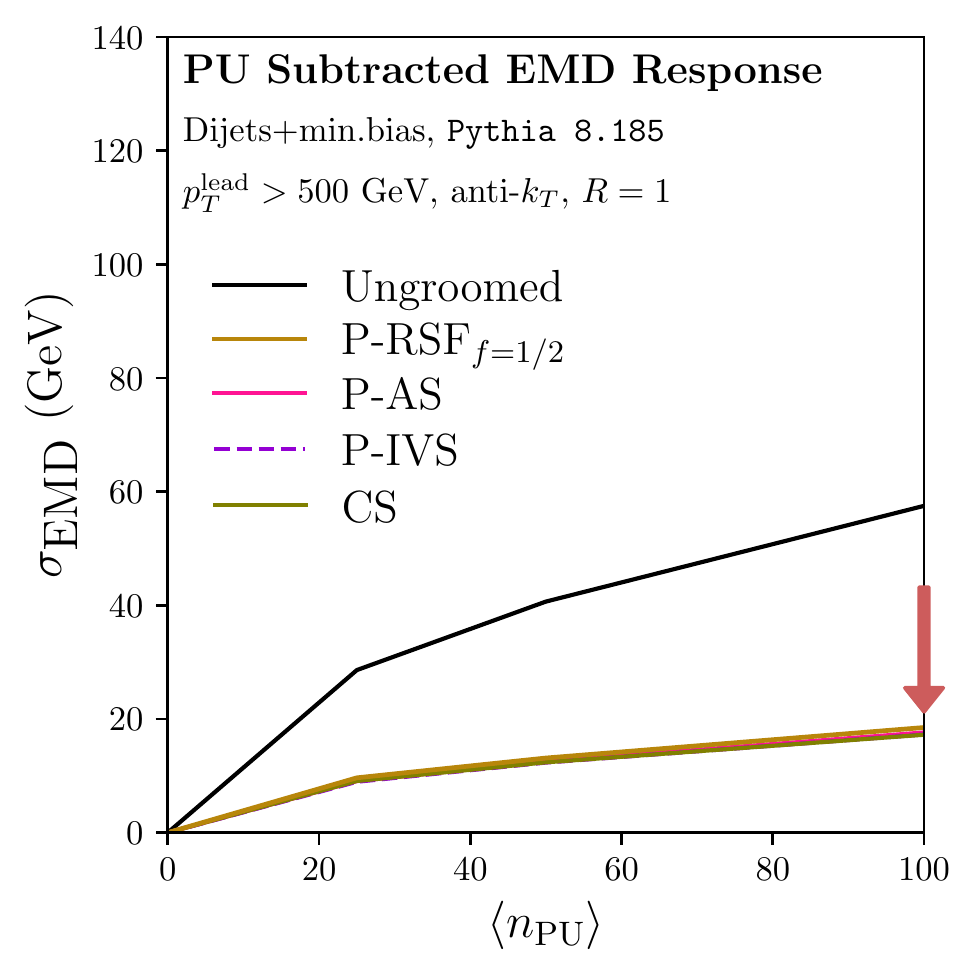}
    }
    \subfloat[]{
        \includegraphics[width=.32\textwidth]
        {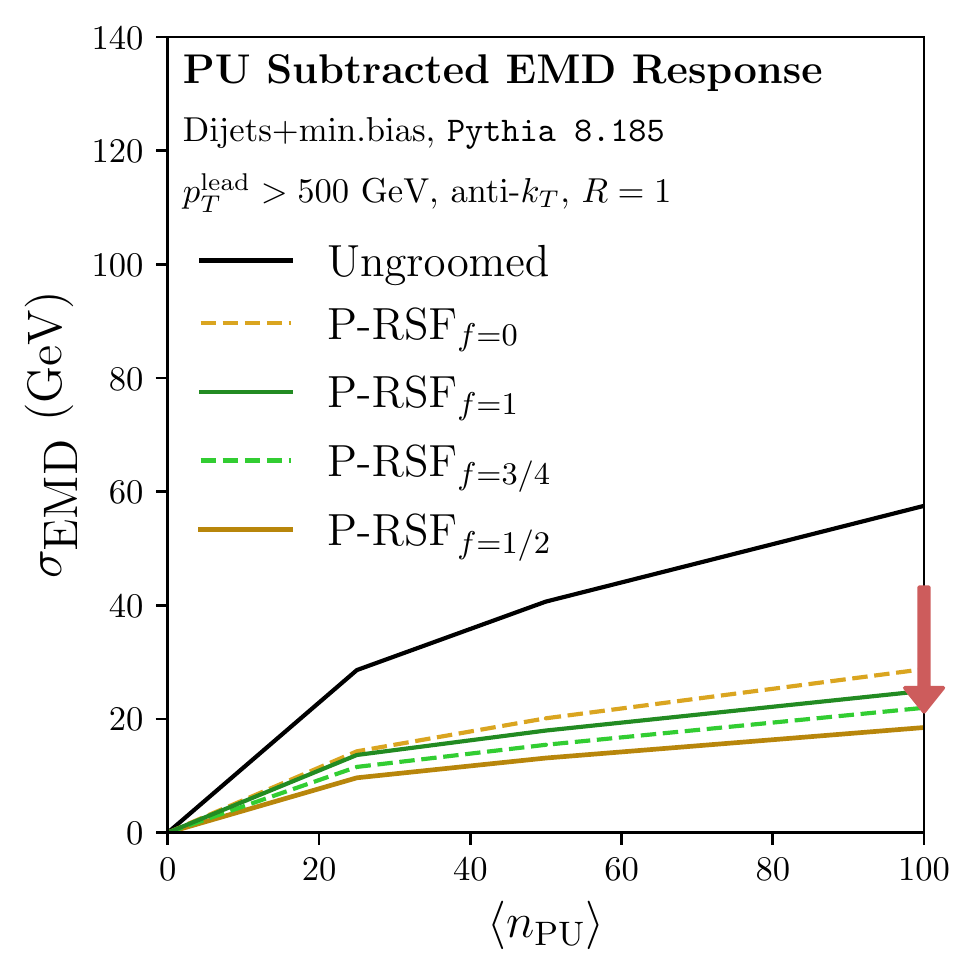}
    }
    \\
    \subfloat[]{
        \includegraphics[width=.32\textwidth]
        {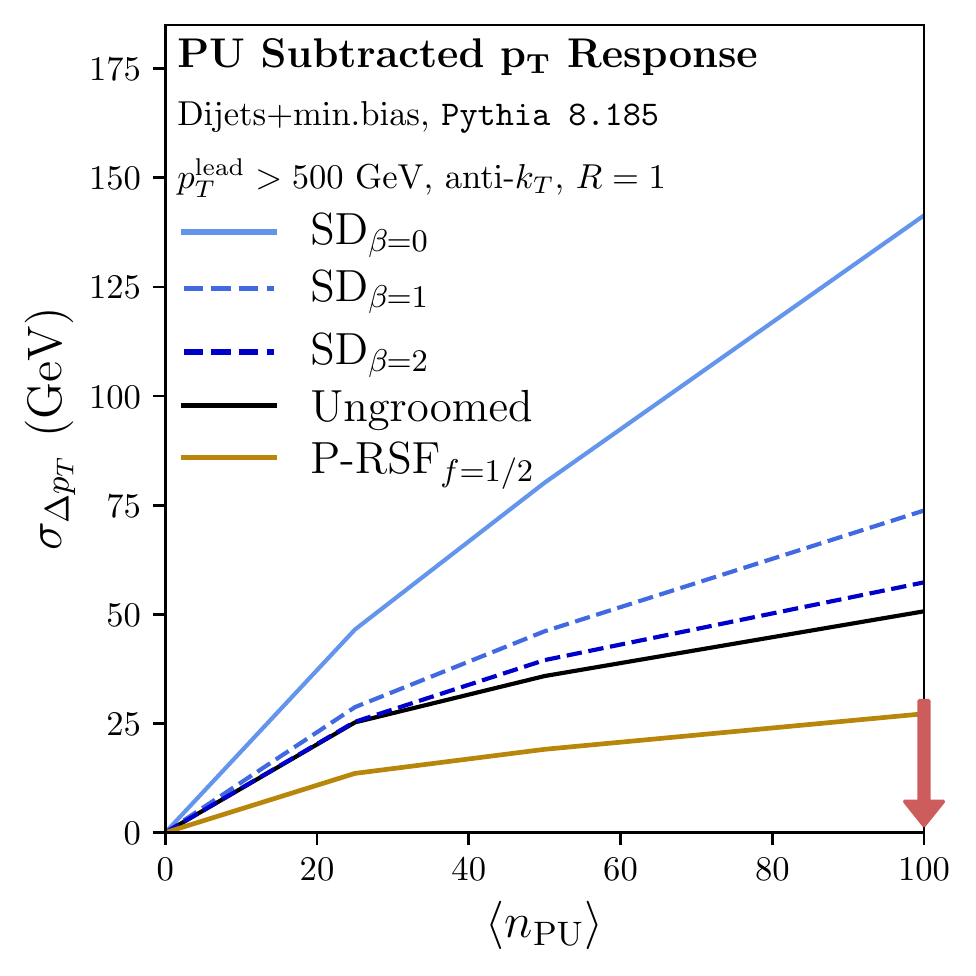}
    }
    \subfloat[]{
        \includegraphics[width=.32\textwidth]
        {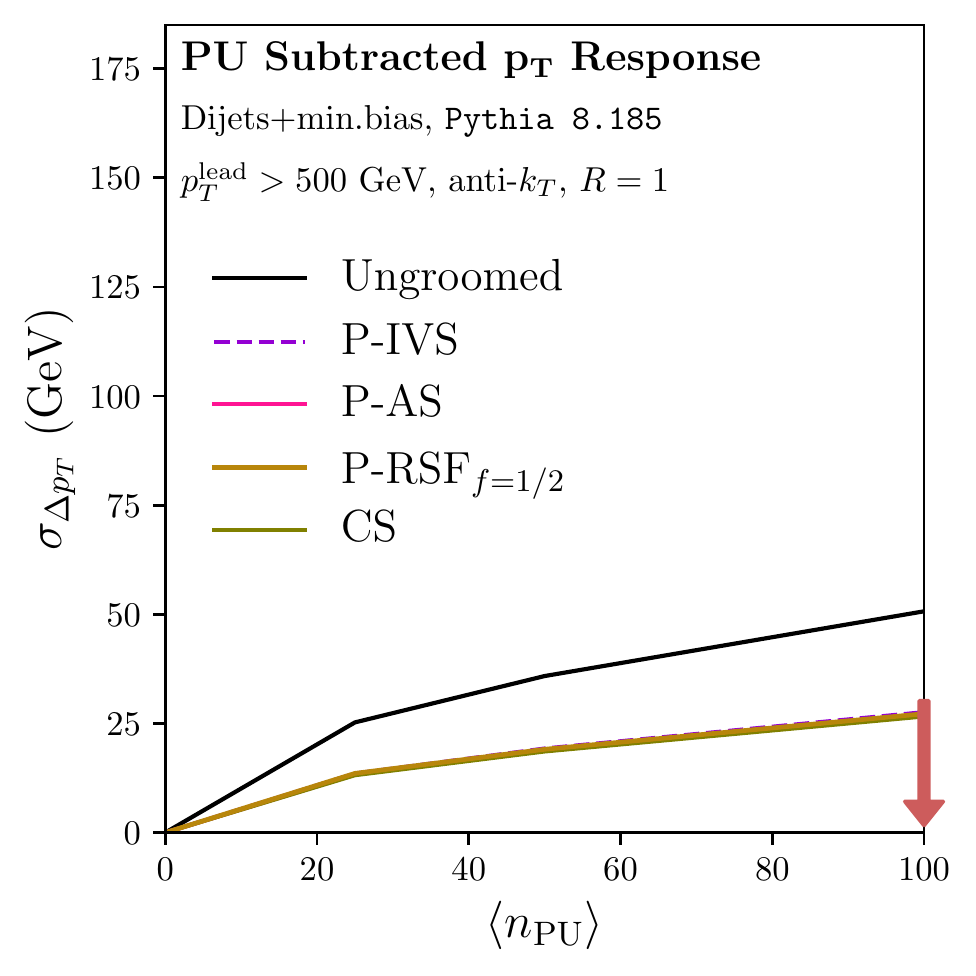}
    }
    \subfloat[]{
        \includegraphics[width=.32\textwidth]
        {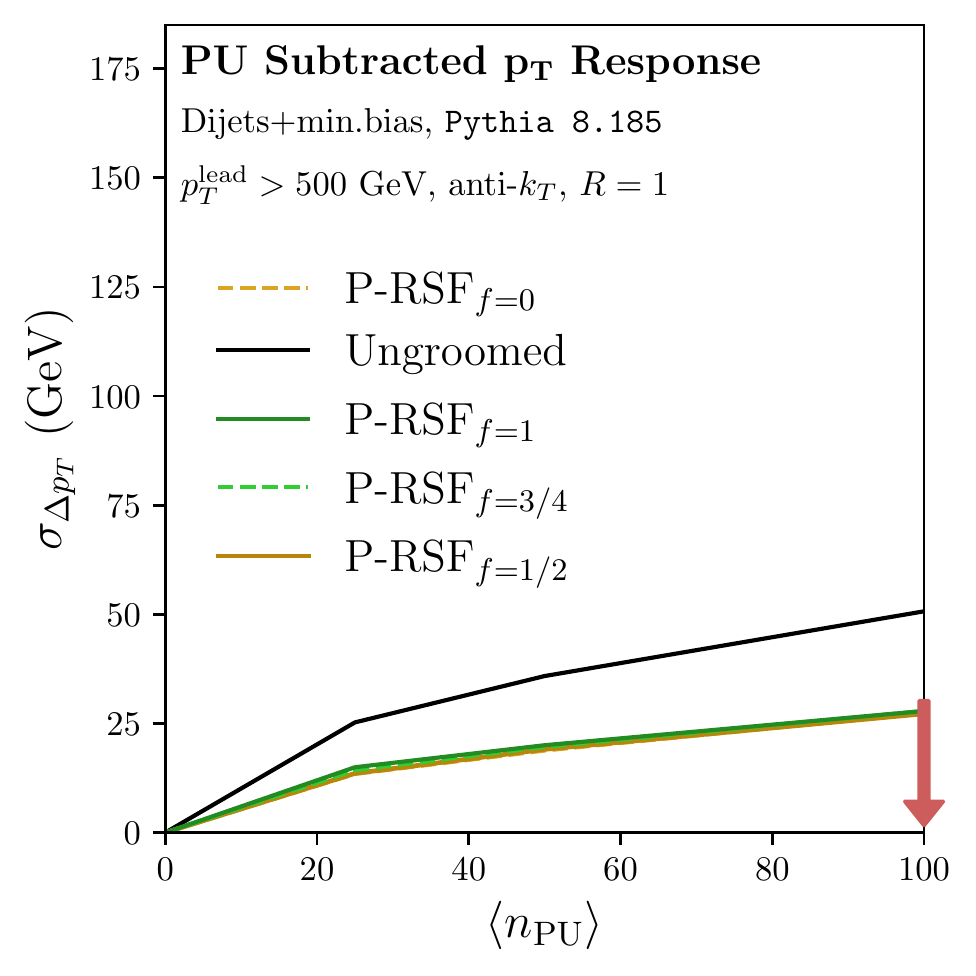}
    }
    \\
    \subfloat[]{
        \includegraphics[width=.32\textwidth]
        {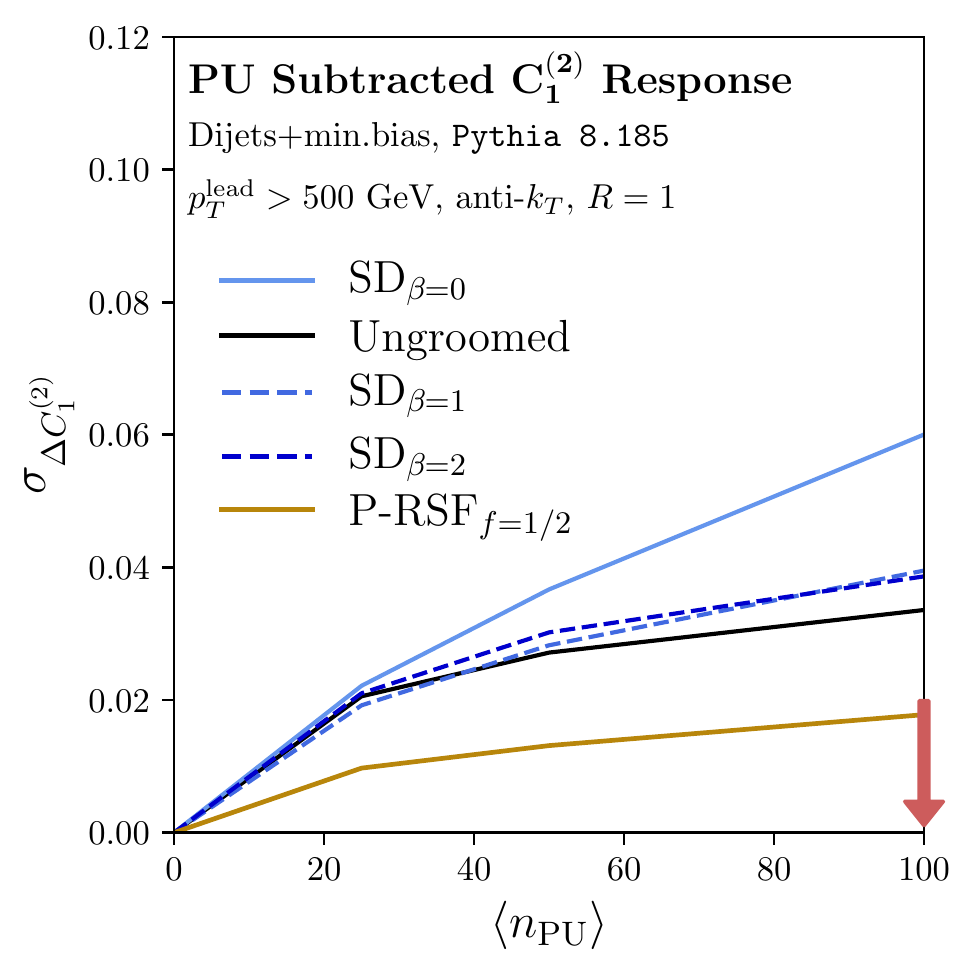}
    }
    \subfloat[]{
        \includegraphics[width=.32\textwidth]
        {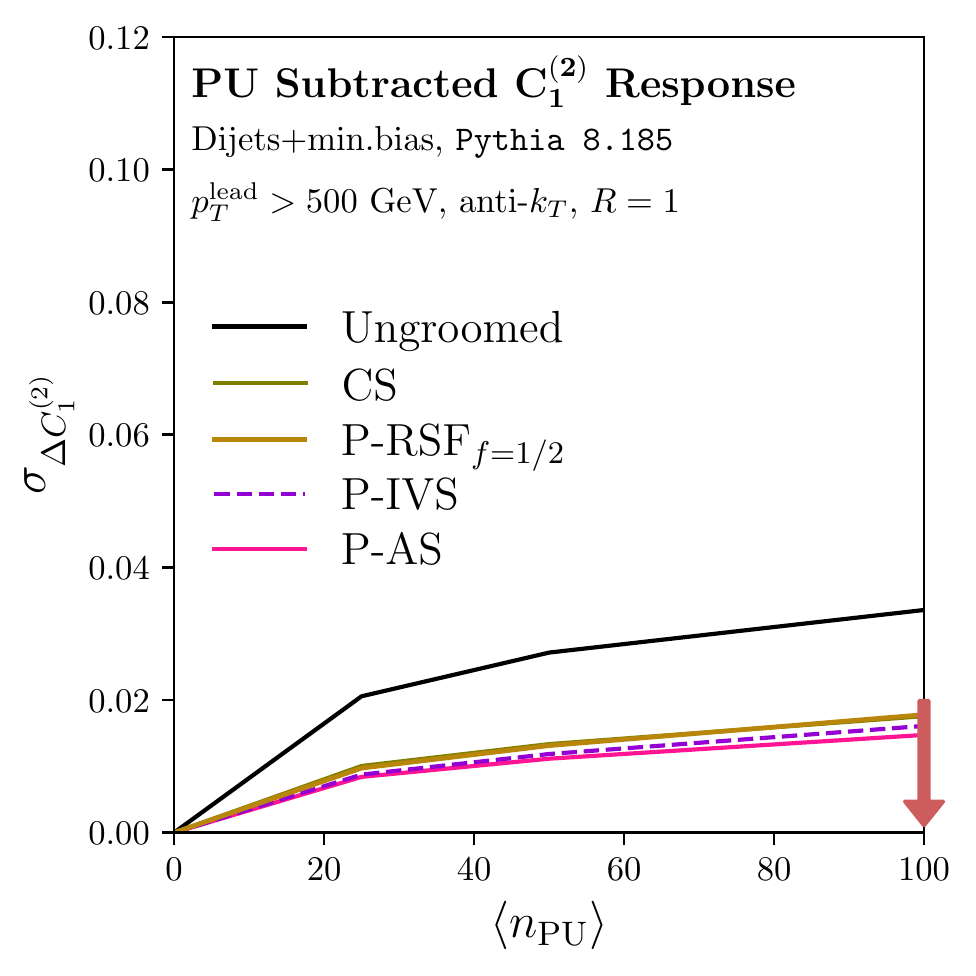}
    }
    \subfloat[]{
        \includegraphics[width=.32\textwidth]
        {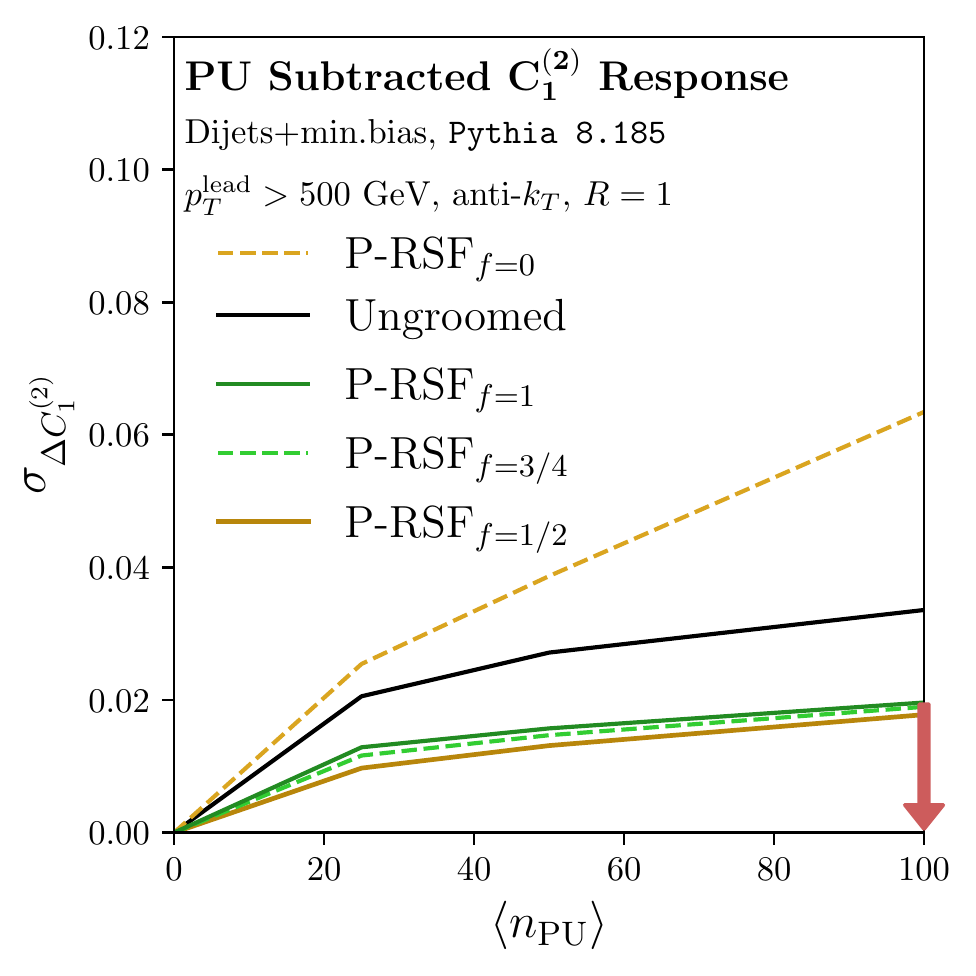}
    }
\caption{
    Same as \Fig{pufrenzy_ave}, but for the standard deviation of the pileup-induced shifts.
}
\label{fig:pufrenzy_stddev}
\end{figure}

In \Figs{pufrenzy_ave}{pufrenzy_stddev}, we present results on PU mitigation using the optimization procedure described above for a variety of groomers, as an extension of the discussion in \Sec{pileup}.
We focus on the PU-subtracted EMD, the PU-subtracted \(\Delta p_T\), and the PU-subtracted \(\Delta C_1^{(2)}\) as a function of the average number of PU events \(\langle n_{\rm PU} \rangle\).

The \PIRANHA{} groomers displayed in \Fig{pufrenzy_ave}, together with CS, tend to perform only slightly better than the traditional groomers, on average, in removing pileup;
both \PIRANHA{} and traditional groomers have been tuned so that \(\langle \Delta p_T\rangle=0\), and this tuning is reflected by the relatively small changes to the extensive properties and even the substructure of the PU-subtracted jets.
The notable exception is \PRSF{0}.
Since \PRSF{0} preferentially grooms away hard radiation, it is poorly suited to grooming away the low-energy additive contamination that is contributed by pileup, and it performs more poorly than other \PIRANHA{} groomers in reproducing both extensive and substructure properties of PU-subtracted jets.

However, \Fig{pufrenzy_stddev} shows that \PIRANHA{} groomers and CS are more \textit{reliable} PU mitigators than traditional groomers:
the standard deviations in the shifts of PU-subtracted extensive and substructure properties tend to be significantly smaller for \PIRANHA{} groomers than for traditional groomers.
Again, \PRSF{0} is a noticeable exception and is less reliable as a PU mitigator than either \PIRANHA{} or traditional grooming methods.

\begin{figure}[t!]
    \subfloat[]{
        \includegraphics[width=.5\textwidth]{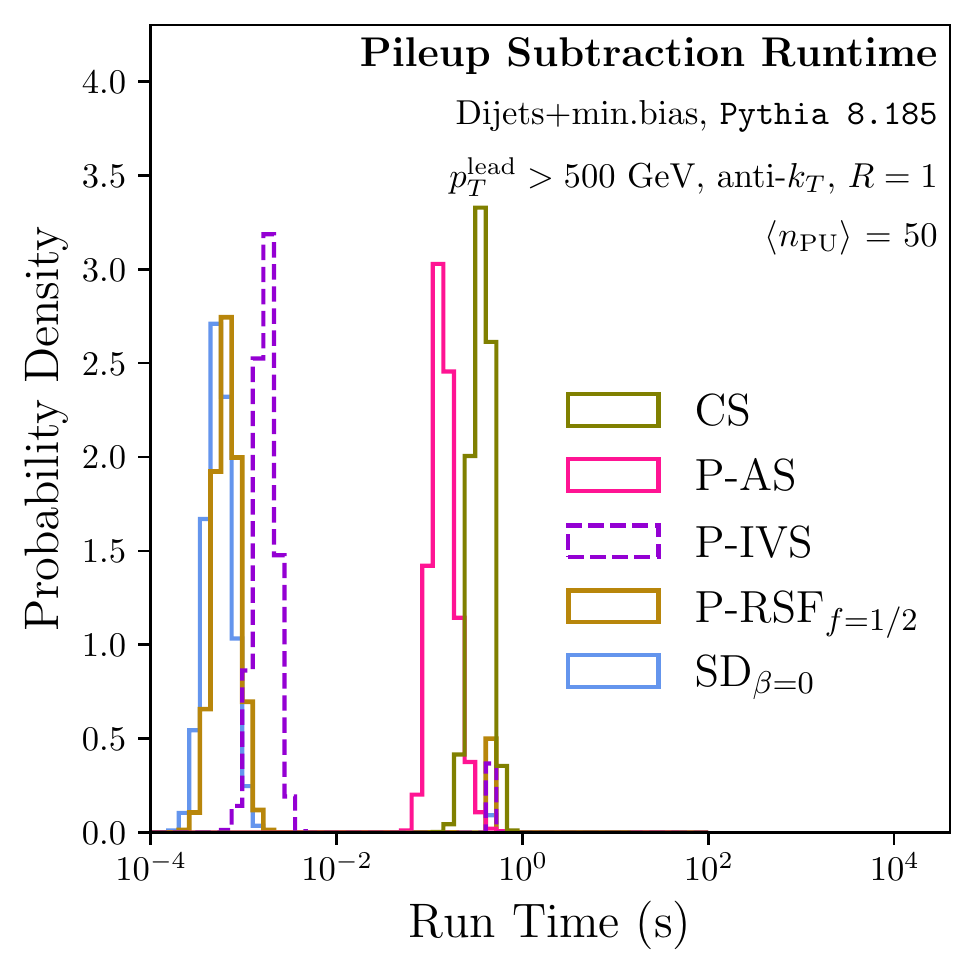}
    }
    \subfloat[]{
        \includegraphics[width=.5\textwidth]{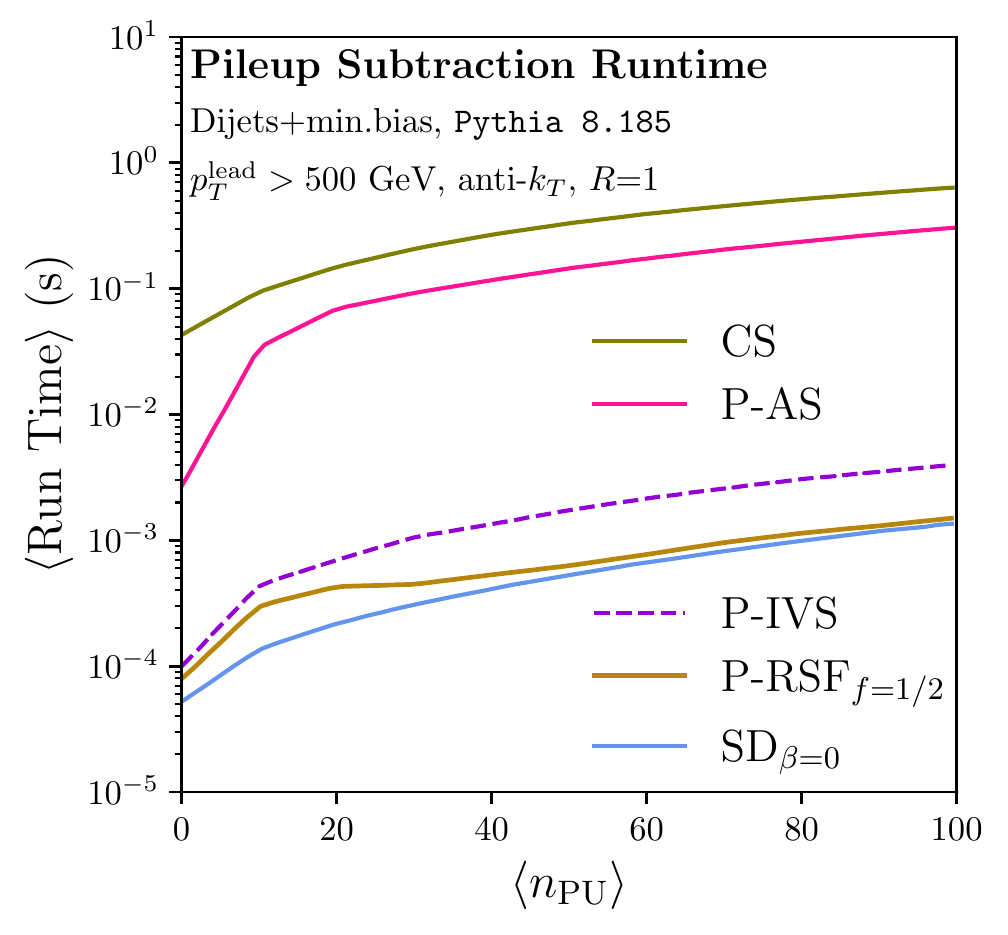}
    }
    \caption{
    Runtimes for \PIRANHA{} algorithms, and comparison to some traditional grooming and pileup mitigation algorithms.
    (a) shows the probability density for the logarithm of the runtime for each algorithm when \(\langle n_{\rm PU} = 50\rangle\), where \(n_{\rm PU}\) is the average number of pileup events.
    (b) shows the average runtime for each algorithm as a function of \(\langle n_{\rm PU} \rangle\).
    Runtimes are shown for the \texttt{Pythia 8.185} datasets of \Reff{Soyez:2018opl}, and we simulate pileup by layering a Poisson-distributed number of minimum-bias events over a hard dijet event.
    }
    \label{fig:runtimes}
\end{figure}

We also briefly present results for the runtimes of the grooming algorithms discussed in this paper, including the Constituent Subtraction (CS) pileup mitigation algorithm, as a function of the amount of additive contamination in events with PU.
In particular, \Fig{runtimes} displays the properties of runtimes for grooming algorithms functioning as pileup mitigators as we vary the average number of layered minimum bias events, \(\langle n_{\rm PU} \rangle\).
The tree-based grooming strategies, such as Soft Drop and P-RS, are the fastest.
P-IVS is an order of magnitude slower.
Finally, P-AS and CS are the slowest, with runtimes of \(\mathcal O({\rm 1s})\) even for a reasonable number of pileup events.
The combination of high speed and high performance of balanced P-RS provides simple preliminary evidence for their utility in the removal of pileup.
\\~\\

\begin{figure}[p]
    \centering
    \subfloat[]{
        \includegraphics[width=.32\textwidth]{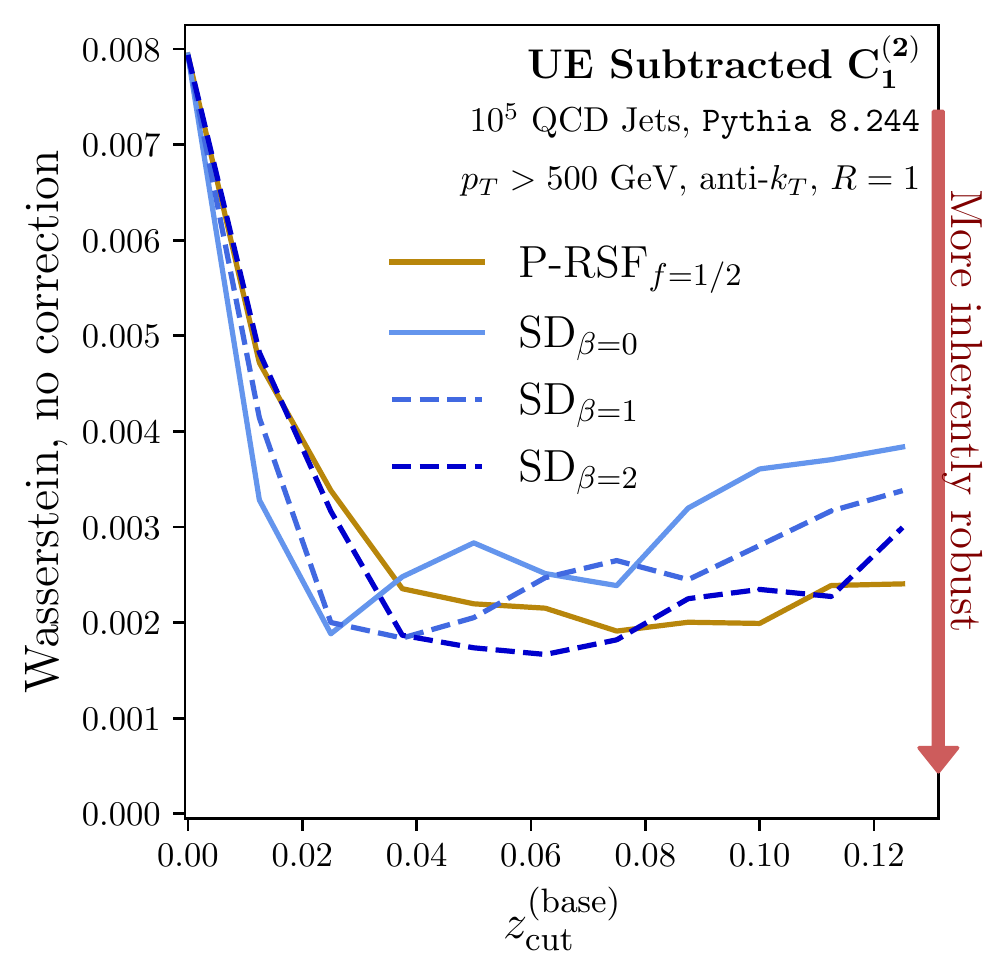}
        \label{fig:uefrenzy_wasserstein_sd}
    }
    \subfloat[]{
        \includegraphics[width=.32\textwidth]{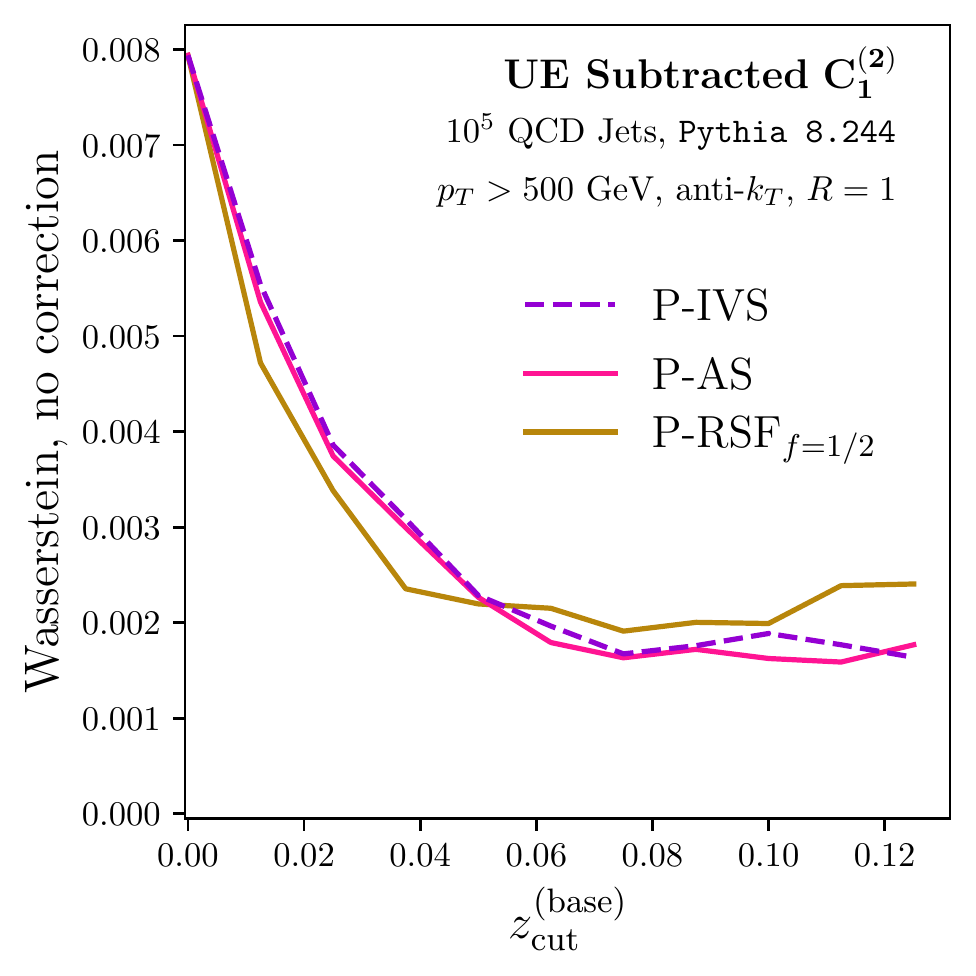}
        \label{fig:uefrenzy_wasserstein_pira}
    }
    \subfloat[]{
        \includegraphics[width=.32\textwidth]{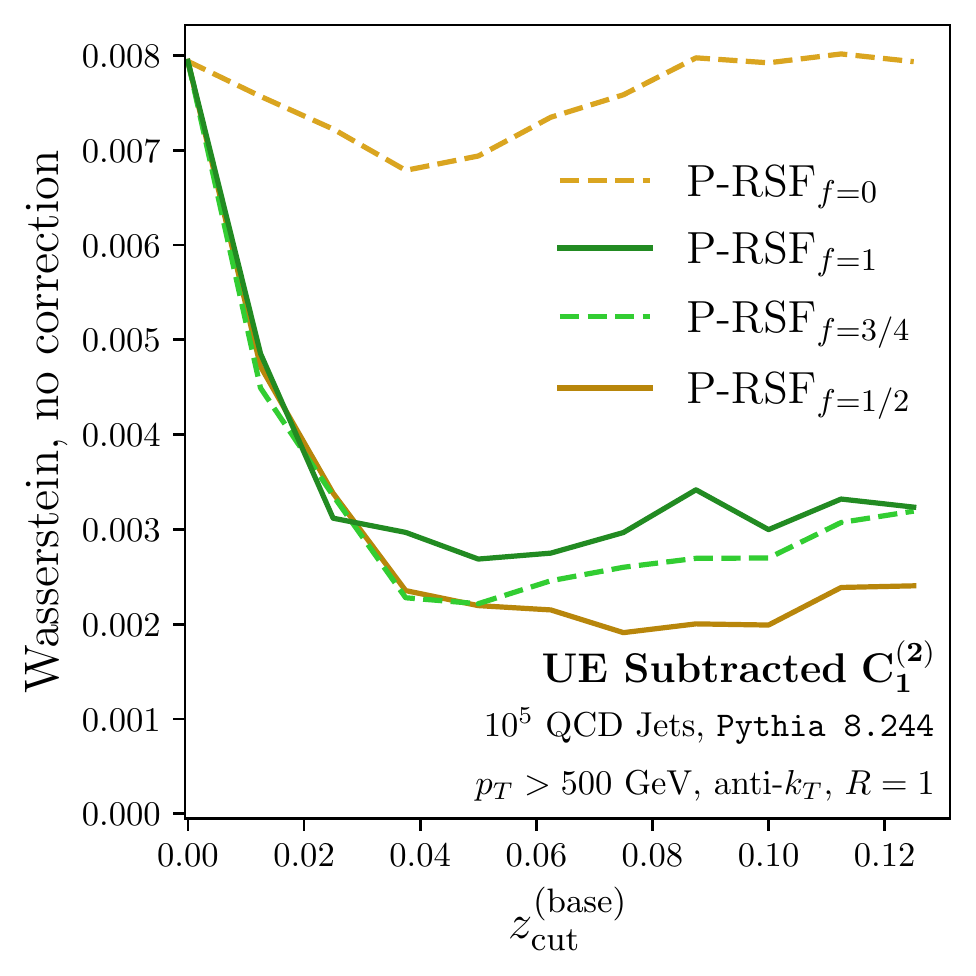}
    }
    \\
    \subfloat[]{
        \includegraphics[width=.32\textwidth]{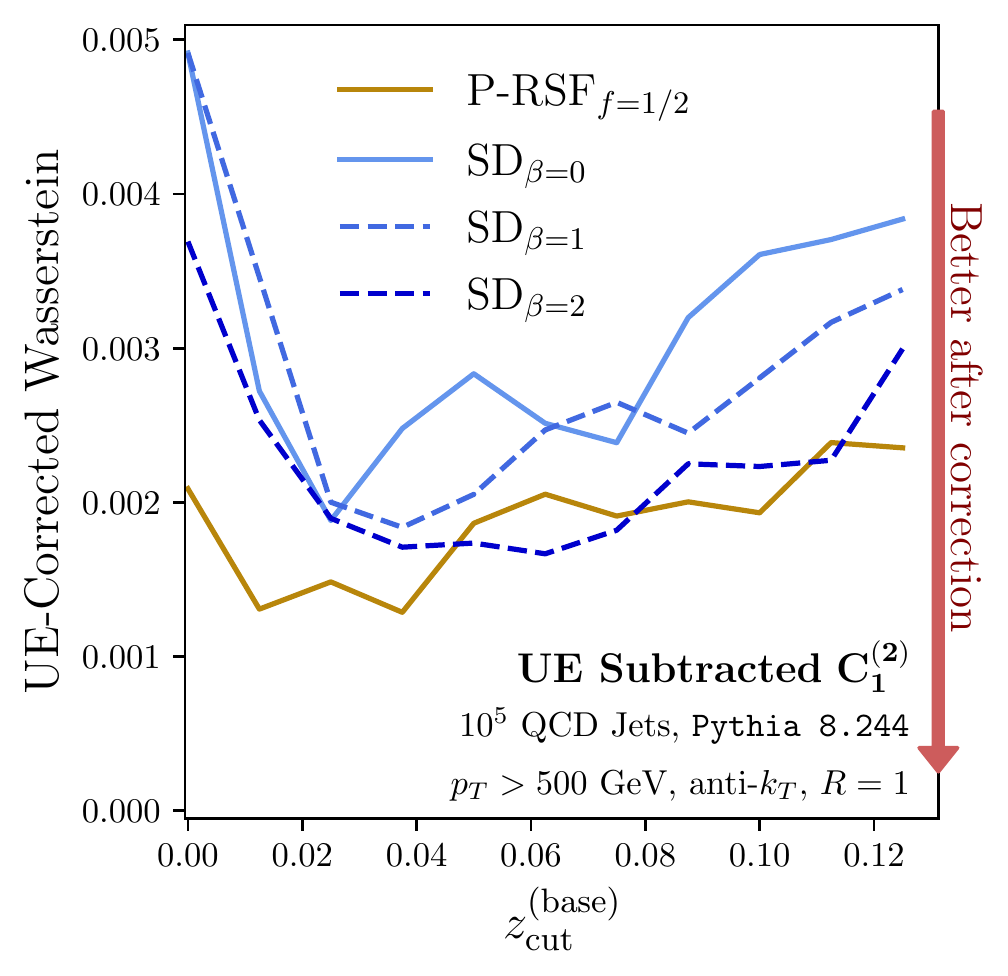}
        \label{fig:uefrenzy_correctedwasserstein_sd}
    }
    \subfloat[]{
        \includegraphics[width=.32\textwidth]{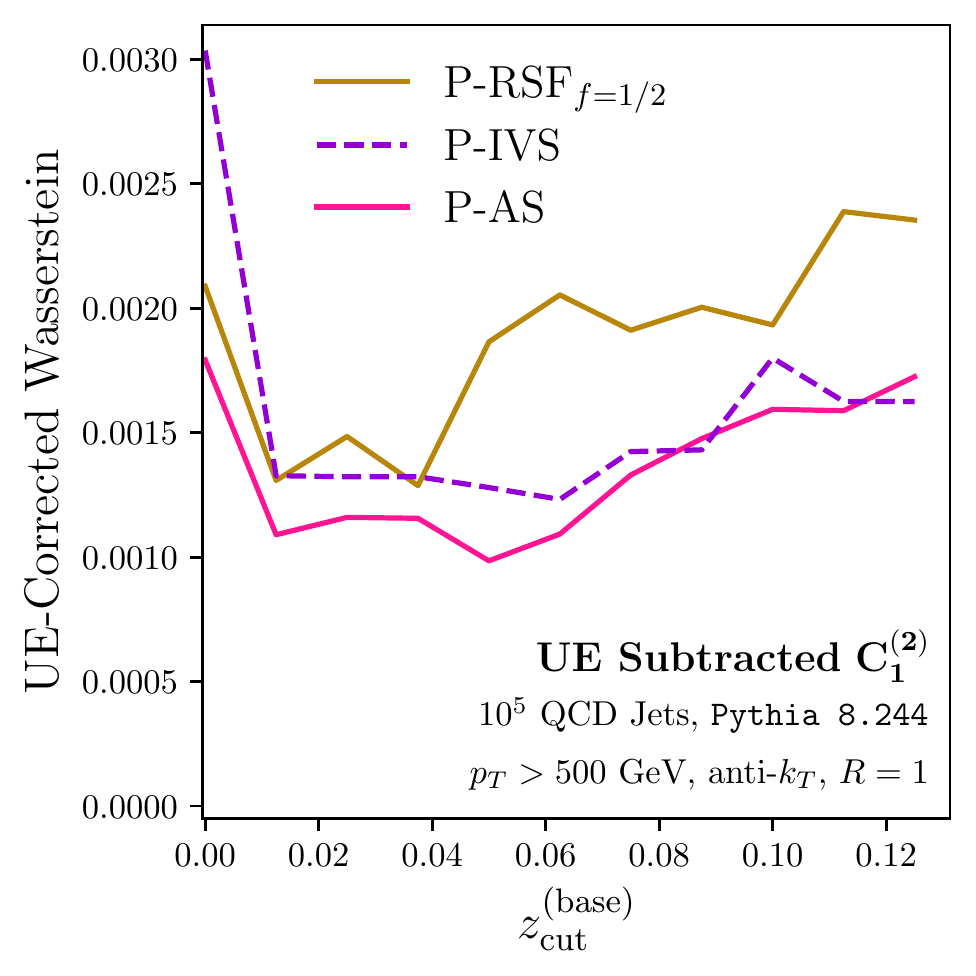}
        \label{fig:uefrenzy_correctedwasserstein_pira}
    }
    \subfloat[]{
        \includegraphics[width=.32\textwidth]{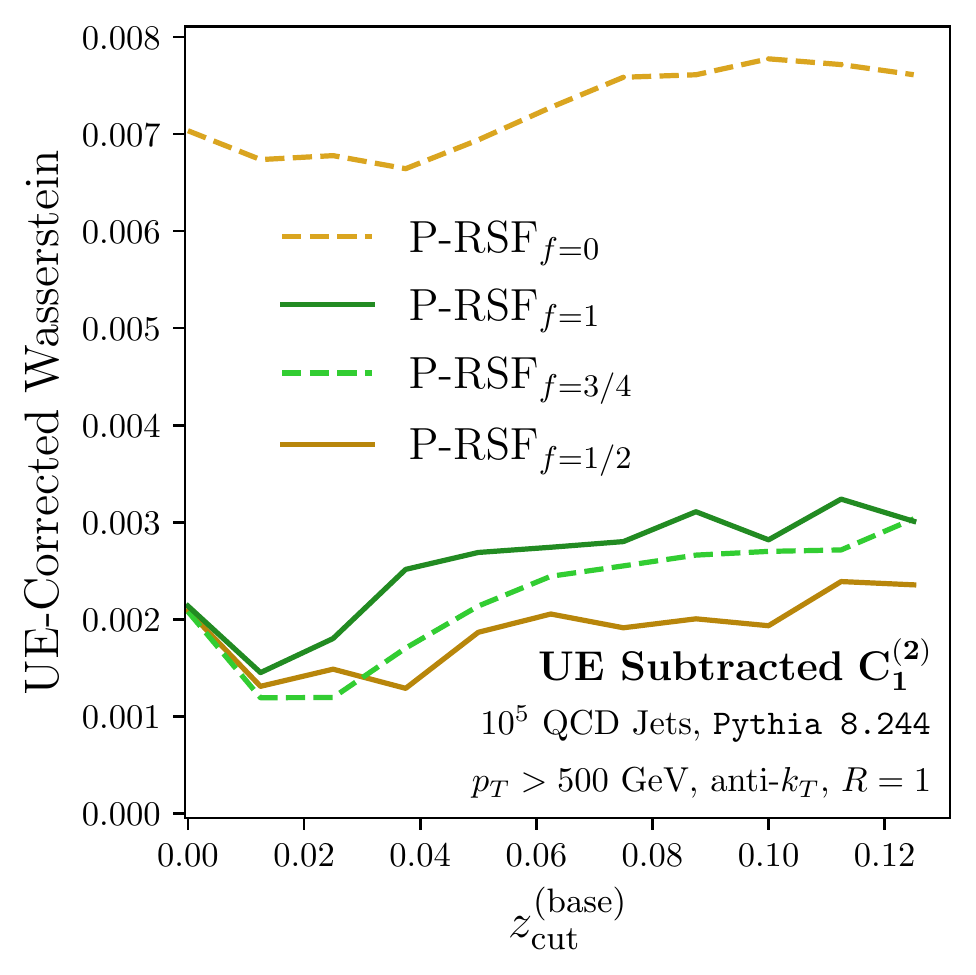}
    }
    \\
    \subfloat[]{
        \includegraphics[width=.32\textwidth]{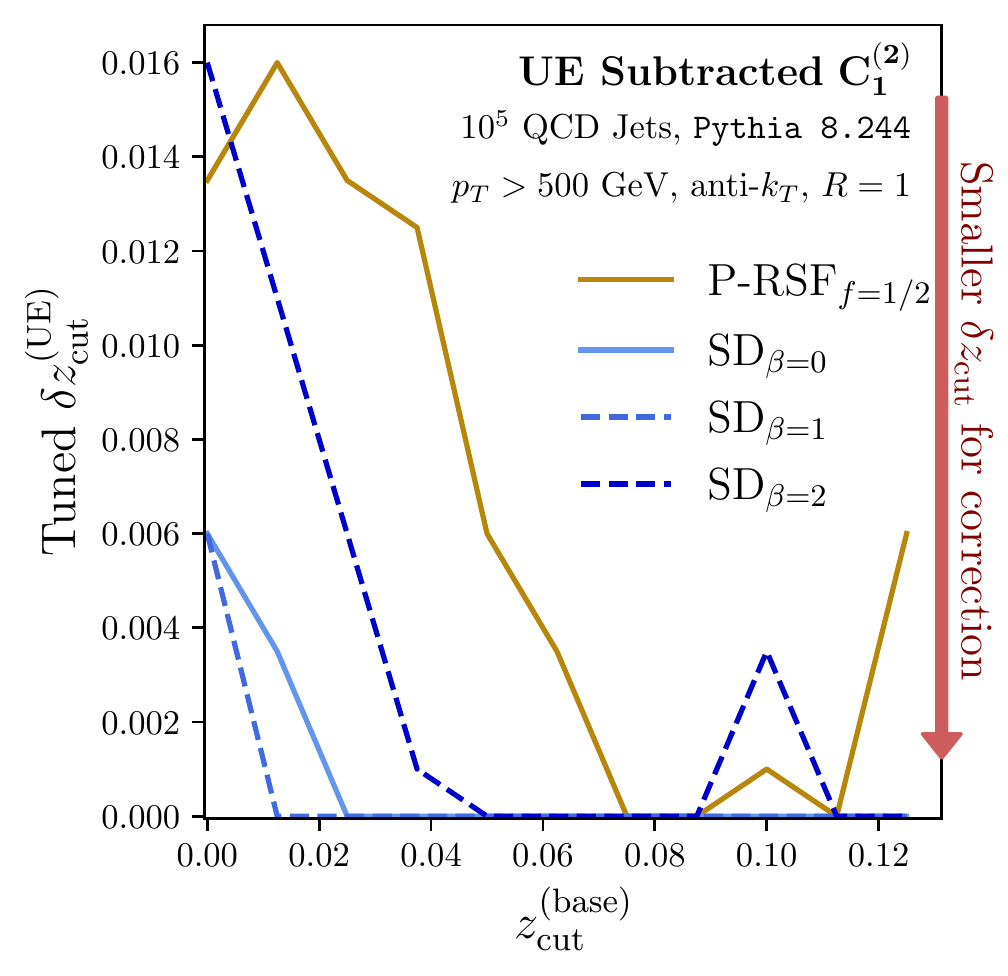}
    \label{fig:uefrenzy_deltazcut_sd}
    }
    \subfloat[]{
        \includegraphics[width=.32\textwidth]{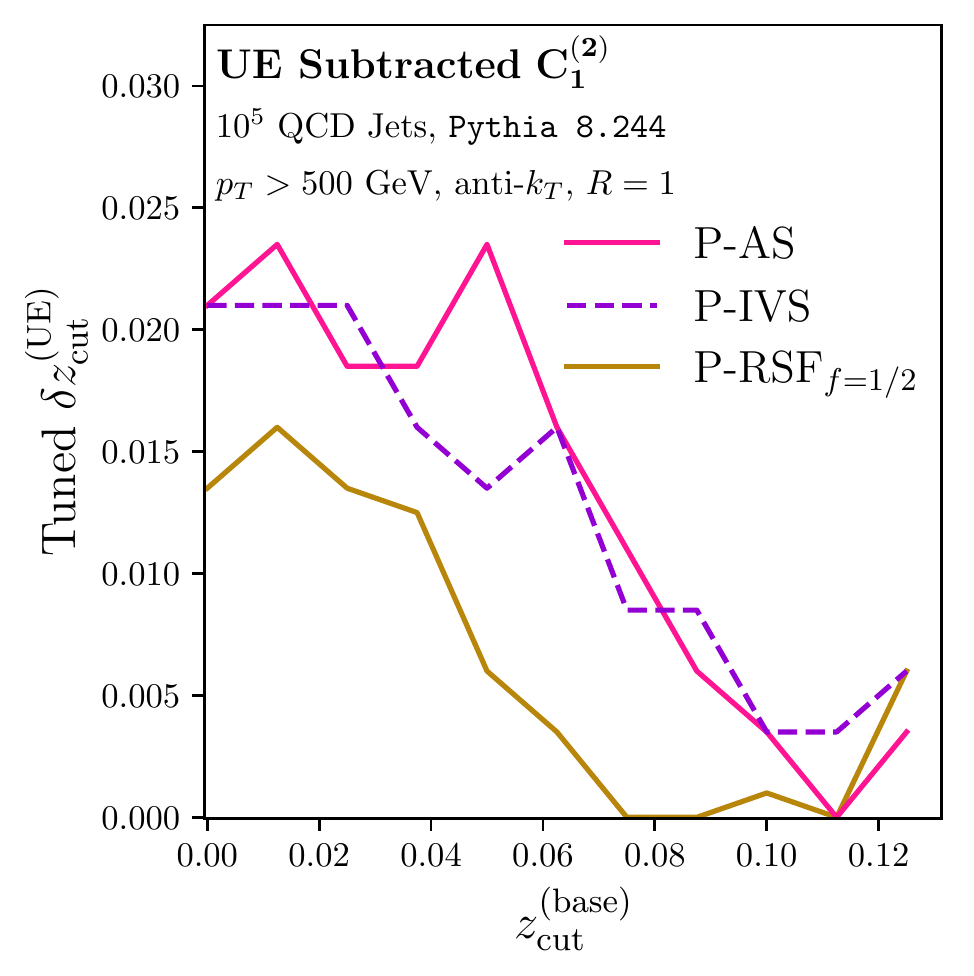}
    \label{fig:uefrenzy_deltazcut_pira}
    }
    \subfloat[]{
        \includegraphics[width=.32\textwidth]{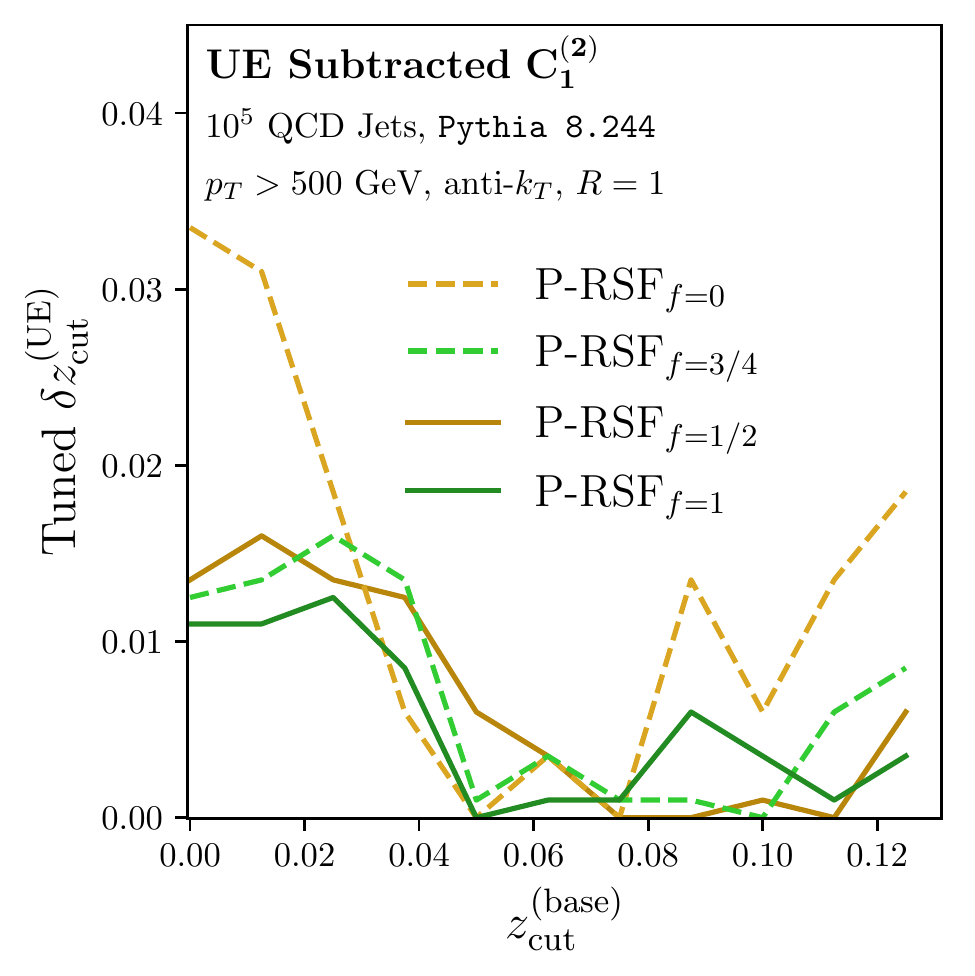}
    }
\caption{
    Wasserstein distance between jet \(C_1^{(2)}\) distributions (top row) without UE correction and (middle row) with UE correction, and (bottom row) the optimal value of \(\delta \zcut\) for UE correction.
}
\label{fig:uefrenzy}
\end{figure}

\subsection{Underlying Event}
\label{app:uefrenzy}

For the underlying event, we present results for the Wasserstein distances between substructure distributions with and without UE, discussed in \Sec{ue}, as well as the additional grooming \(\delta\zcut\) required for optimal UE correction.
We collect these results in \Fig{uefrenzy}.

In each of the plots, a lower value along the $y$-axis -- of either the Wasserstein distance or the optimal additional \(\delta\zcut\) -- indicates a desirable property for UE correction.
A lower Wasserstein distance without UE correction indicates inherent robustness of the groomer to the presence of UE.
A lower UE-corrected Wasserstein distance indicates that the groomer is able to better correct for the presence of UE.
Finally, a lower \(\delta\zcut\) indicates greater robustness against UE:
less additional grooming must be applied for UE correction.

Notably, \Fig{uefrenzy_correctedwasserstein_sd} shows that the UE-corrected Wasserstein distance of \PRSF{1} is smaller than that of Soft Drop for several values of \(\beta_{\rm SD}\) and a range of \(z_{\rm cut}^{\rm (base)}\), indicating that \PRSF{1} is a more delicate tool for the removal of additive contamination due to UE.
However, P-IVS and P-AS perform even better at UE removal, as shown in \Fig{uefrenzy_correctedwasserstein_pira}.
It may be that simple improvements to P-RS that emulate the features of P-IVS and P-AS, such as the ``Hard-Balanced Recursive Subtractors'' mentioned briefly in \Sec{rsf_discont}, may lead to even better performance in UE removal.
Finally, while the P-RSF groomers generally exhibit similar performance in UE correction, \PRSF{0} performs poorly at UE correction and is extremely sensitive to the effects of UE;
this is unsurprising, as \PRSF{0} deliberately grooms away hard radiation and leaves in soft radiation, such as the radiation contributed by UE.

Since the procedure we use to study the ability of each groomer to remove UE is based on a computationally expensive optimization of \(\delta\zcut\), the plots we show in \Fig{uefrenzy_deltazcut_sd} have some unphysical, jagged behavior, especially in the second two rows.
We expect that with a dedicated study with better statistics that the jagged behavior of these plots will be smoothed, but that the qualitative nature of our conclusions will remain unchanged.

\section{Towards Resummed Recursive Subtraction}
\label{app:calc}

We would like to gain theoretical insight, where possible, into the properties of \PIRANHA{}-groomed jets.
To move towards a perturbative understanding of \PIRANHA{}, we begin by studying the behavior of P-RSF at leading order (LO), deriving the results for P-RSF substructure presented in \Sec{sd_discont_lo}, and describe why this is sufficient for generic \PIRANHA{} groomers at LO.
Beyond LO, we find difficulties in using existing resummation techniques because of the intricate correlations between subtractions of different emissions.
Nonetheless, we can study the specific example of \PRSF{1} to obtain analytic resummed results up to leading logarithmic accuracy (LL) and numerical results beyond LL.

The angular-ordered, tree-based structure of P-RSF is an important feature that helps facilitate perturbative calculation.
In particular, we recall that the branching structure of an angular-ordered tree of emissions is comparable to the history of coherent parton branchings in the parton model~\cite{Collins:2011zzd}, and that the philosophy of local parton-hadron duality \cite{Azimov:1984np,Andersson:1989ww,Shifman:2003de,Neill:2020tzl} asserts that energy flows are not dramatically changed in the presence of hadronization.
We may therefore approximate the probability of finding a physical particle within an angular-ordered jet tree by the probability of finding an unphysical, unconfined QCD parton with the same energy and angle.
Each partonic emission in a jet is accompanied by a factor of the strong coupling \(\alpha_s\), enabling a perturbative analysis of the behavior of groomed observables.

We focus on quantifying the substructure of \PIRANHA{}-groomed jets using the two-prong generalized energy correlation functions (ECFs) of \Reff{Larkoski:2013eya}.
These were described by \Eqs{ECFdefn_lo}{ECFdefn_pheno}, and we repeat their approximate form in the central (\(y=0\)) and narrow (\(R_0 \ll 1\)) limit here for convenience:
\begin{equation}
    C_1^{(\varsigma)} \simeq \frac{1}{2}\sum_{i=1}^M\sum_{j=1}^M z_i z_j \left(\frac{\theta_{ij}}{R_0}\right)^\varsigma
    \label{eqn:ECFdefn_repeat}
    .
\end{equation}
As before, \(z_i\) indicates the energy fraction of particle \(i\), \(\theta_{ij}\) indicates the opening angle between particles \(i\) and \(j\), and \(R_0\) is the jet radius.

The \(C_1^{(\varsigma)}\) furnish a set of interesting but relatively simple observables to study in the context of \PIRANHA{}.
The behavior of the \PIRANHA{} groomed jet \(p_T\), for example, is relatively uninteresting:
since the \PIRANHA{} groomers discussed in this paper operate by subtracting a fixed amount of transverse momentum
\begin{align}
    \Delta p_T = \zcut\,p_T = \rho\,A_{\rm jet}
    ,
\end{align}
the changes to the transverse momentum of a jet due to the P-RSF grooming procedure are set trivially by the parameter \zcut{}.
However, as discussed in \Sec{sd_discont_lo}, refining calculations of observables involving the energy fraction \(z_g\) or angle \(r_g\) of the first emission to pass the grooming is a nuanced task that we will defer to future work.

We discuss the leading behavior of the distributions for P-RSF groomed ECFs in \App{LO_RSF}, with a brief discussion of the LO behavior of \PIRANHA{} groomers in general.
We discuss the behavior of P-RSF\(_{1}\) groomed ECFs at resummed accuracy, comparing results obtained with perturbative QCD, leading logarithmic parton showers, and \texttt{Pythia 8.244}, in \App{resumresults}.
We introduce the mathematical technology we use for calculations in perturbative QCD in \App{caesar_basics}, and our leading logarithmic parton shower in \App{partonshower}.

\subsection{Substructure at Fixed Order}
\label{app:LO_RSF}

For convenience, we first briefly repeat some comments made in \Sec{sd_discont_lo} while discussing the behavior of P-RSF at lowest order in \(\alpha_s\) (LO).
At LO, the emission of a single parton with energy fraction \(z\) and angle \(\theta\) is described approximately by the pseudo-probability distribution
\begin{align}
    \frac{\alpha_s}{\pi} p_i(z) \frac{1}{\theta}
    \approx
    \frac{2 C_{R_i} \alpha_s}{\pi}~\frac{1}{z}~\frac{1}{\theta},
    \label{eqn:dglap_approx}
\end{align}
where \(p_i(z)\) is a Dokshitzer-Gribov-Lipatov-Altarelli-Parisi (DGLAP) splitting function \cite{Gribov:1972ri,Dokshitzer:1977sg,Altarelli:1977zs} and \(C_{R_i}\) is the quadratic Casimir for the SU(3) color representation \(R_i\) of the mother parton, with \(C_F = \frac{N_C^2 - 1}{2 N_C} = \frac{4}{3}\) for quarks and \(C_A = N_C = 3\) for gluons.\footnote{
By focusing on the singular pieces of the splitting function in \Eq{dglap_approx}, we implicitly focus on the probability that a \textit{gluon} is emitted from the mother parton.
For example, a quark may emit a gluon, so that the resulting final state consists of both a gluon and a quark, while a gluon may split into two gluons.
In the language of splitting functions, a gluon may in principle split into two quarks;
however, the associated splitting function does not have any singularities as \(z \to 0\) and is therefore sub-leading to the singular behavior reflected in \Eq{dglap_approx}.
}
We use reduced splitting functions,
\begin{equation}
    \overline{p}_i(z) = p_i(z) + p_i(1-z),
\end{equation}
with \(z \in (0, 1/2)\), which describe the probability distribution of the softest parton produced in the splitting.

The singular nature of the DGLAP splitting functions in the limits \(z,\,\theta \to 0\) indicates that the phase space of partonic emissions is dominated by emissions with small energy fraction \(z\) and angle \(\theta\).
The problems posed by these singularities are addressed systematically by resummation, and we review the resummation method we apply here (often called the CAESAR method \cite{Banfi:2001bz,Banfi:2004yd}) in \App{caesar_basics}.
The non-singular pieces of these splitting functions are written explicitly in \Eqs{quark_splitting}{gluon_splitting}.

Because each emission is suppressed by a factor of \(\alpha_s\), the behavior of the P-RSF groomed ECFs of \Eq{ECFdefn_repeat}, away from \(C_1^{(\varsigma)} = 0\), is determined by the phase space distribution of two-parton jets whose particles survive the P-RSF grooming procedure, up to corrections proportional to \(\alpha_s^2\).
If the two partons survive the grooming procedure, they can be described with the modified energy fractions
\begin{subequations}
\begin{align}
    z_{\rm soft}' &= z - f\,\zcut > 0,
    \\
    z_{\rm hard}' &= 1 - z - (1-f)\zcut > 0,
\end{align}
\end{subequations}
where \(z\) is the original energy fraction of the softer parton.
The groomed jet ECF then takes the form
\begin{equation}
    C_1^{(\varsigma)}(z,\theta)
    =
    \left(z - f\,\zcut\right) \left(\frac{\theta}{R_0}\right)^\varsigma \frac{\left(1 - z - (1-f)\zcut\right)}{\left(1 - \zcut\right)^2}
    \approx
    \left(z - f\,\zcut\right)\left(\frac{\theta}{R_0}\right)^\varsigma
    \,
    \frac{\left(1 - (1-f)\zcut\right)}{\left(1 - \zcut\right)^2}
    \label{eqn:oneEm_groomedECFdefn}
    .
\end{equation}
In the last step, we use that the phase space of the emission is dominated by regions with \(z,\,\theta \ll 1\).
The factors of \((1-\zcut)^2\) that appear in the denominators of \Eq{oneEm_groomedECFdefn} emerge from the normalization of the groomed \(C_1^{(\varsigma)}\) by the groomed jet \(p_T\), as in the definition of \Eq{ECFdefn_repeat}.

We also note that if either of the partons in our two-parton configuration at LO is entirely groomed away, the groomed value of \(C_1^{(\varsigma)}\) is zero.
The region of phase space where one parton is groomed away at LO conspires with virtual contributions and the one-parton configuration with no splittings to ensure that the \(C_1^{(\varsigma)}\) distribution integrates to one.
These contributions may quickly be calculated once we find the distribution away from the origin, \(C_1^{(\varsigma)} > 0\), and we include them at the end of the following discussion.

In the DGLAP splitting functions, the energy fraction \(z\) takes values from \(0\) to \(1\).
Since the \(C_1^{(\varsigma)}\) are sensitive to the softest emission in the jet, we are instead interested in restricting \(z\) to take values from \(0\) to \(1/2\).
We are therefore interested in the \textit{reduced} splitting functions, \(\Bar{p}(z) = p(z) + p(1-z)\);
these have the same singular behavior as the unmodified DGLAP splitting functions near \(z = 0\), but are only used in the range \(z\in(0, 1/2)\).
From these considerations, the fixed-order distribution of \(C_1^{(\varsigma)}\) away from zero becomes
\begin{align}
    \rho^{\rm(LO)}_{i,~\varsigma}(C>0)
    =
    \frac{\alpha_s}{\pi}\int_0^{R_0}\frac{\dd\theta}{\theta}
    \int_0^{1/2} \dd z\, \Bar{p}_i(z)
    \delta\left(C - C_1^{(\varsigma)}(z,\theta)\right),
    \label{eqn:LO_dist_setup}
\end{align}
where  \(C_1^{(\varsigma)}(z,\theta)\) is the functional form of the groomed ECF given in \Eq{oneEm_groomedECFdefn}.

Up to terms that are power-suppressed in \(C\), \(\zcut\), or both, the fixed-order distribution for the groomed observable away from \(C = 0\) is
\begin{subequations}
\label{eqn:piranha_LO}
\begin{align}
    \rho^{\rm(LO)}_{i,~\varsigma}(C>0)
    &=
    \frac{2\alpha_s C_{R_i}}{\varsigma~\pi}
    \frac{1}{C}
    \left(
        2\tanh^{-1}\left(1  -2 \tilde C - 2\,f\,\zcut\right)
        + B_i
        +
        \mathcal{O}\left(C,\,\zcut\right)
    \right)
    \\
    &=
    \frac{2\alpha_s C_{R_i}}{\varsigma~\pi}
    \frac{1}{C}
    \left(
        -\log\left(\tilde C + f\,\zcut\right)
        + B_i
        +
        \mathcal{O}\left(C,\,\zcut\right)
    \right)
\end{align}
\end{subequations}
where \(\tilde C = C \times (1-\zcut)^2 / (1 - (1-f)\zcut)\), \(C_{R_i}\) is the quadratic Casimir of the color representation of the initiating parton, and we have defined the factor \(B_i\) to capture behavior from non-singular components of the splitting functions:
\(B_q = -3/4\) for quark initiated jets and \(B_g = -11/12 + n_f/(6 C_A)\) for gluon initiated jets.

Including the \(\mathcal{O}(\alpha_s^0)\) jet configuration with a single parton for which \(C_1^{(\varsigma)} = 0\), the two-parton contributions for which \(z < f \zcut\) and the softer emission is completely eliminated, and the virtual contributions at \(\mathcal{O}(\alpha_s^1)\), we can write the full LO probability distribution for \(C_1^{(\varsigma)}\) as
\begin{align}
    \label{eqn:plusdist_prsf_lo}
    \rho^{\rm(LO)}_{i,~\varsigma}(C)
    &\approx
    \delta(C)
    +
    \frac{2\alpha_s C_{R_i}}{\varsigma~\pi}
    \left[\frac{1}{C}
    \left(
        -\log\left(\tilde C + f\,\zcut\right)
        + B_i
        +
        \mathcal{O}\left(C,\,\zcut\right)
    \right)
    \right]^{(C_{\rm max})}_+
    ,
\end{align}
where \({C_{\rm max}} = (1/2 - f \zcut)(1 - (1-f)\zcut)/(1 - \zcut)^2\) is the maximum value of \(C_1^{(\varsigma)}\), and we have fixed the normalization of the distribution with \textit{plus-function regularization}.\footnote{
The plus-function regulation of the function \(q(x)\) is a distribution \([q(x)]^{(x_0)}_+\) defined by its integral against any function \(f(x)\) in a suitable space of test functions,
\begin{align}
    \int_0^{x_0} \dd x [q(x)]^{(x_0)}_+ f(x)
    &=
    \int_0^{x_0} \dd x ~ q(x) \left(f(x)  - f(0)\right)
    \label{eqn:plusdist_def}
    .
\end{align}
See Appendix B of \Reff{Ligeti:2008ac} and the presentation of \Reff{Ebert:2016gcn} for satisfying discussions.
}
The plus-function regularized expression of \Eq{plusdist_prsf_lo} may be derived by replacing \(\overline{p}(z)\) in \Eq{LO_dist_setup} with its plus-function analog.
Alternatively, one can recognize that the full distribution for \(C_1^{(\varsigma)}\) must integrate to one even at LO, and that additional contributions to the LO distribution in \Eq{piranha_LO} can only come from the regions of phase space where \(C_1^{(\varsigma)}=0\);
this is enough to uniquely specify the result of \Eq{plusdist_prsf_lo} at LO.

Unlike the LO distributions for traditionally groomed jet ECFs, such as those of Soft Drop \cite{Larkoski:2014wba}, the LO distributions for \PRSF{1/2} and P-RSF groomed \(C_1^{(\varsigma)}\) do not exhibit piece-wise behavior.
While the piece-wise behavior of Soft Drop observable distributions is smoothed out by all-orders effects \cite{Benkendorfer:2021unv}, the smoothness of \PIRANHA{}-groomed distributions at LO is a manifestation of their continuity.

\subsection{The Groomed Energy Fraction for a Single Emission}
\label{app:groomedenergyfraction}
Next, we discuss the \textit{groomed energy fraction}, \(z_g\), of the first emission to survive the grooming procedure.
In Soft Drop, \(z_g\) is the energy fraction of the first emission to pass the Soft Drop criterion \(z > \zcut \theta^{\beta_{\rm SD}}\).
In P-RSF, \(z_g\) is the groomed energy fraction of the first emission that is not entirely removed by the grooming, which we call the \textit{critical emission}.

The behavior of \(z_g\) for Soft Drop was derived to double-logarithmic accuracy by \Reff{Larkoski:2015lea} with the technology of Sudakov safety.
Expanding their double-logarithmic result to \(\mathcal{O}(\alpha_s)\), the \(z_g\) distribution for Soft Drop takes the approximate form
\begin{align}
    \rho_{i,\,\,\rm SD}(z_g; \beta_{\rm SD}) = 
    \begin{dcases}
       \frac{2 \alpha_s C_{R_i}}{\pi |\beta|} \, 
       \overline{p}_i(z_g) \, 
       \log(z_g /z_c)
       \, 
       \Theta(\zcut < z_g)
       +
       \mathcal{O}(\alpha_s^2),
       &
       \beta_{\rm SD} < 0;
       \\
       \frac{\overline{p}_i(z_g)}{\int_{\zcut}^{1/2} \dd z \,\, \overline{p}_i(z)}
       \, 
       \Theta(\zcut < z_g),
       &
       \beta_{\rm SD} = 0;
       \\
       \sqrt{\frac{\alpha_s C_{R_i}}{\beta}} \,\, \overline{p}_i(z_g)
       +
       \frac{2 \alpha_s C_{R_i}}{\beta} \,\, \overline{p}_i(z_g)
        \log\min(2\zcut, 2z_g)
       +
       \mathcal{O}(\alpha_s^{3/2}),
       &
       \beta_{\rm SD} > 0,
    \end{dcases}
    \label{eqn:sd_zg}
\end{align}
where \(i\) again indicates the flavor (quark or gluon) of the initiating parton, and we note that the expression for \(\beta_{\rm SD} = 0\) receives no \(\mathcal{O}(\alpha_s)\) corrections at double-logarithmic accuracy.
If \(\beta_{\rm SD} \leq 0\), the probability distribution has no support when \(z_g < \zcut\) because \(z_g\) is determined by the Soft Drop criterion;
the vanishing of the \(z_g\) distribution when \(z_g < \zcut\) for \(\beta_{\rm SD} \leq 0\) therefore reflects the soft discontinuous behavior of Soft Drop.
If \(\beta_{\rm SD} > 0\) there is instead a kink at \(z_g = \zcut\) that similarly reflects Soft Drop's soft discontinuity.

The behavior of the P-RSF \(z_g\) distribution is more subtle, for reasons we discuss below.
At the level of a single emission, however, the \(z_g\) distribution for Soft Drop may be translated directly to P-RSF because of the similarities between Soft Drop with \(\beta = 0\) and P-RSF:\footnote{
In \Eq{prsf_zg_lo}, we choose \(z_g\) to be the groomed energy fraction of what had been the softer emission before grooming, regardless of which emission is softer after grooming.
This avoids additional corrections if \(f_{\rm soft} < 1/2\), where more grooming is applied to the harder emission than the softer emission at each branch and it becomes possible that the harder of two ungroomed emissions becomes the softer of the two emissions after grooming.
If we instead define \(z_g\) to be the softer \textit{groomed} emission, there are corrections to this formula for \(z_g > 1/2 - \mathcal{O}(\zcut)\) when \(f_{\rm soft} < 1/2\).
We also assume that \(\zcut < 1/2\);
if \(f_{\rm soft} < 1/2\) and \(\zcut > 1/2\), there are additional \(\mathcal{O}(\zcut)\) corrections to \Eq{prsf_zg_lo} due to the constraint that the harder ungroomed emission is not entirely groomed away.
}
\begin{align}
    \rho_{i,\,\,\text{P-RSF}}(z_g; f)
    = 
    (1-\zcut)
    \frac{
    \overline{p}_i\left[
        \left(1 - \zcut\right)z_g\,+\,f \zcut
    \right]
    }
    {\int_{f \zcut}^{1/2} \dd z' \,\, \overline{p}(z')}
    +
    \mathcal{O}(\alpha_s)
    \label{eqn:prsf_zg_lo}
    .
\end{align}
\(\rho_{i,\,\,\text{P-RSF}}(z_g; f)\) has support in the domain \(z_g \in \left(0,\,(1/2 - f\zcut)\,/\,(1-\zcut)\right)\) at this order of accuracy.
The additional factors of \((1-\zcut)\) relative to \Eq{sd_zg} capture the fact that \(z_g\) is normalized further after grooming:
\(z_g = (z_0 - f\zcut)/(1-\zcut)\), where \(z_0\) denotes the ungroomed softer energy fraction.
The lower limit \(f \zcut\) on the integration variable \(z'\) reflects that we consider configurations for which the softer emission survives the grooming procedure.

Much like Soft Drop with \(\beta_{\rm SD} = 0\), the LO behavior of the P-RSF \(z_g\) distribution is determined by the fact that the first surviving emission must have an energy fraction greater than the assigned grooming.
At LO, there is a single emission, and therefore the assigned grooming is simply \(f \zcut\).
Unlike Soft Drop, however, the \(z_g\) distribution for P-RSF has support down to \(z_g = 0\):
a two-parton event with \(z = f \zcut + \delta\), \(\delta \ll 1\), is mapped to a groomed two-parton event with \(z_g = \delta\), while the two-parton event with \(z = f \zcut - \delta\) is mapped to a one-parton event;
though \(z_g\) is undefined in the latter case, where the groomed jet consists of only a single parton, the fact that \(z_g\) can become infinitesimally small is a mark of the continuity of P-RSF.

Improving the accuracy of the calculation of \(z_g\) for P-RSF will require a more detailed analysis which we leave to future work.
An important complication emerges because there may be emissions before the critical emission that can soak up some of the grooming;
the amount by which the critical emission is groomed is therefore not \(\zcut\) but a smaller effective value, \(\zcut^{\rm(eff)} = \zcut - \Delta \zcut\), where \(\Delta \zcut\) depends on the energy fractions of earlier emissions.
This leads to problems when following the approach of \Reff{Larkoski:2015lea}, suited to the calculation of \(z_g\) for Soft Drop.
\Reff{Larkoski:2015lea} calculates the distribution of \(z_g\) in the framework of Sudakov safety by first marginalizing over the distribution of the angle of the critical emission, \(r_g\).
However, the distribution of \(r_g\) depends on the amount by which the critical emission is groomed, \(\zcut^{(\rm eff)}\), while a leading logarithmic calculation of \(\zcut^{(\rm eff)}\) depends on the value of \(r_g\).
The circular dependence of the distributions of \(r_g\) and \(\zcut^{\rm(eff)}\) prevents us from applying the formalism of \Reff{Larkoski:2015lea} without more careful modification.

\subsection{Substructure at Leading Logarithmic Accuracy and Beyond}
\label{app:resumresults}
The global nature of \PIRANHA{} grooming algorithms leads to intricate correlations between the grooming of different final-state particles, and thus to complications in perturbative calculations beyond LO.
For \PRSF{1/2}, P-IVS, and P-AS, every particle in a jet is groomed by some amount, and the grooming of any particular final-state particle is correlated with the grooming of every other.
This subtlety renders existing techniques for multiple emissions calculations less effective.
\PIRANHA{}-groomed observables depend on global information, and a more detailed understanding of \PIRANHA{}-groomed correlators will require techniques that can elucidate global information associated with a jet.

While we do not get around this subtlety for the \PIRANHA{} grooming procedures presented in this paper, we may use P-RSF to at least gain some intuition for the resummed behavior of recursive subtractors.
In particular, we navigate around the subtleties associated with \PIRANHA{} by studying the substructure of P-RSF with \(f_{\rm soft} = 1\), or \PRSF{1}.
Crucial to our analysis is that \PRSF{1} eventually terminates, such that it does not necessarily affect every particle within an event.\footnote{
\PRSF{0} is another subtraction algorithm that eventually terminates, but it subtracts away hard radiation instead of soft radiation.
Therefore, we focus on \PRSF{1} in the discussion below.
We leave a more detailed survey of the analytic properties of \PRSF{1/2}, \PRSF{0}, and Recursive Subtraction more generally to future work.
}
Due to this additional simplicity, we can study \PRSF{1} at leading logarithmic (LL) accuracy in slightly more detail, and find analytic expressions for observables related to the first emission to survive the grooming procedure, or \textit{critical} emission.

The calculation of these resummed distributions is discussed in more detail in \App{caesar_basics}, whose methods may be used to show that the LL distribution for the energy fraction \(z\) and the angle \(\theta\) of the first emission to survive the P-RSF\(_1\) procedure is
\begin{equation}
    \rho^{\rm(LL)}_{\rm crit}(\zcrit, \thetacrit)
    =
    \frac{2 C_R \alpha_s}{\pi}
    \frac{1}{\zcrit}\frac{1}{\thetacrit}
    \exp\left[-\frac{2 C_R \alpha_s}{\pi}
    \log\thetacrit
    \log 2 \zcut
    \right]
    .
    \label{eqn:LLcrit_ztheta}
\end{equation}
In \Eq{LLcrit_ztheta} and in the remainder of our discussion, we suppress the index \(i\) denoting the flavor of the initiating parton because our LL results depend unambiguously on the representation of the initiating parton through \(C_R\).

Using this resummed distribution, the singular pieces of the splitting function, and dropping subleading terms in \(\zcut\), we use the cumulative analog of \Eq{LO_dist_setup} at LL to find the cumulative distribution \(\Sigma^{\rm(LL)}_{\varsigma,~{\rm crit}}(C)\) for the contribution of the critical emission to the groomed \(C_1^{(\varsigma)}\)  for general \(f_{\rm soft}=f\):


\begin{equation}
\begin{aligned}
    \Sigma^{\rm(LL)}_{\varsigma,~{\rm crit}} (C)
    &=
    C^{\eta_f}
    + \frac{2C_R\alpha_s}
    {\varsigma \pi}
    \left(
        \frac{C}{f\,\zcut}
    \right)
    ^{\eta_f}
    \frac{1}{\eta_f(\eta_f-1)}
    \\
    &~~~~~~
    \times
    \Bigg[
        \left(
        \frac{f\,\zcut}{C}
        \right)
        ^{\eta_f - 1}
        {}_{~~2}F_1
        \left(
            1, 1-\eta_f, 2-\eta_f,
            -\frac{C}{f\,\zcut}
        \right)
        \\
        &~~~~~~~~~~~~
        +
        \left(
        \frac{f\,\zcut}{C}
        \right)
        ^{\eta_f}
        \left(\eta_f-1\right)
        \log\left(
        1 + \frac{C} {f\,\zcut}
        \right)
        \\
        &~~~~~~~~~~~~
        -\left(f\,\zcut\right)
        ^{\eta_f - 1}
        {}_{~~2}F_1
        \left(
            1,1-\eta_f,2-\eta_f,
            1-1/(2\,f\,\zcut)
        \right)
        \\
        &~~~~~~~~~~~~
        -
        \left(f\,\zcut\right)
        ^{\eta_f}
        \left(\eta_f-1\right)
        \log\left(1/(2\,f\,\zcut)\right)
    \Bigg]
    ,
    \label{eqn:ECF_LL}
\end{aligned}
\end{equation}
where we have defined
\begin{align}
    \eta_f
    &=
    -\frac{2 C_R \alpha_s}{\varsigma \pi}
    \log(2\,f\,\zcut)
    .
\end{align}
To gain some intuition for this intimidating expression, we may also expand this to first order in our fixed coupling \(\alpha_s\) to retrieve
%
\begin{equation}
\begin{aligned}
    \Sigma^{\rm(LL)}_{\varsigma~{\rm crit}} (C)
    &\approx
    1 + \frac{2 C_R \alpha_s} {\varsigma \pi}
    \left(
        -\log(2\,f\,\zcut)
        \log C
        +
        {\rm Li}_2
        \left(
        -\frac{C}{f\,\zcut}
        \right)
        -
        {\rm Li}_2(1 - 1/(2\,f\,\zcut))
    \right)
    .
    \label{eqn:convolDistExpansion}
\end{aligned}
\end{equation}
Taking a derivative gives
\begin{equation}
    \rho^{\rm(LL)}_{\varsigma~{\rm crit}} (C \,>\, 0)
    =
    \frac{\dd}{\dd C}\Sigma^{\rm(LL)}_{\varsigma~{\rm crit}} (C \,>\, 0)
    \approx
    -\frac{2 C_R \alpha_s} {\varsigma~\pi}\,\frac{1}{C}\,\log\left(2C + 2f \zcut\right)
    \label{eqn:ll_pdf_expansion}
    ,
\end{equation}
which more clearly reveals the LL contribution of the critical emission to the groomed value of \(C_1^{(\varsigma)}\).
\Eq{ll_pdf_expansion} differs from the LO result of \Eq{piranha_LO} by a factor of \(2\) inside the logarithm simply because we have integrated \(z\) from \(0\) to \(1/2\) and used only the singular parts of the splitting function in computing the LL result, while the LO result includes non-singular pieces of the splitting function.
By noting that the probability distribution must integrate to 1, one can also produce an expression with end-point contributions at \(C = 0\) which can then be compared to \Eq{plusdist_prsf_lo}.

Even for P-RSF with \(f_{\rm soft} = 1\), however, the subtractive nature of the grooming procedure leads to additional subtleties.
Usually, the distribution of partonic emissions in the \(\log z\)--\(\log \theta \) plane is approximately uniform, so that a particular emission will tend to have an exponentially higher contribution to the two-pronged substructure of a jet than any other.
Since the first emission to survive an angular ordered hard-cutoff grooming procedure such as Soft Drop will have a non-negligible energy fraction \(z > \zcut\) and a large angle, it is this first surviving emission, the critical emission, that contributes dominantly to \(C_1^{(\varsigma)}\).
This trick made the original calculation of Soft Drop groomed substructure quite simple \cite{Larkoski:2014wba}.

In the case of \PRSF{1}, however, the critical emission is partially groomed and may have a groomed energy fraction \(z_{\text{P-RSF}_1} \sim z - \zcut \ll 1\).
It is therefore possible for other emissions -- the ungroomed \textit{subsequent emissions} which are narrower than the critical emission and therefore untouched by the \PRSF{1} algorithm -- to contribute greater values to \(C_1^{(\varsigma)}\).
The ungroomed \textit{subsequent} emissions after the \PRSF{1} algorithm terminates can lead to larger contributions to substructure observables than they would in the case of Soft Drop.

\begin{figure}[t!]
\centering
\subfloat[]{
\includegraphics[width=.48\textwidth]{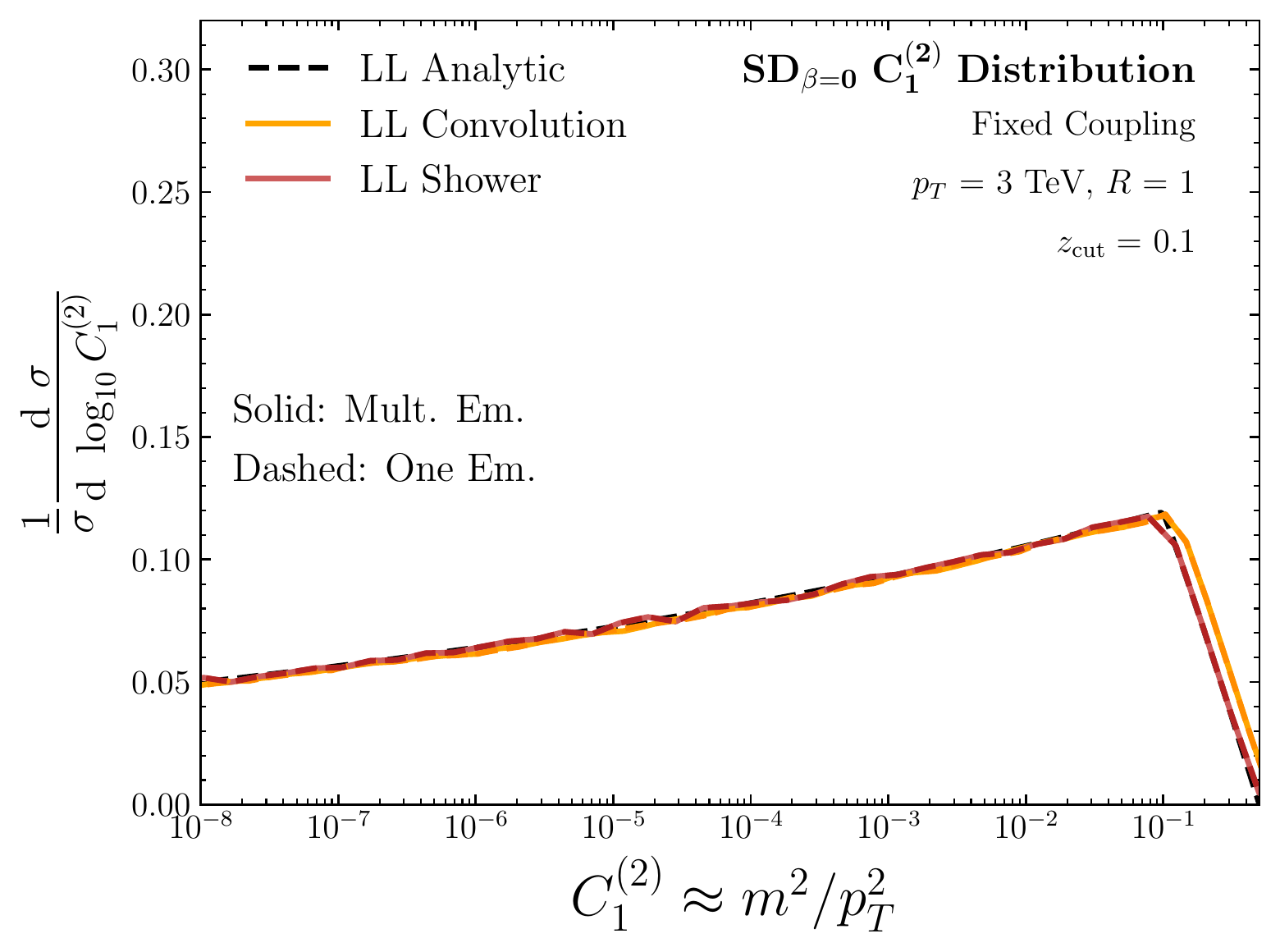}
\label{fig:LL_SD0}
}
%
\subfloat[]{
\includegraphics[width=.48\textwidth]{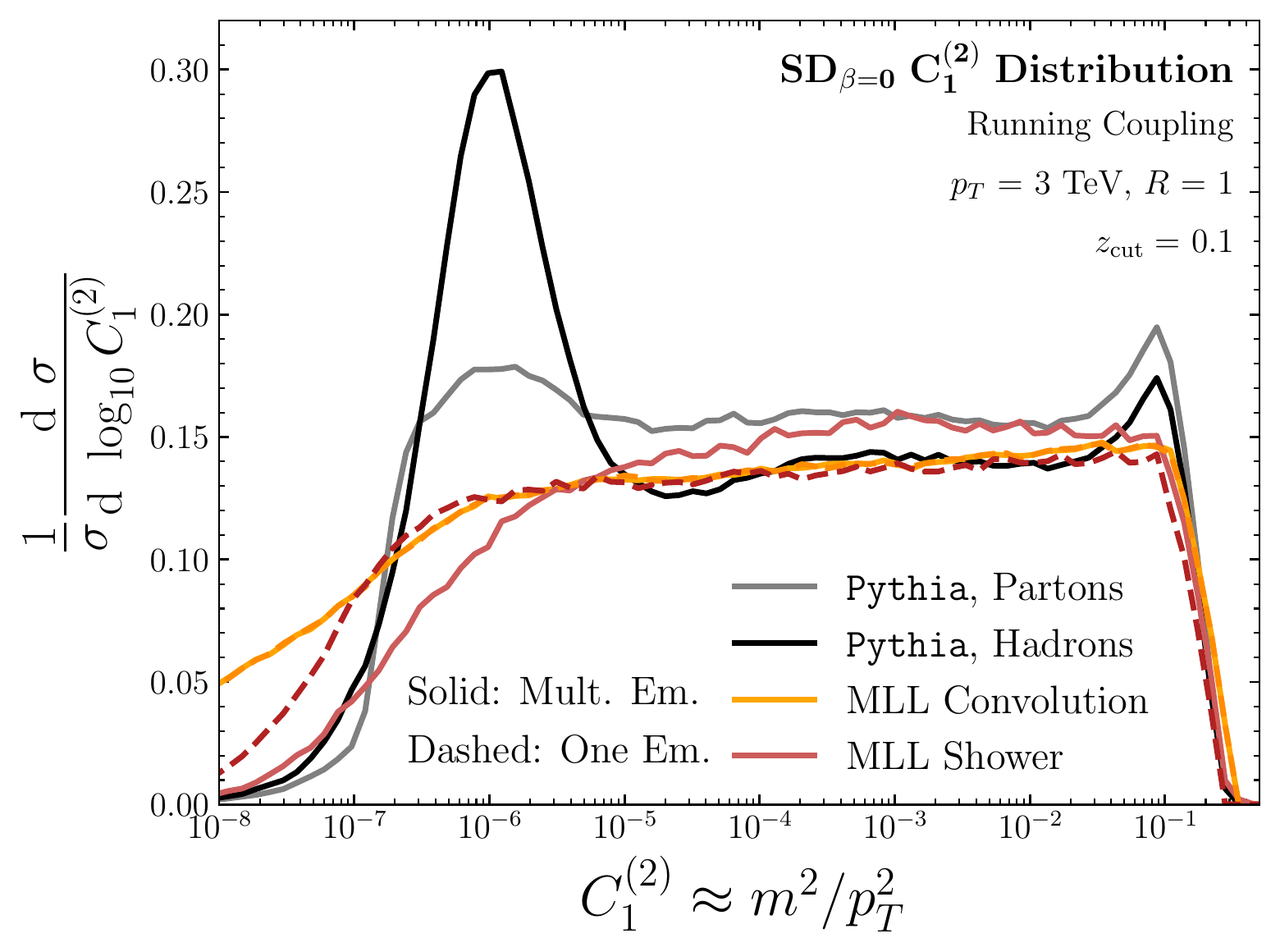}
\label{fig:MLL_SD0}
}
\\
%
\subfloat[]{
\includegraphics[width=.48\textwidth]{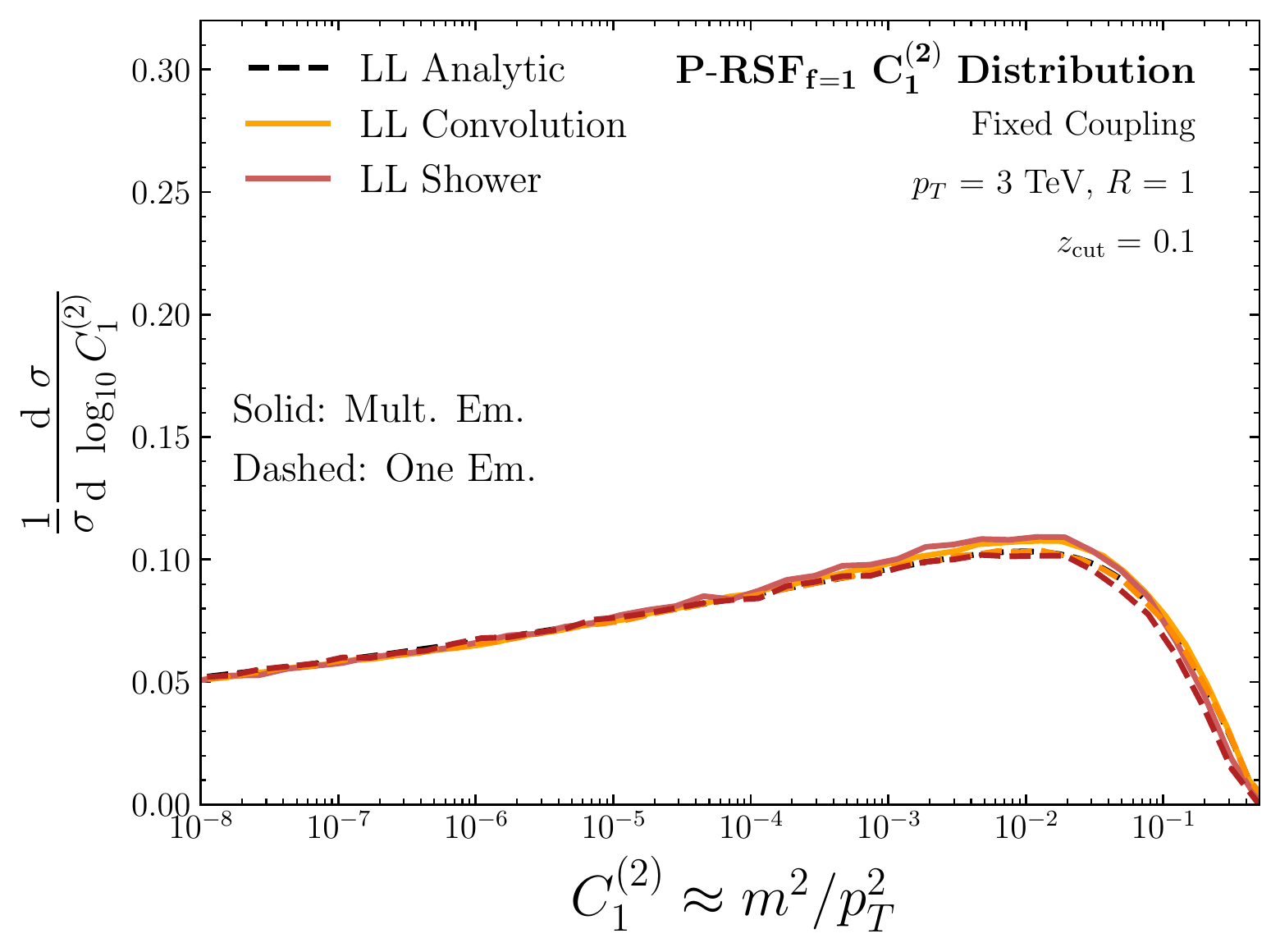}
\label{fig:LL_PRSF1}
}
%
\subfloat[]{
\includegraphics[width=.48\textwidth]{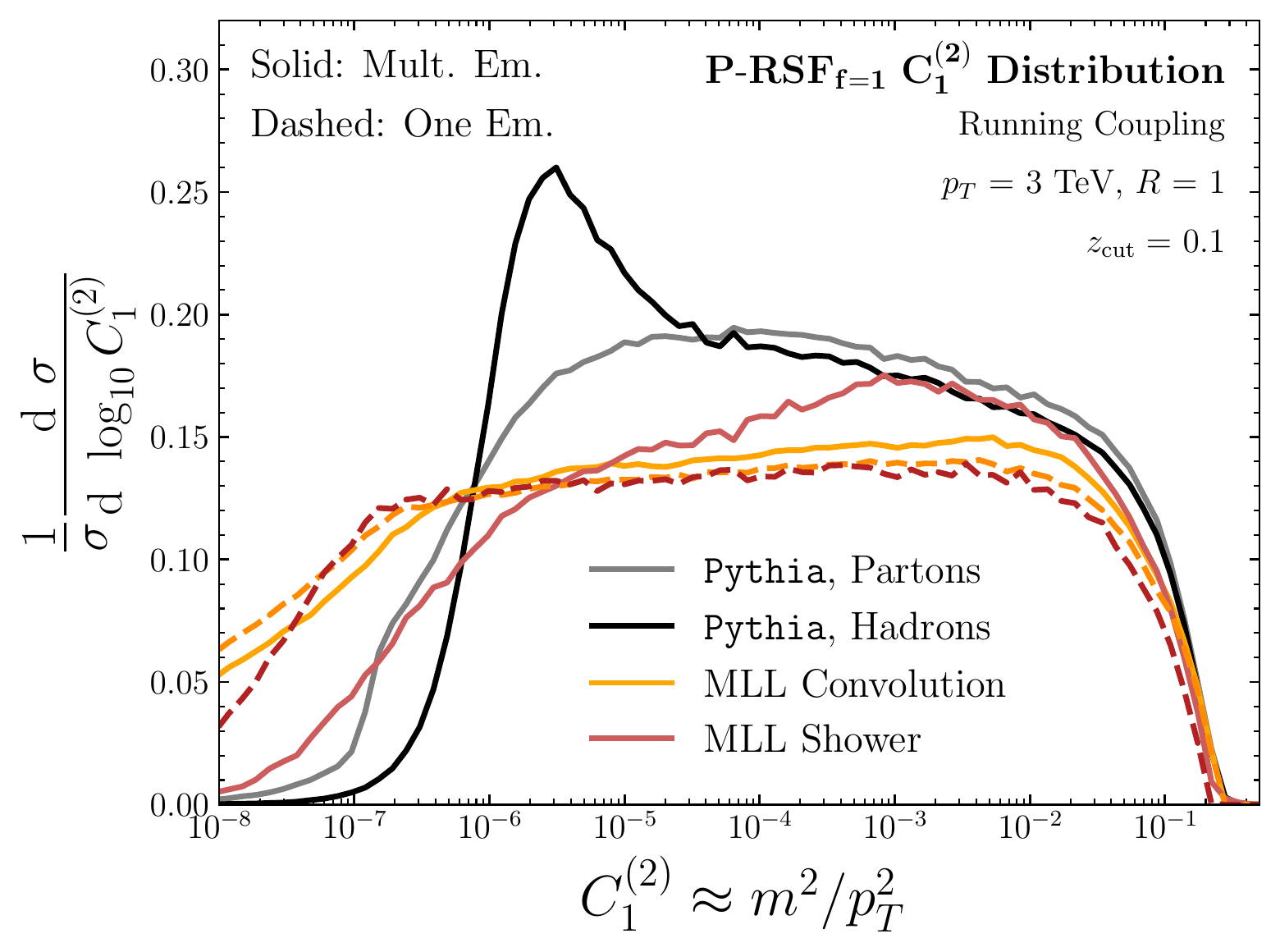}
\label{fig:MLL_PRSF1}
}
\caption{
Distributions of the observable \(C_1^{(2)}\approx m^2 / p_T^2\) of \Eq{ECFdefn_repeat} for quark jets groomed with \SD{0} (top row) and \PRSF{1} (bottom row), with and without multiple emissions, at (left column) LL accuracy and (right column) MLL accuracy.
For each plot, we display the results of the convolution method of \Eq{convolution} using numerical integration and perturbative QCD with \(10^6\) phase space samples as well as the parton shower algorithm of \App{partonshower}.
At LL, we compare these results to the analytic prediction of \Eq{ECF_LL};
at MLL, we compare them to parton-level and hadron-level results obtained with \texttt{Pythia 8.244} using the anti-\(k_t\) algorithm.
The figures demonstrate that \PRSF{1} is less sensitive to hadronization corrections in the large observable regime, near \(C_1^{(2)} \approx 10^{-2}\), and that \PRSF{1} groomed substructure is more sensitive to the effects of multiple emissions than that of Soft Drop.
}
\label{fig:Calculations}
\end{figure}

As a result, performing resummed calculations of Soft Drop groomed substructure distributions that use only the critical emissions are more accurate than calculations for \PRSF{1} that only use the critical emission.
The relative accuracy of the LL, single-emission result for Soft Drop and \PRSF{1} is demonstrated in \Figs{LL_SD0}{LL_PRSF1}, which compares the fixed-order behavior of \(C_1^{(2)}\) distributions for jets groomed with Soft Drop or \PRSF{1}, respectively, using either the critical emission or all emissions in the jet.
Our parton shower procedure is inspired by \Reff{Larkoski:2013paa} and is described in more detail in \App{partonshower}.
Including multiple emissions does not dramatically change the Soft Drop substructure distribution, but leads to non-negligible changes in \PRSF{1} substructure distributions.

We may include effects from the running of the strong coupling constant and multiple emissions within the ungroomed jet to determine the behavior of the cumulative distribution of \(C_1^{(\varsigma)}\) at an even higher level of accuracy, modified leading logarithmic (MLL) accuracy, as described in \App{caesar_basics}.
While including these features in our calculation does not fully achieve next-to-leading-logarithmic (NLL) accuracy, the MLL calculation is an important piece of the full NLL result.
To consider the effects of several emissions on the distribution of the groomed ECFs, we convolve the probability distributions of several emissions that contribute to the total observable \(C_1^{(\varsigma)}\):

\begin{equation}
\begin{aligned}
    \rho^{\rm (MLL)}_\varsigma(C_{\rm tot})
    &=
    \int_{f\,\zcut}^{1/2} \dd \zcrit
    \int_0^1 \dd \thetacrit
    ~
    \rho_{\rm crit}(\thetacrit)
    \rho_{\rm crit}(\zcrit | \zpre)
    \\
    &~~~
    \int \dd \zpre
    ~
    \rho_{\rm pre}(\zpre|\thetacrit)
    \int \dd c_{\rm sub}
    ~
    \rho_{\rm sub}(c_{\rm sub}|\zcrit,\thetacrit)
    \\
    &~~~~~~~~~~~~~~~
    \delta\left(
        C_{\rm tot}
        -
        C^{(\varsigma)}_{1,~\rm crit}
        (\thetacrit, \zcrit, \zpre)
        -
        c_{\rm sub}
    \right)
    \label{eqn:convolution}
    .
\end{aligned}
\end{equation}

Each factor of \(\rho\) in \Eq{convolution} indicates a resummed probability distribution corresponding to a particular emission, discussed in greater detail in \App{caesar_basics}.
At the MLL level of accuracy, these resummed probability distributions are all calculated with a running coupling \(\alpha_s\).
We model non-perturbative effects by freezing the value of \(\alpha_s\) below the non-perturbative scale of \(\mu_{\rm freeze} = 1\) GeV, as in \Eq{frozencoupling}.
Variations in our choice of non-perturbative scale lead only to small changes to our results.

The expressions that describe multiple emissions within the groomed jet are:
\begin{itemize}
\item
\(\rho_{\rm crit}(\thetacrit) \rho(\zcrit | \zpre)\) represents the probability distribution for the emission angle \(\thetacrit\) and energy fraction \(\zcrit\) of the \textit{critical} emission, or the first emission to survive the \PRSF{1/2} procedure.
This distribution is conditioned on the presence of earlier emissions that modify the grooming procedure.

\item
\(\rho_{\rm pre}(\zpre | \thetacrit)\) encodes the probability that emissions that are wider than the critical emission, which we call \textit{pre-critical} emissions, have an energy fraction \(\zpre\).
%
%
Any non-zero value of \(\zpre\) mitigates the grooming of the critical emission.

\item
\(\rho_{\rm sub}(c_{\rm sub} | \zcrit, \thetacrit)\) is the probability distribution for emissions narrower than the critical emission to contribute a value \(c_{\rm sub}\) to the overall observable \(C^{(\varsigma)}_1\).
We call these emissions \textit{subsequent} emissions, and they are unmodified by the grooming procedure.

\item
Finally, \(C^{(\varsigma)}_{1,~\rm crit}(\thetacrit,\zcrit,\zpre)\) is the contribution of the critical emission to \(C^{(\varsigma)}_{1,~\rm tot}\), in the presence of pre-critical emissions with energy fraction \(\zpre\).
\end{itemize}

We evaluate the convolution of \Eq{convolution} through numerical integration and compare the resulting distributions to the results from running coupling parton showers and \texttt{Pythia 8.244} in \Fig{Calculations}.
In particular, \Figs{MLL_SD0}{MLL_PRSF1} compare our numerical MLL results to hadron-level results for anti-\(k_t\) jets in \texttt{Pythia 8.244} and demonstrate that this trend of comparatively large multiple-emissions effects in \PRSF{1} continues at MLL.

First, let us note the difference between our MLL results obtained with Monte Carlo integration, described in \App{caesar_basics}, and via parton showers, described in \App{partonshower}.
Our Monte Carlo results use a frozen coupling below a non-perturbative scale \(\mu_{\rm freeze} = 1\) GeV (see, for example, \Eq{frozencoupling}). Our parton shower results both freeze the coupling at \(\mu_{\rm freeze} = 1\) GeV and stop the shower below a non-perturbative cutoff \(\mu_{\rm cutoff} = 1\) MeV, tuned by hand to match results from \texttt{Pythia 8.244}.
The shower cutoff effectively sets \(\alpha_s\) to zero below the scale set by \(\mu_{\rm cutoff}\).
Therefore, the plots in \Figs{MLL_SD0}{MLL_PRSF1} effectively show the results obtained with three different models of non-perturbative effects, with \texttt{Pythia 8.244} the most accurate.
Our MLL parton shower results in \Figs{MLL_SD0}{MLL_PRSF1} are still normalized, but the normalization is not immediately evident due to the presence of zero bins that are not shown in the logarithmically-scaled figures.
In passing from our rough parton shower model to the more accurate hadronization model of \texttt{Pythia}, we expect that it is these zero bins that contribute to the additional hadron-level peak shown in the \texttt{Pythia} data.

We also point out that in \Fig{MLL_SD0}, the parton-level \texttt{Pythia 8.244} distribution for \(C_1^{(2)}\) also has an additional peak near the discontinuity in the parton-level LL result, in the regime of relatively large \(C_1^{(2)} \approx 10^{-1}\).
There is a comparable additional peak in the hadron-level \texttt{Pythia} distribution for \(C_1^{(2)}\) at this relatively large value for the observable.

We also note that, while our numerical results for \PRSF{1} in the MLL figures agree roughly with \texttt{Pythia 8.244} in the regime of large \(C_1^{(2)}\), the hadron-level \texttt{Pythia} distributions have an additional peak near \(C_1^{(2)} \approx 3\times 10^{-6}\) due to hadronization corrections to the observable at that scale.
There is a similar additional peak in the hadron-level \texttt{Pythia} distribution for mMDT (\SD{0}) groomed \(C_1^{(2)}\).

\subsection{Summary of Resummation: Convolution}
\label{app:caesar_basics}
We now review the resummation procedure we apply in our LL and MLL results.
In particular, we introduce the basics of the CAESAR formalism, named due to its use in the context of the Computer Automated Expert Semi-Analytical Resummer (CAESAR) program \cite{Banfi:2004yd}.\footnote{
The results of our presentation below are also a special case of the more general, observable-independent formalism of jet calculus \cite{Konishi:1979cb}
--
closely connected to the resummation procedure used by parton showers
--
for which pedagogical introductions may be found in e.g.\ \Reffs{Dokshitzer:1991wu,vanBeekveld:2023lsa}.
}
To the best of our knowledge, this formalism was first introduced in the context of semi-analytic resummation of event shapes \cite{Banfi:2001bz}.
While we include an introduction here for completeness, a presentation of the concepts that appear in the following discussion can be found in more elegant or complete forms throughout existing literature \cite{Larkoski:2017fip, Larkoski:2021aav, Cohen:2020afv, Banfi:2001bz, Banfi:2004yd, Banfi:2004nk, Banfi:2010xy, Luisoni:2015xha, Bauer:2018svx, Baron:2020xoi}.

The distribution of energy fraction and angle of a partonic emission in a jet is described at fixed order by
\begin{align}
    \rho_{i,~\rm f.o.}(z,\theta)
    &=
    \frac{\alpha_s}{\pi} \bar{p}_i(z) \frac{1}{\theta}
    ,
\end{align}
where \(z\in(0, 1/2)\) is the energy fraction of the softer branch of the emission and \(\bar{p}(z) = p(z) + p(1-z)\) indicates a reduced splitting function, modified from the splitting function \(p(z)\) to describe the softer branch.
We would like to regulate the singularities of this distribution by using techniques of resummation in perturbative QCD and thus gain greater predictive power over groomed jet observables.
In this appendix, we discuss the resummed distributions used in the text and give some quick derivations.

To begin, we provide explicit expressions for the DGLAP splitting functions, which describe the probability for the emission of partonic radiation from quarks and gluons, at one loop accuracy:
\begin{align}
&p_{q\to q g}(z) = C_F \frac{1+(1-z)^2}{z},
\label{eqn:quark_splitting}
\\
&p_{g\to g g}(z) = 2C_A\frac{1-z}{z} + C_A z(1-z) + T_F n_f (z^2 + (1-z)^2).
\label{eqn:gluon_splitting}
\end{align}
Here, \(z\) is the energy fraction of an emitted gluon relative to the total energy of the emitted gluon and an initial parton.
\(p_q\) describes the distribution of \(z\) in the case that a gluon emitted by a quark, \(q \to q g\), and \(p_g\) describes the distribution of \(z\) in the case \(g \to gg\).
\(T_F = \frac{1}{2}\) is the Dynkin index of the fundamental representation of the standard model SU(3) color gauge group, and \(n_f\) indicates the number of quark flavors.
In the following analysis, we take \(n_f = 5\), including all quark flavors except for the top quark.
As in the text, \(C_R\) is the quadratic Casimir for the representation \(R\) of SU(3).
\(C_R = C_F = \frac{4}{3}\) for quarks and \(C_R = C_A = 3\) for gluons.

We also introduce some graphical notation to represent the probability of a splitting whose angle \(\theta\) and softer energy fraction \(z\) lie within some region of phase space.
In particular, we describe the probability of finding a splitting in some region in the \(\log(z^{-1})\)-\(\log(\theta^{-1})\) plane, or Lund plane, with a diagram of the Lund plane with the corresponding region filled:
\begin{align}
    \iint_{
    \begin{tikzpicture}[scale=.06]
    \begin{axis}
    [xmin=0, xmax=5,
    ymin=0, ymax=5,
    axis line style = {draw=none},
    ticks=none]
    	\draw[blue, line width=9pt, fill=blue,fill opacity=0.3] \pgfextra{
    	  \pgfpathellipse{\pgfplotspointaxisxy{2.5}{2.5}}
    		{\pgfplotspointaxisdirectionxy{0}{1.8}}
        	{\pgfplotspointaxisdirectionxy{1.2}{.5}}
    	};
    \end{axis}
    \end{tikzpicture}
    }
    \frac{\alpha_s}{\pi}~
    \frac{\dd\theta}{\theta}~
    \dd z~
    [\bar{p}_i(z)]^{(1/2)}_+
    ~~~
    \triangleq
    ~~~
    \begin{tikzpicture}[
    baseline={([yshift=-.8ex]current bounding box.center)},
    vertex/.style={anchor=base,
    circle,fill=black!25,minimum size=18pt,inner sep=2pt},
    scale=.3]
    \begin{axis}
    [
    xlabel=\scalebox{2.5}
    {\(\log(\theta^{-1})\)},
    ylabel=\scalebox{2.5}
    {\(\log(z^{-1})\)},
    xmin=0, xmax=5,
    ymin=0, ymax=5,
    axis line style = {line width=3pt},
    axis y line*=left,
    axis x line*=bottom,
    axis lines = middle,
    ticks=none]
    	\draw[blue, line width=3pt, fill=blue,fill opacity=0.3]
    	\pgfextra{
    	  \pgfpathellipse{\pgfplotspointaxisxy{1.1}{1.6}}
    		{\pgfplotspointaxisdirectionxy{-.2}{.9}}
    		{\pgfplotspointaxisdirectionxy{.7}{.5}}
    	};
    \end{axis}
    \end{tikzpicture}
    ~~~
    \triangleq
    ~~~
    \begin{tikzpicture}[
    baseline={([yshift=-.8ex]current bounding box.center)},
    vertex/.style={anchor=base,
    circle,fill=black!25,minimum size=18pt,inner sep=2pt},
    scale=.3]
    \begin{axis}
    [xmin=0, xmax=5,
    ymin=0, ymax=5,
    axis line style = {line width=3pt},
    axis y line*=left,
    axis x line*=bottom,
    axis lines = middle,
    ticks=none]
    	\draw[blue, line width=3pt, fill=blue,fill opacity=0.3]
    	\pgfextra{
    	  \pgfpathellipse{\pgfplotspointaxisxy{1.1}{1.6}}
    		{\pgfplotspointaxisdirectionxy{-.2}{.9}}
    		{\pgfplotspointaxisdirectionxy{.7}{.5}}
    	};
    \end{axis}
    \end{tikzpicture}
    ,
\end{align}
where the shaded oval in the Lund plane above represents an arbitrary shape, and \([\bar{p}(z)]^{(1/2)}_+\) is a plus-function regulated \textit{reduced} splitting function, \(\bar{p}_i(z) = p_i(z) + p_i(1-z)\), using the definition of plus-function regularization of \Eq{plusdist_def}.\footnote{We may use \([\bar{p}(z)]^{(1/2)}_+\) for the applications presented in this work because we are interested in observables whose dependence on energy is captured entirely by the energy fraction of the softer parton at any branching.
If we would like to deal with observables that require more information than just the softer energy fraction, such as the flavor of the emitted partons, we would need to work with a plus-function regulated splitting function \([p(z)]^{(1)}_+\).
}
It follows, for example, that
\begin{align}
    \begin{tikzpicture}[
    baseline={([yshift=-.8ex]current bounding box.center)},
    vertex/.style={anchor=base,
    circle,fill=black!25,minimum size=18pt,inner sep=2pt},
    scale=.3]
    \begin{axis}
    [xmin=0, xmax=5,
    ymin=0, ymax=5,
    axis line style = {line width=3pt},
    axis y line*=left,
    axis x line*=bottom,
    axis lines = middle,
    ticks=none]
    	\draw[blue, line width=3pt, fill=blue,fill opacity=0.3]
    	\pgfextra{
    	  \pgfpathellipse{\pgfplotspointaxisxy{0}{0}}
    		{\pgfplotspointaxisdirectionxy{0}{100}}
    		{\pgfplotspointaxisdirectionxy{100}{0}}
    	};
    \end{axis}
    \end{tikzpicture}
    =
    0
    ~~~~~~~~~
    {\rm and}
    ~~~~~~~~~
    \begin{tikzpicture}[
    baseline={([yshift=-.8ex]current bounding box.center)},
    vertex/.style={anchor=base,
    circle,fill=black!25,minimum size=18pt,inner sep=2pt},
    scale=.3]
    \begin{axis}
    [xmin=0, xmax=5,
    ymin=0, ymax=5,
    axis line style = {line width=3pt},
    axis y line*=left,
    axis x line*=bottom,
    axis lines = middle,
    ticks=none]
    	\draw[blue, line width=3pt, fill=blue,fill opacity=0.3]
    	\pgfextra{
    	  \pgfpathellipse{\pgfplotspointaxisxy{0}{0}}
    		{\pgfplotspointaxisdirectionxy{0}{100}}
    		{\pgfplotspointaxisdirectionxy{100}{0}}
    	};
    	\draw[blue, line width=3pt, fill=white,fill opacity=1.0]
    	\pgfextra{
    	  \pgfpathellipse{\pgfplotspointaxisxy{1.1}{1.6}}
    		{\pgfplotspointaxisdirectionxy{-.2}{.9}}
    		{\pgfplotspointaxisdirectionxy{.7}{.5}}
    	};
    \end{axis}
    \end{tikzpicture}
    ~~
    =
    ~~
    \begin{tikzpicture}[
    baseline={([yshift=-.8ex]current bounding box.center)},
    vertex/.style={anchor=base,
    circle,fill=black!25,minimum size=18pt,inner sep=2pt},
    scale=.3]
    \begin{axis}
    [xmin=0, xmax=5,
    ymin=0, ymax=5,
    axis line style = {line width=3pt},
    axis y line*=left,
    axis x line*=bottom,
    axis lines = middle,
    ticks=none]
    	\draw[blue, line width=3pt, fill=blue,fill opacity=0.3]
    	\pgfextra{
    	  \pgfpathellipse{\pgfplotspointaxisxy{0}{.9}}
    		{\pgfplotspointaxisdirectionxy{0}{200}}
    		{\pgfplotspointaxisdirectionxy{2.5}{.9}}
    	};
    	\draw[blue, line width=3pt, fill=white,fill opacity=1.0]
    	\pgfextra{
    	  \pgfpathellipse{\pgfplotspointaxisxy{1.1}{1.6}}
    		{\pgfplotspointaxisdirectionxy{-.2}{.9}}
    		{\pgfplotspointaxisdirectionxy{.7}{.5}}
    	};
    \end{axis}
    \end{tikzpicture}
    ~~
    =
    ~~
    -~
    \begin{tikzpicture}[
    baseline={([yshift=-.8ex]current bounding box.center)},
    vertex/.style={anchor=base,
    circle,fill=black!25,minimum size=18pt,inner sep=2pt},
    scale=.3]
    \begin{axis}
    [xmin=0, xmax=5,
    ymin=0, ymax=5,
    axis line style = {line width=3pt},
    axis y line*=left,
    axis x line*=bottom,
    axis lines = middle,
    ticks=none]
    	\draw[blue, line width=3pt, fill=blue,fill opacity=0.3]
    	\pgfextra{
    	  \pgfpathellipse{\pgfplotspointaxisxy{1.1}{1.6}}
    		{\pgfplotspointaxisdirectionxy{-.2}{.9}}
    		{\pgfplotspointaxisdirectionxy{.7}{.5}}
    	};
    \end{axis}
    \end{tikzpicture}
    .
\end{align}
As depicted above, we may replace any vertical line -- representing an integral over \(z\) at fixed \(\theta\) -- or any vertical strip -- representing an integral over \(z\) and some range of \(\theta\) -- with zero, since the integral of a plus-function regulated distribution is zero.

In this paper, we are concerned with the jet ECFs \(C_1^{(\varsigma)}\), which can be roughly calculated by adding contributions from all of the emissions within a jet, such that the contribution from each emission increases as \(z\) or \(\theta\) increases and vanishes in the limit that \(z\) or \(\theta\) vanishes.
We call such observables \textit{additive} observables, and we may treat the \(C_1^{(\varsigma)}\) as additive observables in our leading order computations.

The calculation of distributions of additive observables is simplified by the approximately uniform distribution of partonic emissions within the Lund plane.
In particular, the uniformity of emissions in the Lund plane implies that there is a single emission with splitting parameters \(z\) and \(\theta\) that we expect to be exponentially higher than the values of \(z\) and \(\theta\) of other emissions.
Additive observables are therefore generically dominated by the contribution from a single such emission, and we may approximate the probability that an observable \(\mathcal{O}\) is less than the sum of contributions from different emissions, \(\mathbb{P}(\mathcal{O} < \sum_i x_i)\), by the probability that it is less than the dominant value, \(\mathbb{P}(\mathcal{O} < X)\), with \(X = \max_j x_j\).
We refer to this as the \textit{leading logarithmic} (LL) approximation.

Mathematically, the LL approximation for an additive observable \(\mathcal{O}\), with contributions \(x_i > 0\) from emissions \(i\), asserts that there is a dominant contribution \(X = \max_i x_i\) such that \(\mathcal{O} = \sum_i x_i \approx X\).
This approximation holds up to effects that are sub-leading in \(\alpha_s\) and logarithms of the observable.
In the calculation of cumulative distributions for \(\mathcal{O}\), we may write without approximation that \(\Theta(\mathcal{O} - \sum_i x_i) \leq \Theta(\mathcal{O} - X)\), where equality emerges when we have at most a single non-zero \(x_j = X\).
The LL approximation notes that \(x_i / X \) is exponentially small, and therefore that \(\Theta(\mathcal{O} - \sum_i x_i) \approx \Theta(\mathcal{O} - X)\).
We should also note that typically, calculations applying the LL approximation will also drop terms that are suppressed by powers of \(X\) since these are sub-dominant to the leading logarithmic terms in observable distributions.

We therefore define the LL cumulative probability distribution \(\Sigma^{\rm(LL)}_{\mathcal{O}}(X)\) of an additive observable \(\mathcal{O}\) as the probability that no emission contributes a value greater than \(X\) to \(\mathcal{O}\), up to sub-leading behavior in \(\alpha_s\) and logarithms or powers of the observable.
The region of phase space in which an observable contributes a value greater than \(X\) to the observable is called a \textit{veto region}.
We sum and integrate over all possible final-state emissions outside the veto region when calculating \(\Sigma^{\rm(LL)}_{\mathcal{O}}(x)\).

As an example, the approximate mass of a jet due to a single emission is given by \(m^2 = Q^2 z\theta^2\), where \(Q\) is the total energy of the jet.
Therefore, the jet mass distribution is given at leading logarithmic accuracy by
\begin{equation}
\begin{aligned}
    \Sigma^{\rm(LL)}_{m^2}(m^2)
    &=
    \sum_{n = 0}^{\infty}\frac{1}{n!}
    \left(
    \frac{\alpha_s}{\pi}
    \int_0^1 \frac{\dd\theta}{\theta}
    \int_0^{\frac{1}{2}} \dd z\,[\bar{p}_i(z)]_+^{[1/2]}
    \Theta\left(
    \frac{m^2}{Q^2} - z\theta^2
    \right)
    \right)^n
    \\
    &=
    \exp\left[
    -\frac{\alpha_s}{\pi}
    \int_0^1 \frac{\dd\theta}{\theta}
    \int_0^{\frac{1}{2}} \dd z\,\bar{p}_i(z)
    \Theta\left(
    z \theta^2 - \frac{m^2}{Q^2}
    \right)
    \right]
    .
\end{aligned}
\end{equation}
In graphical notation, the analogous derivation is
\begin{equation}
\begin{aligned}
    \Sigma^{\rm(LL)}_{m^2}(m^2)
    &=
    \sum_{n = 0}^{\infty}\frac{1}{n!}
    \left(
    \begin{tikzpicture}[
    baseline={([yshift=-.8ex]current bounding box.center)},
    vertex/.style={anchor=base,
    circle,fill=black!25,minimum size=18pt,inner sep=2pt},
    scale=.3]
    \begin{axis}
    [xmin=0, xmax=5,
    ymin=0, ymax=5,
    axis line style = very thick,
    axis y line*=left,
    axis x line*=bottom,
    axis lines = middle,
    ticks=none]
    	\addplot[name path=f, domain=0:5,
        style=very thick, blue]
        {4.5-1.5*x};
        \path[name path=axis]
        (axis cs:0,5) -- (axis cs:5,5);
        \addplot [
            thick,
            color=blue,
            fill=blue,
            fill opacity=0.3
        ]
        fill between[
            of=f and axis
        ];
        \node [rotate=-55] at (axis cs:  3.2,  2.8)
        {${\scriptstyle \frac{m^2}{Q^2} > z\theta^2}$};
    \end{axis}
    \end{tikzpicture}
    \right)^n
    =
    \exp\left[
    ~-~
    \begin{tikzpicture}[
    baseline={([yshift=-.8ex]current bounding box.center)},
    vertex/.style={anchor=base,
    circle,fill=black!25,minimum size=18pt,inner sep=2pt},
    scale=.3]
    \begin{axis}
    [xmin=0, xmax=5,
    ymin=0, ymax=5,
    axis line style = very thick,
    axis y line*=left,
    axis x line*=bottom,
    axis lines = middle,
    ticks=none]
    	\addplot[name path=f, domain=0:5,
        style=very thick, blue]
        {4.5-1.5*x};
        \path[name path=axis]
        (axis cs:0,0) -- (axis cs:1,0);
        \addplot [
            thick,
            color=blue,
            fill=blue,
            fill opacity=0.3
        ]
        fill between[
            of=f and axis
        ];
        \node [rotate=-50] at (axis cs:  1.25,  1.25)
        {${\scriptstyle \frac{m^2}{Q^2} < z\theta^2}$};
    \end{axis}
    \end{tikzpicture}
    ~~
    \right]
    ,
\end{aligned}
\end{equation}
where in the final equality, we see that the LL cumulative distribution is one over the exponent of the veto region area.
The area of the veto region for an observable is also called the \textit{radiator} \(R_\mathcal{O}\) for the observable.

A similar analysis yields the cumulative probability distribution function of the angle \(\theta\) for the first emission in the \PRSF{1/2} procedure to satisfy \(z > \zcut\), which we call the \textit{critical emission}:
\begin{align}
    \Sigma^{\rm(LL)}_{{\rm crit}, \theta}(\theta_{\rm crit})
    &=
    \exp\left[
        ~-~
        \begin{tikzpicture}[
        baseline={([yshift=-.8ex]current bounding box.center)},
        vertex/.style={anchor=base,
        circle,fill=black!25,minimum size=18pt,inner sep=2pt},
        scale=.3]
        \begin{axis}
        [xmin=0, xmax=5,
        ymin=0, ymax=5,
        axis line style = very thick,
        axis y line*=left,
        axis x line*=bottom,
        axis lines = middle,
        ticks=none]
        	\addplot[name path=f, domain=0:2.3,
            style=very thick, blue]
            {2.5};
            \addplot[very thick, samples=50, smooth,domain=0:6,blue, name path=three] coordinates {(2.3,-1)(2.3,2.5)};
            \path[name path=axis]
            (axis cs:0,0) -- (axis cs:2.3,0);
            \addplot [
                thick,
                color=blue,
                fill=blue,
                fill opacity=0.3
            ]
            fill between[
                of=f and axis
            ];
            \node [rotate=0] at (axis cs:  2.0,  4.0)
            {${\scriptstyle z < \zcut}$};
            \coordinate (I)  at (3.3, 3.4);
            \coordinate (F)  at (3.3, 4.9);
            \draw[-Stealth,line width=.2mm] (I) -> (F);
            \node [rotate=0] at (axis cs:  1.18,  1.38)
            {${\scriptstyle \theta > \theta_{\rm crit}}$};
            \coordinate (I)  at (1.8, .5);
            \coordinate (F)  at (.3, .5);
            \draw[-Stealth,line width=.2mm] (I) -> (F);
        \end{axis}
        \end{tikzpicture}
        ~~
    \right]
    =
    \exp\left[
    -\frac{\alpha_s}{\pi}
    \int_{\theta_{\rm crit}}^1
    \frac{\dd\theta}{\theta}
    \int_{\zcut}^{1/2}
    \dd z\,
    \bar{p}_i(z)
    \right],
\end{align}
with a corresponding probability density
\begin{equation}
   \rho^{\rm (LL)}_{{\rm crit}, \theta}(\theta_{\rm crit})
    =
    \frac{1}{\theta_{\rm crit}}
    ~
    \frac{\alpha_s}{\pi}
    \int_{\zcut}^{1/2}
    \dd z\,
    \bar{p}(z)
    \exp\left[
    -\frac{\alpha_s}{\pi} \int_{\theta_{\rm crit}}^1
    \frac{\dd\theta}{\theta}
    \int_{\zcut}^{1/2} \dd z\,
    \bar{p}_i(z)
    \right]
    .
\end{equation}

The resummed, joint probability distribution for the energy fraction and angle of the critical emission is therefore
\begin{equation}
    \rho_{\rm crit}(z,\theta)
    =
    \rho_{{\rm crit}, z}(z | \theta)
    \rho_{{\rm crit}, \theta}(\theta)
    =
    \alpha_s(\kappa_{\rm crit})
    ~
    \bar{p}(z)
    \frac{1}{\theta}
    \exp\left[
    - \frac{1}{\pi}
    \int_{\theta}^1
    \frac{\dd\theta'}{\theta'}
    \int_{\zcut}^{1/2} \dd z'
    \bar{p}_i(z')~
    \alpha_s(\kappa')
    \right]
    ,
\end{equation}
where the argument \(\kappa\) of the strong coupling accounts for the possible inclusion of running coupling effects.
In particular, at LL we do not include effects due to the running of the coupling, and \(\kappa = p_T R\) is a scale set only by the dynamics and parameters of the jet.
When including running coupling effects, \(\kappa \approx p_T R z \theta \approx p_\perp\) is approximately related to \(p_\perp\), the momentum of the softer emission that is transverse to the momentum of the harder emission.
This is because QCD correlations and cross sections that use \(\alpha_s\) evaluated at a scale \(\kappa\) generically include logarithms of the form \(\log(p_{\perp} / \kappa)\).
To make these logarithms small, one may choose \(\kappa \sim p_\perp\), which then requires that we evaluate the strong coupling at the scale \(p_\perp\) as well.
At one loop, the running of \(\alpha_s\) is captured by
\begin{align}
    \alpha^{\rm 1-loop}_s(\kappa)
    =
    \frac{\alpha_s(Q)}{1 + 2\alpha_s(Q) \beta_0 \log(\frac{\kappa}{Q})}
    ,
\end{align}
where \(\beta_0 = (\frac{11}{3}C_A - \frac{4}{3} T_f n_f)/4\pi\) is a coefficient appearing in the QCD beta function: \(\beta(\alpha_s) = -2\beta_0 \alpha_s^2 + \mathcal{O}(\alpha_s^3)\).
Calculations accounting for effects from the running of the QCD coupling and multiple emissions within a jet are called \textit{modified leading logarithmic} (MLL) calculations.

In the numerical MLL results we present in \Fig{Calculations}, we also model non-perturbative QCD effects by \textit{freezing} the coupling below a non-perturbative scale \(\mu_{\rm freeze}\),
\begin{align}
    \alpha^{\rm 1-loop,\,frozen}_s(\kappa; \mu_{\rm freeze})
    =
    \alpha^{\rm 1-loop}_s(\kappa)
    \Theta(\kappa > \mu_{\rm freeze})
    +
    \alpha^{\rm 1-loop}_s(\mu_{\rm freeze})
    \Theta(\mu_{\rm freeze} > \kappa)
    .
    \label{eqn:frozencoupling}
\end{align}
\Eq{frozencoupling} is a simplified analog of several motivated approaches to modeling the infrared structure of QCD for which \(\alpha_s\) also approximately freezes below a non-perturbative scale.\footnote{The freezing of the effective value of \(\alpha_s\) in the low-energy regime can be argued in several complementary contexts.
The freezing of \(\alpha_s\) is discussed by \Reff{Mattingly:1992ud} in the context of optimized perturbation theory \cite{Stevenson:1981vj}, whose goal is to perform perturbative calculations that are minimally sensitive to changes in renormalization scheme, by \Reffs{Stevenson:1994jd,Caveny:1997yr} in the context of expanding perturbative results around the Caswell-Banks-Zaks fixed point \(N_f = N_f^* \lesssim 11 N_c / 2\) of the QCD beta function \cite{Caswell:1974gg,Banks:1981nn}, and by \Reff{Brodsky:2010ur} in the context of holographic QCD.
The freezing of \(\alpha_s\) may be also seen qualitatively by comparing experimental data to field theoretic calculations \cite{Deur:2008rf,Deur:2009tj,Binosi:2016nme}, and is a special case of the more general program of using effective couplings to describe non-perturbative infrared effects \cite{Dokshitzer:1995qm,Dokshitzer:1995zt,Dokshitzer:1997ew,Dokshitzer:1997iz,Korchemsky:1999kt}.
See \Reff{Deur:2016tte} for a recent review of models of the low-energy behavior of QCD with and without a frozen coupling.
}
We take \(\mu_{\rm freeze}\) = 1 GeV in our numerical calculations, and we note that changing our choice of \(\mu_{\rm freeze}\) leads only to small effects in substructure distributions such as those presented in \Fig{Calculations}.

Let us now apply the technology reviewed in this appendix to distributions of \PRSF{1} groomed substructure observables.
An efficient way to include the multiple emission effects that contribute to MLL accuracy is to consider P-RSF\(_1\) so that the grooming affects only soft, wide-angle emissions and terminates before it affects narrow emissions forming the jet core.
For \PRSF{1}, we may divide up the emissions into three qualitative categories:

\begin{enumerate}
    \item[a)]
    soft, wide angle \textit{pre-critical emissions}, which are considered first in an angular-ordered grooming procedure and thus completely removed by \PRSF{1};

    \item[b)]
    a single \textit{critical emission}, which is the first soft emission to survive the grooming procedure but is nonetheless modified by the grooming; and

    \item[c)]
    \textit{subsequent emissions}, which are left unmodified by the grooming procedure.
\end{enumerate}

If we have the radiator for an additive observable \(\mathcal O\), it is simple to evaluate corrections due to the presence of multiple emissions.
These corrections can be evaluated with Laplace transform methods \cite{Banfi:2004yd} to find that at MLL accuracy,
\begin{align}
    \Sigma^{\rm (MLL)}_{\mathcal O}(x)
    &=
   \frac{
   e^{-R_{\mathcal O}(x)
   -
   \gamma_E R'_{\mathcal O}(x)}
   }
   {\Gamma
   \left(1 + R'_{\mathcal O}(x)\right)
   }
   \label{eqn:mll_general_dist}
   ,
\end{align}
where \(\gamma_E\) is the Euler-Mascheroni constant, and the prime denotes a derivative with respect to \(\log(1/x)\).\footnote{
We note that additivity is a sufficient condition for \Eq{mll_general_dist}, but is not strictly necessary.
For example, a weaker but still sufficient condition, named \textit{recursive infrared-collinear (rIRC) safety}, was introduced by \Reff{Banfi:2004yd}.
The definition of rIRC safety is given in Eqs.\ (3.4) and (3.5) of \Reff{Banfi:2004yd}, and an example of additivity is furnished by the thrust observable, discussed in Section 3.2 of \Reff{Banfi:2004yd}.
}
%
%
These considerations lead us to the MLL expressions for the effects of multiple pre-critical and subsequent emissions on jet ECFs.

The cumulative distribution for the contribution of the subsequent emissions to the observable \(C_1^{(\varsigma)}\) takes the form
\begin{align}
    \Sigma^{\rm(MLL)}_{C^{(\varsigma)}_{\rm sub}}
    (C  |  \thetacrit, \zcrit)
    &=
   \frac{
   e^{-R_{\rm sub}(C)
   -
   \gamma_E R'_{\rm sub}(C)}
   }
   {\Gamma
   \left(1 + R'_{\rm sub}(C)\right)
   }
   \approx
   e^{-R_{\rm sub}(C)}
   =
   \Sigma^{\rm(LL)}_{C^{(\varsigma)}_{\rm sub}}
   (C  |  \thetacrit)
   \label{eqn:subsequentDist}
   ,
\end{align}
where \(R_{\rm sub}(C)\) is shorthand for \(R_{C^{(\varsigma)}_{\rm sub}}(C | \thetacrit)\), defined as
\begin{align}
    R_{C^{(\varsigma)}_{\rm sub}}
    (C | \thetacrit)
    &=
    \begin{tikzpicture}[
    baseline={([yshift=-.8ex]current bounding box.center)},
    vertex/.style={anchor=base,
    circle,fill=black!25,minimum size=18pt,inner sep=2pt},
    scale=.3]
    \begin{axis}
    [xmin=0, xmax=5,
    ymin=0, ymax=5,
    axis line style = very thick,
    axis y line*=left,
    axis x line*=bottom,
    axis lines = middle,
    ticks=none]
    	\addplot[name path=f,domain=0:5,
        style=very thick, blue]
        {4.5-1.5*x};
        \path[name path=axis]
        (axis cs:0,5) -- (axis cs:5,5);
        \addplot [
            thick,
            color=blue,
            fill=blue,
            fill opacity=0.3
        ]
        fill between[
            of=f and axis
        ];
        \node at (axis cs:  2.9,  3.8)
        {${\scriptstyle C_{\rm sub}^{(\varsigma)} > z\theta^\varsigma}$};
        \node at (axis cs:  3.4,  1.8)
        {${\scriptstyle \theta > \theta_{\rm crit}}$};
    \end{axis}
    \end{tikzpicture}
    =
    \int_{C/\thetacrit^{\varsigma}}^{1/2}
    \frac{\dd c}{c}
    \int_c^{1/2} \dd z\,
    \bar{p}_i(z)
    \frac{\alpha_s(\kappa)}{k \varsigma}
    .
\end{align}
Of course, the value of \zcrit{} will also have an impact on the subsequent emissions, but we may safely neglect these corrections, which contribute only by scaling \(C_{\rm sub}^{(\varsigma)}\) by factors of the form \(1 - \zcrit\).

The cumulative distribution for the energy fraction carried by pre-critical emissions is calculated similarly and takes the form
\begin{align}
    \Sigma^{\rm(MLL)}_{\zpre}
    (z | \thetacrit)
    &=
   \frac{
   e^{-R_{\rm pre}(z | \thetacrit)
   -
   \gamma_E R'_{\rm pre}(z | \thetacrit)}
   }
   {\Gamma
   \left(1 + R'_{\rm pre}(z | \thetacrit)\right)
   }
   \approx
   e^{-R_{\rm pre}(z | \thetacrit)}
   =
   \Sigma^{\rm(LL)}_{\zpre}
   (z  |  \thetacrit)
   \label{eqn:preDist}
   ,
\end{align}
where \(R_{\rm pre}(z | \thetacrit) = R_{\zpre}(z | \thetacrit)\) is given by
\begin{align}
    R_{\zpre}
    (z | \thetacrit)\
    &=
    \int_{\thetacrit}^{1}
    \frac{\dd\theta}{\theta}
    \int_{z}^{\zcut}\dd z'\,
    \bar{p}_i(z')
    \frac{\alpha_s(\kappa)}{\pi}
    .
\end{align}

We need only use the singular pieces of the splitting functions to achieve LL accuracy, but, by definition, we must include the non-singular pieces of the splitting functions to achieve MLL accuracy.
While the MLL analysis modifies the LL analysis by including additional, higher-order behavior, it is not quite sufficient to produce results at next-to-leading logarithmic (NLL) accuracy;
a full NLL result requires a more intricate treatment including the effects of logarithms due to the constraint that radiation within the jet must lie within the jet radius.
These logarithms emerge because jet observables depend on only subsets of the full phase (global) space of the outgoing radiation, and are known as non-global logarithms \cite{Dasgupta:2001sh,Dasgupta:2002dc,Dasgupta:2002bw,Banfi:2002hw,Appleby:2002ke,Weigert:2003mm,Rubin:2010fc,Banfi:2010pa,Kelley:2011tj,Hornig:2011iu,Kelley:2011aa,Hatta:2013iba,Schwartz:2014wha,Khelifa-Kerfa:2015mma,Larkoski:2015zka,Larkoski:2016zzc,Banfi:2021owj}.
Regardless, the MLL results presented in the paper provide a sense of the physics contained within NLL results.

\subsection{Summary of Resummation: Parton Shower Algorithm}
\label{app:partonshower}

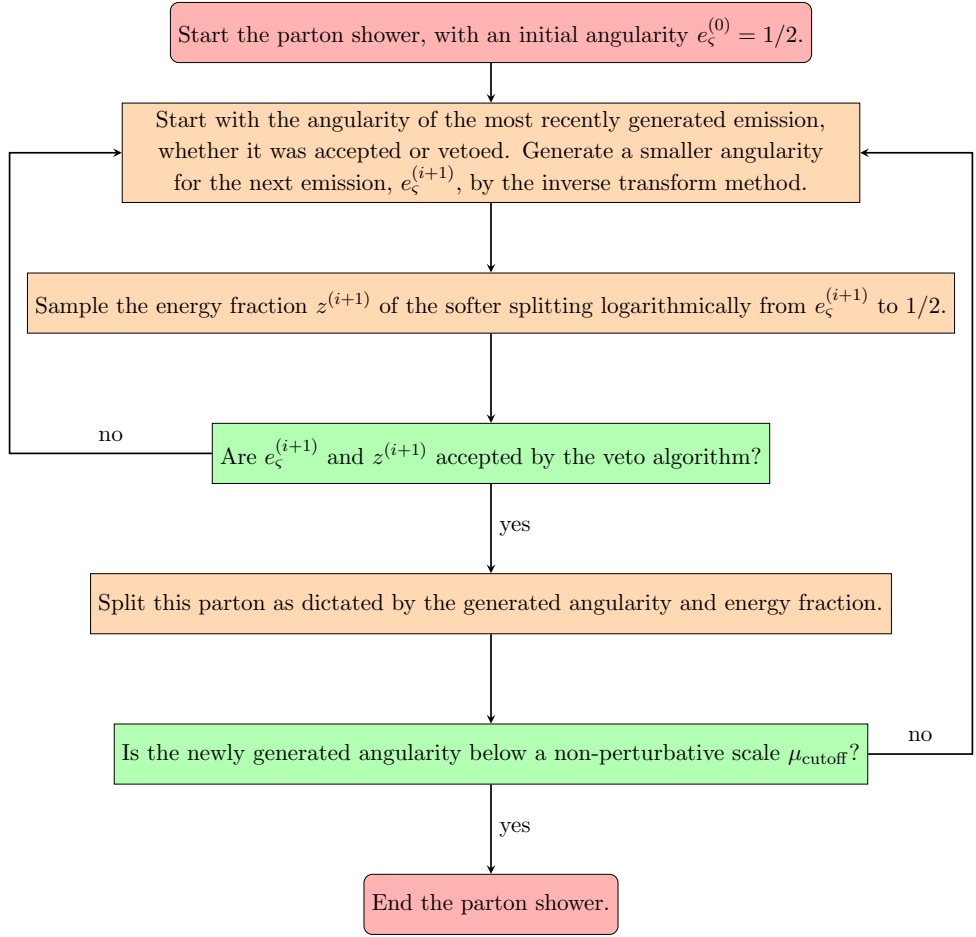
\begin{figure}[t!]
    \centering
    \scalebox{.8}{
    \begin{tikzpicture}[node distance = 2cm, auto]
    \node (start) [startstop] {Start the parton shower, with an initial angularity \(e_{\varsigma}^{(0)} = 1/2\).};
    \node (pro1) [process, below of=start, text width=12cm]
    {Start with the angularity of the most recently generated emission, whether it was accepted or vetoed. Generate a smaller angularity for the next emission, \(e_{\varsigma}^{(i+1)}\), by the inverse transform method.};
    \node (pro2) [process, below of=pro1, yshift=-0.5cm] {Sample the energy fraction \(z^{(i+1)}\) of the softer splitting logarithmically from \(e_{\varsigma}^{(i+1)}\) to 1/2.};
    \node (dec1) [decision, below of=pro2, yshift=-0.5cm] {Are \(e_{\varsigma}^{(i+1)}\) and \(z^{(i+1)}\) accepted by the veto algorithm?};
    \node (pro3) [process, below of=dec1, yshift=-0.5cm] {Split this parton as dictated by the generated angularity and energy fraction.};
    \node (dec2) [decision, below of=pro3, yshift=-0.5cm] {Is the newly generated angularity below a non-perturbative scale \(\mu_{\rm cutoff}\)?};
    \node (stop) [startstop, below of=dec2, yshift=-0.5cm] {End the parton shower.};

    \draw [arrow] (start) -- (pro1);
    \draw [arrow] (pro1) -- (pro2);
    \draw [arrow] (pro2) -- (dec1);
    \draw [arrow] (dec1) -- node[]{yes}(pro3);
    \draw [arrow] (pro3) -- (dec2);
    \draw [arrow] (dec2) -- node[]{yes}(stop);
    \draw [arrow] (dec1) --  node[anchor=west, above=2pt ] {no} ++(-8,0) |- (pro1);    \draw [arrow] (dec2) --  node[anchor=east, above=2pt ] {no} ++(8,0) |- (pro1);
\end{tikzpicture}
}
    \caption{The basic code flow of the parton shower algorithm described in \App{partonshower}.
    The veto algorithm is only used when considering MLL effects and is skipped at LL.
    The computations of the algorithm are shown in orange, the decisions of the algorithm in green, and the start and stop of the algorithm in red.}
    \label{fig:ps_codeflow}
\end{figure}

In this appendix, we describe the parton shower we use in the text to provide an additional check of our analytic results.
Our parton shower, motivated by Section 5 of \Reff{Larkoski:2013paa}, is available on GitHub \cite{samgithub}.
A concise flow of code for our parton shower at MLL accuracy is shown in \Fig{ps_codeflow}.

In broad strokes, we order our parton shower by the angularity \(e_\varsigma \approx z \theta^\varsigma\) of its emissions.
The angularity \(e_\varsigma\) associated with a single emission \(i\) with softer energy fraction \(z\) and opening angle \(\theta\) takes the form
\(
    e^{(i)}_\varsigma = z \left(\theta/R\right)^\varsigma
    ,
\)
where \(R\) indicates the jet radius, which we set to 1.

To obtain a parton shower valid at LL accuracy, we implement the following procedure:
\begin{enumerate}
    \item
    \label{item:LL_ps_start}
    Initialize a jet consisting of a single parton.
    The splittings of this parton will give rise to all of the final-state particles in the jet at the parton level.
    We take the angularity \(e^{(0)}_\varsigma = R^\varsigma / 2\) at this stage to be the maximum possible angularity for a single splitting within the jet radius \(R\).

    \item
    \label{item:cutoff}
    If the angularity \(e^{(i)}_\varsigma\) of the parton under consideration is below a pre-determined cutoff scale, label it as a final-state parton.
    In our MLL parton shower, for example, the cutoff scale is taken to be a non-perturbative scale \(\mu_{\rm cutoff}/ p_{T,\,{\rm jet}} = \) 1 MeV/\(p_{T,\,{\rm jet}}\), where we have tuned the cutoff roughly by hand in order to get better agreement with substructure distributions obtained with \texttt{Pythia 8.244}.
    Otherwise, the parton splits into two daughter partons, and we continue to the next step.

    \item
    \label{item:LL_angularity_gen}
    The splitting of a parton is described by its angularity \(e^{(f)}_\varsigma\).
    Using the techniques of the previous section, we may write the LL probability that there is no emission with angularity greater than \(e^{(f)}_\varsigma\), given that there has been an emission with angularity \(e^{(i)}_\varsigma\), as
    \begin{align}
        \label{eqn:alg_LLang_ratio}
        \mathbb{P}
        \left(
            {\rm no~emission~with~}e_\varsigma>e^{(f)}_\varsigma
            ~|~
            e_\varsigma < e^{(i)}_\varsigma
        \right)
        &=
        \frac{\Sigma(e^{(f)}_\varsigma)}{\Sigma(e^{(i)}_\varsigma)}
        \\
        \notag
        &=
        \exp\left[
            -\frac{C_R \alpha_s}{\varsigma~\pi}
            \left(
                \log^2(2 e^{(f)}_\varsigma) - \log^2(2 e^{(i)}_\varsigma)
            \right)
        \right]
        .
    \end{align}
    We can therefore sample from the corresponding probability distribution by the inverse transform method:
    pick a random number \(r\) uniformly distributed between 0 and 1, and set it equal to the above cumulative distribution.
    The corresponding value of \(e^{(f)}_\varsigma\) is
    \begin{align}
        \label{eqn:alg_LLang}
        e^{(f)}_\varsigma
        =
        \frac{1}{2}
        \exp\left[
            -\sqrt{\log^2(2 e^{(i)}_\varsigma) - \frac{\varsigma~\pi}{C_R \alpha_s} \log(r)}
        \right]
        ,
    \end{align}
    which is taken to be the angularity of the splitting from this mother parton.

    \item
    \label{item:LL_ztheta_sampling}
    Given the angularity \(e^{(f)}_\varsigma\) of our splitting, we may now sample from the associated distributions of the energy fraction \(z\) of the softer parton and the angle \(\theta\) of the splitting.
    In particular, at leading logarithmic order, these are both logarithmically distributed.
    At this level of accuracy, we may therefore pick a uniformly distributed random variable \(r'\) and write
    \begin{align}
        z = (2e^{(f)}_\varsigma)^{r'}/2
        ,
        ~~~~~~~~
        \theta = (2e^{(f)}_\varsigma)^{(1-r')/\varsigma}
        .
    \end{align}

    \item
    We may now continue this method recursively by returning to Step \ref{item:cutoff}, allowing the partons we produced in Steps \ref{item:LL_angularity_gen} and \ref{item:LL_ztheta_sampling} to split.
    We can achieve results at LL and MLL accuracy if we allow only the harder of the two partons at any branch to split.
\end{enumerate}

We can also extend this procedure to include running coupling effects and non-singular pieces of the splitting function, which contribute to MLL accuracy.
We produce these higher-order effects by using veto method sampling \cite{Sjostrand:2007gs,Larkoski:2013paa,Sjostrand:2014zea}:
\begin{enumerate}
	\item
	\label{item:MLL_ps_start}
	Generate an angularity, angle, and energy fraction with Steps \ref{item:LL_ps_start}--\ref{item:LL_ztheta_sampling} of the LL parton shower algorithm, with the coupling \(\alpha_s\) of \Eqs{alg_LLang_ratio}{alg_LLang} replaced by a large fixed coupling \(\hat\alpha = \frac{1}{2}\).

	\item
	As dictated by the veto method, accept the generated variables with a probability given by the ratio \(\rho^{\rm(MLL)}(z,\theta)/\hat\rho^{\rm(LL)}(z,\theta)\), where \(\rho^{\rm(MLL)}\) indicates the MLL probability distribution function and \(\hat\rho^{\rm(LL)}\) indicates the LL probability distribution function with the coupling \(\hat\alpha\).
	In particular, generate a new random number \(r''\) and accept the generated splitting if
	\begin{align}
	    r'' < \mathbb{P}_{\rm no~veto}
	    =
	    \frac{\bar{p}_i(z)}{p_{i,\,\rm gen}(z)}\,\frac{\alpha_s(\kappa)}{\hat\alpha}
	    ,
	\end{align}
	where \(p_{i,\,\rm gen}(z) = 2 C_{R_i} / z\) is the singular piece of the splitting function, used implicitly in drawing the angularity \(e_\varsigma^{(f)}\) from an LL distribution in \Eq{alg_LLang}.
    The choice \(\hat \alpha = 1/2\) ensures that this probability is less than 1 if we freeze our coupling below the non-perturbative scale \(\mu_{\rm freeze}\) = 1 GeV.

	\item
	If the veto method accepts the splitting, split the parton under consideration and continue the parton shower recursively by returning to Step \ref{item:MLL_ps_start}.
	If the veto method rejects the splitting, we do \textit{not} split the parton under consideration, but still return to Step \ref{item:MLL_ps_start} to generate a new angularity, this time starting the splitting algorithm at the scale of the rejected angularity.
	In particular, continue as dictated by \Eq{alg_LLang} with the coupling \(\hat\alpha\), by setting \(e_\varsigma^{(i)}\) to the angularity rejected by the veto algorithm in Step 2.
\end{enumerate}
The final step is the key to the veto algorithm, and correctly takes into account the exponentiation of multiple emission contributions in the MLL case.

Additionally, \Figs{LL_SD0}{LL_PRSF1} compare the LL ECF \(C_1^{(2)} \approx m^2 / p_T^2\) of \Eq{ECFdefn_repeat} for quark jets groomed using mMDT (Soft Drop with \(\beta_{\rm SD} = 0\)) and P-RSF\(_1\), obtained analytically, numerically, and using our parton shower procedure.
The agreement between these calculations provides a simple validation for the results we obtain analytically, using numerical integration, and with parton shower methods for P-RSF\(_1\) in the appendices.

\bibliography{piranha_paper}

\end{document}

%% file: includes/utils/paper_preamble.tex
\usepackage{mathtools,amssymb,amsmath,amsthm,mathrsfs}
\usepackage{physics}

\usepackage{hhline}

\pdfoutput=1
\usepackage{subfig}

\usepackage{graphbox,xspace,float,xcolor,color}

\usepackage{tikz, tikzscale, tikz-cd, pgf, pgfplots}  
\usetikzlibrary{angles,quotes,shapes}

\pgfplotsset{compat=newest}
\usepgfplotslibrary{fillbetween}

\tikzset{
  every overlay node/.style={
    draw=black,fill=white,rounded corners,anchor=north west,
  },
}

\tikzstyle{startstop} = [rectangle, rounded corners, minimum width=3cm, minimum height=1cm,text centered, draw=black, fill=red!30]
\tikzstyle{process} = [rectangle, minimum width=3cm, minimum height=1cm, text centered, draw=black, fill=orange!30]
\tikzstyle{decision} = [rectangle, minimum width=3cm, minimum height=1cm, text centered, draw=black, fill=green!30]
\tikzstyle{arrow} = [thick,->,>=stealth]

\DeclareMathSymbol{\mhyphen}{\mathord}{AMSa}{"39}

\usepackage{pifont}
\newcommand{\cmark}{\textcolor{green!60!black}{\ding{51}}}  
\newcommand{\xmark}{\textcolor{red}{\ding{55}}}  

\usepackage{contour}

\newcommand\dangerthin{%
 \makebox[1.4em][c]{%
 \makebox[0pt][c]{\raisebox{-0.05em}{\textbf{\tiny\textcolor{yellow!75!black}!}}}%
 \makebox[0pt][c]{\raisebox{-0.2em}{\color{yellow!75!black}\Large$\mathbf{\bigtriangleup}$}}}}%

\newcommand\danger{\protect\contourlength{.13pt}\protect\contour{yellow!75!black}{\dangerthin}} 

\definecolor{darkpastelred}{rgb}{0.76, 0.23, 0.13}
\definecolor{ballblue}{rgb}{0.13, 0.67, 0.8}
\definecolor{azure}{rgb}{0.0, 0.5, 1.0}
\definecolor{darkspringgreen}{rgb}{0.09, 0.45, 0.27}
\definecolor{ashgrey}{rgb}{0.7, 0.75, 0.71}
\definecolor{aurometalsaurus}{rgb}{0.43, 0.5, 0.5}
\definecolor{babyblueeyes}{rgb}{0.63, 0.79, 0.95}
\definecolor{cornflowerblue}{rgb}{.53, .80, .93}  
\definecolor{royalblue}{RGB}{99,149,236}  
\definecolor{dodgerblue}{rgb}{0.12, 0.56, 1.0}
\definecolor{forestgreen(traditional)}{rgb}{0.0, 0.27, 0.13}
\definecolor{forestgreen(web)}{rgb}{0.13, 0.55, 0.13}
\definecolor{darkgoldenrod}{rgb}{0.72, 0.53, 0.04}
\definecolor{goldenrod}{rgb}{0.85, 0.65, 0.13}
\definecolor{ochre}{rgb}{0.8, 0.47, 0.13}
\definecolor{plum}{RGB}{221,160,221}

\DeclareRobustCommand{\Sec}[1]{Sec.~\ref{sec:#1}}
\DeclareRobustCommand{\Secs}[2]{Secs.~\ref{sec:#1} and \ref{sec:#2}}

\DeclareRobustCommand{\App}[1]{App.~\ref{app:#1}}

\DeclareRobustCommand{\Appss}[3]{Apps.~\ref{app:#1},~\ref{app:#2}, and~\ref{app:#3}}
\DeclareRobustCommand{\Tab}[1]{Table~\ref{tab:#1}}

\DeclareRobustCommand{\Fig}[1]{Fig.~\ref{fig:#1}}
\DeclareRobustCommand{\Figs}[2]{Figs.~\ref{fig:#1} and \ref{fig:#2}}
\DeclareRobustCommand{\Figss}[3]{Figs.~\ref{fig:#1}, \ref{fig:#2}, and \ref{fig:#3}}
\DeclareRobustCommand{\Eq}[1]{Eq.~(\ref{eqn:#1})}
\DeclareRobustCommand{\Eqs}[2]{Eqs.~(\ref{eqn:#1}) and (\ref{eqn:#2})}

\DeclareRobustCommand{\Reff}[1]{Ref.~\cite{#1}}
\DeclareRobustCommand{\Reffs}[1]{Refs.~\cite{#1}}

\newtheorem{definition}{Definition}

%% file: includes/utils/piranha_utils.tex
\newcommand{\zcut}{\ensuremath{z_{\rm cut}}}
\newcommand{\zcrit}{\ensuremath{z_{\rm crit}}}
\newcommand{\thetacrit}{\ensuremath{\theta_{\rm crit}}}
\newcommand{\zpre}{\ensuremath{z_{\rm pre}}}

\newcommand{\rhoC}{\ensuremath{\rho\,\mathcal{C}}}
\newcommand{\rhoU}{\ensuremath{\rho\,\mathcal{U}}}

\newcommand{\PIRANHA}{\textsc{Piranha}}

\newcommand{\PRSF}[1]{P-RSF\(_{f = #1}\)}
\newcommand{\SD}[1]{SD\(_{\beta = #1}\)}

\definecolor{Epluscolor}{rgb}{0.0, 0.62, 0.38}

\definecolor{Eminuscolor}{rgb}{0.76, 0.23, 0.13}

\definecolor{E1color}{rgb}{0.76, 0.13, 0.28}

\definecolor{E2color}{rgb}{0.2, 0.2, 0.6}

\definecolor{softdrop}{rgb}{0.36, 0.54, 0.66}
\definecolor{rss}{rgb}{1.0, 0.49, 0.0}

%% file: includes/utils/comments.tex
\newif\ifstorecomments
\storecommentsfalse

\ifstorecomments
    \newwrite\commentfile
    \immediate\openout\commentfile=includes/aux/paper_comments.tex

    \AtEndDocument{
        \immediate\closeout\commentfile
    }
\fi

\definecolor{samcolor}{rgb}{0.0, 0.5, 0.0} 

\newcommand{\samtodo}[1]{
    \textbf{\textcolor{plum}{(Sam to-do: #1)}}
    \ifstorecomments
        \write\commentfile{Page \thepage: \textbf{\textcolor{plum}{(Sam to-do: #1)}}}
        \write\commentfile{}
    \fi
}
\newcommand{\ws}[1]{\samtodo{wordsmith!}}

\definecolor{jdtcolor}{rgb}{0.8392,0.1529,0.1569}

\usepackage{tcolorbox}

\newtcolorbox{sambox}[2][]{
    colback=samcolor!5!white,
    colframe=samcolor!75!black,
    colbacktitle=samcolor!10!white,
    coltitle=samcolor!70!black,
    title={#2},fonttitle=\bfseries,#1}

%% file: includes/aux/paper_comments.tex
Page 0: \textbf  {\leavevmode {\color  {samcolor}(?? --Sam)}}

%% file: includes/aux/final_checks.tex
\begin{sambox}{Final Double Checks:}
\begin{itemize}
    \item
    No comments left (\texttt{ctr-f} comment commands);
    
    \item
    No ``\textbackslash iffalse ... \textbackslash fi'' statements or ``...''s.
    
    \item 
    No \texttt{?}s left in citations, references to Sections, etc. (\texttt{ctr-f} `?' in .pdf);
    
    \item
    No files in temp folders;
    
    \item
    Consistency for
    \begin{itemize}
        \item
        Superscript in \(C_1^{(\varsigma)}\)
        
	\item ``subjet'' to ``sub-jet''.
	
	\item ``nonperturbative'' to ``non-perturbative'' 
		
	\item
	De-cluster, PU-subtracted, and UE-corrected all have hyphens.
	
	\item \(d\) to \(\dd\) in integrals
	
	\item
	EMDs \(\to\) the EMD where appropriate
	
	\item
	Energy Movers Distance, Energy Movers' Distance \(\to\) Energy Mover's Distance
	
	\item
	high energy (low energy) vs. high-energy (low-energy)
	
	\item
	\(\beta\) function \(\to\) beta function (so people can find with ctr-f)
	
    \end{itemize}
\end{itemize}
\end{sambox}